\documentclass{article}

\usepackage{amssymb}
\usepackage{amsmath}
\usepackage{epubtk}
\usepackage{booktabs}
\usepackage{graphicx,epsfig}

\def\mdens{\mathrm{g\ cm}^{-3}}

\begin{document}

\title{Physics of Neutron Star Crusts}

\author{%
\epubtkAuthorData{Nicolas Chamel}{%
Institut d'Astronomie et d'Astrophysique \\
Universit{\'e}\ Libre de Bruxelles \\
CP226 \\
Boulevard du Triomphe
B-1050 Brussels, Belgium}{%
nchamel@ulb.ac.be}{%
http://www-astro.ulb.ac.be/}%
\\
\and \\
\epubtkAuthorData{Pawel Haensel}{%
Nicolaus Copernicus Astronomical Center \\
Polish Academy of Sciences \\
Bartycka 18, \\
00-716 Warszawa, Poland}{%
haensel@camk.edu.pl}{%
http://www.camk.edu.pl}%
}

\date{}
\maketitle

\begin{abstract}
The physics of neutron star crusts is vast, involving many different
research fields, from nuclear and condensed matter physics to general
relativity. This review summarizes the progress, which has been
achieved over the last few years, in modeling neutron star crusts, both
at the microscopic and macroscopic levels. The confrontation of these
theoretical models with observations is also briefly discussed.
\end{abstract}

\epubtkKeywords{astrophysics, neutron stars, pulsars, magnetars, soft
  gamma repeaters, supernova, elasticity, neutron star cooling,
  low-mass X-ray binaries, oscillations, structure, transport
  properties, conductivity, viscosity, neutrino, equation of state,
  neutron star crust, superfluidity, entrainment, two-fluid model,
  accretion, X-ray astronomy, gravitational waves, pulsar glitches,
  accreting neutron stars, deep crustal heating}

\newpage

\tableofcontents


\newpage
\section{Introduction}
\label{sect.intro}

Constructing  models of neutron stars requires knowledge of
the physics of matter with a density significantly exceeding the density of
atomic nuclei. The simplest picture of the atomic nucleus is a drop of
highly incompressible nuclear matter. Analysis of nuclear masses tells
us that nuclear matter at saturation (i.e.\ at the minimum of the
energy per nucleon) has the density  $\rho_0= 2.8\times
10^{14}~\mdens$, often called \emph{normal nuclear density}. It
corresponds to $n_0=0.16$ nucleons per fermi cubed. 
The density in the cores of massive neutron stars
is expected to be as large as $\sim 5\mbox{\,--\,}10\rho_0$ and in spite of
decades of observations of neutron stars and intense theoretical
studies, the structure of the matter in neutron star cores and in
particular its equation of state remain the well-kept secret of neutron
stars (for a recent review, see the book by Haensel, Potekhin and
Yakovlev~\cite{haensel-06}). The physics of matter with $\rho\sim
5\mbox{\,--\,}10\rho_0$ is a huge challenge to theorists, with observations of
neutron stars being crucial for selecting a correct dense-matter
model. Up to now, progress has been slow and based overwhelmingly
on scant observation~\cite{haensel-06}.

The outer layer of neutron stars with density $\rho<\rho_0$ --
the neutron star crust --  which is the subject of the present
review, represents very different theoretical challenges and
observational opportunities. The elementary constituents of
the  matter are neutrons, protons, and electrons -- like in
the atomic matter around us. The density is ``subnuclear'', so
that the methods developed and successfully
applied in the last decades to terrestrial nuclear physics can 
be applied to neutron star crusts. 
Of course, the physical conditions are extreme and far from terrestrial ones. The compression of matter by gravity crushes atoms
and forces, through electron captures, the neutronization of the
matter. This effect of huge pressure was already predicted in the
1930s (Sterne~\cite{sterne-33}, Hund~\cite{hund-36,
hund-36b}). At densities $\rho\gtrsim 4\times 10^{11}~\mdens$
a fraction of the neutrons is unbound and forms a gas around
the nuclei. For a density approaching $10^{14}~\mdens$, some 90\% of
nucleons are neutrons while nuclei are represented by proton
clusters with a small neutron fraction. 
How far we are taken from terrestrial nuclei with a moderate
neutron excess! Finally, somewhat above $10^{14}~\mdens$
nuclei can no longer exist -- they coalesce into a uniform
plasma of nearly-pure neutron matter, with a few percent
admixture of protons and electrons: we reach the bottom of
the neutron star crust.

The crust contains only a small percentage of a neutron star's mass, but it is
crucial for many astrophysical phenomena involving neutron stars. It
contains matter at subnuclear density, and therefore  there is no
excuse for the theoretical physicists, at least in principle: the
interactions are known, and many-body theory techniques are
available. Neutron star crusts are wonderful cosmic laboratories in
which the full power of theoretical physics can be demonstrated and
hopefully confronted with neutron star observations.

To construct neutron star crust models we have to employ atomic and
plasma physics, as well as the theory of condensed matter, the physics of
matter in strong magnetic fields, the theory of nuclear structure, nuclear
reactions, the nuclear many-body problem,  superfluidity, physical
kinetics, hydrodynamics, the physics of liquid crystals, and the theory of
elasticity. Theories have to be applied under extreme physical
conditions, very far from the domains where they were originally
developed and tested. Therefore, caution is a must!

Most of this review is devoted to theoretical descriptions of
various aspects of neutron star crusts. The plural ``crusts'' in the
title is well justified; depending on the scenario of their
formations, the crusts may be very different in their composition and
structure as sketched in Figure~\ref{fig.sect.intro.surfaces}. In
Section~\ref{sect.plasma} we briefly describe the basic plasma
parameters relevant to crust physics and delineate various plasma
regimes in the temperature-density plane. We also address the
important question of the magnetic field.

\epubtkImage{neutron_star_surfaces.png}{%
\begin{figure}[htbp]
  \centerline{\includegraphics[scale=0.8]{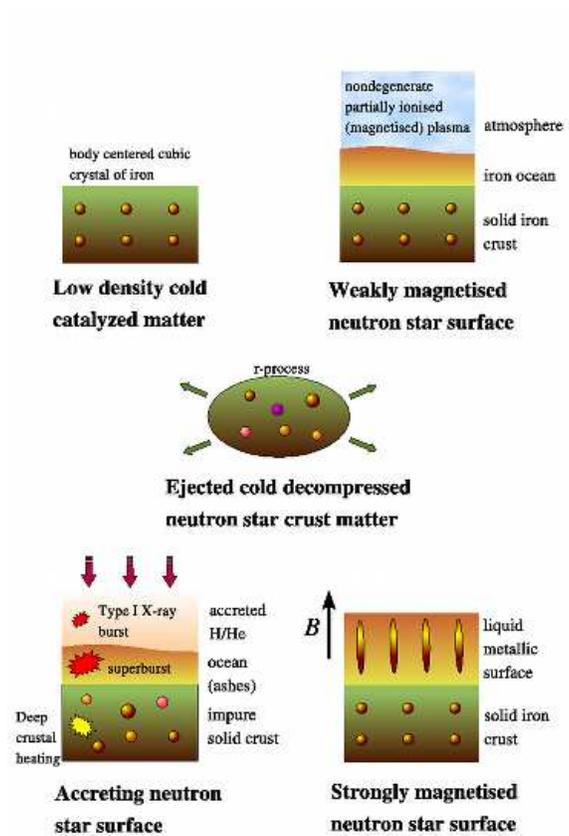}}
  \caption{Schematic pictures of various neutron star surfaces.}
\label{fig.sect.intro.surfaces}
\end{figure}}

The ground state of the crust (one of the possible ``crusts'') is
reviewed in Section~\ref{sect.groundstate}. We also discuss there the
uncertainties concerning the densest bottom layers of the crust and we 
mention possible deviations from the ground
state. As we describe in Section~\ref{sect.accretion}, a crust formed
via accretion is expected to be very different from that formed during the
aftermath of a supernova explosion. We show how it can be a
site for nuclear reactions. We study its thermal structure during
accretion, and briefly review the phenomenon of X-ray bursts. We
quantitatively analyze the phenomenon of deep crustal heating.

To construct a neutron star model one needs the equation of state (EoS)
of the crust, reviewed in Section~\ref{sect.eos}. We consider
separately the ground-state crust, and the accreted crust. For the
sake of comparison, we also describe another EoS of matter at
subnuclear density -- that relevant for the collapsing type~II
supernova core.

Section~\ref{sect.structure} is devoted to the stellar-structure aspects of neutron star crusts. We start with the
simplest case of a spherically-symmetric static neutron star and
derive approximate formulae for crust mass and moment of
inertia. Then we study the deformation of the crust in a rotating
star. Finally, we consider the effects of magnetic fields on the
crust structure. Apart from isotropic stress resulting from
pressure, a solid crust can support an elastic strain. Elastic
properties are reviewed in Section~\ref{sect.elast}. A
separate subsection is devoted to the elastic parameters of
the so-called ``pasta'' layers, which behave like liquid
crystals. The inner crust is permeated by a neutron
superfluid. Various aspects of crustal superfluidity are
reviewed in Section~\ref{sect.super}. After a brief
introduction to superconductivity and its relevance for
neutron star crusts, we start with the static properties of
neutron superfluidity, considering first a uniform neutron
gas, and then discussing the effects of the presence of the
nuclear crystal lattice. In the following, we consider superfluid hydrodynamics. 
We stress those points, which have
been raised only recently. We consider also the important
problem of the critical velocity above which superfluid flow
breaks down. The interplay of superfluid flow and vortices is
reviewed. The section ends with a discussion of entrainment
effects. Transport phenomena are reviewed in
Section~\ref{sect.cond}. We present calculational methods and
results for electrical and thermal conductivity, and shear
viscosity of neutron star crusts. Differences between accreted
and ground-state crusts, and the potential  role of
impurities are illustrated by examples. Finally, we discuss
the very important effects of magnetic fields on transport
parameters. In Section~\ref{sect.hydro}, we review macroscopic 
models of the crust, and we describe in particular a two-fluid model,
which takes into account the stratification of crust
layers, as well as the presence of a neutron superfluid. We
show how entrainment effects between the superfluid and the
charged components can be included using the variational
approach developed by Brandon Carter~\cite{carter-89}. 
Section~\ref{sect.neutrino} is devoted to a description of
the wealth of neutrino emission processes associated with crusts. We limit ourselves to the basic mechanisms, which, according
to existing calculations, are the most important ones at
subsequent stages of neutron-star cooling. 

The confrontation of theory with observations is presented in
Section~\ref{sect.obs}. Neutron stars are born very hot, and
we briefly describe in Section~\ref{sect.obs.supernova} the
present status of the theory of hot dense matter at subnuclear
densities; this layer of the proto-neutron star will eventually
become the neutron star crust. The crust is crucial for neutron star cooling, as observed by a distant observer. Namely, the
crust separates the neutron star core from its surface, where the
observed X-ray radiation is produced. The relation between crust
physics and observations of cooling neutron stars is studied
in Section~\ref{sect.obs.cooling}.  In
Section~\ref{sect.obs.rproc} we briefly consider possible
r-processes associated with the ejection and subsequent
decompression of the neutron star crusts. Pulsar glitches are
thought to originate in neutron star crust and glitch models
are confronted with observations of glitches in pulsar timing
data in Section~\ref{sect.obs.glitches}. The asteroseismology of neutron 
stars from their gravitational wave radiation is discussed in
Section~\ref{sect.obs.gw}. Due to its elasticity, the solid
crust can support mountains and shear (torsional) oscillations, both
associated with gravitational wave emission. The crust-core
interface can be crucial for the damping of r-modes, which, if
unstable, could be a promising source of gravitational waves from
rotating neutron stars. Observations of oscillations in the
giant flares from Soft Gamma Repeaters are confronted with
models of torsional oscillations of crusts in
Section~\ref{sect.obs.sgr}. As discussed in
Section~\ref{sect.obs.LMXB}, the modeling of phenomena associated
with low-mass X-ray binaries (LMXB)  requires a rather
detailed knowledge of the physics of neutron star crusts. New
phenomena discovered in the last decade (and some very
recently) necessitate realistic physics of accreted neutron
star crusts, including deep crustal heating  and the correct degree
of purity. All aspects of accreted crusts, relevant for soft
X-ray transients, X-ray superbursts, and persistent X-ray
transients, are discussed in this section.

\newpage


\section{Plasma Parameters}
\label{sect.plasma}

\subsection{No magnetic field}
\label{sect.plasma.noB}

In this section we introduce several parameters that will be used
throughout this review. We follow the notations of the book by
Haensel, Potekhin and Yakovlev~\cite{haensel-06}.

We consider a one-component plasma model of neutron star crusts,
assuming a single species of nuclei at a given density
$\rho$. We restrict ourselves to matter composed of atomic nuclei
immersed in a nearly ideal
and uniform, strongly degenerate electron gas of number density
$n_e$. This model is valid at $\rho>10^{5} \mathrm{\ g\ cm}^{-3}$.
A neutron gas is also present at densities greater than neutron drip density
$\rho_{\mathrm{ND}}\approx 4\times 10^{11} \mathrm{\ g\ cm}^{-3}$.

For an ideal, fully degenerate, relativistic electron gas, the Fermi
energy (using the letter F to denote quantities evaluated at the 
Fermi level and the letter $e$ for electrons) is given by
\begin{equation}
\epsilon_{\mathrm{F}e}=m^*_e c^2 \;, ~~
m_e^*=m_e\left(1+ x_{\mathrm{r}}^2\right)^{1/2} \, ,
\label{eq.sect.plasma.noB.mstar.e}
\end{equation}
where $x_{\mathrm{r}}$ is a dimensionless relativity parameter defined in
terms of the Fermi momentum
\begin{equation}
p_{\mathrm{F}e}= \hbar (3\pi^2 n_e)^{1/3} \, ,
\label{eq.sect.plasma.noB.pF}
\end{equation}
by
\begin{equation}
x_{\mathrm{r}}={p_{\mathrm{F}e}\over m_e c}~
\label{eq.sect.plasma.noB.xr}
\end{equation}
and $m_e^*$ is the electron effective mass at the Fermi surface. The
Fermi velocity is
\begin{equation}
v_{\mathrm{F}e}={p_{\mathrm{F}e}\over m^*_e} =
c {x_{\mathrm{r}}\over \left(1+x_{\mathrm{r}}^2\right)^{1/2}} \, .
\end{equation}
Electrons are strongly degenerate for $T\ll T_{\mathrm{F}e}$, where the
electron Fermi temperature is defined by
\begin{equation}
 T_{\mathrm{F}e}={\epsilon_{\mathrm{F}e}-m_e c^2\over
 k_{\mathrm{B}}}=5.93\times 10^{9}\;
 \left(\gamma_{\mathrm{r}}-1\right)\;\mathrm{K}\;,
\label{sect.plasma.noB.TFe}
\end{equation}
$k_{\mathrm{B}}$ is the Boltzmann constant and
\begin{equation}
\gamma_{\mathrm{r}}={m^*_e\over m_e}=\left(1+x_{\mathrm{r}}^2\right)^{1/2} \, .
\label{eq.sect.plasma.noB.gamma.r.e}
\end{equation}
Let the mass number of nuclei be $A$ and their proton number be $Z$. The
electric charge neutrality of matter implies that the number density
of ions (nuclei) is
\begin{equation}
n_{\mathrm{N}}=n_{\mathrm{e}}/Z \, .
\label{eq.sect.plasma.noB.ni}
\end{equation}
For $\rho<\rho_{\mathrm{ND}}$, the quantities $n_{\mathrm{N}}$ and $A$ are related to
the mass density of the crust by
\begin{equation}
\rho\approx n_{\mathrm{N}}A m_{\mathrm{u}} \, ,
\label{eq.sect.plasma.noB.rho_ni_A}
\end{equation}
where $m_{\mathrm{u}}$ is the atomic mass unit,
$m_{\mathrm{u}}=1.6605\times10^{-24}$~g. For
$\rho>\rho_{\mathrm{ND}}$, one has to replace $A$  in
Equation~(\ref{eq.sect.plasma.noB.rho_ni_A}) by
$A^\prime=A+A^{\prime\prime}$, where  $A$ is the number of nucleons
bound in the nucleus and $A^{\prime\prime}$ is the number of free
(unbound) neutrons per ion. $A^\prime$ is, thus, the number of
nucleons per ion.  The electron relativity parameter can be expressed
as
\begin{equation}
x_{\mathrm{r}}=1.00884
\left({\rho_6 Z\over A^\prime}\right)^{1/3}~\;,
\end{equation}
where $\rho_6\equiv  \rho/10^6 \mathrm{\ g\ cm}^{-3}$.

The ion plasma temperature $T_{\mathrm{pi}}$ is defined, in terms of the
the ion plasma frequency
\begin{equation}
\omega_{\mathrm{pi}}=\left(\frac{4\pi e^2 n_{\mathrm{N}}Z^2}{A m_{\mathrm{u}}}\right)^{1/2} \, ,
\end{equation}
by
\begin{equation}
T_{\mathrm{pi}}={\hbar \omega_{\mathrm{pi}}\over k_{\mathrm{B}}}
=7.832\times 10^6\;\left({\rho_6\over A^\prime}\;
{Z^2\over A}\right)^{1/2}\;\mathrm{K} \, .
\end{equation}
Quantum effects for ions become very important for $T\ll T_{\mathrm{pi}}$.
The electron plasma frequency is
\begin{equation}
\omega_{\mathrm{p}e}=\left(\frac{4\pi e^2 n_e}{m^*_e}\right)^{1/2} \, ,
\label{eq.sect.plasma.noB.omega_pe}
\end{equation}
so that the electron plasma temperature
\begin{equation}
T_{\mathrm{p}e}=\frac{\hbar\omega_{\mathrm{p}e}}{k_{\mathrm{B}}}=3.300\times
10^8 x_{\mathrm{r}}^{3/2}\gamma_{\mathrm{r}}^{-1/2}~\mathrm{K} \, ,
\label{eq.sect.plasma.noB.T_pe}
\end{equation}
which can be rewritten as
\begin{equation}
T_{\mathrm{p}e}= 3.34\times 10^8
~\mathrm{K}\left(\rho_6{Z\over A}\right)^{1/2}\times \left[ 1+
1.02\left(\rho_6 {Z\over A}\right)^{2/3} \right]^{-1/4} \, .
\label{eq.sect.plasma.noB.T_pe2}
\end{equation}

\epubtkImage{crust_domains.png}{%
  \begin{figure}[htbp]
    \centerline{\includegraphics[scale=0.6]{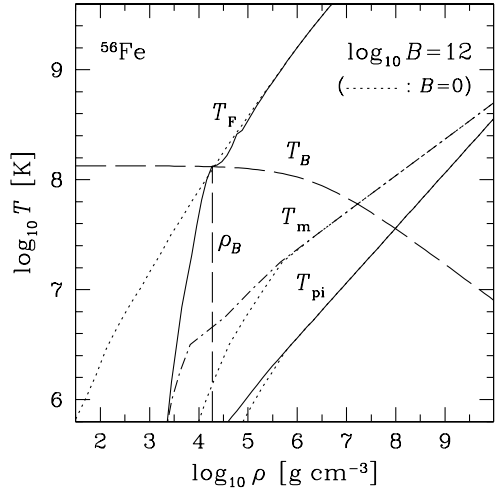}}
    \caption{Different parameter domains in the $\rho-T$ plane for
    $^{56}$Fe plasma with magnetic field $B=10^{12}$~G. Dash-dot line:
    melting temperature $T_{\mathrm{m}}$. Solid lines: $T_{\mathrm{F}}$ -- Fermi
    temperature for the electrons (noted $T_{\mathrm{F}e}$ in the main
    text); $T_{\mathrm{pi}}$ -- ion plasma temperature. Long-dash lines:
    $T_B$ and $\rho_B$ relevant for the quantized regime of the
    electrons (Section~\ref{sect.plasma.B}); for comparison we also
    show, by dotted lines,  $T_{\mathrm{F}}$, $T_{\mathrm{m}}$ and
    $T_{\mathrm{pi}}$ for $B=0$. For further explanation see the
    text. From~\cite{haensel-06}.}
    \label{fig.sect.plasma.crust_domains}
\end{figure}}

\epubtkImage{Tm.png}{%
\begin{figure}[htbp]
  \centerline{\includegraphics[scale=0.4]{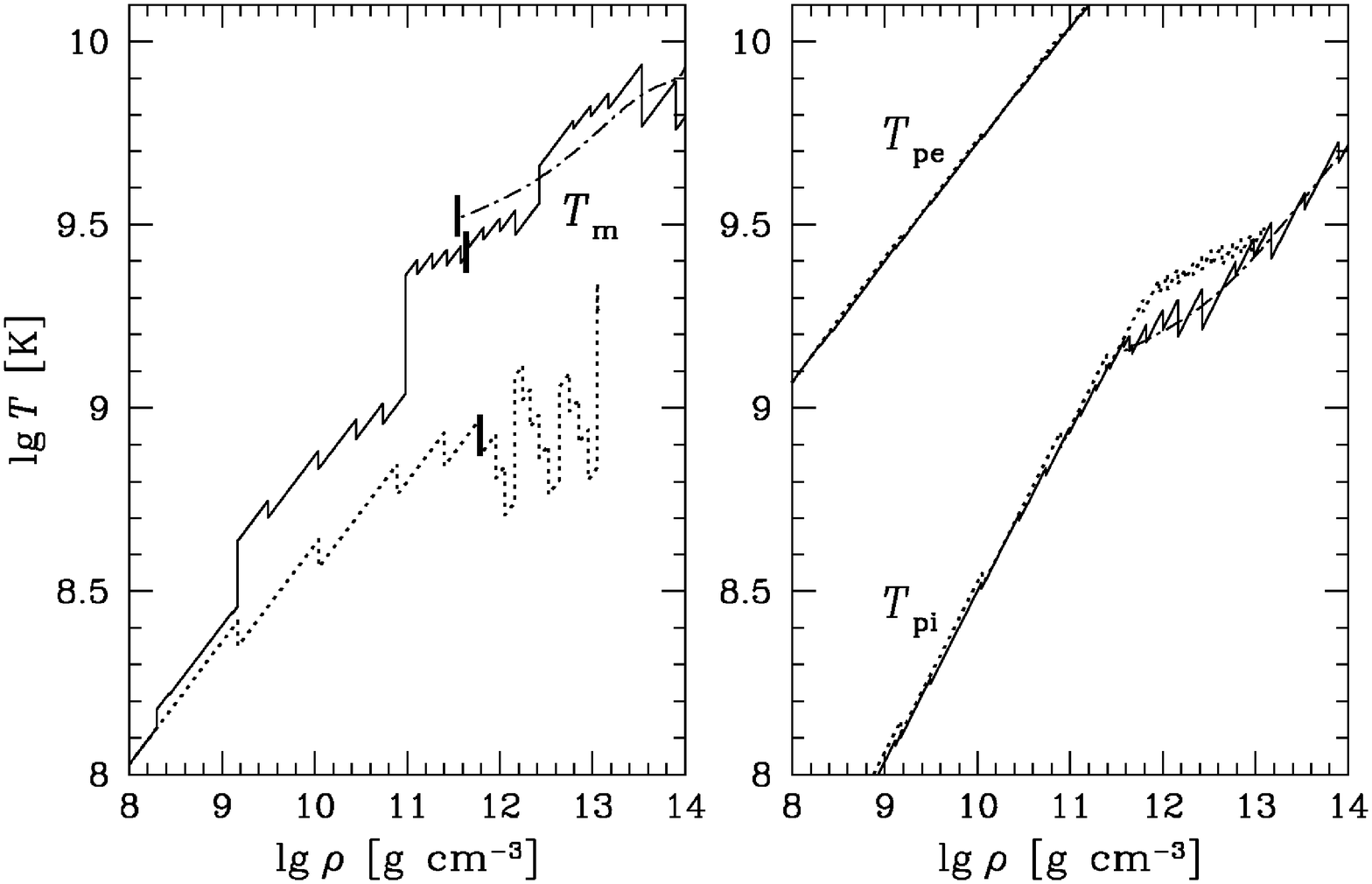}}
  \caption{Left panel: melting temperature versus
  density. Right panel: electron and ion plasma temperature versus
  density. Solid lines: the ground-state composition of the
  crust is assumed: Haensel \& Pichon~\cite{haensel-94} for the outer
  crust, and Negele \& Vautherin~\cite{nv-73} for the inner crust. Dot
  lines:  accreted crust, as calculated by Haensel \&
  Zdunik~\cite{haensel-90b}. Jumps result from discontinuous
  changes of $Z$ and $A$. Dot-dash line: results obtained for the
  compressible liquid drop model of Douchin \&
  Haensel~\cite{douchin-01} for the ground state of the inner crust; a
  smooth behavior (absence of jumps) results from the approximation
  inherent in the compressible liquid drop model. Thick vertical
  dashes: neutron drip point for a given crust model. Figure made by
  A.Y.\ Potekhin.}
  \label{fig.sect.plasma.Tm}
\end{figure}}

The crystallization of a Coulomb plasma of ions occurs at the
temperature
\begin{equation}
\label{eq.sect.plasma.noB.Tm}
 T_{\mathrm{m}}={Z^2 e^2\over
 R_{\mathrm{cell}}k_{\mathrm{B}}\Gamma_{\mathrm{m}}}\approx 1.3\times
 10^5\;Z^2\left({\rho_6\over A^\prime}\right)^{1/3}\; {175\over
 \Gamma_{\mathrm{m}}}\;\mathrm{K} \, ,
\end{equation}
where the ion sphere (also called the unit cell or Wigner--Seitz cell) radius is
\begin{equation}
R_{\mathrm{cell}}=\left(\frac{4\pi n_{\mathrm{N}}}{3}\right)^{-1/3} \, .
\label{eq.sect.plasma.Rcell}
\end{equation}
For classical ions ($T \gtrsim T_{\mathrm{pi}}$) one has
$\Gamma_{\mathrm{m}}\approx 175$. For $T \lesssim T_{\mathrm{pi}}$ the
zero-point quantum vibrations of ions become important and lead
to crystal melting at lower $\Gamma_{\mathrm{m}}$.

\subsection{Effects of magnetic fields}
\label{sect.plasma.B}

Typical pulsars have surface magnetic fields $B\sim
10^{12}$~G. Magnetars  have much higher magnetic fields, $B\sim
10^{14}\mbox{\,--\,}10^{15}$~G. The properties of the outer envelope of neutron
stars can be drastically modified by a sufficiently strong magnetic
field $\pmb B$. It is convenient to introduce the ``atomic'' magnetic
field $B_0$
\begin{equation}
B_0={m_e^2 e^3 c\over \hbar^3} = 2.35\times 10^9 \mathrm{\ G} \, .
\label{eq:ground-B0}
\end{equation}
It is the value of $B$ for which the electron cyclotron energy is equal to
$e^2/a_0=2\times 13.6 \mathrm{\ eV}$ ($a_0$ is the Bohr radius). Putting it
differently, at $B=B_0$ the characteristic magnetic length
$a_{\mathrm{m}}=(\hbar c/eB)^{1/2}$ equals the Bohr radius. For
typical pulsars and magnetars the surface magnetic field is
significantly stronger than $B_0$. As a result, the atomic structure
at low pressure is expected to be strongly modified. The motion of electrons
perpendicular to $\pmb{B}$ is quantized into Landau orbitals. Assuming
that $\pmb{B}=[0,0,B]$, the electron energy levels are given by
$\epsilon_n (p_z)=c(m_e^2c^2+2\hbar\omega_{\mathrm{c}e}m_e n +p_z^2)^{1/2}$,
where $n$ is the Landau quantum number and $p_z$ the z-component of the electron 
momentum. The ground state Landau level
$n=0$ is nondegenerate with respect to the spin (the spin is antiparallel to
$\pmb{B}$, with spin quantum number $s=-1$), while the higher levels $n>0$
are doubly degenerate ($s=\pm1$). The cyclotron frequency for
electrons is $\omega_{\mathrm{c}e}={eB/m_e c}$; it is 1836 times larger
than for protons. The Coulomb binding of electrons by the atomic
nucleus is significantly less effective along $\pmb B$, while in the
plane perpendicular to $\pmb B$ the electron motion is confined to the 
$n=0$ Landau level. Therefore atoms get a cylindrical shape and can
form linear chains along $\pmb B$. The attraction between these
chains can lead to a phase transition into a ``magnetically
condensed'' phase (for a recent review on this topic, see
Medin \& Lai~\cite{medin-07}). The density of the condensed phase at
zero pressure $P=0$ (i.e.\ at the stellar surface) and zero
temperature $T=0$ is
\begin{equation}
\rho_{\mathrm{s}}\simeq 4.4\times 10^3\;{A\over 56}
\left({Z\over 26}\right)^{-3/5}
B_{12}^{6/5}~\mdens \, ,
\label{eq:ground-rho_s-B}
\end{equation}
where $B_{12}=B/10^{12} \mathrm{\ G}$. For each element, there is a
critical temperature at given $B$, below which a magnetically condensed
phase exists at $P=0$. The values of $T_{\mathrm{crit}}$ at several $B$
for $^{56}$Fe, $^{12}$C, and $^{4}$He are given in
Table~\ref{sect.plasma.Tc.B}.

\begin{table}[htbp]
  \caption[Critical temperature (in K) below which a condensed phase
  exists at $P=0$, for several magnetic field strengths
  $B_{12}=B/10^{12} \mathrm{\ G}$ and matter compositions.]{Critical temperature (in K) below which a condensed phase
  exists at $P=0$, for several magnetic field strength
  $B_{12}=B/10^{12} \mathrm{\ G}$ and matter
  composition. From~\cite{medin-07}.}
  \label{sect.plasma.Tc.B}
  \vskip 4mm

  \centering
    \begin{tabular}{c|ccc}
      \toprule
       $B_{12}=$  &   10  &  100  &  1000 \cr
      \midrule
      $^{56}$Fe & $7\times 10^5$ & $3\times 10^6$ & $2\times 10^7$ \\
      $^{12}$C & $3\times 10^5$ & $3\times 10^6$ & $2\times 10^7$ \\
      $^{4}$He & $3\times 10^5$ & $2\times 10^6$ & $9\times 10^7$ \\
      \bottomrule
    \end{tabular}
\end{table}

We now briefly consider the effects of the magnetic field on plasma properties at finite pressure $P>0$. The magnetic field {\it
  strongly quantizes} the motion of electrons, if it confines most of
them to the ground Landau state $n=0$. Parameters relevant to a strong
quantization regime are
\begin{equation}
T_{\mathrm{c}e}={\hbar \omega_{\mathrm{c}e}\over k_{\mathrm{B}}}\approx
1.343\times 10^8\;B_{12}~\mathrm{K} \, ,
~~ \rho_B=7.045\times 10^3{A^\prime\over
Z}\;B_{12}^{3/2} \mathrm{\ g\ cm}^{-3} \, ,
\label{eq.sect.plasma.B.Tce.rhoB}
\end{equation}
and
\begin{equation}
T_B= T_{\mathrm{c}e}~\mathrm{if}~\rho<\rho_B \, , ~~~\mathrm{or}~~~
T_B= T_{\mathrm{c}e}/\sqrt{1+x_{\mathrm{r}}^2}~~\mathrm{if}~\rho>\rho_B \, .
\label{eq.sect.plasma.B.TB}
\end{equation}
The field $\pmb{B}$ is strongly quantizing if $\rho<\rho_B$ {\it and }
$T\ll T_{\mathrm{c}e}$. On the contrary, a magnetic field is called
{\it weakly quantizing} if  many Landau levels are occupied, but
still $T\ll T_B$. Finally, $\pmb{B}$ is nonquantizing
if  $T\gg T_B$. The temperature $T_B$ and density $\rho_B$ are
 shown in Figure~\ref{fig.sect.plasma.crust_domains}.

The magnetic field can strongly modify transport properties
(Section~\ref{sect.cond.mag}) and neutrino emission
(Section~\ref{sect.neutrino.synchrotron}). Its effect on the equation
of state is significant only if it is strongly quantizing (see
Section~\ref{sect.structure.B}).

\newpage


\section{The Ground State Structure of Neutron Star Crusts}
\label{sect.groundstate}

According to the cold catalyzed matter hypothesis, the matter inside
cold non-accreting neutron stars is assumed to be in complete
thermodynamic equilibrium with respect to \emph{all} interactions at
zero temperature and is therefore supposed to be in its ground state
with the lowest possible energy. The validity of this assumption is
discussed in Section~\ref{sect.groundstate.imp}.

The ground state structure of a neutron star crust is sketched in
Figure~\ref{fig.sect.groundstate.crust}. The outer crust
(Section~\ref{sect.groundstate.outer}) consists of a body-centered
cubic lattice of iron $^{56}$Fe. At $\rho\sim 10^4~\mdens$ the atoms
are fully ionized owing to the high density. At densities above $10^7~\mdens$, the composition of the nuclei becomes more neutron rich as a
result of electron captures. The inner crust
(Section~\ref{sect.groundstate.inner}), which extends from $\rho_{\mathrm{ND}}\simeq
4\times 10^{11}~\mdens$ to $\sim \rho_0/3 \simeq 10^{14}~\mdens$, is characterized
by the presence of free neutrons, which may condense into a superfluid
phase in some layers (see Section~\ref{sect.super}). At the bottom of the crust, 
some calculations predict various ``pasta'' phases of non-spherical nuclei, such as slabs
or cylinders as discussed in Section~\ref{sect.groundstate.pasta}.

The ground state of a neutron star crust is obtained by minimizing the
total energy density $\varepsilon_{\mathrm{tot}}$ for a given baryon
density $n_{\mathrm{b}}$ under the assumption of $\beta$-equilibrium and
electric charge neutrality. For simplicity, the crust is assumed to be
formed of a perfect crystal with a single nuclear species at lattice
sites (see Jog \& Smith~\cite{jog-82} and references therein for the
possibility of heteronuclear compounds).

\epubtkImage{neutron_star_crust_structure.png}{%
\begin{figure}[htbp]
  \centerline{\includegraphics[scale=0.8]{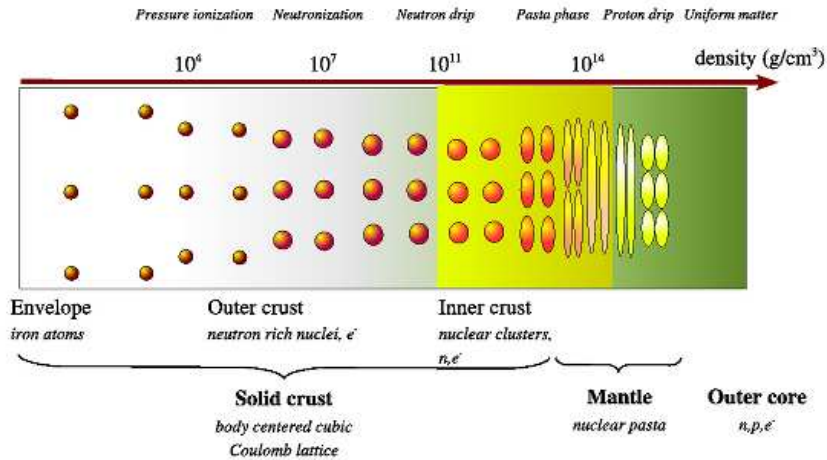}}
  \caption{Schematic picture of the ground state structure of
  neutron stars along the density axis. Note that the main part of this figure 
  represents the solid crust since it covers about 14 orders of magnitude 
  in densities.}
  \label{fig.sect.groundstate.crust}
\end{figure}}

\subsection{Structure of the outer crust}
\label{sect.groundstate.outer}

Matter at densities below neutron drip $\rho_{\mathrm{ND}}$ is not
only relevant for neutron star crusts but also for white
dwarfs. Following the classical paper of Baym, Pethick and
Sutherland~\cite{bps-71}, the total energy density in a given layer
can be written as
\begin{equation}
\label{eq.sect.groundstate.outer.total_energy}
\varepsilon_{\mathrm{tot}} = n_{\mathrm{N}} E\{A,Z\} +\varepsilon_e+\varepsilon_L,
\end{equation}
where $n_{\mathrm{N}}$ is the number density of nuclei, $E\{A,Z\}$ is the
energy of a nucleus with $Z$ protons and $A-Z$ neutrons,
$\varepsilon_e$ is the electron kinetic energy density and
$\varepsilon_L$ is the lattice energy density, which accounts for the
electron-electron, electron-ion and ion-ion Coulomb interactions.

In dense, cold, neutron star crust, electron-charge--screening effects are
negligible and the electron density is essentially
uniform~\cite{watanabe-03}. The reason is that the electron Thomas--Fermi 
screening length is larger than the lattice spacing~\cite{pethick-95}. Charge screening
effects are much more important in neutron star matter with large
proton fractions, such as encountered in supernovae and newly born
hot neutron stars~\cite{maruyama-05}. At densities $\rho\gg 10 A Z
\mathrm{\ g\ cm}^{-3}$ ($\sim 10^4 \mathrm{\ g\ cm}^{-3}$ for iron), 
the electrons can be treated as a quasi-ideal Fermi gas
so that
\begin{equation}
\label{eq.sect.groundstate.outer.electron_energy}
\varepsilon_e = \frac{m_e^4 c^5}{8 \pi^2\hbar^3}\left( x_r (2x_r^2+1)\sqrt{x_r^2+1} - \ln\{ x_r+\sqrt{x_r^2+1}\} \right) \, ,
\end{equation}
with $x_r$ defined by Equation~(\ref{eq.sect.plasma.noB.xr}).

The lattice energy density can be estimated from the Wigner--Seitz
approximation illustrated on
Figure~\ref{fig.sect.groundstate.outer.ws_app}. The crust is
decomposed into a set of independent spheres centered around
each nucleus, with a radius $R_{\mathrm{cell}}$ defined by
Equation~(\ref{eq.sect.plasma.Rcell}). Each sphere is electrically
neutral and therefore contains $Z$ protons and $Z$ electrons. The
lattice energy density is then given by the density of nuclei times
the Coulomb energy of one such sphere (excluding the Coulomb energy of
the nucleus, which is already taken into account in
$E\{A,Z\}$). Assuming point-like nuclei since the lattice spacing is
much larger than the size of the nuclei\epubtkFootnote{Lattice energy 
including finite size effects is given by
  Equation~(\ref{eq.sect.groundstate.inner.LDM.lat_energy}). Even at
  the bottom of the outer crust finite size effects represent a small
  correction to the lattice energy, less than 1\%.}, the lattice
energy can be expressed as (see, for instance,
Shapiro \& Teukolsky~\cite{shapiro-83}, p30-31)
\begin{equation}
\label{eq.sect.groundstate.outer.lattice_energy}
\varepsilon_L = -\frac{9}{10} \left(\frac{4\pi}{3}\right)^{1/3} Z^{2/3} e^2 n_e^{4/3}  \, .
\end{equation}

\epubtkImage{ws_app.png}{%
\begin{figure}[htbp]
  \centerline{\includegraphics[scale=0.5]{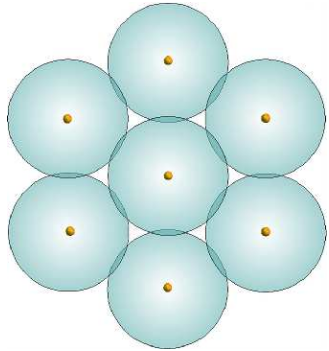}}
  \caption{In the Wigner--Seitz approximation the crystal
  (represented here as a two-dimensional hexagonal lattice) is
  decomposed into independent identical spheres, centered around each
  site of the lattice. The radius of the sphere is chosen so that the
  volume of the sphere is equal to $1/n_{\mathrm{N}}$, where $n_{\mathrm{N}}$ is
  the density of lattice sites (ions).}
  \label{fig.sect.groundstate.outer.ws_app}
\end{figure}}

An exact calculation of the lattice energy for cubic lattices yields
similar expressions except for the factor 9/10, which is replaced by
0.89593 and 0.89588 for body-centered and face-centered
cubic-lattices, respectively (we exclude simple cubic lattices since they 
are generally unstable; note that polonium is the only known element on Earth 
with such a crystal structure under normal conditions~\cite{legut-07}). 
This shows that the equilibrium structure of the crust is expected to be 
a body-centered cubic lattice, since this gives the smallest lattice energy. 
Other lattice types, such as hexagonal closed packed for instance, might be realized in
neutron star crusts. Nevertheless, the study of Kohanoff and
Hansen~\cite{kohanoff-96} suggests that such noncubic lattices may
occur only at small densities, meaning that $r_e \sim a_0$, while in
the crust $r_e \ll a_0$, where $r_e\equiv(3/4\pi n_e)^{1/3}$ and $a_0=
\hbar^2/m_e e^2$ is the Bohr
radius. Equation~(\ref{eq.sect.groundstate.outer.lattice_energy})
shows that the lattice energy is negative and therefore reduces the
total Coulomb energy. The lattice contribution to the total energy
density is small but large enough to affect the equilibrium structure
of the crust by favoring large nuclei. Corrections due to
electron-exchange interactions, electron polarization and quantum zero
point motion of the ions are discussed in the book by Haensel,
Potekhin and Yakovlev~\cite{haensel-06}.

The main physical input is the energy $E\{A,Z\}$, which has been
experimentally measured for more than 2000 known
nuclei~\cite{audi-03}. Nevertheless, this quantity has not been
measured yet for the very neutron rich nuclei that could be present in
the dense layers of the crust and has therefore to be calculated. The
most accurate theoretical microscopic nuclear mass tables, using
self-consistent mean field methods, have been calculated by the
Brussels group and are available on line~\cite{iaa-bruslib}.

According to the first law of thermodynamics, the total pressure $P$
is given by
\begin{equation}
\label{sect.groundstate.outer.P}
P=n_{\mathrm{b}}^2 \frac{\mathrm{d}}{{\mathrm{d}}n_{\mathrm{b}}}\left(\frac{\varepsilon_{\mathrm{tot}}}{n_{\mathrm{b}}}\right)\, .
\end{equation}
Using Equation~(\ref{eq.sect.groundstate.outer.total_energy}), the
total pressure can be expressed as
\begin{equation}
\label{eq.sect.groundstate.outer.eos}
P=P_e+\frac{1}{3} \varepsilon_L \, .
\end{equation}
The electron pressure $P_e$ is defined by
\begin{equation}
\label{eq.sect.groundstate.outer.Pe}
P_e=\varepsilon_e - \mu_e n_e,
\end{equation}
where the electron chemical potential $\mu_e$ is simply given by the
electron Fermi energy $\epsilon_{\mathrm{F}e}$,
Equation~(\ref{eq.sect.plasma.noB.mstar.e}), and $p_{\mathrm{F}e}$, the
electron Fermi momentum given by
Equation~(\ref{eq.sect.plasma.noB.pF}). The electrons make the
dominant contribution to the total pressure in the outer crust.

The structure of the ground state crust is determined by minimizing the total
energy density $\varepsilon_{\mathrm{tot}}$ for a given baryon density
$n_{\mathrm{b}}=A n_{\mathrm{N}}$ imposing charge neutrality, $n_p=n_e$. However
$n_{\mathrm{b}}$ (or the average mass density $\rho$) can suffer jumps at
some values of the pressure. The pressure, on the contrary, should be continuous and
monotonically increasing with increasing depth below the
stellar surface. Therefore we will look for a ground state at $T=0$
and at a fixed $P$. This  corresponds to minimization of the Gibbs free
energy per nucleon,
$g(P)=\left(\varepsilon_{\mathrm{tot}}+P\right)/n_{\mathrm{b}}$, under
the condition of electric charge neutrality. For a completely-ionized
one-component plasma, one constructs a table $g(P;A,Z)$ and then finds
an absolute minimum in the $(A,Z)$ plane. The procedure, based on the
classical paper of Baym, Pethick and Sutherland~\cite{bps-71}, is
described in detail in the book by Haensel, Potekhin and
Yakovlev~\cite{haensel-06}. Every time that the ground state shifts to
a new nucleus with a smaller proton fraction, $(A,Z)\longrightarrow
(A^\prime,Z^\prime)$, there is a few percent jump of density at the
$(A,Z)$ and $(A^\prime,Z^\prime)$ shell interface,
\begin{equation}
{\Delta n_{\mathrm{b}}\over n_{\mathrm{b}}}\approx
{\Delta \rho\over \rho}\approx {Z\over A}{A^\prime\over
Z^\prime}-1 \, ,
\label{eq.sect.groundstate.outer.jump}
\end{equation}
which stems from a strict continuity of the pressure. It should be
stressed that these density discontinuities are the direct consequence
of the one-component plasma approximation. Jog \& Smith~\cite{jog-82}
have shown that the transition between two adjacent layers with a
single nuclear species is much smoother due to the existence of mixed
lattices. In particular, they have found that between a layer with a
pure body-centered cubic (bcc) lattice of $(Z,A)$ nuclei (lower
density) and a layer with a pure bcc lattice of $(Z^\prime,A^\prime)$
nuclei, a bcc lattice with $(Z,A)$ nuclei at the corners of the conventional 
cube and $(Z^\prime,A^\prime)$ at the center is energetically favored.

The structure of the crust is completely determined by the
experimental nuclear data up to a density of the order $\rho \sim 6
\times 10^{10} \mathrm{\ g\ cm}^{-3}$. At higher densities the nuclei are so
neutron rich that they have not yet been experimentally studied, and
the energy $E\{A,Z\}$ must be extrapolated. Consequently the
composition of the nuclei in these dense layers is model
dependent. Nevertheless most models show the predominance of nuclei
with the magic neutron numbers $N=50, 82$, thus revealing the crucial
role played by the quantum shell effects. The structure of the outer
crust is shown in Table~\ref{table.sect.groundstate.outer.struct} for
one particularly representative recent model. Up-to-date theoretical
mass tables are available online~\cite{iaa-bruslib}.

\begin{table}[htbp]
\caption[Sequence of nuclei in the ground state of the outer crust of
  a neutron star calculated by R\"uster et al.]{Sequence of nuclei in
  the ground state of the outer crust of neutron star calculated by
  R\"uster et al.~\cite{ruster-06} using experimental nuclear data
  (upper part), and the theoretical mass table of the Skyrme model
  BSk8 (lower part).}
\label{table.sect.groundstate.outer.struct}
\vskip 4mm

\centering
\begin{tabular}{c c c c c}
\toprule
$\rho_\mathrm{max}$ [$\mdens$] & Element & $Z$ & $N$ & $R_\mathrm{cell}$ [fm]\\
\midrule
$8.02 \times 10^{6}$   &  ${}^{56}$Fe & 26 & 30 & $1404.05$\\
$2.71 \times 10^{8}$   &  ${}^{62}$Ni & 28 & 34 & $449.48$ \\
$1.33 \times 10^{9}$   &  ${}^{64}$Ni & 28 & 36 & $266.97$ \\
$1.50 \times 10^{9}$ &  ${}^{66}$Ni & 28 & 38 & $259.26$ \\
$3.09 \times 10^{9}$   &  ${}^{86}$Kr & 36 & 50 & $222.66$ \\
$1.06 \times 10^{10}$ &  ${}^{84}$Se & 34 & 50 & $146.56$ \\
$2.79 \times 10^{10}$ &  ${}^{82}$Ge & 32 & 50 & $105.23$ \\
$6.07 \times 10^{10}$ &  ${}^{80}$Zn & 30 & 50 & $80.58$ \\
\midrule
$8.46 \times 10^{10}$ &  ${}^{82}$Zn & 30 & 52 & $72.77$ \\
$9.67 \times 10^{10}$ & ${}^{128}$Pd & 46 & 82 & $80.77$ \\
$1.47 \times 10^{11}$ & ${}^{126}$Ru & 44 & 82 & $69.81$ \\
$2.11 \times 10^{11}$ & ${}^{124}$Mo & 42 & 82 & $61.71$ \\
$2.89 \times 10^{11}$ & ${}^{122}$Zr & 40 & 82 & $55.22$\\
$3.97 \times 10^{11}$ & ${}^{120}$Sr & 38 & 82 & $49.37$\\
$4.27 \times 10^{11}$ & ${}^{118}$Kr & 36 & 82 & $47.92$\\
\bottomrule
\end{tabular}
\end{table}

\subsection{Structure of the inner crust}
\label{sect.groundstate.inner}

With increasing density, the ground-state value of $Z/A$ decreases and
neutrons become less and less bound. Let us define the ``net neutron
chemical potential'' of a neutron in a nucleus
\begin{equation}
\mu^\prime_n\equiv \mu_n-m_n c^2=
\left(
{\partial E\{ A,Z \}\over \partial N}
\right)_Z-m_n c^2 \, .
\label{eq.sect.groundstate.inner.munprime}
\end{equation}
As long as $\mu^\prime_n<0$, all neutrons are bound within nuclei. The
neutron drip point corresponds to $\mu^\prime_n=0$; beyond this point
neutrons ``drip out of nuclei'', i.e.\ they begin to fill states
in the continuous part of the energy spectrum. We can roughly localize
the neutron drip point using the approximate mass formula for
$E^\prime\{A,Z\}=E\{A,Z\}-A m c^2$, where, for simplicity, we neglect
the neutron-proton mass difference, putting $m_n\approx m_p\approx m
=939 \mathrm{\ MeV}/c^2$. Neglecting surface and Coulomb terms, we have
\begin{equation}
E^\prime\{A,Z\}\approx A\;
\left(E_{\mathrm{vol}}+E_{\mathrm{sym}}\delta^2\right) \, ,
\label{eq.sect.groundstate.inner.Emass.vol}
\end{equation}
where $\delta\equiv (N-Z)/A$, and $E_{\mathrm{vol}}$ and $E_{\mathrm{sym}}$ are
nuclear volume and symmetry energies, respectively. Experimentally,
$E_{\mathrm{vol}}\simeq-16$~MeV and $E_{\mathrm{sym}}\simeq 32$~MeV. Using
Equation~(\ref{eq.sect.groundstate.inner.Emass.vol}), we can easily
show that the value of $\delta$ at which $\mu^\prime_n=0$ is
\begin{equation}
\delta_{\mathrm{ND}}=\sqrt{1-E_{\mathrm{vol}}/E_{\mathrm{sym}}}-1.
\label{eq.sect.groundstate.inner.deltaND}
\end{equation}
Using experimental values of $E_{\mathrm{vol}}$ and $E_{\mathrm{sym}}$, we
find $\delta_{\mathrm{ND}}=0.225$.

Neglecting neutron-proton mass difference, the beta equilibrium
condition
\begin{equation}
\label{eq.sect.groundstate.outer.beta}
p+e \rightarrow n+\nu_e\, 
\end{equation}
reads
\begin{equation}
\mu_e=\mu_n-\mu_p\approx 4 E_{\mathrm{sym}} \delta \, .
\label{eq.sect.groundstate.inner.betaND}
\end{equation}
At the density under consideration, electrons are ultrarelativistic, so
that $\mu_e\simeq 5.16\;\left(\rho_9 Z/A\right)^{1/3}$~MeV, where
$\rho_9=\rho/10^9~\mdens$. It is now easy to show that the neutron
drip density is roughly $\rho_{\mathrm{ND}}\approx 2\times 10^{11}~\mdens$,
which is about half of the value obtained in complete
calculations~\cite{bps-71, haensel-94, ruster-06}. For
$\rho>\rho_{\mathrm{ND}}\approx 4\times 10^{11}~\mdens$ a fraction of
neutrons thus forms a gas outside the nuclei.

The inner crust of a neutron star is a unique system, which is not
accessible in the laboratory due to the presence of this neutron
gas. In the following we shall thus refer to the ``nuclei'' in the
inner crust as ``clusters'' in order to emphasize these
peculiarities. The description of the crust beyond neutron drip
therefore relies on theoretical models only. Many-body calculations
starting from the realistic nucleon-nucleon interaction are out of
reach at present due to the presence of spatial inhomogeneities of
nuclear matter. Even in the simpler case of homogeneous nuclear
matter, these calculations are complicated by the fact that nucleons
are strongly interacting via two-body, as well as three-body, forces,
which contain about twenty different operators. As a result, the inner
crust of a neutron star has been studied with phenomenological
models. Most of the calculations carried out in the inner crust rely
on purely classical (compressible liquid drop) and semi-classical
models (Thomas--Fermi approximation and its extensions). The
state-of-the-art calculations performed so far are based on
self-consistent mean field methods, which have been very successful in
predicting the properties of heavy laboratory nuclei.

\subsubsection{Liquid drop models}
\label{sect.groundstate.inner.LDM}

We will present in detail the liquid drop model because this approach
provides very useful insight despite its simplicity. As in
Section~\ref{sect.groundstate.outer}, we first start by writing the
total energy density including the contribution $\varepsilon_n$ of the
neutron gas
\begin{equation}
\label{eq.sect.groundstate.inner.LDM.total_energy}
\varepsilon_{\mathrm{tot}} = n_{\mathrm{N}} E\{A,Z\} +\varepsilon_e+\varepsilon_L+\varepsilon_n \, .
\end{equation}
The nuclear clusters are treated as liquid drops of nuclear matter
whose energy can be decomposed into volume, surface and Coulomb terms
\begin{equation}
\label{eq.sect.groundstate.inner.LDM.cluster_energy}
E\{A,Z\} = E_{\mathrm{N, vol}}+E_{\mathrm{N, surf}}+E_{\mathrm{N, Coul}} \, .
\end{equation}
In the simplest version, the drop is supposed to be
incompressible with a density on the order of $\rho_0$ 
corresponding to the density inside heavy nuclei. This implies that the volume and surface
terms in Equation~(\ref{eq.sect.groundstate.inner.LDM.cluster_energy})
are proportional to $A$ and $A^{2/3}$, respectively.
Each contribution to the energy,
Equation~(\ref{eq.sect.groundstate.inner.LDM.cluster_energy}),
 can then be parameterized in terms of the numbers $A$ and $Z$.
The parameters are adjusted to the known experimental masses
of nuclei, with $Z/A \sim 0.5$. The first models of neutron
star crust were based on such semi-empirical mass formulae, see
for instance the book by Haensel, Potekhin and
Yakovlev~\cite{haensel-06}. However, such formulae can not be
reliably extrapolated to the neutron rich nuclear clusters in
neutron star crusts, where $Z/A$ varies from $\sim 0.3$ at the
neutron drip threshold to $\sim 0.1$ at the bottom of the
crust. Besides, the presence of the neutron liquid has a
profound effect on the clusters. First, it reduces the surface
energy of the clusters as compared to that of isolated nuclei.
Second, it exerts a pressure on the clusters. A major
breakthrough was reached by Baym, Bethe and
Pethick~\cite{bbp-71},  who applied a compressible liquid-drop
model, which included the results of microscopic many-body
calculations, to describe consistently both the nucleons in the
clusters and the ``free'' neutrons. The energy
Equation~(\ref{eq.sect.groundstate.inner.LDM.cluster_energy})
of the cluster then depends not only on $A$ and $Z$ but on a
few additional parameters, such as, for instance, the size of
the cluster and the density of the neutrons and protons inside it.

The volume contribution in
Equation~(\ref{eq.sect.groundstate.inner.LDM.cluster_energy}) is given
by
\begin{equation}
\label{eq.sect.groundstate.inner.LDM.bulk_energy}
E_{\mathrm{N, vol}} = \varepsilon\{ n_{n\mathrm{i}}, n_{p\mathrm{i}} \} {\cal V}_{\mathrm{N}},
\end{equation}
where $\varepsilon\{ n_n, n_p\}$ is the energy density of homogeneous
nuclear matter and $n_{n\mathrm{i}}$ and $n_{p\mathrm{i}}$ are respectively the
neutron and proton densities inside the clusters. ${\cal V}_{\mathrm{N}}$
is the volume of the cluster. For consistency the energy density of
the surrounding neutron gas is expressed in terms of the same function
$\varepsilon\{ n_n, n_p\}$ as
\begin{equation}
\label{eq.sect.groundstate.inner.LDM.drip_energy}
\varepsilon_n =\varepsilon\{ n_{n\mathrm{o}}, 0\}(1-w )  \, ,
\end{equation}
where $n_{n\mathrm{o}}$ is the number density of free neutrons outside the
clusters and
\begin{equation}
w=\frac{{\cal V}_{\mathrm{N}}}{{\cal V}_{\mathrm{cell}}}=\left(\frac{r_p}{R_{\mathrm{cell}}}\right)^3
\end{equation}
is the volume fraction of the cluster. Let us define the surface
thermodynamic potential per unit area $\sigma$ and the chemical
potential $\mu_{n\mathrm{s}}$ of neutrons adsorbed on the surface of the
drop (forming a neutron skin) by
\begin{equation}
\label{eq.sect.groundstate.inner.LDM.surf_param}
\sigma = \frac{ \partial E_{\mathrm{N, surf}} }{\partial {\cal A}}\biggr\vert_{N_{\mathrm{s}}}\, , \ \ \mu_{n\mathrm{s}}= \frac{ \partial E_{\mathrm{N, surf}} }{\partial N_{\mathrm{s}}}\biggr\vert_{\cal A}  \, ,
\end{equation}
where $\cal A$ is the surface area of the cluster and $N_{\mathrm{s}}$ is
the number of adsorbed neutrons. Since energy is an extensive
thermodynamic variable, it follows from Euler's theorem about
homogeneous functions that
\begin{equation}
\label{eq.sect.groundstate.inner.LDM.surf_energy}
E_{\mathrm{N, surf}} = \sigma {\cal A} + N_{\mathrm{s}} \mu_{n\mathrm{s}}  \, .
\end{equation}
The Coulomb energy of a uniformly-charged spherical drop of radius
$r_p$ is given by
\begin{equation}
\label{eq.sect.groundstate.inner.LDM.Coul_energy}
E_{\mathrm{N, Coul}} = \frac{3}{5} \frac{Z^2 e^2}{r_p} \, .
\end{equation}
Corrections due to the diffuseness of the cluster surface and due to
quantum exchange can be found, for instance, in
reference~\cite{mackie-77}. The electron energy density is
approximately given by
Equation~(\ref{eq.sect.groundstate.outer.electron_energy}). In the
Wigner--Seitz approximation, assuming uniformly-charged spherical
clusters of radius $r_p$, the lattice energy is given by~\cite{bbp-71}
\begin{equation}
\label{eq.sect.groundstate.inner.LDM.lat_energy}
\varepsilon_L = -\frac{9}{10} \left(\frac{4\pi}{3}\right)^{1/3} Z^{2/3} e^2 n_e^{4/3} \left(1-\frac{1}{3}w^{2/3}\right) \, .
\end{equation}
The physical input of the liquid drop model is the energy density of
homogeneous nuclear matter $\varepsilon\{ n_n,n_p \}$, the surface
potential $\sigma$ and the chemical potential $\mu_{n\mathrm{s}}$. For
consistency, these ingredients should be calculated from the same
microscopic nuclear model. The surface properties are usually
determined by considering two semi-infinite phases in equilibrium
(nucleons in clusters and free neutrons outside) separated by a plane
interface (for curvature corrections, see, for
instance, \cite{lorenz-91,douchin-00}). In this approximation, the
surface energy is proportional to the surface area
\begin{equation}
\label{eq.sect.groundstate.inner.LDM.app_surf_energy}
E_{\mathrm{N, surf}} = \biggl(\sigma_{\mathrm{s}} + s_n (n_{n\mathrm{i}} - n_{n\mathrm{o}}) \mu_{n\mathrm{s}} \biggr)  {\cal A} \, ,
\end{equation}
where $\sigma_{\mathrm{s}}$ is the surface tension and $s_n$ the neutron
skin thickness. Unlike the surface area $\cal A$, the proportionality
coefficient does not depend on the actual shape of the nuclear
clusters.

The structure of the inner crust is determined by minimizing the total
energy density $\varepsilon_{\mathrm{tot}}$ for a given baryon density
$n_{\mathrm{b}}$ imposing electric charge neutrality $n_p=n_e$. The
conditions of equilibrium are obtained by taking the partial derivative
of the energy density $\varepsilon_{\mathrm{tot}}$ with respect to the
free parameters of the model. In the following we shall neglect curvature
corrections to the surface energy. In this approximation, the surface
potential $\sigma_{\mathrm{s}}$ is independent of the shape and size of the
drop. The variation of the surface tension with the neutron excess is
illustrated in
Figure~\ref{fig.sect.groundstate.inner.LDM.DH_surface_tension}.

\epubtkImage{douchin3.png}{%
\begin{figure}[htbp]
  \centerline{\includegraphics[scale=0.4]{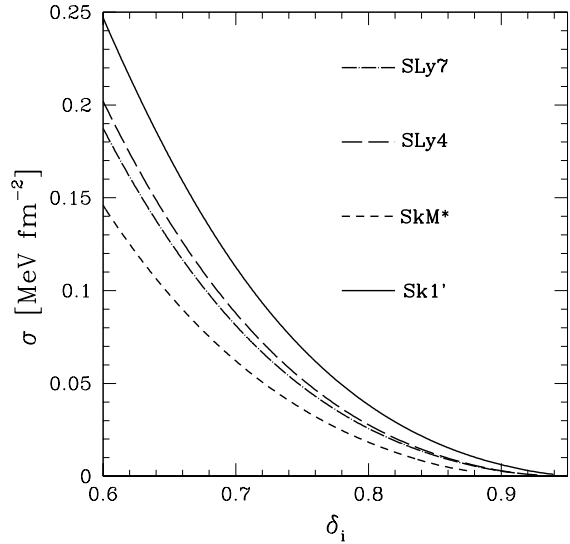}}
  \caption{Surface tension of the nuclei in neutron star crusts
  versus neutron excess parameter $\delta_{\mathrm{i}}$ inside the nuclear
  cluster, in the plane interface approximation with different Skyrme
  models~\cite{douchin-99}.}
  \label{fig.sect.groundstate.inner.LDM.DH_surface_tension}
\end{figure}}

The conditions of chemical equilibrium are
\begin{equation}
\label{eq.sect.groundstate.inner.LDM.equi1}
\mu^{\mathrm{bulk}}_{n\mathrm{i}}=\mu^{\mathrm{bulk}}_{n\mathrm{o}}=\mu_{n\mathrm{s}} \, ,
\end{equation}
\begin{equation}
\label{eq.sect.groundstate.inner.LDM.equi2}
\mu^{\mathrm{bulk}}_{n\mathrm{i}}-\mu^{\mathrm{bulk}}_{p\mathrm{i}}-\mu_e=
\frac{8 \pi}{5} e^2 n_{p\mathrm{i}} r_p^2 f_3\{ w \} \, ,
\end{equation}
where
\begin{equation}
\label{eq.sect.groundstate.inner.LDM.mux}
\mu^{\mathrm{bulk}}_{X}\equiv \frac{\partial \varepsilon}{\partial n_X} \, ,
\end{equation}
and  $X=n\mathrm{i},n\mathrm{o},n\mathrm{s}$, ...
\begin{equation}
\label{eq.sect.groundstate.inner.LDM.func3}
f_3\{ w \} \equiv 1 - \frac{3}{2} w^{1/3}+\frac{1}{2} w \, .
\end{equation}
The mechanical equilibrium of the cluster is expressed by
\begin{equation}
\label{eq.sect.groundstate.inner.LDM.equi3}
P^{\mathrm{bulk}}_{\mathrm{i}}-P^{\mathrm{bulk}}_{\mathrm{o}}=\frac{2 \sigma_{\mathrm{s}}}{r_p}-\frac{4\pi}{15}e^2 n_{p\mathrm{i}}^2 r_p^2 (1-w) \, ,
\end{equation}
which can be easily recognized as a generalization of Laplace's
law for an isolated drop. We have introduced bulk nuclear pressures by
\begin{equation}
\label{eq.sect.groundstate.inner.LDM.Px}
P^{\mathrm{bulk}}_{X}\equiv n_X^2 \frac{\partial (\varepsilon/n_X)}{\partial n_X} =  n_X \mu^{\mathrm{bulk}}_X - \varepsilon \, .
\end{equation}
Note that $P^{\mathrm{bulk}}_{\mathrm{i}}$ means the total pressure
inside the drop, $P^{\mathrm{bulk}}_{\mathrm{i}}\equiv
P^{\mathrm{bulk}}_{n\mathrm{i}}+P^{\mathrm{bulk}}_{p\mathrm{i}}$. The
last term in Equation~(\ref{eq.sect.groundstate.inner.LDM.equi3})
comes from the pressure due to Coulomb forces between protons inside
the cluster and between clusters of the lattice.

The mechanical equilibrium of the crystal lattice
can be expressed as
\begin{equation}
\label{eq.sect.groundstate.inner.LDM.virial}
E_{\mathrm{N, surf}} =2 E_{\mathrm{Coul}} \, ,
\end{equation}
where $E_{\mathrm{Coul}}$ is the \emph{total} Coulomb energy
\begin{equation}
\label{eq.sect.groundstate.inner.LDM.tot_Coul_energy}
E_{\mathrm{Coul}}=E_{\mathrm{N, Coul}}+{\cal V}_{\mathrm{cell}} \varepsilon_L\, .
\end{equation}
Relation (\ref{eq.sect.groundstate.inner.LDM.virial}) is
referred to
as a ``virial'' theorem~\cite{bbp-71}.

Combining Equations~(\ref{eq.sect.groundstate.inner.LDM.Coul_energy})
and (\ref{eq.sect.groundstate.inner.LDM.lat_energy}), the total
Coulomb energy can be written as
\begin{equation}
\label{eq.sect.groundstate.inner.LDM.tot_Coul_energy2}
E_{\mathrm{Coul}}=E_{\mathrm{N, Coul}}f_3\{ w \} \, ,
\end{equation}
where the dimensionless function $f_3\{ w \}$ is given by
Equation~(\ref{eq.sect.groundstate.inner.LDM.func3}). Let us emphasize
that the above equilibrium conditions are only valid if nuclear
surface curvature corrections are neglected.

Equation~(\ref{eq.sect.groundstate.inner.LDM.virial}) shows that the
equilibrium composition of the cluster is a result of the
competition between Coulomb effects, which favor small clusters, and
surface effects, which favor large clusters. 
This also shows that the lattice energy is very important for
determining the equilibrium shape of the cluster, especially at the
bottom of the crust, where the size of the cluster is of the same order
as the lattice spacing. Even at the neutron drip, the lattice energy
reduces  the total Coulomb energy by  about 15\%.

The structure of the inner crust, as calculated from a compressible
liquid drop model by Douchin \& Haensel~\cite{douchin-01}, is
illustrated in
Figures~\ref{fig.sect.groundstate.inner.LDM.DH_inner_crust1}
and~\ref{fig.sect.groundstate.inner.LDM.DH_inner_crust2}. One
remarkable feature,  which is confirmed by more realistic models, is
that the number $Z \sim 40$ of protons in the clusters is almost
constant throughout the inner crust. It can also be seen that, as the
density increases, the clusters get closer and closer, while their
size $r_p$ varies very little. Let us also notice that at the bottom
of the crust the number $N_{\mathrm{s}}$ of neutrons,  adsorbed on the
surface of the clusters,  decreases with increasing density, because the
properties of the matter inside and outside the clusters become more and more
alike. The results of different liquid drop models are compared in
Figure~\ref{fig.sect.groundstate.inner.LDM.charge}.

\epubtkImage{douchin1b.png}{%
\begin{figure}[htbp]
  \centerline{\includegraphics[scale=0.4]{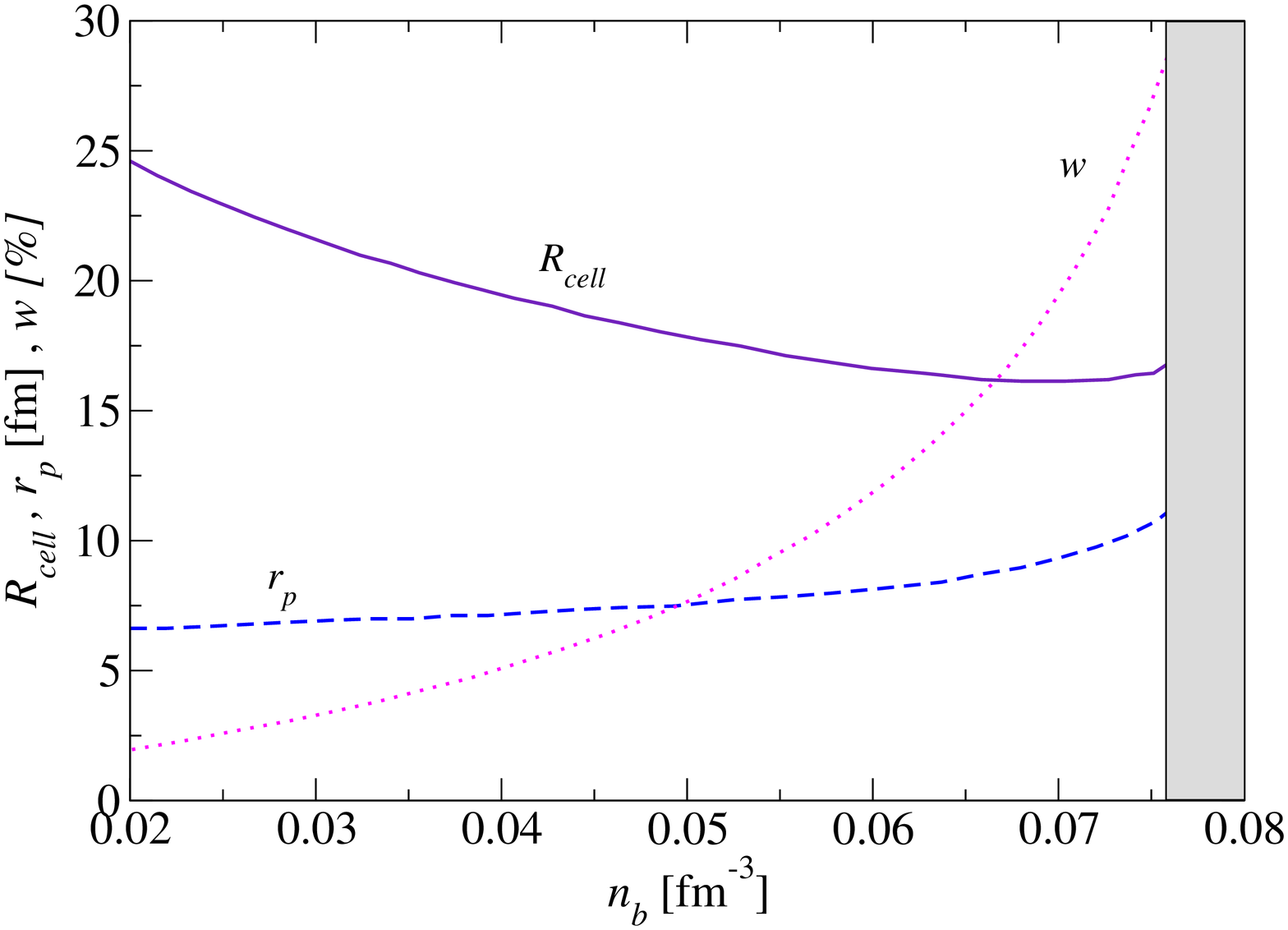}}
  \caption{Structure of the ground state of the inner crust.
   Radius $R_{\mathrm{cell}}$ of the Wigner--Seitz cell,
  proton radius $r_p$ of spherical nuclei, and fraction $w$ of volume
  filled by nuclear clusters (in percent), versus average baryon
  number density $n_{\mathrm{b}}$ as calculated by Douchin \&
  Haensel~\cite{douchin-00b}. }
  \label{fig.sect.groundstate.inner.LDM.DH_inner_crust1}
\end{figure}}

\epubtkImage{douchin2b.png}{%
\begin{figure}[htbp]
  \centerline{\includegraphics[scale=0.4]{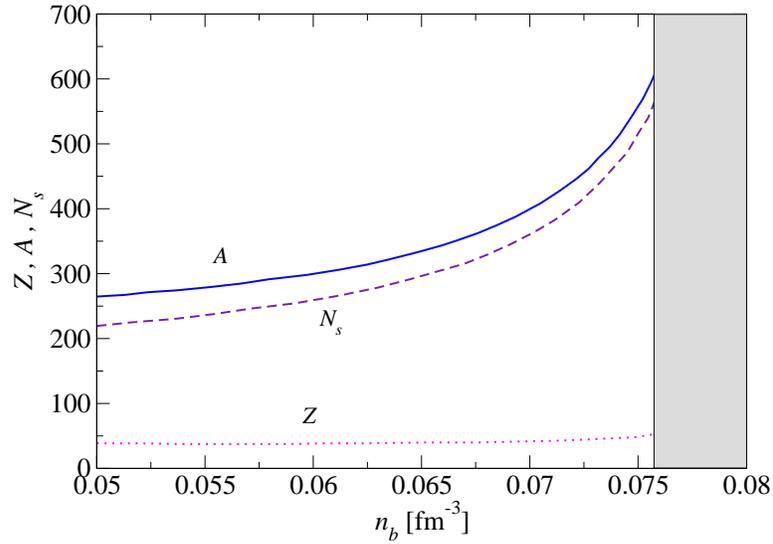}}
  \caption{Composition of nuclear clusters in the ground state
  of the inner crust. Baryon number A of spherical
  clusters and their proton number Z, versus average baryon number
  density $n_{\mathrm{b}}$ as calculated by Douchin \&
  Haensel~\cite{douchin-00b}. $N_{\mathrm{s}}$ is the number of neutrons
  adsorbed on the surface of the clusters.}
  \label{fig.sect.groundstate.inner.LDM.DH_inner_crust2}
\end{figure}}

\epubtkImage{charge_cldm.png}{%
\begin{figure}[htbp]
  \centerline{\includegraphics[scale=0.4]{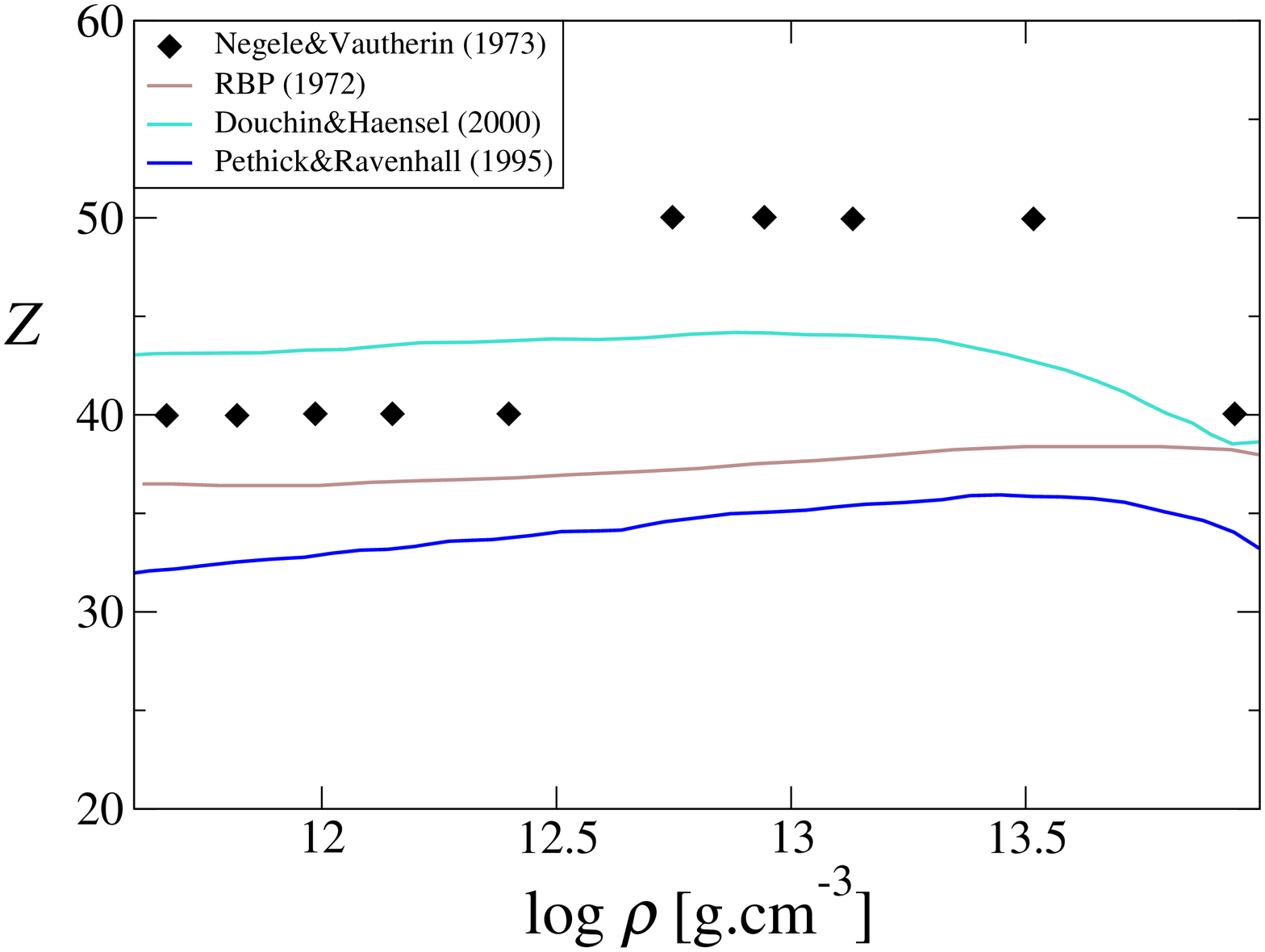}}
  \caption{Proton number $Z$ of the nuclear clusters vs.
  density $\rho$ in the ground state of the inner crust
  of neutron stars, calculated by various authors from different
  liquid drop models based on many-body theories  with
  effective interactions: RBP(Ravenhall, Bennett and
  Pethick)~\cite{rbp-72}, Douchin \& Haensel~\cite{douchin-00b},
  Pethick \& Ravenhall~\cite{pethick-95}. For comparison, the results of the quantum calculations of Negele \&
  Vautherin~\cite{nv-73} (diamonds) are also shown.}
  \label{fig.sect.groundstate.inner.LDM.charge}
\end{figure}}

The liquid drop model is very instructive for understanding the
contribution of different physical effects to the structure of the
crust. However this model is purely classical and consequently
neglects quantum effects. Besides,  the assumption of
clusters with a sharp cut surface is questionable, especially in the
high density layers where the nuclei are very neutron rich.

\subsubsection{Semi-classical models}
\label{sect.groundstate.inner.TF}

Semi-classical models have been widely applied to study the structure
of neutron star crusts. These models assume that the number of
particles is so large that the quantum numbers describing the system
vary continuously and instead of wave functions one can use the number
densities of the various constituent particles. In this approach, the
total energy density is written as a functional of the number
densities of the different particle species
\begin{equation}
\label{eq.sect.groundstate.inner.TF}
\varepsilon( \pmb{r} ) = \varepsilon_{\mathrm{N}}\{ n_n(\pmb{r}),n_p(\pmb{r}) \} + \varepsilon_e\{ n_e(\pmb{r}) \} + \varepsilon_{\mathrm{Coul}}\{ n_e(\pmb{r}), n_p(\pmb{r}) \}\, ,
\end{equation}
where $\varepsilon_{\mathrm{N}}$, $\varepsilon_e$ and $\varepsilon_{\mathrm{Coul}}$ are the nuclear, electron and Coulomb contributions,
respectively.

The idea for obtaining the energy functional is to assume that the
matter is \emph{locally} homogeneous: this is known as the
Thomas--Fermi or local density approximation. This approximation is
valid when the characteristic length scales of the density variations
are much larger than the corresponding interparticle spacings. The
Thomas--Fermi approximation can be improved by including density
gradients in the energy functional.

As discussed in Section~\ref{sect.groundstate.outer}, the electron
density is almost constant so that the local density approximation is
very good with  the electron energy functional given by
Equation~(\ref{eq.sect.groundstate.outer.electron_energy}).

The Coulomb part in Equation~(\ref{eq.sect.groundstate.inner.TF}) can
be decomposed into a classical and a quantum contribution.  The classical
contribution is given by
\begin{equation}
\label{eq.sect.groundstate.inner.TF.Coulomb_energy}
 \varepsilon^{\mathrm{class}}_{\mathrm{Coul}}\{ n_e(\pmb{r}), n_p(\pmb{r}) \} = \frac{1}{2} e\left( n_p(\pmb{r})- n_e(\pmb{r}) \right) \phi(\pmb{r})\, ,
\end{equation}
where $e$ is the proton electric charge and $\phi$ is the
electrostatic potential, which obeys Poisson's equation,
\begin{equation}
\label{eq.sect.groundstate.inner.TF.Poisson}
\Delta \phi = -4 \pi e\left( n_p(\pmb{r}) - n_e(\pmb{r}) \right) \,.
\end{equation}
The quantum contribution $\varepsilon_{\mathrm{Coul}}^{\mathrm{corr}}$ accounts
for the quantum correlations. For instance, the local part of the
Coulomb exchange correlations induced by the Pauli exclusion principle
is given by the Slater--Kohn--Sham functional~\cite{kohnsham-65}
\begin{equation}
\label{eq.sect.groundstate.inner.TF.corr_Coulomb_energy}
\varepsilon_{\mathrm{Coul}}^{\mathrm{corr}}\{ n_e(\pmb{r}), n_p(\pmb{r}) \}=-\frac{3}{4} \left(\frac{3}{\pi}\right)^{1/3} e^2 \left(n_p(\pmb{r})^{4/3}+n_e(\pmb{r})^{4/3}\right)\, .
\end{equation}
The nuclear functional $\varepsilon_{\mathrm{N}}\{
n_n(\pmb{r}),n_p(\pmb{r}) \}$ is less certain. Its local part
is just a function of $n_n$ and $n_p$, and can, in principle, be
inferred using the results of many-body calculations of the
ground state of uniform asymmetric nuclear matter. However,  the
many-body calculations for the nonlocal part of the nuclear functional
are  much more difficult and have never been done in a fully satisfactory
way. A simpler procedure is to postulate a purely phenomenological
expression of the nonlocal part of the nuclear functional. The free
parameters are then determined to reproduce some nuclear
properties, for instance, the experimental atomic
masses. Alternatively, the nuclear functional can be obtained from
effective theories. In this case, the bare nucleon-nucleon interaction
is replaced by an effective phenomenological interaction. It is then
possible to deduce the nuclear functional in a systematic way using
the extended Thomas--Fermi approximation (see for
instance~\cite{brack-85}). This approach has been developed for
neutron star crust matter by Onsi and
collaborators~\cite{onsi-97,onsi-08}.

The total energy density is equal to the energy density of one unit cell of the
lattice times the number of cells. The minimization of the total
energy density under the constraints of a fixed total baryon  density
$n_{\mathrm{b}}$ and \emph{global} electro-neutrality
\begin{equation}
\label{eq.sect.groundstate.inner.TF.electroneutrality}
n_{\mathrm{b}} = \frac{1}{{\cal V}_{\mathrm{cell}}} \int_{\mathrm{cell}} \mathrm{d}^3r \left( n_n(\pmb{r}) + n_p(\pmb{r}) \right)\, , \ \ \int_{\mathrm{cell}} \mathrm{d}^3 r \left( n_p(\pmb{r}) -n_e(\pmb{r}) \right) = 0 \, ,
\end{equation}
leads to Euler--Lagrange equations for the nucleon densities. In
practice the unit cell is usually approximated by a sphere of the same
volume ${\cal V}_{\mathrm{cell}}$.  The boundary conditions are that the
gradients of the densities and of the Coulomb potential vanish at the
origin $r=0$ and on the surface of the sphere $r=R_{\mathrm{cell}}$. Instead of solving the Euler--Lagrange equations, the nucleon
densities are usually parameterized by  some simple analytic functions
with correct boundary behavior. Free parameters are then
determined by minimizing the energy as in the compressible liquid drop
models discussed in Section~\ref{sect.groundstate.inner.LDM}. The
proton number of the nuclear clusters in the inner crust vs. mass density 
is shown in Figure~\ref{fig.sect.groundstate.inner.TF.charge} for different models. 
In those semiclassical models (as
well as in liquid drop models discussed in
Section~\ref{sect.groundstate.inner.LDM}), the number of bound
nucleons inside the clusters in neutron star crusts varies continuously
with depth. However, the nuclear clusters are expected to exhibit
specific magic numbers of nucleons as similarly observed in isolated
terrestrial nuclei, due to the clustering of quantum single-particle
energy levels. The scattering of the unbound neutrons on the nuclear
inhomogeneities leads also to ``shell'' (Casimir or band)
effects~\cite{bulgac-01, magierski-02, chamel-05}. The energy
corrections due to shell effects have been studied perturbatively in
semiclassical models~\cite{oyamatsu-94, dutta-04} and in Hartree--Fock
calculations~\cite{magierski-02}. They have been found to be
small. However, since the energy differences between different nuclear
configurations are small, especially at high densities, these shell
effects are important for determining the equilibrium structure of the
crust. Calculations of the ground state structure of the crust, including proton shell 
effects, have recently been carried out by Onsi et al.~\cite{onsi-08}. 
As can be seen in Figure~\ref{fig.sect.groundstate.inner.QM.charge}, these
shell effects significantly change the composition of the clusters predicting 
proton magic numbers $Z=20,40,50$. 

\epubtkImage{charge.png}{%
  \begin{figure}[htbp]
    \centerline{\includegraphics[scale=0.4]{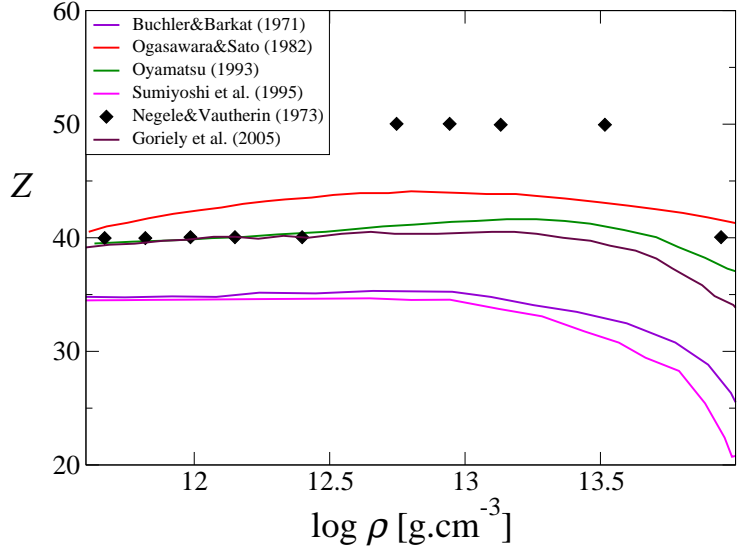}}
    \caption{Proton number $Z$ of the nuclear clusters vs.\ the mass
    density $\rho$ in the ground state of the inner crust, calculated
    by different semi-classical models: Buchler \&
    Barkat~\cite{buchler-71}, Ogasawara \& Sato~\cite{ogasawara-82},
    Oyamatsu~\cite{oyamatsu-93}, Sumiyoshi et al.~\cite{sumiyoshi-95},
    Goriely et al.~\cite{goriely-05b}. For comparison the results of the quantum calculations of Negele \&
    Vautherin~\cite{nv-73} (diamonds) are also shown.}
    \label{fig.sect.groundstate.inner.TF.charge}
\end{figure}}

\subsubsection{Quantum calculations}
\label{sect.groundstate.inner.quantum}

Quantum calculations of the structure of the inner crust were
pioneered by Negele \& Vautherin~\cite{nv-73}. These types of calculations
have been improved only recently by Baldo and
collaborators~\cite{baldo-05, baldo-06}. Following the Wigner--Seitz
approximation~\cite{ws-33}, the inner crust is decomposed into
independent spheres, each of them centered at a nuclear cluster,
whose radius is defined by Equation~(\ref{eq.sect.plasma.Rcell}), as 
illustrated in Figure~\ref{fig.sect.groundstate.outer.ws_app}. 
The determination of the equilibrium structure of the crust thus 
reduces to calculating the composition of one of the spheres. Each sphere 
can be seen as an exotic ``nucleus''. The methods developed in nuclear
physics for treating isolated nuclei can then be directly applied.

Starting from many-body calculations of uniform nuclear matter
with realistic nucleon-nucleon interaction, and expanding
the nucleon density matrix in relative and center of mass coordinates,
Negele \& Vautherin~\cite{nv-72} derived a set of nonlinear equations
for the single particle wave functions of the nucleons,
$\varphi_\alpha^{(q)}(\pmb{r})$,  where $q=n,p$ for neutron and
proton, respectively, and $\alpha$ is the set of quantum numbers characterizing
each single particle state. Inside the Wigner--Seitz sphere, these
equations take the form
\begin{equation}
\label{eq.sect.groundstate.inner.quantum.HF}
-\pmb{\nabla} \cdot \frac{\hbar^2}{2 m_q^\oplus(r)} \pmb{\nabla}
\varphi_\alpha^{(q)}(\pmb{r}) + U_q( r) \varphi_\alpha^{(q)}(\pmb{r})
+ \frac{W_q(r)}{r} \pmb{\ell} \cdot \pmb{\sigma} \,
\varphi_\alpha^{(q)}(\pmb{r})  = \epsilon^{(q)}_\alpha
\varphi_\alpha^{(q)}(\pmb{r}) \, ,
\end{equation}
where $\epsilon^{(q)}_\alpha$ is the single particle energy,
$\pmb{\ell}\equiv - \mathrm{i}\pmb{r}\times\pmb{\nabla}$ is the dimensionless
orbital--angular-momentum operator and $\pmb{\sigma}$ is a vector
composed of Pauli spin matrices. The effective masses
$m_q^\oplus(r)$, the mean fields $U_q(r)$ and the spin-orbit
potentials $W_q(r)$ depend on wave functions of all nucleons
inside the sphere through the particle number densities
\begin{equation}
\label{eq.sect.groundstate.inner.quantum.nq}
n_q(\pmb{r}) = \sum_\alpha |\varphi_\alpha^{(q)}(\pmb{r})|^2 \, ,
\end{equation}
the kinetic energy densities (in units of $\hbar^2/2m$, where $m$ is
the nucleon mass)
\begin{equation}
\label{eq.sect.groundstate.inner.quantum.tauq}
\tau_q(\pmb{r}) = \sum_\alpha |\pmb{\nabla} \varphi_\alpha^{(q)}(\pmb{r})|^2 \, ,
\end{equation}
and the spin-orbit densities
\begin{equation}
\label{eq.sect.groundstate.inner.quantum.Jq}
\pmb{J_q}(\pmb{r}) = \pmb{r} \sum_\alpha \varphi_\alpha^{(q)}(\pmb{r})^*
\frac{\pmb{\ell} \cdot \pmb{\sigma}}{r^2}
\varphi_\alpha^{(q)}(\pmb{r}) \,.
\end{equation}

Equations~(\ref{eq.sect.groundstate.inner.quantum.HF}) reduce to
\emph{ordinary differential equations} by expanding a wave function
on the basis of the total angular momentum. Apart from the nuclear
central and spin-orbit potentials, the protons also feel a Coulomb
potential. In the Hartree--Fock approximation, the Coulomb potential is
the sum of a direct part $e\phi(\pmb{r})$, where $\phi(\pmb{r})$ is
the electrostatic potential, which obeys Poisson's
Equation~(\ref{eq.sect.groundstate.inner.TF.Poisson}), and an exchange
part, which is nonlocal in general. Negele \& Vautherin adopted the
Slater approximation for the Coulomb exchange, which
leads to a \emph{local} proton Coulomb potential. As a remark, the
expression of the Coulomb exchange potential used nowadays was
actually suggested by Kohn \& Sham~\cite{kohnsham-65}. It is
smaller by a factor $3/2$ compared to that initially proposed by
Slater~\cite{slater-51} before the formulation of the density
functional theory. It is obtained by taking the 
derivative of Equation~(\ref{eq.sect.groundstate.inner.TF.corr_Coulomb_energy})
 with respect to the proton density $n_p(r)$. Since the clusters in the
crust are expected to have a very diffuse surface and a thick neutron skin
(see Section~\ref{sect.groundstate.inner.TF}), the spin-orbit coupling
term for the neutrons (which is proportional to the gradient of the
neutron density) was neglected.

Equations~(\ref{eq.sect.groundstate.inner.quantum.HF}) have to
be solved self-consistently. For a given number $N$ of neutrons and
$Z$ of protons and some initial guess of the effective masses and
potentials, the equations are solved for the wave functions of
 $N$ neutrons and $Z$ protons, which correspond to
  the lowest energies
 $\epsilon^{(q)}_\alpha$. These wave functions are then used
to recalculate the effective masses and potentials. The process is
iterated until the convergence is achieved.

\epubtkImage{charge_qm.png}{%
\begin{figure}[htbp]
  \centerline{\includegraphics[scale=0.4]{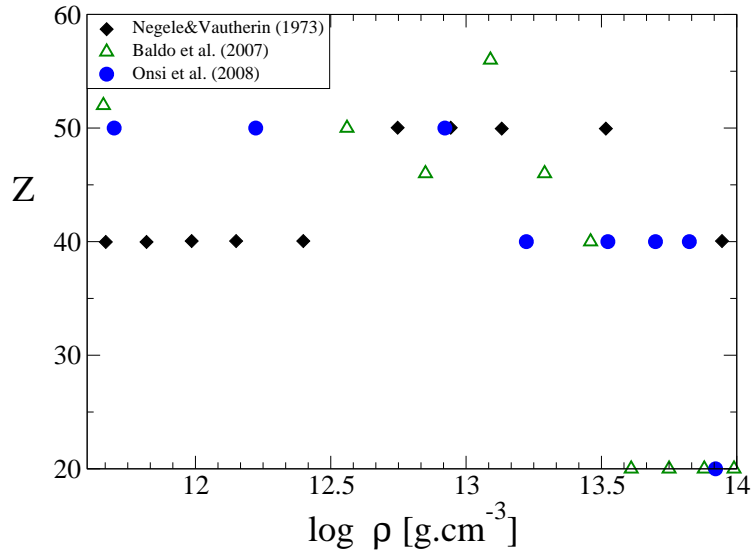}}
  \caption{Proton number $Z$ of the nuclear clusters vs. the
  mass density $\rho$ in the ground state of the inner crust,
  calculated by different quantum models:
  Negele \& Vautherin~\cite{nv-73}, the model P2 
  from Baldo et al.~\cite{baldo-07, baldo-07b}
  and the recent calculations of Onsi~et~al~\cite{onsi-08}.}
  \label{fig.sect.groundstate.inner.QM.charge}
\end{figure}}

Negele \& Vautherin~\cite{nv-73} determined the structure of the inner
crust by minimizing the total energy per nucleon in a Wigner--Seitz
sphere, and thus treating the electrons as a relativistic Fermi gas. Since the
sphere is electrically neutral, the number of electrons is equal to
$Z$ and the electron energy is easily evaluated from
Equation~(\ref{eq.sect.groundstate.outer.electron_energy}) with
$n_e=Z/{\cal V}_{\mathrm{cell}}$, where ${\cal V}_{\mathrm{cell}}$ is the volume
of the sphere. As for the choice of boundary conditions, Negele \&
Vautherin imposed that wave functions with even parity (even $\ell$)
and the radial derivatives of wave functions with odd parity (odd
$\ell$) vanish on the sphere $r=R_{\mathrm{cell}}$. This prescription leads
to a roughly constant neutron density outside the nuclear
clusters. However, the densities had still to be averaged in the
vicinity of the cell edge in order to remove unphysical
fluctuations. The structure and the composition of the inner crust is
shown in
Table~\ref{table.sect.groundstate.inner.quantum.struct}. These results
are qualitatively similar to those obtained with liquid drop models
(see Figure~\ref{fig.sect.groundstate.inner.LDM.charge} in
Section~\ref{sect.groundstate.inner.LDM}) and semiclassical models
(see Figure~\ref{fig.sect.groundstate.inner.TF.charge} in
Section~\ref{sect.groundstate.inner.TF}). The remarkable distinctive
feature is the existence of strong proton quantum-shell effects with a
predominance of nuclear clusters with $Z=40$ and $Z=50$. The same magic 
numbers have been recently found by Onsi et al.~\cite{onsi-08} using 
a high-speed approximation to the Hartree--Fock method with an 
effective Skyrme force that was adjusted on essentially all nuclear data. 
Note however that the predicted sequence of magic numbers differs from 
that obtained by Negele and Vautherin as can be seen in 
Figure~\ref{fig.sect.groundstate.inner.QM.charge}. 
Neutron quantum effects are also important (while not obvious from the table)
as can be inferred from the spatial density fluctuations inside the clusters
in Figure~\ref{fig.sect.groundstate.inner.quantum.NV_cells}. This
figure also shows that these quantum effects disappear at high densities,
where the matter becomes nearly homogeneous. The quantum shell
structure of nuclear clusters in neutron star crusts is very different
from that of ordinary nuclei owing to a large number of neutrons
(for a recent review on the shell structure of very neutron-rich
nuclei, see, for instance, \cite{dobaczewski-07}). For instance,
clusters with $Z=40$ are strongly favored in neutron star crusts, while
$Z=40$ is not a magic number in ordinary nuclei (however, it
corresponds to a filled proton subshell).

\begin{table}[htbp]
  \caption[Sequence of nuclear clusters in the ground state of the
  inner crust calculated by Negele \& Vautherin.]{Sequence of nuclear
  clusters in the ground state of the inner crust calculated by Negele
  \& Vautherin~\cite{nv-73}. Here $N$ is the \emph{total} number of
  neutrons in the Wigner--Seitz sphere (i.e., it is a sum of the
  number of neutrons bound in nuclei and of those forming a neutron
  gas, per nucleus). Isotopes are labelled with the \emph{total}
  number of nucleons in the Wigner--Seitz sphere.}
  \label{table.sect.groundstate.inner.quantum.struct}
  \vskip 4mm

  \centering
    \begin{tabular}{c c c c c}
      \toprule
      $\rho$ [g~cm$^{-3}$] & Element & $Z$ & $N$ & $R_\mathrm{cell}$ [fm]\\
      \midrule
      $4.67 \times 10^{11}$ & $^{180}$Zr & 40 & 140 & 53.60 \\
      $6.69 \times 10^{11}$ & $^{200}$Zr & 40 & 160 & 49.24 \\
      $1.00 \times 10^{12}$ & $^{250}$Zr & 40 & 210 & 46.33 \\
      $1.47 \times 10^{12}$ & $^{320}$Zr & 40 & 280 & 44.30 \\
      $2.66 \times 10^{12}$ & $^{500}$Zr & 40 & 460 & 42.16 \\
      $6.24 \times 10^{12}$ & $^{950}$Sn & 50 & 900 & 39.32 \\
      $9.65 \times 10^{12}$ & $^{1100}$Sn & 50 & 1050 & 35.70 \\
      $1.49 \times 10^{13}$ & $^{1350}$Sn & 50 & 1300 & 33.07 \\
      $3.41 \times 10^{13}$ & $^{1800}$Sn & 50 & 1750 & 27.61 \\
      $7.94 \times 10^{13}$ & $^{1500}$Zr & 40 & 1460 & 19.61 \\
      $1.32 \times 10^{14}$ & $^{982}$Ge & 32 & 950 & 14.38 \\
      \bottomrule
    \end{tabular}
\end{table}

\begin{table}[htbp]
  \caption[Sequence of nuclear clusters in the ground state of the
  inner crust calculated by Baldo et al.]{Sequence of nuclear clusters
  in the ground state of the inner crust calculated by
  Baldo et al.~\cite{baldo-07, baldo-07b} including pairing
  correlations (their P2 model). The boundary conditions are the same
  as those of Negele and Vautherin~\cite{nv-73}. Similarly, $N$ is the
  \emph{total} number of neutrons in the Wigner--Seitz sphere. The
  isotopes are labelled with the \emph{total} number of nucleons in
  the  Wigner--Seitz sphere, as in
  Table~\ref{table.sect.groundstate.inner.quantum.struct}.}
  \label{table.sect.groundstate.inner.quantum.Baldo_struct}
  \vskip 4mm

  \centering
    \begin{tabular}{c c c c c}
      \toprule
      $\rho$ [g~cm$^{-3}$] & Element & $Z$ & $N$ & $R_\mathrm{cell}$ [fm]\\
      \midrule
      $4.52 \times 10^{11}$ & $^{212}$Te  & 52 & 160 & 57.19 \\
      $1.53 \times 10^{12}$ & $^{562}$Xe  & 54 & 508 & 52.79 \\
      $3.62 \times 10^{12}$ & $^{830}$Sn  & 50 & 780 & 45.09 \\
      $7.06 \times 10^{12}$ & $^{1020}$Pd & 46 & 974 & 38.64 \\
      $1.22 \times 10^{13}$ & $^{1529}$Ba & 56 & 1473 & 36.85 \\
      $1.94 \times 10^{13}$ & $^{1351}$Pd & 46 & 1305 & 30.31 \\
      $2.89 \times 10^{13}$ & $^{1269}$Zr & 40 & 1229 & 25.97 \\
      $4.12 \times 10^{13}$ & $^{636}$Cr  & 20 & 616 & 18.34 \\
      $5.65 \times 10^{13}$ & $^{642}$Ca  & 20 & 622 & 16.56 \\
      $7.52 \times 10^{13}$ & $^{642}$Ca  & 20 & 622 & 15.05 \\
      $9.76 \times 10^{13}$ & $^{633}$Ca  & 20 & 613 & 13.73 \\
      \bottomrule
    \end{tabular}
\end{table}

\epubtkImage{NV_cells.png}{%
\begin{figure}[htbp]
  \centerline{\includegraphics[scale=0.6]{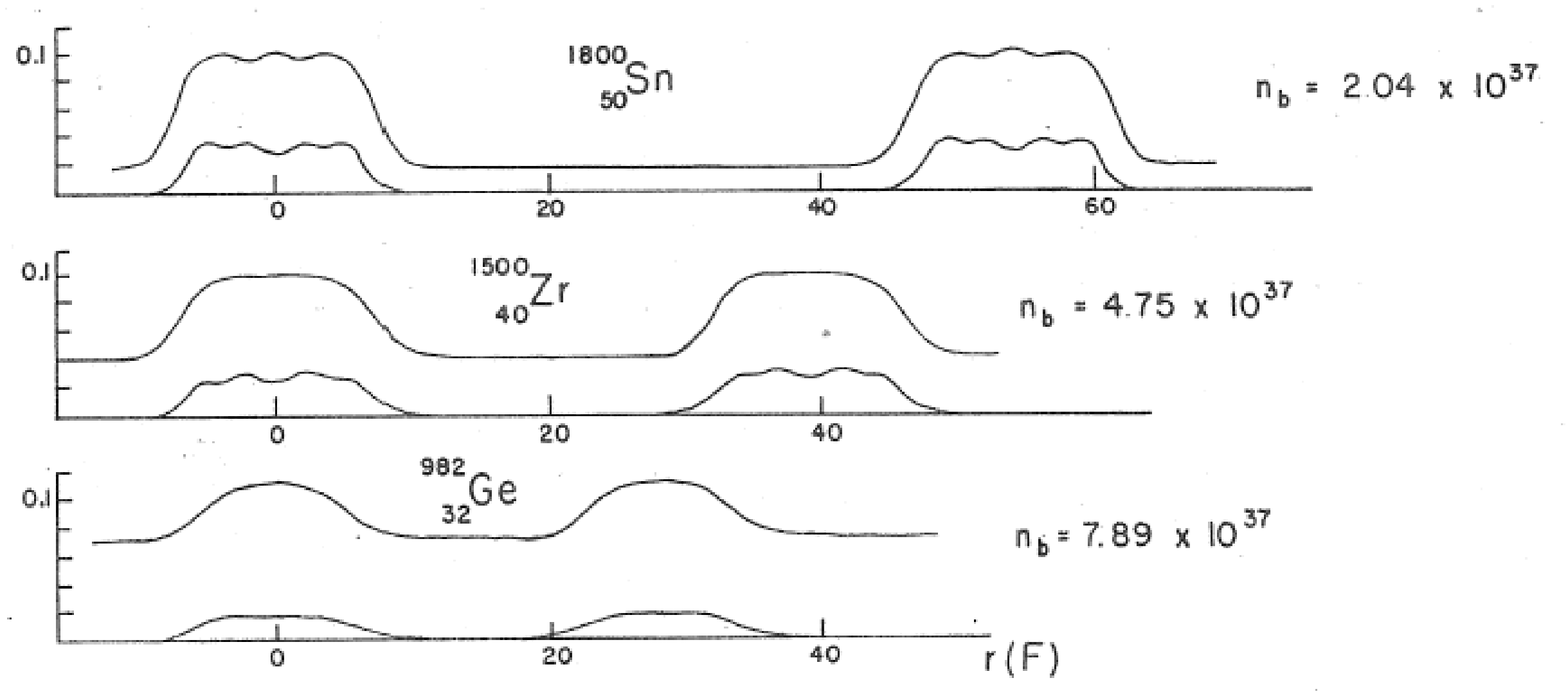}}
  \caption{Nucleon number densities (in fm$^{-3}$) along the axis
  joining two adjacent Wigner--Seitz cells of the ground state of the
  inner crust,  for a few  baryon densities $n_{\mathrm{b}}$
  (in cm$^{-3}$), as calculated by Negele \& Vautherin~\cite{nv-73}.}
  \label{fig.sect.groundstate.inner.quantum.NV_cells}
\end{figure}}

Negele \& Vautherin assume that nucleons can be described as
independent particles in a mean field induced by all other
particles. However, neutrons and protons are expected to form bound
pairs due to the long-range attractive part of the nucleon-nucleon
interaction, giving rise to the property of superfluidity (Section~\ref{sect.super}). 
Baldo and collaborators~\cite{baldo-05b, baldo-07b, baldo-05, baldo-06, baldo-07}
have recently studied the effects of these pairing correlations on the
structure of neutron star crusts, applying the generalized energy-density--functional theory in the Wigner--Seitz approximation. They
found that the composition of the clusters differs significantly from
that obtained by Negele \& Vautherin~\cite{nv-73}, as can be seen
from  Table~\ref{table.sect.groundstate.inner.quantum.Baldo_struct}
and Figure~\ref{fig.sect.groundstate.inner.QM.charge}.
However, Baldo et al. stressed that these results depend on the more-or-less
arbitrary choice of boundary conditions imposed on the Wigner--Seitz
sphere, especially in the deepest  layers of the inner crust.
Therefore, above $2\times 10^{13}~\mdens$
results of Baldo et al. (and those of Negele \& Vautherin)
 should be taken with a grain of salt. Another limitation of the Wigner--Seitz  approximation is that it does not allow the calculation of transport properties,
since neutrons are artificially confined inside the sphere. 
A more realistic treatment has been recently proposed by applying the band theory of 
solids~(see \cite{chamel-08} and references therein).

\subsubsection{Going further: nuclear band theory}
\label{sect.groundstate.inner.beyond}

The unbound neutrons in the inner crust of a neutron star are closely
analogous to the ``free'' electrons in an ordinary (i.e.\ under
terrestrial conditions) metal\epubtkFootnote{``free'' means here that
  the electrons are not bound. However, they are interacting with
  other electrons and with the atomic lattice.}. Assuming that the
ground state of cold dense matter below saturation density possesses
the symmetry of a perfect crystal, which is usually taken for granted,
it is therefore natural to apply the band theory of solids to neutron
star crusts (see Carter, Chamel \& Haensel~\cite{cchII-05} for the application 
to the pasta phases and Chamel~\cite{chamel-05, chamel-06} for the application to the 
general case of 3D crystal structures).

The band theory is explained in standard solid-state physics
textbooks, for instance in the book by Kittel~\cite{kittel-96}. Single
particle wave functions of nucleon species $q=n,p$ in the crust are
characterized by a wave vector $\pmb{k}$ and obey the Floquet--Bloch
theorem
\begin{equation}
\label{eq.sect.groundstate.inner.quantum.beyond.Bloch}
\varphi_{\pmb{k}}^{(q)}(\pmb{r}+\pmb{T}) = e^{ \mathrm{i} \pmb{k} \cdot \pmb{T} } \varphi_{\pmb{k}}^{(q)}(\pmb{r}) \, ,
\end{equation}
where $\pmb{T}$ is any lattice translation vector (which transforms
the lattice into itself). This theorem implies that the wave functions
are modulated plane waves, called simply Bloch waves
\begin{equation}
\label{eq.sect.groundstate.inner.quantum.beyond.Bloch.waves}
\varphi_{\pmb{k}}^{(q)}(\pmb{r}) = e^{ \mathrm{i} \pmb{k} \cdot \pmb{r} } u_{\pmb{k}}^{(q)}(\pmb{r}) \, ,
\end{equation}
with $u_{\pmb{k}}^{(q)}(\pmb{r})$ having  the full periodicity of the
lattice,
$u_{\pmb{k}}^{(q)}(\pmb{r}+\pmb{T})=u_{\pmb{k}}^{(q)}(\pmb{r})$.

In the approach of Negele \& Vautherin~\cite{nv-72} (see
Section~\ref{sect.groundstate.inner.quantum}), or in the more popular 
mean field method with effective Skyrme nucleon-nucleon interactions~\cite{bender-03, stone-07}, 
single particle states are solutions of the 
equations
\begin{equation}
\label{eq.sect.groundstate.inner.beyond.HF}
 -\nabla \cdot \frac{\hbar^2}{2 m_q^\oplus(\pmb{r})} \nabla \varphi_{\pmb{k}}^{(q)}(\pmb{r}) + U_q(\pmb{r}) \varphi_{\pmb{k}}^{(q)}(\pmb{r}) - \mathrm{i} \pmb{W_q}(\pmb{r}) \cdot \nabla\times\pmb{\sigma}\varphi_{\pmb{k}}^{(q)}(\pmb{r})  = \epsilon^{(q)}({\pmb{k}})  \varphi_{\pmb{k}}^{(q)}(\pmb{r}) \, ,
\end{equation}
neglecting pairing correlations (the application of band theory including pairing
correlations has been discussed in~\cite{cch-05}).
Despite their apparent simplicity, these equations are highly
nonlinear, since the various quantities depend on the wave functions (see
Section~\ref{sect.groundstate.inner.quantum}). 

As a result of the lattice symmetry, the crystal can be partitioned into identical
primitive cells, each containing exactly one lattice site. The
specification of the primitive cell is not unique. A particularly
useful choice is the Wigner--Seitz or Voronoi cell, defined by the set
of points that are closer to a given lattice site than to any
other. This cell is very convenient since it reflects the local
symmetry of the crystal. The Wigner--Seitz cell of a crystal lattice is
a complicated polyhedron in general. For instance, the Wigner--Seitz
cell of a body-centered cubic lattice (which is the expected ground state structure
of neutron star crusts), shown in Figure~\ref{fig.sect.groundstate.inner.beyond.bcc.WS}, 
is a truncated octahedron.

\epubtkImage{bcc_ws.png}{%
\begin{figure}[htbp]
  \centerline{\includegraphics[scale=0.8]{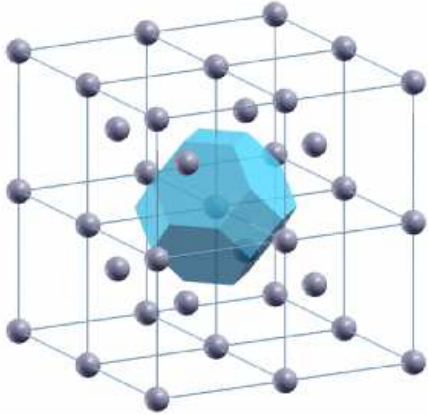}}
  \caption{Wigner--Seitz cell of a body-centered cubic lattice.}
  \label{fig.sect.groundstate.inner.beyond.bcc.WS}
\end{figure}}

Equations~(\ref{eq.sect.groundstate.inner.beyond.HF}) need to be
solved inside only one such cell. Indeed once the wave function in one
cell is known, the wave function in any other cell can be deduced from
the Floquet--Bloch
theorem~(\ref{eq.sect.groundstate.inner.quantum.beyond.Bloch}). This
theorem also determines the boundary conditions to be imposed at the
cell boundary.

For each wave vector $\pmb{k}$, there exists only a
discrete set of single particle energies
$\epsilon^{(q)}_\alpha(\pmb{k})$, labeled by the principal quantum
number $\alpha$, for which the boundary conditions~(\ref{eq.sect.groundstate.inner.quantum.beyond.Bloch})
 are fulfilled. The
energy spectrum is thus formed of ``bands'', each of them being a
\emph{continuous} (but in general not analytic) function of the wave vector
$\pmb{k}$ (bands are
  labelled by increasing values of energy, so that
  $\epsilon^{(q)}_\alpha(\pmb{k})\leq\epsilon^{(q)}_\beta(\pmb{k})$ if
  $\alpha<\beta$). The band index $\alpha$ is associated with the
rotational symmetry of the nuclear clusters around each lattice site, 
while the wave vector $\pmb{k}$ accounts for the translational
symmetry. Both local and global symmetries are therefore properly taken
into account. Let us remark that the band theory  includes uniform
matter as a limiting case of an ``empty'' crystal.

In principle, Equations~(\ref{eq.sect.groundstate.inner.beyond.HF}) have to be
solved for \emph{all} wave vectors $\pmb{k}$. Nevertheless, it can be
shown by symmetry that the single particle states (and, therefore, the
single particle energies) are periodic in $\pmb{k}$-space
\begin{equation}
\label{eq.sect.groundstate.inner.beyond.k-period}
\varphi^{(q)}_{\pmb{k}+\pmb{G}}(\pmb{r})=\varphi^{(q)}_{\pmb{k}}(\pmb{r}) \, ,
\end{equation}
where the reciprocal lattice vectors $\pmb{G}$ are defined by
\begin{equation}
\pmb{G}\cdot\pmb{T}=2\pi N\, ,
\end{equation}
$N$ being any positive or negative integer. The discrete set of all possible
reciprocal vectors $\pmb{G}$ defines a reciprocal lattice in
$\pmb{k}$-space.
Equation~(\ref{eq.sect.groundstate.inner.beyond.k-period}) entails
that only the wave vectors $\pmb{k}$ lying inside the first
Brillouin zone (i.e.\ Wigner--Seitz cell of the reciprocal lattice) are
relevant. The first Brillouin zone of a body-centered cubic lattice is
shown in Figure~\ref{fig.sect.groundstate.inner.beyond.bcc.BZ}.

\epubtkImage{bz_bcc.png}{%
\begin{figure}[htbp]
  \centerline{\includegraphics[scale=0.2]{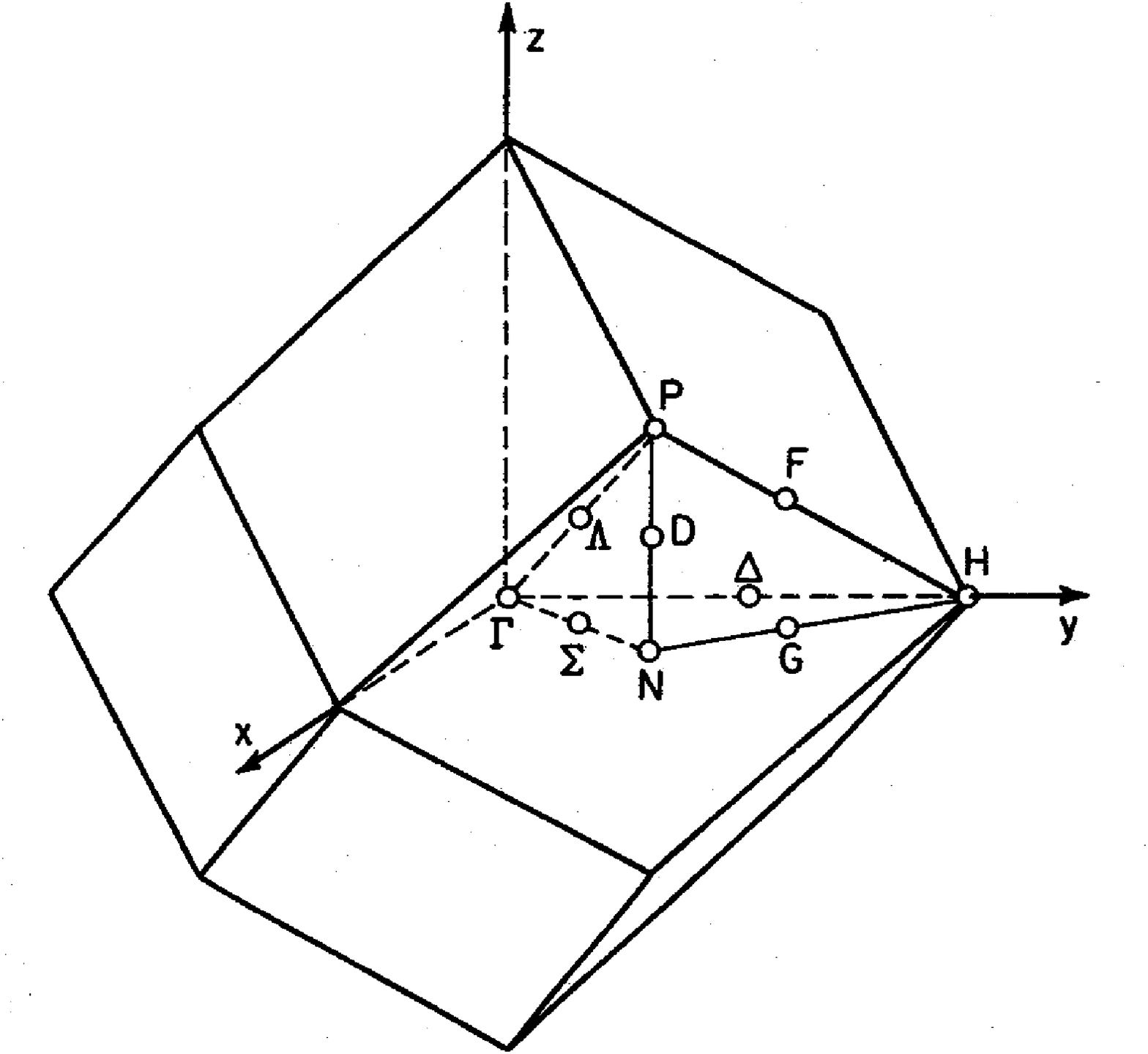}}
  \caption{First Brillouin zone of the body-centered cubic lattice
  (whose Wigner--Seitz is shown in
  Figure~\ref{fig.sect.groundstate.inner.beyond.bcc.WS}). The
  directions $x$, $y$ and $z$ denote the Cartesian axis in
  $\pmb{k}$-space.}
  \label{fig.sect.groundstate.inner.beyond.bcc.BZ}
\end{figure}}

An example of neutron band structure is shown in the right panel of 
Figure~\ref{fig.sect.groundstate.inner.beyond.bands_spec} from 
Chamel et al.\cite{chamel-07}.
The figure also shows the energy spectrum obtained by removing the nuclear clusters 
(empty lattice), considering a uniform gas of unbound neutrons. For comparison, 
the single particle energies, given in this limiting case by an expression of the form
$\epsilon(\pmb{k}) = \hbar^2 k^2/(2 m_n^\oplus)+U_n$, 
have been folded into the first Brillouin zone (reduced zone scheme). It can, thus, be seen that 
the presence of the nuclear clusters leads to distortions of the parabolic energy spectrum, especially at 
wave vectors $\pmb{k}$ lying on Bragg planes (i.e., Brillouin zone faces, see 
Figure~\ref{fig.sect.groundstate.inner.beyond.bcc.BZ}). 

The (nonlinear) three-dimensional partial differential
Equations~(\ref{eq.sect.groundstate.inner.beyond.HF}) are numerically
very difficult to solve (see Chamel~\cite{chamel-05, chamel-06} for a
review of some numerical methods that are applicable to neutron star
crusts). Since the work of Negele \& Vautherin~\cite{nv-73}, the usual
approach has been to apply the Wigner--Seitz
approximation~\cite{ws-33}. The complicated Wigner--Seitz cell 
(shown in Figure~\ref{fig.sect.groundstate.inner.beyond.bcc.WS}) 
is replaced by a sphere of equal volume. It is also assumed that the clusters are spherical so that
Equations~(\ref{eq.sect.groundstate.inner.beyond.HF}) reduce to 
ordinary differential Equations~(\ref{eq.sect.groundstate.inner.quantum.HF}). 
The Wigner-Seitz approximation has been used to predict the structure of the crust, the
pairing properties, the thermal effects, and the low-lying energy-excitation spectrum of the clusters ~\cite{nv-73, bonche-82,
  sandulescu-04b, khan-05, vigezzi-05, baldo-07, monrozeau-07}.

However, the Wigner--Seitz approximation overestimates the importance of neutron
shell effects, as can be clearly seen in Figure~\ref{fig.sect.groundstate.inner.beyond.bands_spec}. 
The energy spectrum is discrete in the Wigner--Seitz 
approximation (due to the neglect of the $\pmb{k}$-dependence of the states), 
while it is continuous in the full band theory. 
The spurious shell effects depend on a particular choice of boundary
conditions, which are not unique. Indeed as pointed out by Bonche \&
Vautherin~\cite{bonche-81}, two types of boundary conditions are
physically plausible yielding a more-or-less constant neutron density
outside the cluster: either the wave function or its radial derivative
vanishes at the cell edge, depending on its parity. Less physical 
boundary conditions have also been applied, like the vanishing of the 
wave functions. Whichever boundary conditions
are adopted, they lead to unphysical spatial fluctuations of the 
neutron density, as discussed in detail by Chamel et al.~\cite{chamel-07}. Negele \&
Vautherin~\cite{nv-73} average the neutron density in the vicinity of
the cell edge in order to remove these fluctuations, but it is not
clear whether this {\it ad hoc} procedure did remove all the spurious
contributions to the total energy. As shown in Figure~\ref{fig.sect.groundstate.inner.beyond.bands_spec}, 
shell energy gaps are on the order of $\Delta \epsilon\sim 100$~keV, at 
$\rho\simeq 7\times 10^{11}~\mdens$. Since these gaps scale
approximately like $\Delta\epsilon\propto\hbar^2/(2 m_n R_{\mathrm{cell}}^2)$ 
(where $m_n$ is the neutron mass), they increase with density $\rho$ 
and eventually become comparable to the 
total energy difference between neighboring configurations. As a consequence, 
the predicted equilibrium structure of the crust becomes very sensitive to 
the choice of boundary conditions in the bottom layers~\cite{baldo-06, chamel-07}. 
One way of eliminating the boundary condition problem without carrying out full band structure
calculations, is to perform semi-classical calculations including only proton shell effects 
with the Strutinsky method, as discussed by Onsi et al.\cite{onsi-08}. 

\epubtkImage{WS_spec-bands_spec.png}{%
\begin{figure}[htbp]
  \centerline{\includegraphics[scale=0.3]{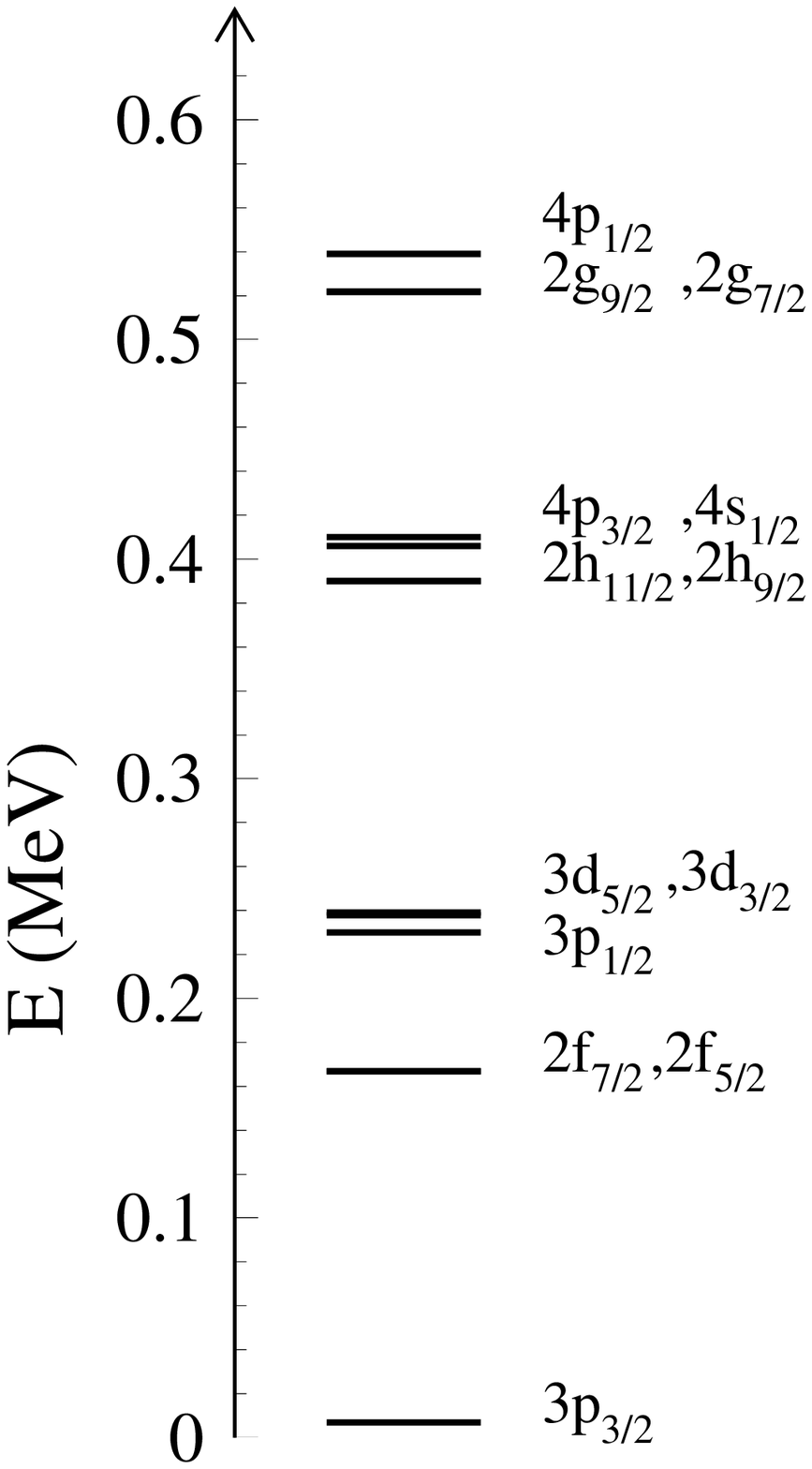}\includegraphics[scale=0.35]{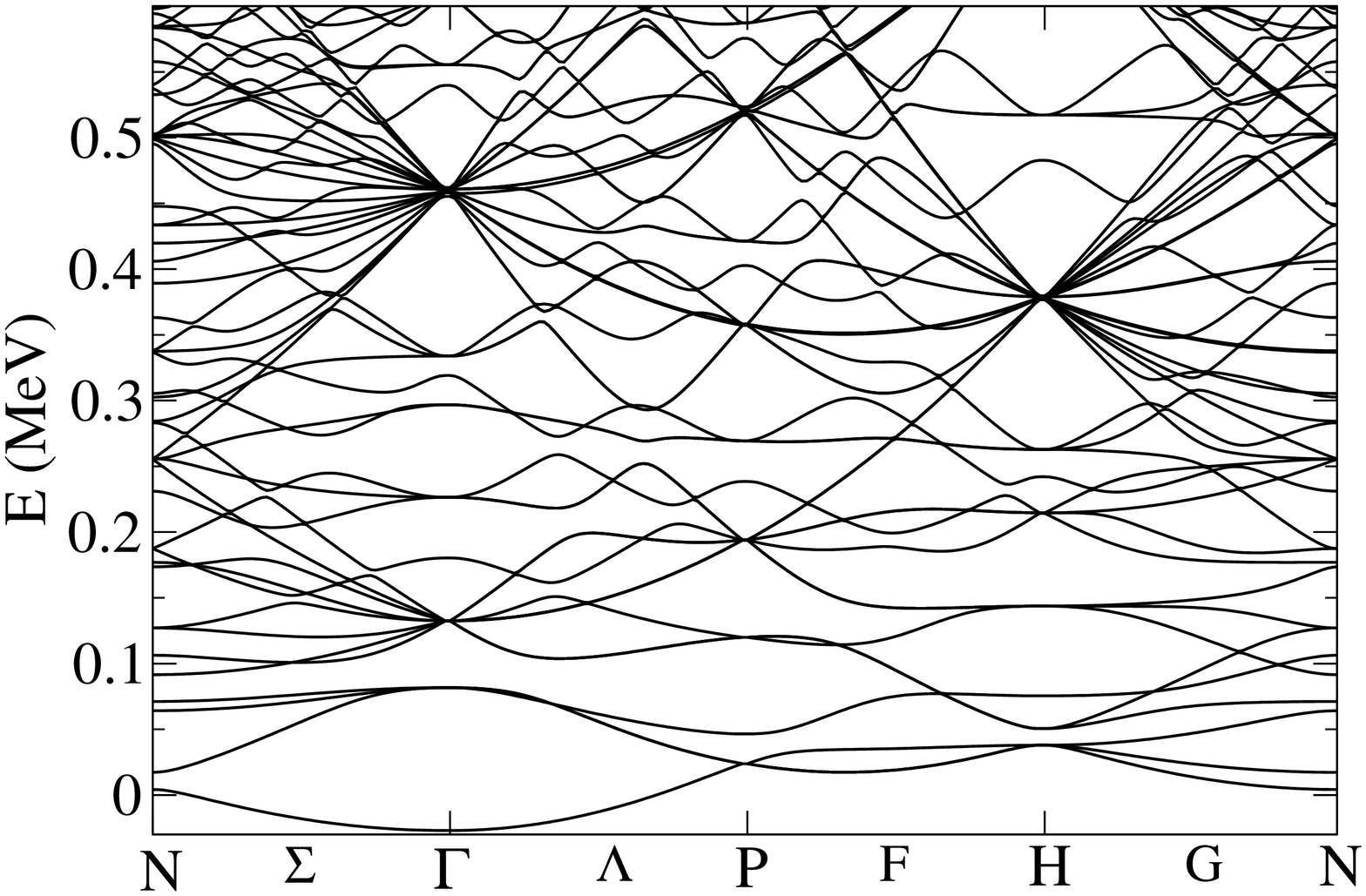}}
  \caption{Single particle energy spectrum of unbound neutrons in
  the ground state of the inner crust composed of
  $^{200}$Zr, at $\rho\simeq 7\times 10^{11}~\mdens$, obtained by Chamel et al.~\cite{chamel-07}. 
  Left: calculation in the Wigner--Seitz approximation. Right:
  full band structure calculation (reduced zone scheme) assuming that the crust is a body-centered
  cubic lattice with nuclear clusters (solid line) and without (dashed line). 
  The capital letters on the horizontal axis refer to lines or points 
  in \pmb{k}-space, as indicated in Figure~\ref{fig.sect.groundstate.inner.beyond.bcc.BZ}.}
  \label{fig.sect.groundstate.inner.beyond.bands_spec}
\end{figure}}

In recent calculations~\cite{magierski-02, newton-06, gogelein-07} the
Wigner--Seitz cell has been replaced by a cube with periodic boundary
conditions instead of Bloch boundary 
conditions~(\ref{eq.sect.groundstate.inner.quantum.beyond.Bloch}).
Although such calculations allow for possible deformations of
the nuclear clusters, the lattice periodicity is still not properly
taken into account, since such boundary conditions are associated with 
only one kind of solutions with $\pmb{k}=0$. Besides the Wigner--Seitz cell is only cubic for a simple
cubic lattice and it is very unlikely that the equilibrium structure
of the crust is of this type (the structure of the crust is expected
to be a body centered cubic lattice as discussed in
Section~\ref{sect.groundstate.outer}). Let us also remember that a simple cubic lattice is unstable. 
It is, therefore, not clear whether these calculations, which require much more computational time
than those carried out in the spherical approximation, are more
realistic. This point should be clarified in future work by a detailed
comparison with full band theory. 
Let us also mention that
recently B\"urvenich et al.~\cite{buervenich-07} have considered
axially-deformed spheroidal W-S cells to account for deformations of
the nuclear clusters.

Whereas the Wigner--Seitz approximation is reasonable at not too high
densities for determining the equilibrium crust structure, full
band theory is indispensable for studying transport properties (which
involve obviously  translational symmetry and, hence, the
$\pmb{k}$-dependence of the states). Carter, Chamel \& Haensel~\cite{cchII-05} using this novel approach have shown that the unbound neutrons move in the
crust as if they had an effective mass much larger than the bare mass
(see Sections~\ref{sect.super.dyn.entr}
and~\ref{sect.super.dyn.entr.ns}). This dynamic effective neutron mass has
been calculated by Carter, Chamel \& Haensel~\cite{cchII-05} in the
pasta phases of rod and slab-like clusters (discussed in
Section~\ref{sect.groundstate.pasta}) and by Chamel~\cite{chamel-05,
  chamel-06} in the general case of spherical clusters. By taking
consistently into account both nuclear clusters, which form a
solid lattice, and the neutron liquid, band theory provides a
unified scheme for studying the structure and properties of neutron star crusts.

\subsection{Pastas}
\label{sect.groundstate.pasta}

The equilibrium structure of nuclear clusters in neutron star
crusts results from the interplay between the \emph{total} Coulomb
energy and the surface energy of the nuclei according to the virial
Equation~(\ref{eq.sect.groundstate.inner.LDM.virial}). At low
densities, the lattice energy
Equation~(\ref{eq.sect.groundstate.inner.LDM.lat_energy}) is a small
contribution to the total Coulomb energy and nuclear clusters are
therefore spherical. However, at the bottom of the crust the nuclei
are very close to one  another. Consequently, the lattice energy
represents a large reduction of the total Coulomb energy, which
vanishes in the liquid core when the nuclear clusters fill all space,
as can be seen from
Equation~(\ref{eq.sect.groundstate.inner.LDM.tot_Coul_energy2}). In
the densest layers of the crust the Coulomb energy is comparable in
magnitude to the net nuclear binding energy (this situation also occurs
in the dense and hot collapsing core of a supernova and in heavy ion
collisions leading to multifragmentation~\cite{botvina-05,
  botvina-06}). The matter thus becomes frustrated and can arrange
itself into various exotic configurations as observed in complex
fluids. For instance, surfactants are organic compounds composed of a
hydrophobic ``tail'' and a hydrophilic ``head''. In solutions,
surfactants aggregate into ordered structures, such as spherical or
tubular micelles or lamellar sheets. The transition between the
different phases is governed by the ratio between the volume of the
hydrophobic and hydrophilic parts as shown in
Figure~\ref{fig.sect.groundstate.pasta.surfactants}. One may expect by
analogy that similar structures could occur in the inner crust of a
neutron star depending on the nuclear packing.

\epubtkImage{surfactant.png}{%
  \begin{figure}[htbp]
    \centerline{\includegraphics[scale=0.8]{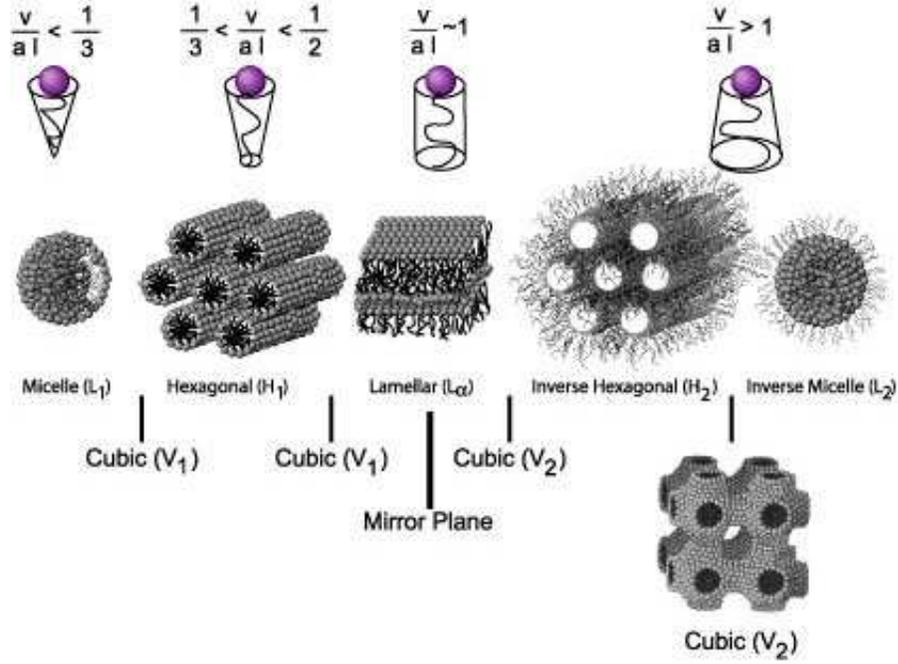}}
    \caption{Structures formed by self-assembled surfactants in
    aqueous solutions, depending on the volume ratio of the
    hydrophilic and hydrophobic parts. Adapted
    from~\cite{jonsson-98}.}
    \label{fig.sect.groundstate.pasta.surfactants}
\end{figure}}

According to the Bohr--Wheeler fission condition~\cite{bohr-39} an
isolated spherical nucleus in vacuum is stable with respect to
quadrupolar deformations if
\begin{equation}
\label{fig.sect.groundstate.pasta.bohrwheeler}
E^{(0)}_{\mathrm{N, Coul}} < 2 E^{(0)}_{\mathrm{N, surf}} \, ,
\end{equation}
where $E^{(0)}_{\mathrm{N, Coul}}$ and $E^{(0)}_{\mathrm{N, surf}}$ are the
Coulomb and surface energies of the nucleus, respectively. The
superscript $(0)$ reminds us that we are considering a nucleus in vacuum. The
Bohr--Wheeler condition can be reformulated in order to be applied in
the inner crust, where both Coulomb and surface energies are modified
compared to the ``in vacuum'' values. Neglecting curvature corrections
and expanding all quantities to the linear order in $w^{1/3}$, where
$w$ is the fraction of volume occupied by the clusters, it is found
\cite{pethick-95} that spherical clusters become unstable
to quadrupolar deformation if $w>w_{\mathrm{crit}}=1/8$ \,.

Reasoning by analogy with percolating networks, Ogasawara \&
Sato~\cite{ogasawara-82} suggest that as the nuclei fill more and
more space, they will eventually deform, touch and merge to form new
structures. A long time ago, Baym, Bethe and Pethick~\cite{bbp-71}
predicted that as the volume fraction exceeds 1/2, the crust will be
formed of neutron bubbles in nuclear matter. In the general framework
of the compressible liquid drop model considering the simplest
geometries, Hashimoto and his
collaborators~\cite{hashimoto-84, oyamatsu-84} show that as the
nuclear volume fraction $w$ increases, the stable nuclear shape
changes from sphere to cylinder, slab, tube and bubble, as illustrated
in Figure~\ref{fig.sect.groundstate.pasta.sequence}. This sequence of
nuclear shapes referred  to as ``pastas'' (the cylinder and slab shaped
nuclei resembling ``spaghetti'' and ``lasagna'' respectively)  was
found independently by Ravenhall et al.~\cite{ravenhall-83} with a
specific liquid drop model. The volume fractions at which the various
phases occur are in good agreement with those predicted by Hashimoto
and collaborators on purely geometrical grounds. This criterion,
however, relies on a liquid drop model, for which curvature corrections
to the surface energy are neglected. This explains why some
authors~\cite{lorenz-93, douchin-00b} find within the liquid drop
model that spherical nuclei remain stable down to the transition to
uniform nuclear matter, despite volume fractions exceeding the critical
threshold (see in particular Figure~\ref{fig.sect.groundstate.inner.LDM.DH_inner_crust1}), 
while other groups found the predicted sequence of pasta
phases~\cite{watanabe-01, watanabe_err-03, iida-01, iida_err-03}. The
nuclear curvature energy is, thus, important for predicting the
equilibrium shape of the nuclei at a given density~\cite{pethick-83}.

\epubtkImage{pastas.png}{%
\begin{figure}[htbp]
  \centerline{\includegraphics[scale=0.8]{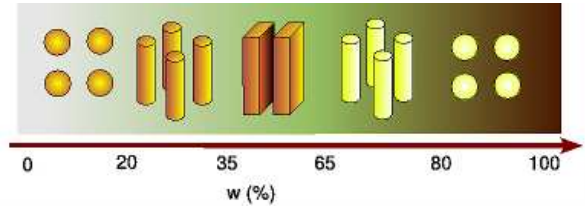}}
  \caption{Sketch of the sequence of pasta phases in the bottom
  layers of  ground-state crusts with an increasing
  nuclear volume fraction, based on the study of Oyamatsu and
  collaborators~\cite{oyamatsu-84}.}
  \label{fig.sect.groundstate.pasta.sequence}
\end{figure}}

These pasta phases have been studied by various nuclear models, from
liquid drop models to semiclassical models, quantum molecular dynamic
simulations and Hartree--Fock calculations (for the current status of
this issue, see, for instance, \cite{watanabe-06}). These models differ
in numerical values of the densities at which the various phases
occur, but they all predict the same sequence of configurations shown in
Figure~\ref{fig.sect.groundstate.pasta.sequence} (see also the discussion 
in Section~5.4 of~\cite{pethick-95}). 
Some models~\cite{lorenz-93, cheng-97, douchin-00b, maruyama-05}
predict that the spherical clusters remain energetically favored throughout
the whole inner crust. 
Generalizing the Bohr--Wheeler
condition to nonspherical nuclei, Iida et al.~\cite{iida-01,
  iida_err-03} showed that the rod-like and slab-like clusters are
stable against fission and proton clustering, suggesting that the
crust layers containing pasta phases may be larger than that predicted
by the equilibrium conditions. It has also been suggested that the
pinning of neutron superfluid vortices in neutron star crusts might
trigger the formation of rod-like
clusters~\cite{mochizuki-99}. Nevertheless, the nuclear pastas may be
destroyed by thermal fluctuations~\cite{watanabe-01,
  watanabe_err-03}. Quite remarkably, Watanabe and
collaborators~\cite{watanabe-06} performed quantum molecular dynamic
simulations and observed the formation of rod-like and slab-like
nuclei by cooling down hot uniform nuclear matter without any
assumption of the nuclear shape. They also found the appearance of
intermediate sponge-like structures, which might be identified with
the ordered, bicontinuous, double-diamond geometry observed in block
copolymers~\cite{matsuzaki-06}. Those various phase transitions
leading to the pasta structures in neutron star crusts are also
relevant at higher densities in neutron star cores, where kaonic or
quark pastas could exist~\cite{maruyama-06}.

The pasta phases cover a small range of densities near the crust-core
interface with $\rho \sim 10^{14}~\mdens$. Nevertheless, by filling
the densest layers of the crust, they may represent a sizable fraction
of the crustal mass~\cite{lorenz-93} and thus may have important
astrophysical consequences. For instance, the existence of nuclear
pastas in hot dense matter below saturation density affects the
neutrino opacity~\cite{horowitz-05, sonoda-07}, which is an important
ingredient for understanding the gravitational core collapse of
massive stars in supernova events and the formation of neutron stars
(see Section~\ref{sect.obs.supernova}). The dynamics of neutron
superfluid vortices, which is thought to underlie pulsar glitches (see
Section~\ref{sect.obs.glitches}), is likely to be affected by the
pasta phase. Besides, the presence of nonspherical clusters in the
bottom layers of the crust influences the subsequent cooling of the
star, hence the thermal X-ray emission by allowing direct Urca
processes~\cite{lorenz-93, gusakov-04} (see
Section~\ref{sect.neutrino}) and enhancing the heat
capacity~\cite{deblasio-95, deblasio-96, elgaroy-96}. The elastic
properties of the nuclear pastas can be calculated using the theory
of liquid crystals~\cite{pethick-98, watanabe-01, watanabe_err-03}
(see Section~\ref{sect.elast.pasta}). The pasta phase could thus
affect the elastic deformations of neutron stars, oscillations,
precession and crustquakes.

\subsection{Impurities and defects}
\label{sect.groundstate.imp}

There are many reasons why the real crust of neutron stars can be
imperfect. In particular, apart from a dominating $(A,Z)$ nuclide at a 
given density $\rho$, it can contain an admixture of different nuclei 
(``impurities''). The
initial temperature at birth exceeds $10^{10}$~K. At such a high $T$,
thermodynamic equilibrium is characterized by a  statistical
distribution of $A$ and $Z$. With decreasing $T$, the $A,Z$ peak
becomes narrower~\cite{bonche-82, burrows-84}. After crystallization
at $T_{\mathrm{m}}$, the composition is basically frozen. Therefore, the
composition at $T<T_{\mathrm{m}}$ reflects the situation at $T\sim T_{\mathrm{m}}$, which can differ from that in the absolute ground state at
$T=0$. For example,  between the neighboring shells, with nuclides
$(A_1,Z_1)$ and $(A_2,Z_2)$, respectively, one might expect a
transition layer composed of a binary mixture of the two
nuclides~\cite{deblasio-98, deblasio-00}. Another way of forming
impurities is via thermal fluctuations of $Z$ and $N_{\mathrm{cell}}$,
which, according to Jones~\cite{jones-99, jones-01},  might be quite significant
at $\rho\gtrsim 10^{12}~\mdens$ and $T\gtrsim 10^9$~K.

The real composition of neutron star crusts can also differ
 from the ground state due to the fallback of material
from the envelope ejected during the supernova explosion and due to
the accretion of matter. In particular, an accreted crust is a site of X-ray
bursts. The ashes of unstable thermonuclear burning at accretion rates
$10^{-8}\,M_\odot \mathrm{\ y}^{-1}\gtrsim \dot{M}\gtrsim
 10^{-9}\,M_\odot \mathrm{\ y}^{-1}$ could be a mixture of $A\simeq
 60\mbox{\,--\,}100$ nuclei and could therefore be relatively ``impure''
 (heterogeneous and possibly amorphous)~\cite{schatz-99,
 schatz-01}. If the initial ashes are a mixture of many nuclides,
 further compression under the weight of accreted matter can keep the
 heterogeneity. If the crust is weakly impure but rather amorphous,
 its thermal and electrical conductivities in the solid phase would be
 \emph{orders of magnitude lower} than in the perfect crystal as
 discussed in Section~\ref{sect.cond}. This would have dramatic
 consequences as far as the rate of the thermal relaxation of the
 crust is concerned (Section~\ref{sect.obs.LMXB.initial-cool-SXTs}).

\newpage


\section{Accreting Neutron Star Crusts}
\label{sect.accretion}

\subsection{Accreting neutron stars in low-mass X-ray binaries}
\label{sect.accretion.LMXB}

In this section we will concentrate on accreting neutron stars
in low-mass X-ray binaries (LMXB), where the mass of the companion is
significantly less than $M_\odot$, and the accretion stage can last as
long as $\sim 10^9$~y.  A binary is sufficiently tight for the
 companion to fill its Roche lobe. The mass transfer proceeds through the inner
Lagrangian point, and the transferred plasma flows in a deep
gravitational potential well, via an accretion disk, towards the
neutron star surface, as illustrated in
Figure~\ref{fig.sect.accretion.LMXB-Piro}.

\epubtkImage{LMXB-Piro.jpg}{%
\begin{figure}[htbp]
  \centerline{\includegraphics[scale=0.7]{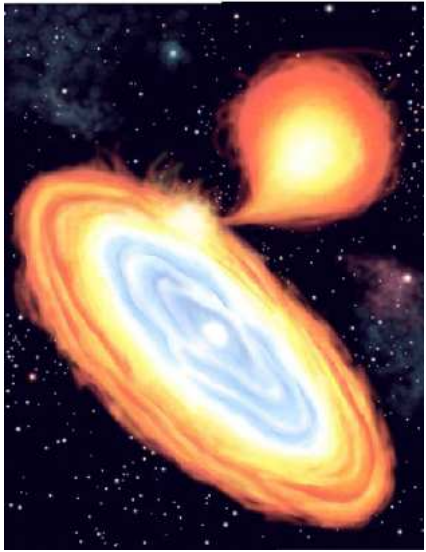}}
  \caption{The artist's view of a low-mass X-ray binary. The
    companion of a neutron star fills its Roche lobe and loses its
    mass via plasma flow through the inner Lagrangian point. Due to
    its angular momentum, plasma orbits around the neutron star, forming
    an accretion disk. Gradually losing angular  momentum due to
    viscosity within the accretion disk, plasma approaches the neutron
    star and eventually falls onto its surface. Figure by T.\
    Piro.}
  \label{fig.sect.accretion.LMXB-Piro}
\end{figure}}

A hydrogen atom falling on a neutron star surface from infinity releases $\sim$~200~MeV of
gravitational binding energy. Therefore, accretion onto
a neutron star releases  $\sim$~200~MeV per accreted nucleon. Most
of this energy is radiated in X-rays, so that the total X-ray
luminosity of an accreting neutron star can be estimated as
$L_{\mathrm{X}}\sim (\dot{M}/10^{-10}\,M_\odot/\mathrm{y})\;10^{36}
\mathrm{\ erg\ s}^{-1}$. Space X-ray observations of accreting neutron stars
were at the origin of X-ray astronomy~\cite{giacconi-62}. Accreted
matter is usually hydrogen rich. It forms the outer envelope of
a neutron star, which contains a hydrogen burning shell, with an energy
release of about 5~MeV/nucleon in a stable burning. Helium ashes from hydrogen burning accumulate in the helium layer, which ignites
under specific density-temperature conditions. For some range of
accretion rate, helium burning is unstable, so that its ignition
triggers a  thermonuclear flash, burning within seconds all the envelope
into nuclear ashes composed of nuclides of the iron group and beyond
it; the energy release in the flash is less than 5~MeV/nucleon. These
flashes are observed as  X-ray bursts, with luminosity rising in a
second to about $10^{38} \mathrm{\ erg s}^{-1}$ ($\approx$ Eddington limit
for neutron stars, $L_{\mathrm{Edd}}$), and then typically decaying in a few
tens of seconds\epubtkFootnote{We restrict ourselves to type~I X-ray
  bursts. There are two X-ray bursters that are of type~II, with
  bursts driven not by thermonuclear flashes on the neutron star
  surface, but originating in the accretion disk itself.}.

Multiplying the burst luminosity by its duration we get an estimate of the
total burst energy $\sim 10^{39}\mbox{\,--\,}10^{40}$~erg. The X-ray bursts  are
quasiperiodic, with typical recurrence time $\sim$ hours-days. Since their
discovery in  1975~\cite{grindlay-76}, about seventy X-ray bursters
have been  found. Many bursters are of transient character, and form a
group of soft X-ray transients (SXTs), with typical active periods of
days -- weeks, separated by periods of quiescence of several months --
years long. During quiescent periods, there is very little or no
accretion, while during much shorter periods of activity there is an
abundant accretion, due  probably to disc flow instability. Some SXTs,
with active periods of years  separated by decades of quiescence, are
called persistent SXTs. In 2000, a special rare type of X-ray {\it
  superbursts} was discovered. Superbursts  last for  a few  to twelve
hours, with recurrence  times of several years. The total energy
radiated in a superburst is $\sim 10^{42}$~erg. Superbursts are explained by
the unstable burning of carbon in deep layers of the outer crust.

In all cases, ignition of the thermonuclear flash takes place in the
neutron star crust, and is sensitive to the crust structure and to the
physical conditions within it. This aspect will be discussed in
Section~\ref{sect.accretion.conditions}. An accreted crust has
a different structure and composition than the ground state one, as
discussed in Section~\ref{sect.accretion.processes}. It has, therefore,
a different equation of state than the ground-state crust (see
Section~\ref{sect.eos.accreted}). Moreover, it is a reservoir of
nuclear energy, which is released in the process of deep crustal
heating, accompanying accretion, reviewed in
Section~\ref{sect.accretion.crust-heating}. Observations of SXTs in
quiescence prove the presence of deep crustal heating
(Section~\ref{sect.obs.LMXB.crust-heat-SXT}). Cooling of the neutron
star surface in quiescence after long periods of accretion (years --
decades) in persistent SXTs also allows one to test physical
properties of the accreted crust
(Section~\ref{sect.obs.LMXB.initial-cool-SXTs}).

\subsection{Nuclear processes and formation of accreted crusts}
\label{sect.accretion.processes}

In this section we will describe the fate of  X-ray burst ashes,
produced at $\lesssim  10^{7}~\mdens$, and then sinking deeper and
deeper under the weight of accreted plasma above them. We start at a few
tens of meters below the surface, and we will end at a depth of $\sim
1$~km, where the density $\gtrsim 10^{13}~\mdens$.  Under conditions
prevailing in accreting neutron star crusts, at $\rho>10^8~\mdens$
matter is strongly degenerate, and is ``relatively cold''  ($T
\lesssim 10^8~\mathrm{K}$, see Figures~\ref{fig.sect.accretion.conditions.MiraldaHP-1}
and~\ref{fig.sect.accretion.conditions.MiraldaHP-2}), so that
thermonuclear processes  are strongly
suppressed because interacting nuclei have to overcome a
large  Coulomb barrier. The structure of an accreted
crust is shown in Figure~\ref{fig.sect.accretion.crust-structure}.

\epubtkImage{accreted-crust.png}{%
\begin{figure}[htbp]
  \centerline{\includegraphics[scale=0.5]{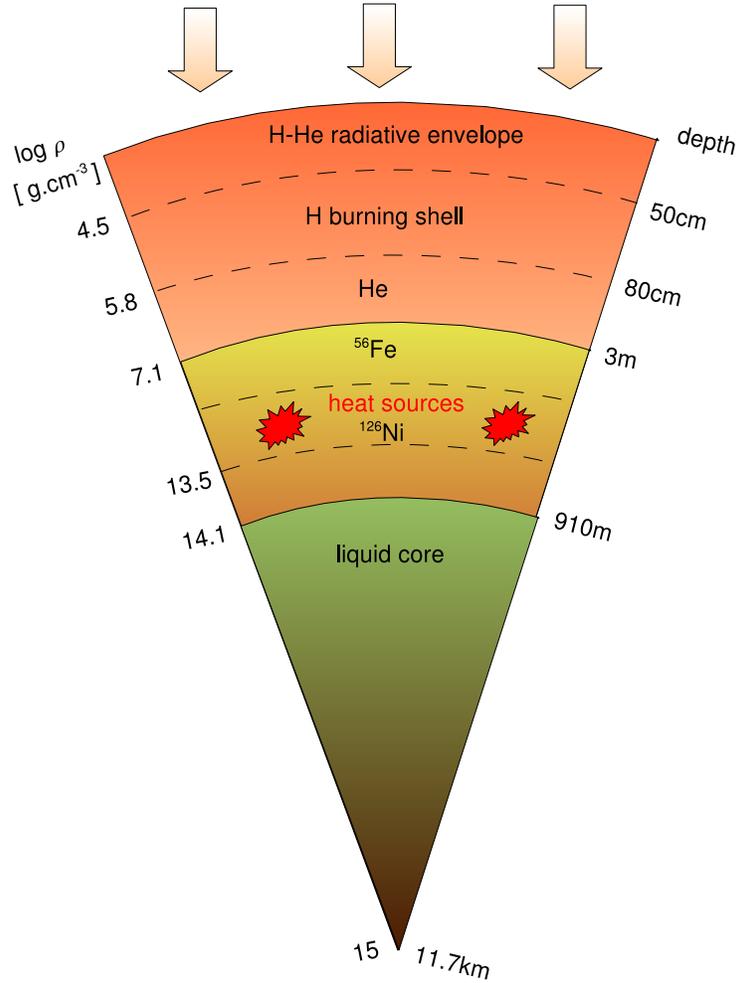}}
  \caption{Model of an accreting neutron star crust. The 
  total mass of the star is $M=1.4\,M_\odot$. Stable
  hydrogen burning takes place in the H-burning shell, and produces
  helium, which accumulates in the He-shell. Helium ignites at
  $\rho\sim10^7~\mdens$, leading to a helium flash and explosive
  burning of all matter above the bottom of the He-layer into
  $^{56}$Ni, which captures electrons to become $^{56}$Fe. After
  $\sim$~1~h, the cycle of accretion, burning of hydrogen and
  explosion triggered by a helium flash repeats again and the layer of
  iron from the previous burst is pushed down. Based on the
  unpublished results of calculations by P.\ Haensel and J.L.\
  Zdunik. Accreted crust model of~\cite{haensel-90a, haensel-90b}. The
  core model of~\cite{douchin-01}.}
  \label{fig.sect.accretion.crust-structure}
\end{figure}}

In what follows we will use a simple model of the accreted crust
formation, based on the one-component plasma approximation at
$T=0$~\cite{haensel-90b, haensel-03}. The (initial) X-burst ashes are
approximated by a one-component plasma with $(A_{\mathrm{i}}, Z_{\mathrm{i}})$
nuclei.

At densities lower than the threshold for {\it pycnonuclear
  fusion}  (which is very uncertain, see
  Yakovlev et al.~\cite{yakovlev-06}), $\rho_{\mathrm{pyc}}\sim
  10^{12}\mbox{\,--\,}10^{13} \mathrm{\ g\ cm}^{-3}$, the number of nuclei in an element
  of matter does not change during the compression resulting from the
  increasing weight of accreted matter. Due to nucleon pairing,
  stable nuclei in dense matter have even $N=A-Z$ and $Z$ (even-even
  nuclides). In the outer crust, in which free neutrons are absent,
  the electron captures proceed in two steps,
\begin{equation}
(A,Z)+e^-\longrightarrow  (A,Z-1)+\nu_e \, ,
\label{eq.sect.accretion.processes.ecap1}
\end{equation}
\begin{equation}
(A,Z-1)+e^-\longrightarrow  (A,Z-2)+\nu_e + Q_j \, .
\label{eq.sect.accretion.processes.ecap2}
\end{equation}
Electron captures lead to a systematic decrease in $Z$ (therefore an
increase in $N=A-Z$) with increasing density.  The first capture,
Equation~(\ref{eq.sect.accretion.processes.ecap1}), proceeds as soon
as $\mu_e > E\left\{A,Z-1\right\}-E\left\{A,Z\right\}$, in a
quasi-equilibrium manner, with negligible energy release. It
produces an odd-odd nucleus, which is strongly unstable in a dense
medium, and captures a second electron in a nonequilibrium manner,
Equation~(\ref{eq.sect.accretion.processes.ecap2}), with energy
release $Q_j$, where $j$ is the label of the nonequilibrium process.

\epubtkImage{ecaptures.png}{%
\begin{figure}[htbp]
  \centerline{\includegraphics[scale=0.6]{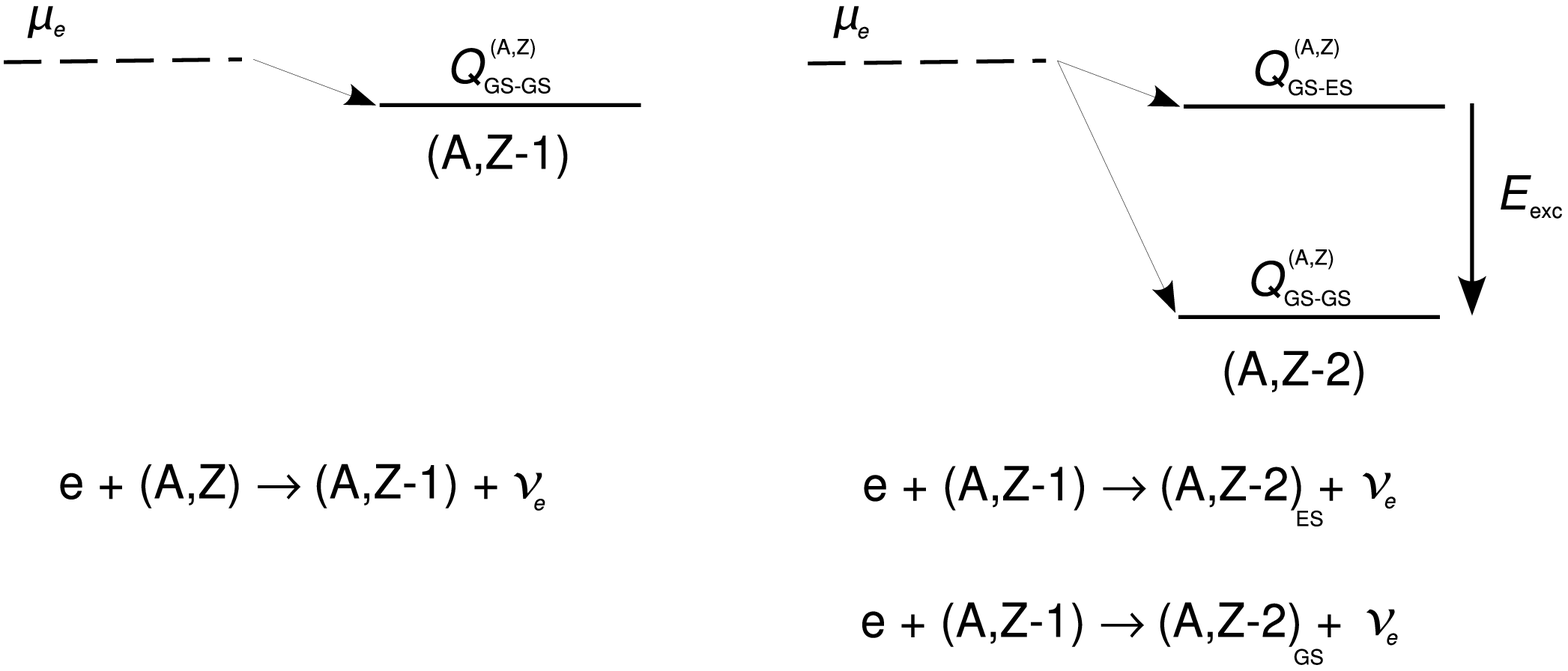}}
  \caption{Electron capture processes.}
  \label{fig.sect.accretion.ecaptures}
\end{figure}}

\epubtkImage{NZ_rho_pyc.png}{%
\begin{figure}[htbp]
  \centerline{\includegraphics[scale=0.5]{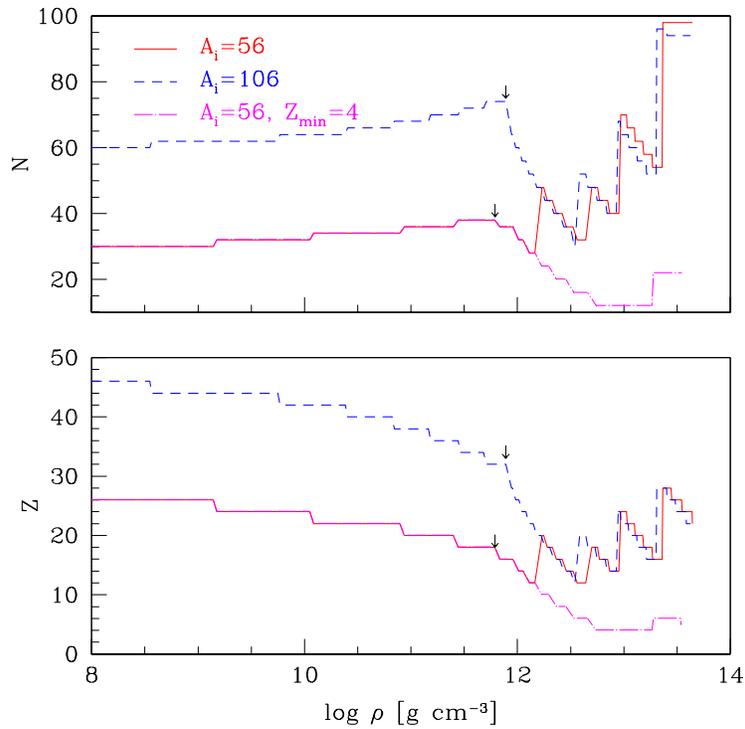}}
  \caption{$Z$ and $N$ of nuclei vs.\ matter density in an
    accreted crust, for different models of dense
    matter. Solid line: $A_{\mathrm{i}}=106$; dotted line: $A_{\mathrm{i}}=56$.
    Every  change of $N$ and $Z$, which takes place at a constant
    pressure, is accompanied by a jump in density: it is represented
    by small steep (but not perpendicular!) segments of the
    curves. These segments connect the top and the bottom
    density of a thin reaction shell. Arrows indicate positions of the
    neutron drip point. From Haensel \& Zdunik~\cite{haensel-03}.}
  \label{eq.sect.accretion.processes.NS-rho-pyc}
\end{figure}}

After the neutron-drip point ($\rho>\rho_{\mathrm{ND}}\simeq6\times10^{11}~\mdens$), electron captures trigger
neutron emissions,
\begin{equation}
(A,Z)+e^-\longrightarrow (A,Z-1)+\nu_e \, ,
\label{eq.sect.accretion.processes.ecapn1}
\end{equation}
\begin{equation}
(A,Z-1)+e^-\longrightarrow (A-\mathrm{k},Z-2)+\mathrm{k}\; n + \nu_e
+ Q_j \, .
\label{eq.sect.accretion.processes.ecapn2}
\end{equation}
Due to the electron captures, the value of $Z$ decreases with
increasing density. In consequence, the Coulomb barrier prohibiting
the nucleus-nucleus reaction lowers. This effect, combined with the
decrease of the mean distance between the neighboring nuclei, and a
simultaneous increase of energy of the quantum zero-point vibrations
around the nuclear lattice sites, opens up a possibility for the
\emph{pycnonuclear  fusion} associated with quantum-mechanical
tunneling through the Coulomb barrier due to zero-point
vibrations  (see, e.g., Section~3.7 of the
book by Shapiro \& Teukolsky~\cite{shapiro-83}). The
pycnonuclear fusion timescale $\tau_{\mathrm{pyc}}$ is a very sensitive
function of $Z$. The chain of the
reactions~(\ref{eq.sect.accretion.processes.ecapn1}) and
(\ref{eq.sect.accretion.processes.ecapn2}) leads to an abrupt decrease
of $\tau_{\mathrm{pyc}}$ typically by 7 to 10 orders of magnitude. In the
one-component plasma approximation, the accreted crust is composed of
spherical shells containing a single nuclide $(A,Z)$. Pycnonuclear
fusion switches on as soon as $\tau_{\mathrm{pyc}}$ is smaller than the
time of the travel  of a piece of matter (due to accretion)
through the considered shell of mass $M_{\mathrm{shell}}(A,Z)$,
$\tau_{\mathrm{acc}}\equiv M_{\mathrm{shell}}/\dot{M}$. The masses of the
shells are on the order of $10^{-5}\,M_\odot$. As a result, in the inner crust
the chain of reactions~(\ref{eq.sect.accretion.processes.ecapn1} and
\ref{eq.sect.accretion.processes.ecapn2}) in several cases is followed
by the pycnonuclear reaction, occurring on a timescale much shorter
than $\tau_{\mathrm{acc}}$. Introducing  $Z'=Z-2$, we then have
\begin{equation}
(A,Z')+(A,Z')\longrightarrow (2A,2Z')+Q_{j,1},
\label{eq.sect.accretion.processes.pyc1}
\end{equation}
\begin{equation}
(2A,2Z')\longrightarrow (2A-\mathrm{k}', 2Z')+\mathrm{k}'\;n + Q_{j,2} \, ,
\label{eq.sect.accretion.processes.pyc2}
\end{equation}
\begin{equation}
\ldots~\ldots\longrightarrow \ldots~\ldots~+Q_{j,3} \, ,
\label{eq.sect.accretion.processes.pyc3}
\end{equation}
where dots in Equation~(\ref{eq.sect.accretion.processes.pyc3})
denote an actual nonequilibrium process, usually  following
reaction~(\ref{eq.sect.accretion.processes.pyc2}). The total heat
deposition in matter, resulting from a chain of reactions involving a
pycnonuclear fusion,
Equations~(\ref{eq.sect.accretion.processes.pyc1}),
(\ref{eq.sect.accretion.processes.pyc2}) and
(\ref{eq.sect.accretion.processes.pyc3}), is
$Q_j=Q_{j,1}+Q_{j,2}+Q_{j,3}$.

The composition of accreted neutron star crusts, obtained by Haensel
\& Zdunik~\cite{haensel-03}, is shown in
Figure~\ref{eq.sect.accretion.processes.NS-rho-pyc}. These results
describe  crusts built of accreted and processed matter up to the
density $5\times 10^{13} \mathrm{\ g\ cm}^{-3}$ (slightly before the
crust-core interface). At a constant accretion
rate $\dot{M}=\dot{M}_{-9}\times 10^{-9}\,M_\odot/\mathrm{yr}$ this
will take $\sim 10^6 \mathrm{\ yr}/\dot{M}_{-9}$. During that time, a shell
of X-ray burst ashes will be compressed from $\sim 10^8 \mathrm{\ g\
  cm}^{-3}$ to $\sim 10^{13} \mathrm{\ g\ cm}^{-3}$.

Two different compositions of X-ray burst ashes at $\lesssim 10^8
\mathrm{\ g\ cm}^{-3}$, $A_{\mathrm{i}}$, $Z_{\mathrm{i}}$, were assumed. In the first
case, $A_{\mathrm{i}}=56$, $Z_{\mathrm{i}}=26$, which is a ``standard
composition''. In the second scenario $A_{\mathrm{i}}=106$, to imitate
nuclear ashes obtained by Schatz et al.~\cite{schatz-01}. The value of
$Z_{\mathrm{i}}=46$ stems then from the condition of beta equilibrium at
$\rho=10^{8} \mathrm{\ g\ cm}^{-3}$. As we see in
Figure~\ref{eq.sect.accretion.processes.NS-rho-pyc}, after
the pycnonuclear fusion region is reached, both curves converge (as
explained in Haensel \& Zdunik~\cite{haensel-03}, this results from
$A_{\mathrm{i}}$ and $Z_{\mathrm{i}}$ in two scenarios).

\subsection{Deep crustal heating}
\label{sect.accretion.crust-heating}

A neutron star crust that is not in full thermodynamic equilibrium
constitutes a reservoir of energy, which can then be released during
the star's evolution. The formation and structure of nonequilibrium neutron
star crusts has been considered by many authors~\cite{vartanyan-76,
  bisnovatyi-79, sato-79, haensel-90a, haensel-03, haensel-07,
  gupta-07}. Such a  crust can be produced by accretion  onto a neutron star in compact LMXB,
   where the original crust built of a catalyzed matter
(see Section~\ref{sect.groundstate}) is  replaced by a
crust with a composition strongly deviating from that of nuclear equilibrium. However, building up the accreted crust takes time. The outer crust
(Section~\ref{sect.groundstate.outer}), containing $\sim
10^{-5}\,M_\odot$, is replaced by the accreted crust in
$(10^4/\dot{M}_{-9})$~y. To replace the whole crust of mass $\sim
10^{-2}\,M_\odot$ by accreted matter requires
$(10^7/\dot{M}_{-9})$~y. After that time has passed, the entire ``old crust'' is pushed down  through the crust-core interface, and is molten
into the liquid core.
The time $(10^7/\dot{M}_{-9})$~y may seem huge.
 However, LMXBs can live for $\sim 10^9$~y, so
that a fully accreted crust on a neutron star is a realistic
possibility.

Heating due to nonequilibrium nuclear processes  in the
outer and inner crust of an accreting neutron star (deep crustal
heating) was calculated, using different scenarios and
models~\cite{haensel-90a, haensel-03, haensel-07}. The effect of
crustal heating on the thermal structure of the interior of an
accreting neutron star can be seen in
Figures~\ref{fig.sect.accretion.conditions.MiraldaHP-1}
and~\ref{fig.sect.accretion.conditions.MiraldaHP-2}. In what follows,
we will describe the most recent calculations of crustal heating by
Haensel \& Zdunik~\cite{haensel-07}. In spite of the model's simplicity
(one-component plasma, $T=0$ approximation), the  heating in the
accreted outer crust obtained by Haensel \& Zdunik~\cite{haensel-07}
agrees nicely with extensive calculations carried out by
Gupta et al.~\cite{gupta-07}. The latter authors considered a
multicomponent plasma, a reaction network of many nuclides, and
included the contribution from the nuclear excited states. They found
that electron captures in the outer crust proceed mostly via the excited
states of the daughter nuclei, which then de-excite, the excitation
energy heating the matter; this strongly reduces neutrino losses, accompanying
nonequilibrium electron captures. The total deep crustal heating
obtained by Haensel \& Zdunik~\cite{haensel-07} is equal to
$Q_{\mathrm{tot}}=1.5$ and 1.9~MeV per accreted nucleon for
$A_{\mathrm{i}}=106$ and $A_{\mathrm{i}}=56$, respectively.

\epubtkImage{H_sources.png}{%
\begin{figure}[htbp]
  \centerline{\includegraphics[scale=0.5]{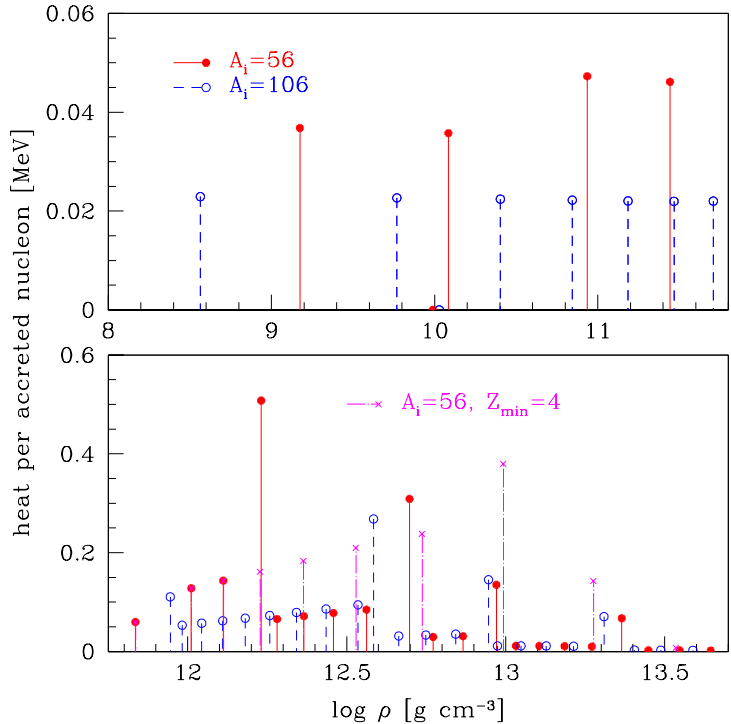}}
  \caption{Heat sources accompanying  accretion, in the outer (upper
  panel) and inner (lower panel) accreted crust. Vertical lines,
  positioned at the bottom of every reaction shell, represent the heat
  per accreted nucleon. Based on Haensel \&
  Zdunik~\cite{haensel-07}. Figure made by J.L.\ Zdunik.}
  \label{fig.sect.accretion.crust-heating.H_sources}
\end{figure}}

In Figure~\ref{fig.sect.accretion.crust-heating.H_sources}, we show
the heat deposited in the matter, per accreted nucleon, in the
thin shells where nonequilibrium nuclear processes occur.
 Actually, reactions proceed at a constant pressure, and there
is a density jump within a thin  ``reaction shell''. The vertical
lines, whose height gives the heat deposited in matter,  are drawn at
the density of the bottom of the reaction shell. The number of heat
sources and the heating
power of a single source depend on the assumed $A_{\mathrm{i}}$. In the case of
$A_{\mathrm{i}}=56$ the number of sources is smaller, and their
heat-per-nucleon values $Q_j$ are larger, than for $A_{\mathrm{i}}=106$.

An important quantity is the integrated heat deposited
in the crust in the outer layer with bottom density $\rho$. It is
given by
\begin{equation}
Q^{(\alpha)}(\rho)= \sum_{j(\rho_j<\rho)}Q^{(\alpha)}_j \, ,
\label{eq.sect.accretion.crust-heating.Qcumul}
\end{equation}
where $(\alpha)$ labels the crustal
heating model (specific $A_{\mathrm{i}},Z_{\mathrm{i}}$, etc.). The quantity
$Q^{(\alpha)}(\rho)$ for two  models of compressional
evolution is plotted in
Figure~\ref{fig.sect.accretion.crust-heating.ecumul}. The second model
illustrates the effect of switching off pycnonuclear
reactions. This was done by artificially blocking pycnonuclear fusion until the nuclear charge went down
to $Z_{\min}=4$, which occurred at
$\rho_{\mathrm{pyc}}=10^{13.25}~\mdens$. And yet, $Q^{(2)}$ for
$\rho>10^{13}~\mdens$ is very similar to that obtained in the first
scenario, which was the most advantageous, as far as crust heating was
concerned. Heating by pycnonuclear fusion at $\rho>10^{13}~\mdens$ is
insignificant. Heat from pycnonuclear fusions at $\rho\sim
10^{12}~\mdens$ is to a large extent replaced by an additional heat
release associated with electron captures and neutron emissions within
the density range $10^{12}\mbox{\,--\,}10^{13}~\mdens$. The values of
$Q^{(1)}$ and $Q^{(2)}$ saturate above $10^{13.6}~\mdens$, where 80\%
of nucleons are in neutron gas phase. All in all, for two scenarios
with $A_{\mathrm{i}}=56$, the total deep crustal heat release is
1.8\,--\,1.9~MeV/nucleon. For $A_{\mathrm{i}}=106$, these
numbers are lowered by about 0.5~MeV/nucleon.

\epubtkImage{ecumul.png}{%
\begin{figure}[htbp]
  \centerline{\includegraphics[scale=0.5]{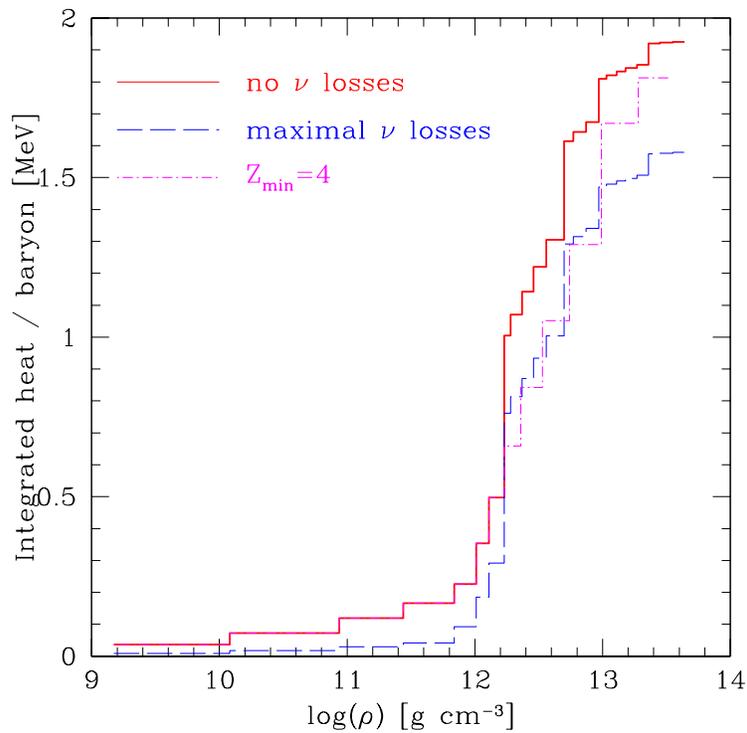}}
  \caption{Integrated heat released in the crust, $Q(\rho)$ (per
    one accreted nucleon) versus $\rho$, assuming initial ashes of
    pure $^{56}$Fe. Solid line: HZ* model of Haensel \&
    Zdunik~\cite{haensel-07}, with $A_{\mathrm{i}}=56$ . Dash-dotted line:
    with pycnonuclear fusion blocked until $Z=Z_{\min}=4$. Based on
    Haensel \& Zdunik~\cite{haensel-07}. Figure made by J.L.\ Zdunik.}
  \label{fig.sect.accretion.crust-heating.ecumul}
\end{figure}}

The quite remarkable weak dependence of the total heat release in the
crust, $Q_{\mathrm{tot}}$, on the nuclear history of an element of matter
undergoing compression from $\sim 10^{8}~\mdens$ to $\sim
10^{13.6}~\mdens$ deserves an explanation~\cite{haensel-07}. One has
to study the most relevant thermodynamic quantity, the Gibbs free
energy per nucleon (baryon chemical potential). Its minimum determines
the state of thermodynamic equilibrium. Moreover, its drop at reaction
surface $P=P_j$ yields the total energy release $Q_j$ per one
nucleon~\cite{prigogine-60}. In the $T=0$ approximation, we have
$\mu_{\mathrm{b}}(P)=[{\cal E}(P)+P]/n_{\mathrm{b}}(P)$ = enthalpy per nucleon.
Minimizing $\mu_{\mathrm{b}}(P)$, at a fixed $P$,
with respect to the independent
thermodynamic variables ($A,Z$, mean free neutron density
$\bar{n}_n$, mean baryon density $n_{\mathrm{b}}$, size of the
Wigner--Seitz cell, etc.), under the constraint of electro-neutrality,
$\bar{n}_p=\bar{n}_e$, we get the ground state
of the crust at a given $P$. This ``cold catalyzed
matter'' (Section~\ref{sect.groundstate}) corresponds to
$\mu^{(0)}_{\mathrm{b}}(P)$. All other $\mu^{(\alpha)}_{\mathrm{b}}(P)$ curves
displaying discontinuous drops due to nonequilibrium reactions
included in a given evolutionary model $(\alpha)$ lie above the
$\mu^{(0)}_{\mathrm{b}}(P)$; see
Figure~\ref{fig.sect.accretion.crust-heating.mup}.

\epubtkImage{mup.png}{%
\begin{figure}[htbp]
  \centerline{\includegraphics[scale=0.5]{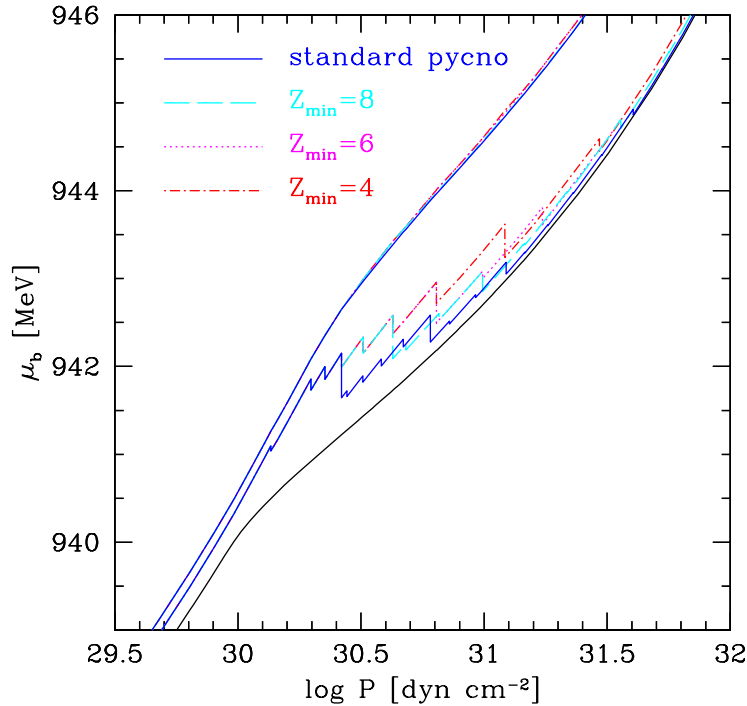}}
  \caption{Baryon chemical potential at $T=0$, $\mu_{\mathrm{b}}=({\cal
      E}+P)/n_{\mathrm{b}}$ vs.\ pressure for different models of neutron
      star crust. Black solid line: $\mu^{(0)}_{\mathrm{b}}(P)$ for the
      ground-state crust. Lines with discontinuous drops:
      $\mu^{(\alpha)}_{\mathrm{b}}(P)$ for four evolution models in
      the accreted crust with $A_{\mathrm{i}}=56$. Upper four
      smooth lines, which nearly coincide:
      $\mu^{(\alpha)}_{\mathrm{b}}(P)+\sum_{j
      (P<P_j)}Q^{(\alpha)}_j\approx \bar{\mu}_{\mathrm{b}}(P)$. For an
      explanation see the text. Based on  Haensel \&
      Zdunik~\cite{haensel-07}. Figure made by J.L.\ Zdunik.}
  \label{fig.sect.accretion.crust-heating.mup}
\end{figure}}

This makes visual the fact that noncatalyzed matter is a reservoir of
energy, released in nonequilibrium processes that move the  matter closer
to the absolute ground state. In spite of dramatic differences between
different $\mu^{(\alpha)}_{\mathrm{b}}(P)$ in the region where the bulk of
the heating occurs, $P=(10^{30}\mbox{\,--\,}10^{31.5}) \mathrm{\ erg\ cm}^{-3}$, the functions
$\mu^{(\alpha)}_{\mathrm{b}}(P)$ tend to $\mu^{(0)}_{\mathrm{b}}(P)$ for
$P\gtrsim 10^{32} \mathrm{\ erg\ cm}^{-3}$. The general structure of different
$\mu^{(\alpha)}_{\mathrm{b}}(P)$ is similar. At the same $P$, their continuous
segments have nearly the same slope. What differs between
$\mu^{(\alpha)}_{\mathrm{b}}(P)$s   are discontinuous drops, by
$Q^{(\alpha)}_j$, at reaction thresholds $P^{(\alpha)}_j$. The
functions $\mu^{(\alpha)}_{\mathrm{b}}(P)$ can therefore be expressed
as (see Haensel \& Zdunik~\cite{haensel-07})
\begin{equation}
\mu^{(\alpha)}_{\mathrm{b}}(P)\approx \bar{\mu}_{\mathrm{b}}(P) -
\sum_{j (P<P_j)}Q^{(\alpha)}_j,
 \label{fig.sect.accretion.crust-heating.decomp-mu}
\end{equation}
where $\bar{\mu}_{\mathrm{b}}(P)$ is a smooth function of
$P$, independent of $(\alpha)$. For $P>10^{33.5} \mathrm{\ erg\ cm}^{-3}$ the
values of $Q_j$ are negligibly small, and  all
$\mu^{(\alpha)}_{\mathrm{b}}(P)$ come quite close to the ground
state line. This implies that the sum $\sum_jQ^{(\alpha)}_j$
must be essentially {\it independent} of ${(\alpha)}$.

\subsection{Thermal structure of accreted crusts and X-ray bursts}
\label{sect.accretion.conditions}

A thermonuclear flash is triggered by an instability in the
thermonuclear burning. The relevant quantities are the local
heating rate due to thermonuclear fusion,
$\dot{\varepsilon}_{\mathrm{nuc}}$, and cooling rate,
$\dot{\varepsilon}_{\mathrm{cool}}$, resulting from heat diffusion,
volume expansion, and neutrino emission. A steady state of an
accreting neutron star corresponds to
$\dot{\varepsilon}_{\mathrm{nuc}}=\dot{\varepsilon}_{\mathrm{cool}}$.
It depends on the accretion rate, composition of accreted plasma, structure and
physical properties (thermal conductivity, neutrino
emissivity, etc.\ of the stellar interior. Examples of the
steady thermal structure of accreting neutron stars are shown
in Figures~\ref{fig.sect.accretion.conditions.MiraldaHP-1}
and~\ref{fig.sect.accretion.conditions.MiraldaHP-2}. 

\epubtkImage{MiraldaHP-1.png}{%
\begin{figure}[htbp]
  \centerline{\includegraphics[scale=0.5]{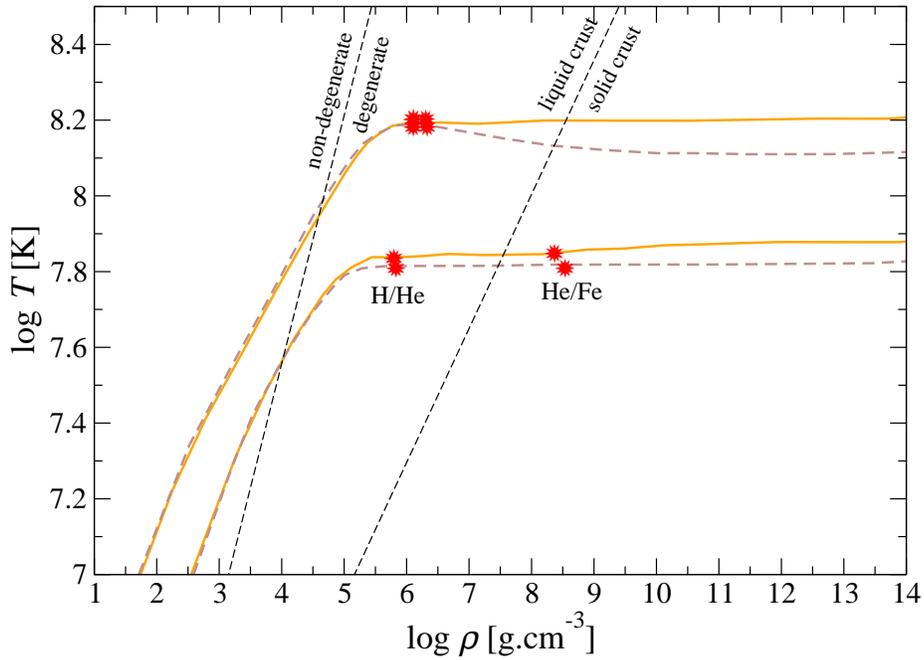}}
  \caption{Temperature (local, in the reference frame of the star)
  vs.\ density  within the crust of an accreting  neutron star (soft
  EoS of the core, $M=1.24\,M_\odot$) in a steady thermal state, with
  standard cooling of the core (no fast cooling of the direct Urca
  type). Upper solid curve -- $\dot{M}=10^{-9.96}\,M_\odot \mathrm{\
  y}^{-1}$. Lower solid curve -- $\dot{M}=10^{-11}\,M_\odot \mathrm{\
  y}^{-1}$. H and He burning shells are indicated by asterisks. Deep
  crustal heating is included. Dashed line -- temperature profile
  without deep crustal heating. Based on Figure~3b of
  Miralda-Escud\'e~el~al.~\cite{miralda-90}.}
  \label{fig.sect.accretion.conditions.MiraldaHP-1}
\end{figure}}

\epubtkImage{MiraldaHP-2.png}{%
\begin{figure}[htbp]
  \centerline{\includegraphics[scale=0.5]{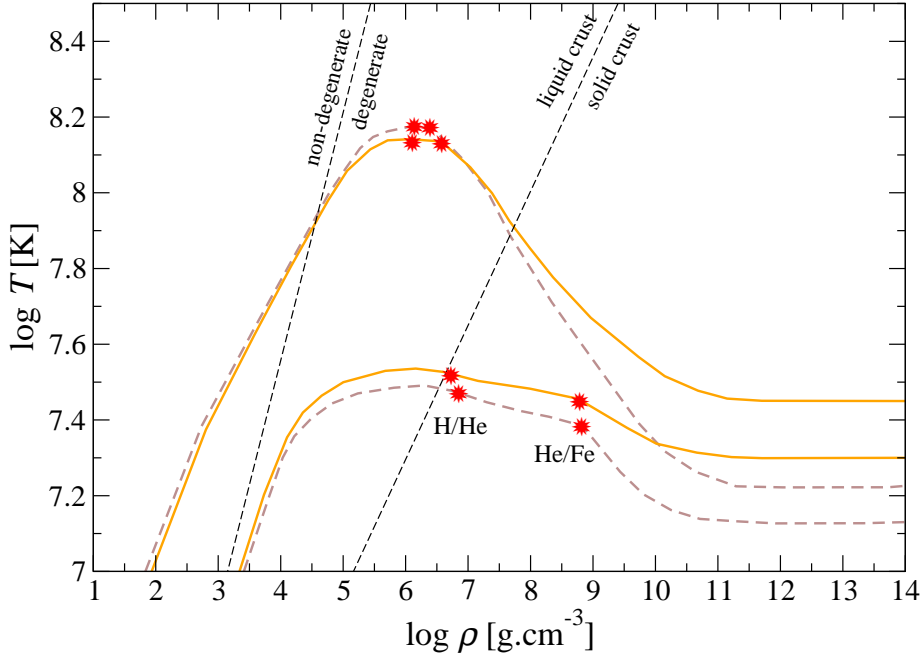}}
  \caption{Same as in
  Figure~\ref{fig.sect.accretion.conditions.MiraldaHP-1}, but for fast
  neutrino cooling due to pion condensation in the
  inner core. Based on Figure~3c of
  Miralda-Escud\'e~et~al.~\cite{miralda-90}.}
  \label{fig.sect.accretion.conditions.MiraldaHP-2}
\end{figure}}

A stable steady state of an accreting neutron star satisfies
\begin{equation}
\dot{\varepsilon}_{\mathrm{nuc}}=\dot{\varepsilon}_{\mathrm{cool}} \, , ~
{\partial\dot{\varepsilon}_{\mathrm{nuc}}\over\partial T}
<{\partial\dot{\varepsilon}_{\mathrm{cool}}\over\partial T} \, .
\label{eq.sect.accretion.conditions.therm-stab}
\end{equation}
The inequality guarantees that any thermal perturbation will be damped
by a self-regulated cooling. Under the increasing weight of accreted
matter, an element of the crust in the He burning shell moves in the
$\rho-T$ plane, in the direction of increasing $\rho$. At some
moment, after crossing the ``ignition line'' characterized by
\begin{equation}
{\partial\dot{\varepsilon}_{\mathrm{nuc}}\over\partial T}
={\partial\dot{\varepsilon}_{\mathrm{cool}}\over\partial T}~
\Longrightarrow \rho_{\mathrm{ign}}(T) \, ,
 \label{eq.sect.accretion.conditions.Tinstab1}
\end{equation}
burning becomes unstable, and a self-accelerating
thermonuclear flash is ignited because
\begin{equation}
{\partial\dot{\varepsilon}_{\mathrm{nuc}}\over\partial T}
>{\partial\dot{\varepsilon}_{\mathrm{cool}}\over\partial T} \, .
 \label{eq.sect.accretion.conditions.Tinstab2}
\end{equation}
In the standard picture, it is the instability in the helium burning
via $3\alpha\longrightarrow ^{12}\mathrm{C}$, which triggers an X-ray
burst (see, e.g., \cite{fujimoto-81}). The total energy release in a
burst can then be easily estimated (neglecting the general
relativistic correction) via
\begin{equation}
E_{\mathrm{burst}}\approx Q_{\mathrm{nuc}}\;M_{\mathrm{burn}}/ m_{\mathrm{u}} \, ,
 \label{eq.sect.accretion.conditions.E-burst}
\end{equation}
where $Q_{\mathrm{nuc}}$ is the mean energy per nucleon released in
the thermonuclear flash (2\,--\,8 MeV, depending on composition of
burnt material), $M_{\mathrm{burn}}=M_{\mathrm{env}}(\rho_{\mathrm{ign}})$
is the mass of the envelope burnt in the flash (determined by
the ignition density $\rho_{\mathrm{ign}}$), and $m_{\mathrm{u}}$ is the
atomic mass unit. The value of $M_{\mathrm{env}}(\rho_{\mathrm{ign}})$
can be read from Figure~\ref{fig.sect.structure.Mcr-z-GSC-ACC}. A good
model of X-ray bursts should yield $E_{\mathrm{burst}}\sim
10^{40}~$erg with a recurrence time of hours. This can be
satisfied with ignition of helium flashes at $\sim
10^7~\mdens$.

\newpage


\section{Equation of State}
\label{sect.eos}

In this section we discuss the Equation of State (EoS) of the neutron star crust.
Three different cases will be considered: cold catalyzed matter in Section~\ref{sect.eos.gs},
accreted crust matter in Section~\ref{sect.eos.accreted} assuming the formation scenario described
in Section~\ref{sect.accretion}, and hot dense matter in supernova
cores in Section~\ref{sect.eos.supernova}.

\subsection{Ground state crust}
\label{sect.eos.gs}

The EoS of the outer crust in the ground
state approximation (see Section~\ref{sect.groundstate.outer})
is rather well established, so that the
pressure at any given density is determined within a
few percent accuracy ~\cite{bps-71, haensel-94, ruster-06} as can be seen in
Figure~\ref{fig.sect.eos.outer}.

\epubtkImage{eos_outer_crust.png}{%
\begin{figure}[htbp]
  \centerline{\includegraphics[scale=0.5]{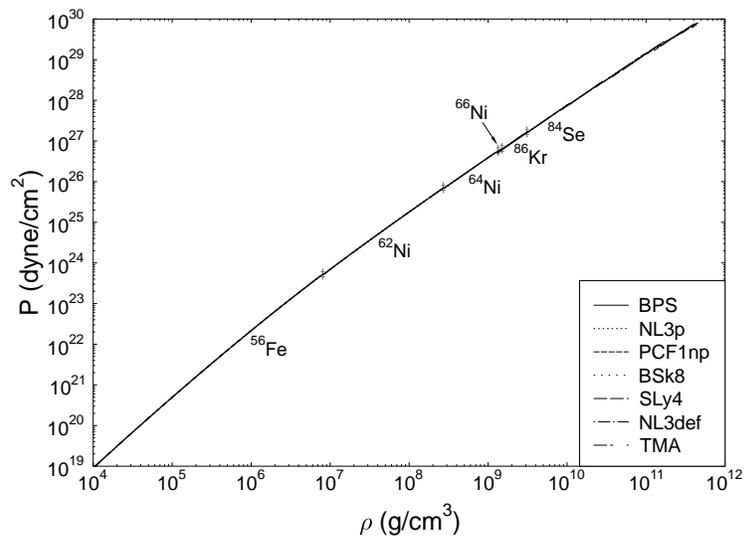}}
  \caption{EoS of the ground state of the outer crust for various
nuclear models. From R\"uster et al.~\cite{ruster-06}. A zoomed-in segment
of the EoS just before the neutron drip can be seen in
Figure \ref{fig.sect.eos.outer.ND}.}
\label{fig.sect.eos.outer}
\end{figure}}

\epubtkImage{eos_outer_crust_zoom.png}{%
\begin{figure}[htbp]
  \centerline{\includegraphics[scale=0.5]{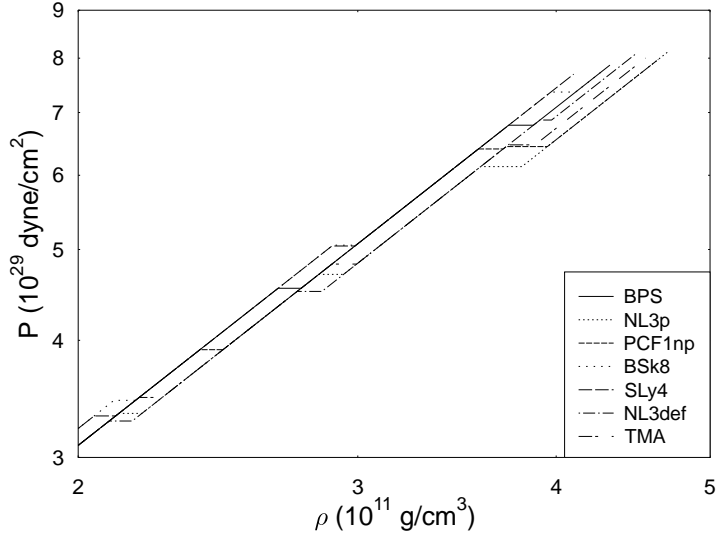}}
  \caption{EoS of the ground state of the outer crust just before
  neutron drip for various
nuclear models. From R\"uster et al.~\cite{ruster-06}}.
\label{fig.sect.eos.outer.ND}
\end{figure}}

Uncertainties arise above the density $\rho>6\times10^{10}~\mdens$ as shown in
Figure~\ref{fig.sect.eos.outer.ND} because experimental data are lacking.
However, one may hope that with the improvement of experimental techniques,
experimental data on very exotic nuclei will become available in the future.

On the contrary, the inner crust nuclei cannot be studied in a laboratory
because their  properties are influenced by the gas of dripped neutrons, as reviewed
in Section~\ref{sect.groundstate.inner}. This means that \emph{only} theoretical models
can be used there and consequently the EoS after neutron drip is much more uncertain
than in the outer layers. The neutron gas contributes more and more to the total
pressure with increasing density. Therefore, the problem of correct modeling of
the EoS of a pure neutron gas at subnuclear densities becomes important. The
true EoS of cold catalyzed matter stems from a true nucleon
Hamiltonian, expected to describe nucleon interactions at
$\rho\lesssim \rho_0$, where $\rho_0$ is the nuclear saturation density.
To make the solution of the many-body
problem feasible, the task is reduced to finding an
\emph{effective nucleon Hamiltonian}, which would enable one to
calculate reliably both the properties of laboratory nuclei and the EoS
of cold catalyzed matter for $10^{11}\,\mdens \lesssim \rho
\lesssim \rho_0$. The task also includes the calculation of
the crust-core transition. We will illustrate the general results
with two examples of the EoS of the inner crust, calculated in the
compressible--liquid-drop model (see Section~\ref{sect.groundstate.inner.LDM})
using the effective nucleon-nucleon interactions
FPS ({\bf F}riedman-{\bf P}anharipande-{\bf S}kyrme~\cite{pandharipande-89}) and
SLy ({\bf S}kyrme-{\bf Ly}on~\cite{chabanat-97, chabanat-98, chabanat-98err}).

\epubtkImage{crust12.png}{%
  \begin{figure}[htbp]
    \centerline{\includegraphics[scale=0.6]{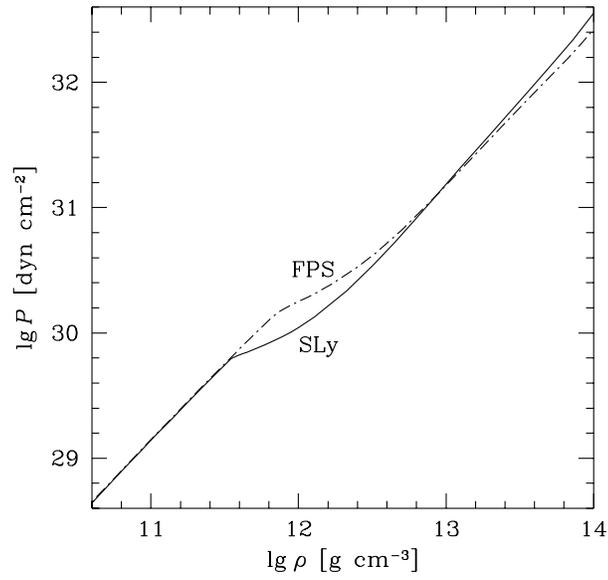}}
    \caption{Comparison of the SLy and FPS
    EoSs. From~\cite{haensel-06}.}
    \label{fig.sect.eos.SLyvsFPS}
\end{figure}}

As one can see in Figure~\ref{fig.sect.eos.SLyvsFPS}, significant
differences between the SLy and FPS EoS are restricted to the density
interval $4\times 10^{11}\mbox{\,--\,}4\times 10^{12}~\mdens$. They result mainly
from the fact that the density at which neutron drip occurs for each is
different: $\rho_{\mathrm{ND}}(\mathrm{SLy})\simeq 4\times 10^{11}~\mdens$ (in
good agreement with the ``empirical EoS'' of Haensel \&
Pichon~\cite{haensel-94}), while $\rho_{\mathrm{ND}}(\mathrm{FPS})\simeq
6\times 10^{11}~\mdens$. For $4\times 10^{12}~\mdens \lesssim \rho
\lesssim 10^{14}~\mdens$ the SLy and FPS EoSs are very similar, with
the FPS EoS being a little softer at the highest densities
considered. The detailed behavior of the two EoSs near the crust-core
transition can be seen in Figure~\ref{fig.sect.eos.SLyvsFPS2}. The FPS
EoS is softer there than the SLy EoS (for pure neutron matter the FPS
model is softer at subnuclear densities; see~\cite{haensel-06}).

\epubtkImage{crust13.png}{%
  \begin{figure}[htbp]
    \centerline{\includegraphics[scale=0.6]{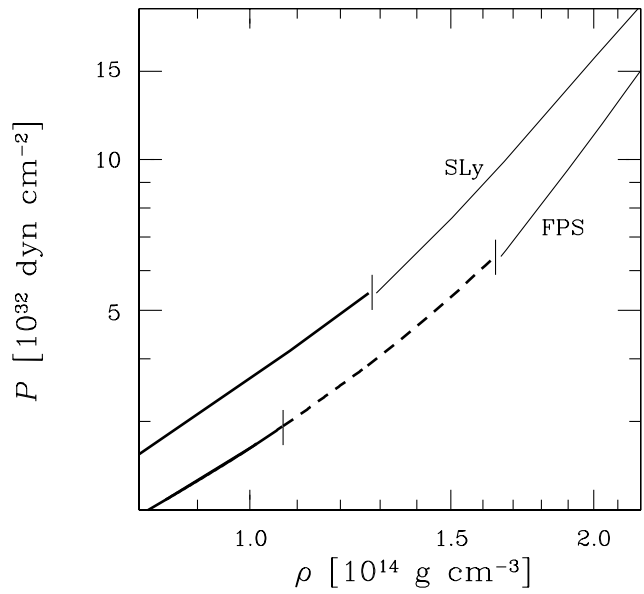}}
    \caption{Comparison of the SLy and FPS EoSs near the crust-core
      transition. Thick solid line: inner crust with spherical
      nuclei. Dashed line corresponds to ``exotic nuclear
      shapes''. Thin solid line: uniform $npe$
      matter. From~\cite{haensel-06}.}
    \label{fig.sect.eos.SLyvsFPS2}
\end{figure}}

In the case of the SLy EoS, the crust-liquid core transition
takes place as a very weak first-order phase transition, with a
relative density jump on the order of one percent. Notice that,
for this model, spherical nuclei persist  to the very bottom
of the crust~\cite{douchin-00}. As seen from Figure~\ref{fig.sect.eos.SLyvsFPS2},
the crust-core transition is
accompanied by a noticeable stiffening of the EoS. For the FPS
EoS the situation is different. Namely, the crust-core
transition takes place through a sequence of phase transitions
with changes of nuclear shapes as discussed in Section~\ref{sect.groundstate.pasta}.
These phase transitions make the crust-core transition smoother than in
the SLy case, with a gradual increase of stiffness (see Figure~\ref{fig.sect.eos.supernova.gamma}).
While the presence of exotic nuclear shapes is expected to have dramatic
consequences for the transport, neutrino emission, and elastic
properties of neutron star matter, their effect on the EoS is
rather small.

\epubtkImage{gamma_outer_crust.png}{%
\begin{figure}[htbp]
  \centerline{\includegraphics[scale=0.5]{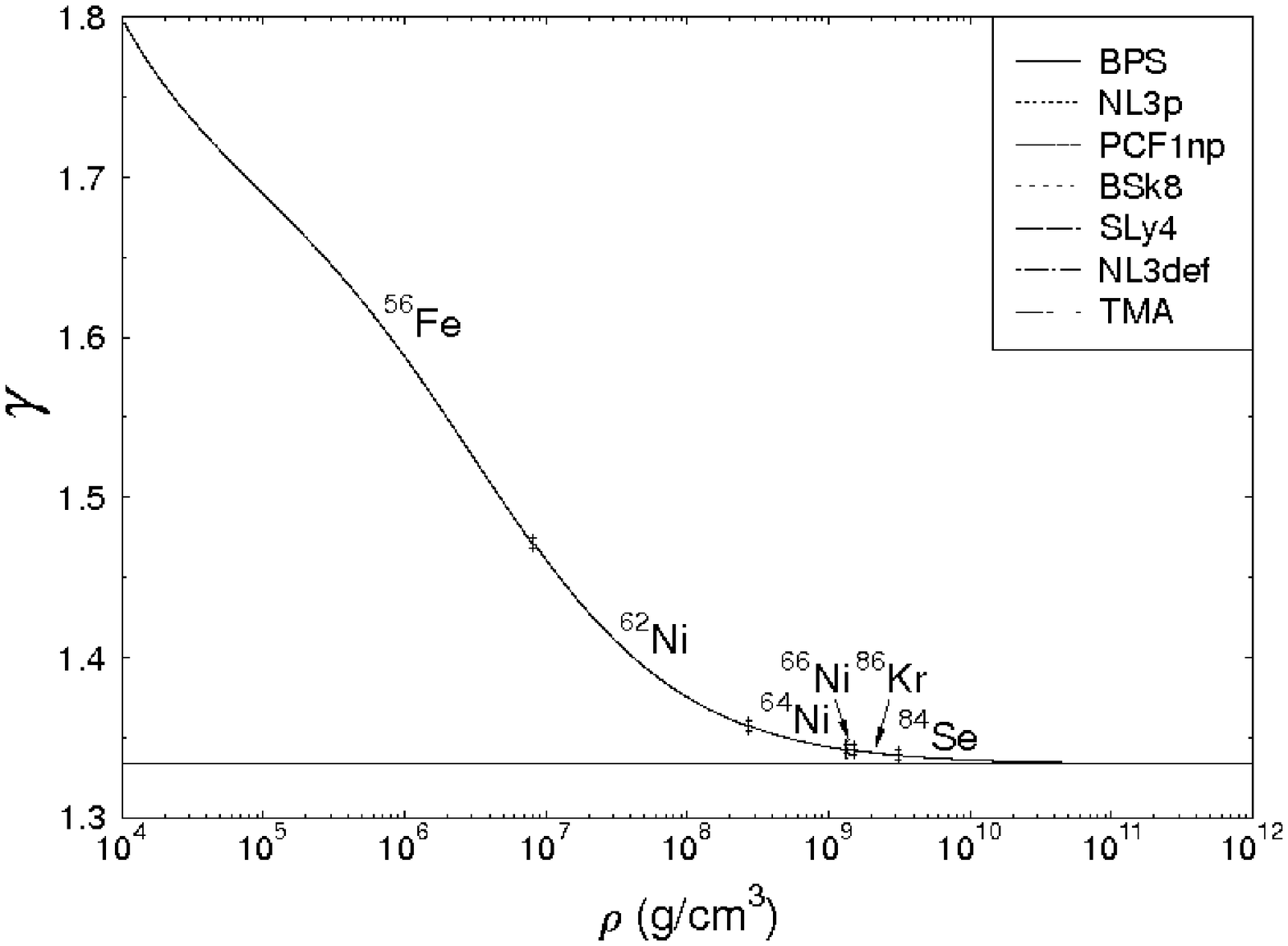}}
  \caption{Adiabatic index $\gamma$ for various EoSs
of the ground-state outer crust below neutron drip. The horizontal
line corresponds to $\gamma=4/3$. The neutron drip point
$\rho_{\mathrm{ND}}\approx 4\times 10^{11}~\mdens$ depends slightly on
the EoS model used and, therefore, is not marked.
 From R\"uster et al.~\cite{ruster-06}.}
\label{fig.sect.eos.gammaGS.outer}
\end{figure}}

\epubtkImage{gamma_inner_crust.png}{%
\begin{figure}[htbp]
  \centerline{\includegraphics[scale=0.5]{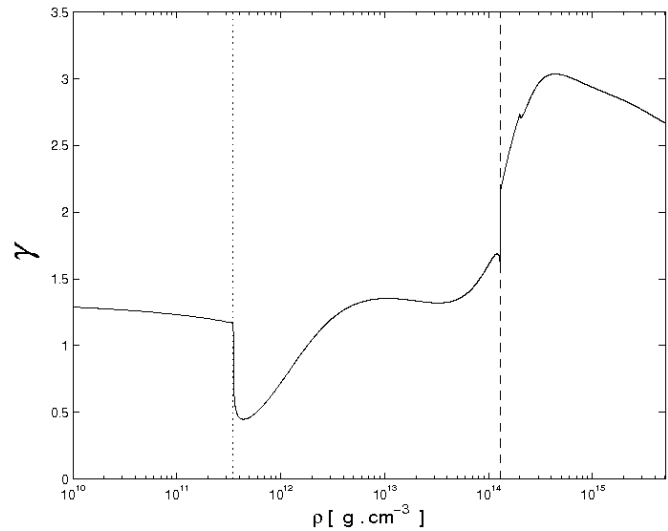}}
  \caption{Adiabatic index $\gamma$ for the EoS of the
ground-state  crust. Dotted vertical lines
correspond to the neutron drip and crust-core interface
points. Calculations performed using the SLy EoS of
Douchin \& Haensel~\cite{douchin-01}.}
\label{fig.sect.eos.gammaGS}
\end{figure}}

\epubtkImage{crust14.png}{%
  \begin{figure}[htbp]
    \centerline{\includegraphics[scale=0.5]{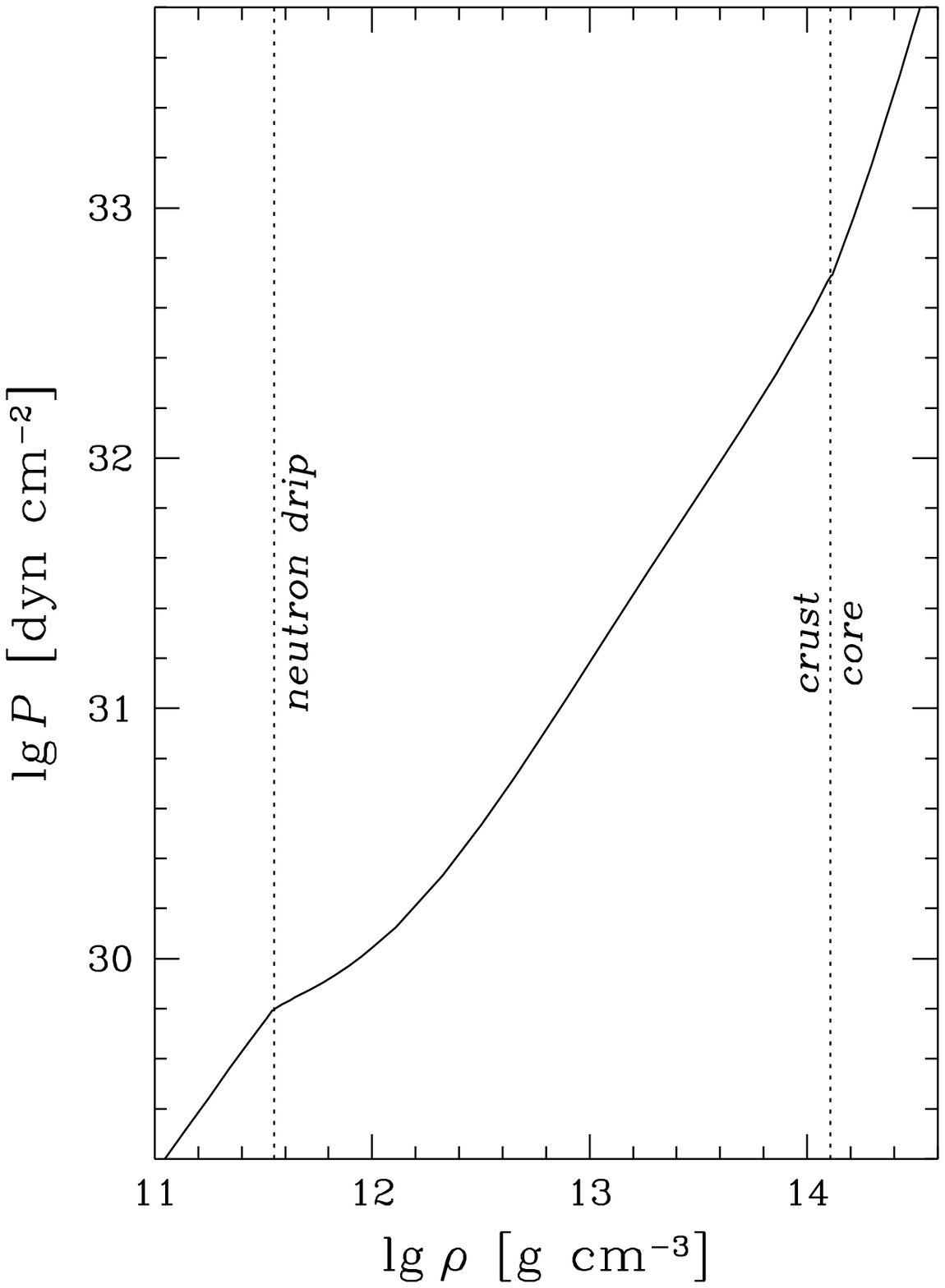}}
    \caption{The SLy EoS. Dotted vertical lines correspond to the
    neutron drip and crust-core transition. From~\cite{haensel-06}.}
    \label{fig.sect.eos.SLy}
\end{figure}}

The overall SLy EoS of the crust, calculated including adjacent
segments of the liquid core and the outer crust, is shown in
Figure~\ref{fig.sect.eos.SLy}. In the outer crust segment, the SLy EoS
cannot be visually distinguished from the EoSs of Haensel \&
Pichon~\cite{haensel-94} or R\"uster et al.~\cite{ruster-06}, which
are  based on experimental nuclear masses. An important dimensionless
parameter, measuring the stiffness of an EoS at a given density, is
the adiabatic index,
\begin{equation}
\label{eq.sect.eos.gamma}
\gamma=\frac{\mathrm{d}\log P} {\mathrm{d}\log
  n_{\mathrm{b}}}=\frac{n_{\mathrm{b}}}{P}\frac{\mathrm{d}P}
  {\mathrm{d}n_{\mathrm{b}}}\, ,
\end{equation}
 which at subnuclear density can be approximated by $\gamma \simeq
(\rho/P)\mathrm{d}P/\mathrm{d}\rho$. The total pressure $P$ is defined by
Equation~(\ref{sect.groundstate.outer.P}). The adiabatic index
$\gamma$ is shown in Figures~\ref{fig.sect.eos.gammaGS.outer} and~\ref{fig.sect.eos.gammaGS}
as a function of mass density $\rho$.
At  $\rho_{\mathrm{ND}}>\rho\gtrsim 10^9~\mdens$, we have
 $\gamma\approx 4/3$. This is because in these outer crust
 layers, pressure is very well
approximated by the sum of the contribution of the ultrarelativistic
electron gas ($P_e$) and of the lattice contribution ($P_{L}$),
which both have
 the same density dependence $\propto \rho^{4/3}$ (see
Section~\ref{sect.groundstate.outer}).
One notices in Figure~\ref{fig.sect.eos.gammaGS} a dramatic softening
in the density region following the neutron drip point, $\rho\gtrsim \rho_{\mathrm{ND}}$.
This means, in particular,  that no stable stars can exist with central densities
around $\rho_{\mathrm{ND}}$ because the compressibility of the matter is too low.
Then the EoS stiffens gradually, with a significant increase of $\gamma$
near the crust-core interface. A jump in $\gamma$ on the core side is connected
with the disappearance of nuclei, and a subsequent stiffening (due
to the nucleon-nucleon interaction) in the uniform $npe$ liquid.
Figures~\ref{fig.sect.eos.gammaGS.outer}
and~\ref{fig.sect.eos.gammaGS} show that the EoS of the inner crust
is very different from the polytropic form $P\propto \rho^{\gamma_0}$, where
${\gamma_0}$ is a constant. Tabulated and analytical EoSs of
the ground state crust are
available online~\cite{hempel-eos, ioffe}.

\subsection{Accreted crust}
\label{sect.eos.accreted}

A model of the EoS of accreted crusts was calculated by
Haensel \& Zdunik~\cite{haensel-90b}. They used the compressible liquid drop model
(see Section~\ref{sect.groundstate.inner.LDM}) with a ``single nucleus''
scenario.

In Figure~\ref{fig.sect.eos.acc} this EoS is compared with the
SLy model of cold catalyzed matter described in Section~\ref{sect.eos.gs}.
At $\rho<\rho_{\mathrm{ND}}$ both EoSs are very similar. The reason is that
for $\rho<\rho_{\mathrm{ND}}$, as discussed in Section~\ref{sect.groundstate.outer},
we have $P\simeq P_e(n_e)$ with
$n_e=(Z/A) n_{\mathrm{b}}$ and the ratio $Z/A$ is quite similar for both
accreted and ground state crusts. Large differences appear for $\rho_{\mathrm{ND}}\lesssim \rho
\lesssim 10\rho_{\mathrm{ND}}$, where the EoS of accreted matter is
stiffer than that of cold catalyzed matter. One also notices well-pronounced density jumps at constant pressure in the EoS of accreted
matter. They are associated with discontinuous changes in nuclear
composition, an artifact of the one-component plasma approximation.
The jumps are particularly large for
$\rho_{\mathrm{ND}}\lesssim \rho \lesssim 10\rho_{\mathrm{ND}}$.

\epubtkImage{crust16.png}{%
\begin{figure}[htbp]
  \centerline{\includegraphics[scale=0.6]{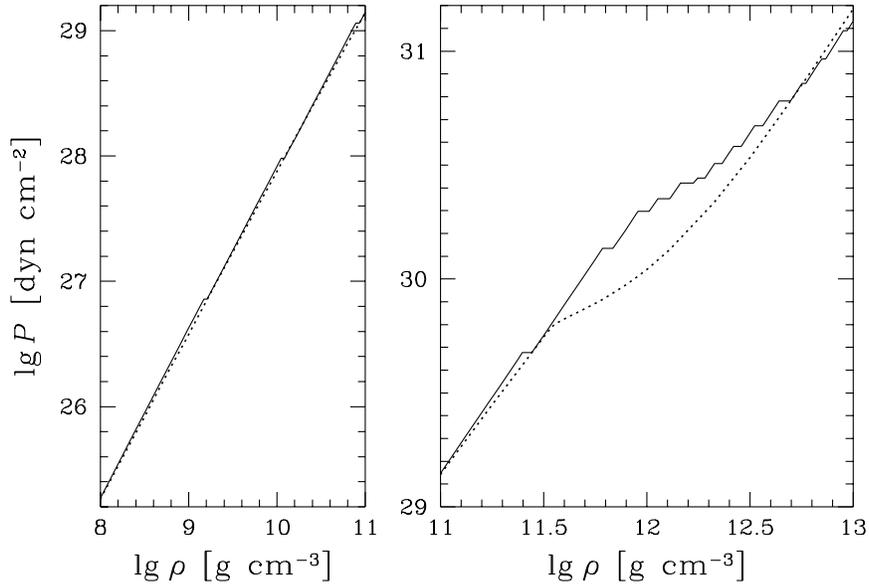}}
  \caption{Comparison of the SLy EoS for cold catalyzed matter (dotted line)
and the EoS of accreted crust (solid line). Figure by A.Y.\ Potekhin.}
\label{fig.sect.eos.acc}
\end{figure}}

The difference between the cold catalyzed  and
accreted matter EoSs decreases for large
density. Both curves are very close to each other
for $\rho>10^{13}~\mdens$. This is because
for such a high density the pressure is mainly produced
by the neutron gas and is not sensitive to the detailed
composition of the nuclear clusters. In view of
this, one can use the EoS of the catalyzed matter for
calculating the hydrostatic equilibrium of the high-density ($\rho >
10^{13} \mathrm{\ g\ cm}^{-3}$) internal layer of the
accreted crust.

\subsection{Effect of magnetic fields on the EoS}
\label{sect.EOS.B}

 Typical values of the surface magnetic field of radio pulsars
 are $B\sim 10^{12}$~g. For magnetars, surface magnetic fields
 can be as high as  $\sim 10^{15}$~G. Effects of such fields
 on the EoS and structure of the crust are briefly reviewed  in
 Section~\ref{sect.structure.B}. A detailed study of the effect
 of $\pmb{B}$ on neutron star envelopes  can be found
 in Chapter~4 of \cite{haensel-06}.

\subsection{Supernova core at subnuclear density}
\label{sect.eos.supernova}

The outer layers of the supernova core, which after a
successful explosion will become the envelope of a
proto-neutron star, display a similar range of densities
$\rho\lesssim 10^{14}~\mdens$  and are governed by the same
nuclear Hamiltonian as the neutron star crust. This is why
we include it in the present review.

The subnuclear density layer of supernova cores shows
 many similarities with that of neutron star crusts. In both cases,
the matter is formed of nuclear clusters embedded in a sea of leptons and hadrons.
 Nevertheless, the conditions are very different since supernova cores
are lepton rich (contain many electrons, positrons as well as trapped neutrinos
and antineutrinos) with lepton fraction
\begin{equation}
Y_{\mathrm{L}}={n_{e^-}+n_{\nu_e}-n_{e^+}-n_{\bar{\nu}_e}\over n_b}\lesssim 0.4 \, ,
\label{eq:Y_L.def}
\end{equation}
and very hot with temperatures typically $T\lesssim 40 \mathrm{\
  MeV}/k_{\mathrm{B}}$ while the matter in neutron star crusts is
cold, neutrino free, and in $\beta$-equilibrium. Neutrinos remain
trapped with the supernova core for several seconds,  which is their
diffusion timescale in dense hot matter. The presence of nuclei in
dense hot matter is of utmost importance for the neutrino opacity of
the supernova core. Namely, the scattering cross section of neutrinos
off nuclei of mass number $A$ is $A$ times larger than the sum of the
individual cross sections for  $A$ nucleons in a nucleon gas (see,
e.g., \cite{shapiro-83}). As long as neutrinos are trapped,
$Y_L=const.$, and collapse is adiabatic.

Of particular importance for supernova simulations is the adiabatic
index defined by
\begin{equation}
  \label{eq.sect.eos.supernova.gamma}
  \gamma = \frac{\partial \log P}{\partial \log
  \rho}\biggr\vert_{s,Y_e} \, ,
  \end{equation}
where $s$ is the entropy per nucleon in units of $k_{\mathrm{B}}$ and $Y_e$ is
the electron fraction. For low entropies per nucleon on the order of
the Boltzmann constant $k_{\mathrm{B}}$, and for densities $\rho \lesssim
5\times 10^{13}~\mdens$ the value of $\gamma$ is mainly determined by
the relativistic electrons and is thus close to 4/3 (adiabatic index
of an ultra-relativistic Fermi gas); see
Figure~\ref{fig.sect.eos.supernova.gamma}. The adiabatic index jumps
to larger values when the nuclear clusters dissolve into a uniform
mixture of nucleons and leptons, as discussed in
Section~\ref{sect.eos.gs}. The adiabatic index is shown in
Figure~\ref{fig.sect.eos.supernova.gamma} for three different EoS that
have been used in supernova simulations. The impact of the presence of
nuclear pasta phases (Section~\ref{sect.groundstate.pasta}) on the
adiabatic index is illustrated in
Figure~\ref{fig.sect.eos.supernova.gamma.pasta}.

The striking differences between the adiabatic index of
supernova matter, $\gamma_{\mathrm{SN}}(\rho)$, and that for cold
catalyzed matter in neutron stars, $\gamma_{\mathrm{NS}}(\rho)$,
deserves additional explanation. In the core collapse,
compression of the matter becomes adiabatic as soon as
$\rho\gtrsim 10^{11}~\mdens$, so that the entropy per nucleon
$s=const.$ Simultaneously, due to neutrino trapping,
the electron-lepton fraction is frozen, $Y_{\mathrm{L}}=const.$ 
The condition $s=const.\approx 1 k_{\mathrm{B}}$ blocks evaporation of
nucleons from the nuclei; the motion of nucleons have to
remain ordered. Therefore, the fraction of free nucleons stays
small and they do not contribute significantly to the
pressure, which is supplied by the electrons, until the density
reaches $10^{14}~\mdens$.

At $\rho\gtrsim 10^{14}~\mdens$, nuclei coalesce forming
uniform nuclear matter. Thus, there are two
density regimes for $\gamma_{\mathrm{SN}}(\rho)$. For
 $\rho\lesssim 10^{14}~\mdens$, pressure is supplied by
 the electrons, while nucleons are confined to the nuclei,
 so that $\gamma_{\mathrm{SN}}\simeq 4/3 \approx 1.3$.
Then, for $\rho\gtrsim 10^{14}~\mdens$ nuclei coalesce into
uniform nuclear matter, and the supernova matter stiffens
violently, with the adiabatic index jumping by a factor of about
two, to $\gamma_{\mathrm{SN}}\approx 2\mbox{\,--\,}3$. This stiffening is
actually responsible for the {\it bounce} of infalling
matter. An additional factor stabilizing nuclei at
$\rho\lesssim 10^{14}~\mdens$   in spite of a high
$T>10^{10}$~K, is a large lepton fraction, $Y_{\mathrm{L}}\approx
0.4$, enforcing a relatively large proton fraction,
$Y_p^{\mathrm{SN}}\approx 0.3$, to be compared with
$Y_p^{\mathrm{NS}}\approx 0.05$ for neutron stars.

Finally, for supernova matter we notice the absence of a neutron-drip
softening, so well pronounced in $\gamma_{\mathrm{NS}}$,
Figure~\ref{fig.sect.eos.gammaGS.outer}. This is because neutron gas
is present in supernova matter also at $\rho<10^{11}~\mdens$, and the
increase of the free neutron fraction at higher density is prevented
by strong neutron binding in the nuclei (large $Y_p^{\mathrm{SN}}$), and
low $s\approx 1 k_{\mathrm{B}}$.

\epubtkImage{gamma_supernova.png}{%
\begin{figure}[htbp]
  \centerline{\includegraphics[scale=0.4]{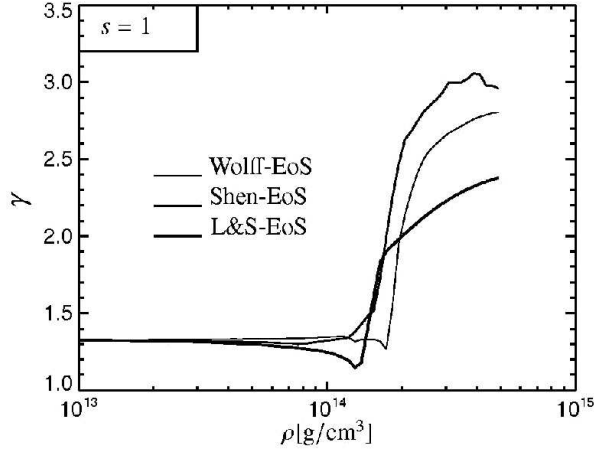}}
  \caption{Variation of the adiabatic index $\gamma$ of  supernova
  matter with   mass density $\rho$ for three different EoSs with trapped neutrinos:
  Lattimer and Swesty~\cite{lattimer-91} (compressible liquid drop model),
  Shen~\cite{shen-98, shen-98b} (relativistic mean field theory
  in the local density approximation) and
  Wolff~\cite{hillebrandt-84} (Hartree--Fock with Skyrme
  nucleon-nucleon interaction).
  The lepton fraction is $Y_L=0.4$ and the entropy
  per nucleon is equal to $1 k_{\mathrm{B}}$.}
  \label{fig.sect.eos.supernova.gamma}
\end{figure}}

\epubtkImage{lassaut.png}{%
\begin{figure}[htbp]
  \centerline{\includegraphics[scale=0.4]{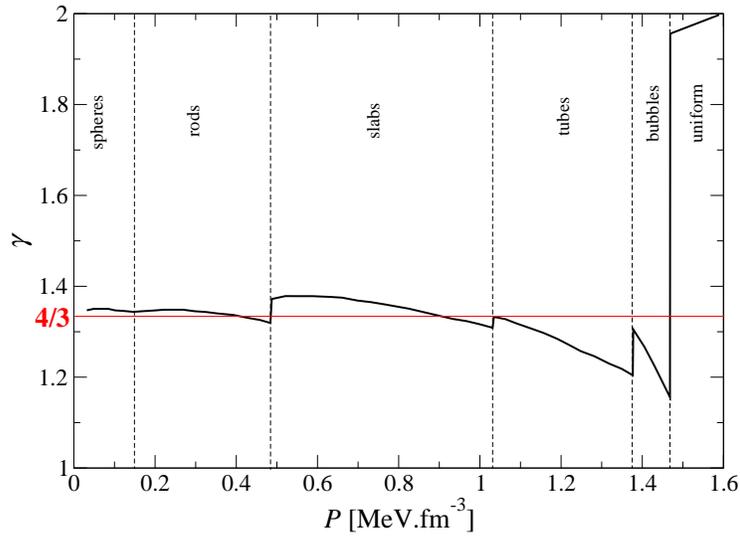}}
  \caption{Variation of the adiabatic index of supernova matter,  $\gamma$,
   with pressure $P$ in the nuclear pasta phases for a fixed
    electron fraction
   $Y_e=0.285$ and the entropy per nucleon $1 k_{\mathrm{B}}$,
    from the Thomas--Fermi calculations with the Skyrme interaction SkM
    of Lassaut et al.~\cite{lassaut-87}.}
  \label{fig.sect.eos.supernova.gamma.pasta}
\end{figure}}

\newpage


\section{Crust in Global Neutron Star Structure}
\label{sect.structure}

The equilibrium structure of the crust of a neutron star results from the
balance of pressure, electromagnetic and elastic stresses, and gravitational pull
exerted by the whole star. The global structure of the crust can be calculated 
by solving Einstein's equations 
\begin{equation}
R_{\mu\nu}-\frac{1}{2} R g_{\mu\nu}=\frac{8\pi G}{c^4} T_{\mu\nu} \, ,
\end{equation}
where $G$ is the gravitational constant and $c$ the speed of light.
The Ricci scalar $R_{\mu\nu}$ and the scalar curvature $R$ are determined from the 
spacetime metric $g_{\mu\nu}$, which represents gravity. Pressure, electromagnetic and elastic 
stresses are taken into account in the energy-momentum tensor $T_{\mu\nu}$.

As a first approximation, treating the star as an ideal fluid, the energy-momentum
tensor is given by (see, e.g., Landau \& Lifshitz~\cite{landau-75})
\begin{equation}
  T^{\mu \, (\mathrm{liq})}_{\ \nu}=(\rho c^2+P)u^\mu u_\nu + P
  \delta^\mu_\nu \, ,
  \label{eq.sect.structure.Tliq}
\end{equation}
where $\rho$ is the mass-energy density and $u^\mu$ is the 4-velocity of the fluid.
Equation~(\ref{eq.sect.structure.Tliq}) can be written in an
equivalent but more general form~\cite{carter-89}
\begin{equation}
  T^{\mu \, (\mathrm{liq})}_{\ \nu}=n^\mu \pi_\nu + P \delta^\mu_\nu \, ,
  \label{eq.sect.structure.Tliq2}
\end{equation}
where $n^\mu=n u^\mu$ is the total 4-current, $n$ is the total particle number density in the fluid rest frame
and $\pi_{\nu}$ is the momentum per particle of the fluid given by
\begin{equation}
  \pi_{\nu}=\mu u_{\nu} \, .
\end{equation}
The quantity $\mu$ is a dynamic effective mass defined by the
relation
\begin{equation}
  P = (n \mu - \rho) c^2 \, ,
\end{equation}
where $c$ is the speed of light. As shown by Carter \&
Langlois~\cite{carter-98}, Equation~(\ref{eq.sect.structure.Tliq2})
can be easily transposed to fluid mixtures (in order to account for
superfluidity inside the star) as follows
\begin{equation}
  T^{\mu \, (\mathrm{mix})}_{\ \nu}=\sum n_{_{\mathrm{X}}}^\mu \pi^{_{\mathrm{X}}}_\nu + \Psi \delta^\mu_\nu \, ,
  \label{eq.sect.structure.Tliq.multi}
\end{equation}
where X labels matter constituents and $\Psi$ is a \emph{generalized}
pressure, which is not simply given by the sum of the partial pressures
of the various constituents (see Section~\ref{sect.hydro}).

The electromagnetic field can be taken into account by including the
following contribution to the stress-energy tensor
\begin{equation}
  T^{\mu \, (\mathrm{em})}_{\ \nu}=\frac{1}{4\pi} F^{\mu\rho} F_{\nu\rho} -
  \delta^\mu_\nu \frac{1}{8\pi} F^{\sigma\rho} F_{\sigma\rho}\, ,
  \label{eq.sect.structure.Tem}
\end{equation}
where $F_{\mu\nu}$ is the electromagnetic 2-form.
Likewise elastic strains in the solid crust contribute through an additional term 
$T^{\mu\, (\mathrm{elast})}_{\ \nu}$. While the components of
$T^{\mu\, (\mathrm{elast})}_{\ \nu}$ are small compared to $P$ (remember that
the shear modulus is $\sim 10^{-2}$ of $P$, Section~\ref{sect.elast}),
it can produce nonaxial deformations in rotating neutron stars,
and nonsphericity in nonrotating stars, as discussed in Section~\ref{sect.obs.gw}.
For the time being, we will use the ideal-fluid approximation,
$T^{\mu}_{\ \nu}\approx T^{\mu \, (\mathrm{liq})}_{\ \nu}+T^{\mu \, (\mathrm{em})}_{\ \nu}$, and consider
the effect of $T^{\mu\, (\mathrm{elast})}_{\ \nu}$ on neutron star structure
in Section~\ref{sect.obs.gw}. We will, therefore, consider purely-hydrostatic equilibrium of the crust, instead of a more general
hydro-elastic equilibrium.

\subsection{Spherical nonrotating neutron stars}
\label{sect.structure.crust-non-rot}

A static (nonrotating) neutron star, built of a perfect fluid, has
spherical symmetry. The spacetime is also spherically
symmetric, with the metric (see, for instance, Landau \& Lifshitz~\cite{landau-75};
here we use the notation of
Haensel, Potekhin and Yakovlev~\cite{haensel-06})
\begin{equation}
\mathrm{d}s^2=c^2 \mathrm{d}t^2 \mathrm{e}^{2\Phi}-
\mathrm{e}^{2\lambda} \mathrm{d}r^2 - r^2 (\mathrm{d}\theta^2 +
\sin^2\theta \mathrm{d}\phi^2) \, ,
\label{eq.sect.structure.metric}
\end{equation}
where $t$ is a time coordinate, $r$ is a radial coordinate called
``circumferential radius'' (see below), and
$\theta$ and $\phi$ are polar and azimuthal angular
coordinates. The dimensionless metric function $\Phi=\Phi(r)$ and 
$\lambda=\lambda(r)$ have to be determined from Einstein's equations. For
a flat (Minkowski) spacetime we would have $\Phi=\lambda=0$.
One can show that $\lambda(r)$ is determined by the
mass-energy contained within radius $r$,  divided by $c^2$,
to be denoted by $m(r)$~\cite{landau-75},
\begin{equation}
\mathrm{e}^{-\lambda(r)}=\left(1-{2Gm(r)\over r c^2}\right)^{1/2} \, .
\label{eq.sect.structure.lambda.m}
\end{equation}

Let us limit ourselves to the static case $t=const$. Fixing $r$, $\theta=\pi/2$, 
and then integrating $\mathrm{d}s$ over $\phi$ from zero to $2\pi$, 
 we find that the proper length of the equator of the star, i.e., its
 circumference, as  measured by a local observer, is equal to $2\pi
 r$. This is why $r$ is called the circumferential radius.
 Notice that Equation~(\ref{eq.sect.structure.metric})
 implies that the infinitesimal proper radial distance (corresponding to
 the infinitesimal difference of radial coordinates $\mathrm{d}r$) is
 given by $\mathrm{d}\ell=\mathrm{e}^\lambda \mathrm{d}r$.

From Einstein's equations, we get the (relativistic) equations of
hydrostatic equilibrium for a static spherically-symmetric star
\begin{equation}
{\mathrm{d}P\over \mathrm{d} r}= -{G\rho m\over r^2}
\left(1 + {P\over \rho c^2}\right)
\left(1 + {4\pi P r^3\over m c^2}\right)
\left(1 - {2Gm\over c^2 r}\right)^{-1} \, ,
\label{eq.sect.structure.TOV1}
\end{equation}
\begin{equation}
{\mathrm{d}m\over \mathrm{d}r}= 4 \pi r^2 \rho \, ,
\label{eq.sect.structure.TOV2}
\end{equation}
\begin{equation}
{\mathrm{d}\Phi\over \mathrm{d} r}= -{1\over \rho c^2}
{\mathrm{d}P\over \mathrm{d} r} \left(1 +{P\over \rho
c^2}\right)^{-1} \, .
\label{eq.sect.structure.TOV3}
\end{equation}
Equation~(\ref{eq.sect.structure.TOV1}) is the famous
Tolman--Oppenheimer--Volkoff equation of hydrostatic equilibrium~\cite{tolman-39, OV-39}.
Equation~(\ref{eq.sect.structure.TOV2}) enables one to calculate  $m(r)$
within a radius $r$. 
Finally, Equation~(\ref{eq.sect.structure.TOV3}) determines the
metric function $\Phi(r)$.
Equations~(\ref{eq.sect.structure.TOV1})--(\ref{eq.sect.structure.TOV3}) have
to be supplemented with an equation of state (EoS) $P=P(\rho)$.

Let us consider the differential Equations~(\ref{eq.sect.structure.TOV1})
and~(\ref{eq.sect.structure.TOV2}), which
determine the global structure of a neutron star. They are integrated
from the star center, $r=0$, with the boundary conditions $\rho(0)
=\rho_{\mathrm{c}}$ [$P(0)=P(\rho_{\mathrm{c}})$] and $m(0)=0$. It is clear from
Equation~(\ref{eq.sect.structure.TOV1}), that pressure is
strictly decreasing with increasing $r$. The integration is
continued until $P=0$, which corresponds to the surface of the
star, with radial coordinate $r=R$, usually called the
\emph{star radius}.

The \emph{gravitational} mass of the star is defined by
$M=m(R)$.  The mass $M$ is the source of the gravitational
field outside the star ($r>R$), and creates an
outer spacetime described by the Schwarzschild metric,
\begin{equation}
r>R~:~~~~\mathrm{d}s^2=c^2\mathrm{d}t^2\;\left(1-{2GM\over r c^2}\right) -
\mathrm{d}r^2\;\left(1-{2GM\over r c^2}\right)^{-1/2}
 - r^2 (\mathrm{d}\theta^2 + \sin^2\theta \mathrm{d}\phi^2) \, .
\label{eq.sect.structure.metric.Schw.}
\end{equation}

The crust corresponds to the layer $r_{\mathrm{cc}}<r<R$, where
$r_{\mathrm{cc}}$ determines the crust-core interface. The depth below
the stellar surface, $z$,  is defined as the proper radial distance
between the star surface and a given surface of radius $r$. It is
given by
\begin{equation}
z(r)=\int_r^R \mathrm{e}^\lambda \mathrm{d}r \, .
\label{eq.sect.structure.z-def}
\end{equation}
The
structure of the crust depends on its EoS, stellar mass, and the EoS of
its liquid core. In Figures~\ref{fig.sect.structure.Mcr-rho-z-SLy}
and~\ref{fig.sect.structure.Mcr-rho-z-FPS} we present the structure of
the crust of a $1.4\,M_\odot$ star, for two EoSs of the neutron star
interior.

An accreted crust has a different composition and thus a different EoS
(stiffer) than the ground-state crust (see
Sections~\ref{sect.accretion} and~\ref{sect.eos}). For the comparison
to be meaningful, however, these two EoSs have to be calculated from
the same nuclear Hamiltonian. We satisfied this by using in both cases
the same compressible liquid drop model of Mackie \&
Baym~\cite{mackie-77}. The plots of $\Delta M(z)$ for the ground state
and accreted crust of a $1.4\,M_\odot$ neutron star are shown in
Figure~\ref{fig.sect.structure.Mcr-z-GSC-ACC}.

\subsection{Approximate formulae}
\label{sect.structure.approximate}

For astrophysically relevant neutron star masses $M>M_\odot$,
the gravitational mass of the crust $M_{\mathrm{cr}}=M-m(r_{\mathrm{cc}})$, where
$r_{\mathrm{cc}}$ is the radial coordinate of the crust-core interface,
 constitutes less than 3\% of $M$. Moreover, for realistic
 EoSs  and  $M>M_\odot$, the difference $\Delta R=R-r_{\mathrm{cc}}$ does not exceed 15\% of $R$. Clearly, $M_{\mathrm{cr}}/M\ll 1$
 and an approximation in which the terms ${\cal O}(M_{\mathrm{cr}}/M)$
 are neglected is usually sufficiently precise. Neglecting
 the terms ${\cal O}(\Delta R / R)$ gives a less accurate but
 still useful approximation.  In what
 follows we will use the above ``light and thin crust
 approximation'' to obtain useful approximate expressions for the
 crustal parameters.

Let us first derive an approximate equation for hydrostatic equilibrium
within the crust. Let $z$ be the proper depth below the neutron star
surface, Equation~(\ref{eq.sect.structure.z-def}). Within the crust,
one can approximate $z$ by
\begin{equation}
z\approx {R-r\over \sqrt{1-r_{\mathrm{g}}/R}} \, ,
\label{eq.sect.structure.z.approx}
\end{equation}
where $r_{\mathrm{g}}=2GM/c^2$ is the Schwarzschild radius of mass
$M$.

In Equation~(\ref{eq.sect.structure.TOV1}) one can use the approximation
$m\simeq M$, and neglect $P/\rho c^2$ and $4\pi P r^3/Mc^2$
as compared to one. Then the equation for hydrostatic equilibrium
can be rewritten in a Newtonian form,
\begin{equation}
{\mathrm{d}P\over \mathrm{d}z}= g_{\mathrm{s}}\rho \, ,
\label{eq.sect.structure.TOV.z}
\end{equation}
where $g_{\mathrm{s}}$ is the surface gravity,
\begin{equation}
g_{\mathrm{s}}={GM\over R^2 \sqrt{1-r_{\mathrm{g}}/R}} \, .
\label{eq.sect.structure.g_s.def}
\end{equation}
 Using $P(z=0)=0$,  we obtain a formula relating the pressure at depth $z$
 to the mass $\Delta M(z)$ of the crust layer above $z$, 
\begin{equation}
\Delta M(z)=4\pi {R^2\over g_{\mathrm{s}}} \sqrt{1 -\frac{r_{\mathrm{g}}}{R}}~ P(z) \, .
\label{eq.sect.structure.P.z}
\end{equation}
In particular, putting $z=z_{\mathrm{cc}}$, and denoting the
pressure at the crust-core interface by $P_{\mathrm{cc}}$, we get
\begin{equation}
M_{\mathrm{cr}}
\approx {4\pi R^2 P_{\mathrm{cc}}\over GM}\left(1-{r_{\mathrm{g}}\over
R}\right) \, .
\label{eq.sect.structure.Mcr}
\end{equation}

\epubtkImage{Mcr-rho-SLy-rho-z-SLy.png}{%
  \begin{figure}[htbp]
    \centerline{\includegraphics[scale=0.3]{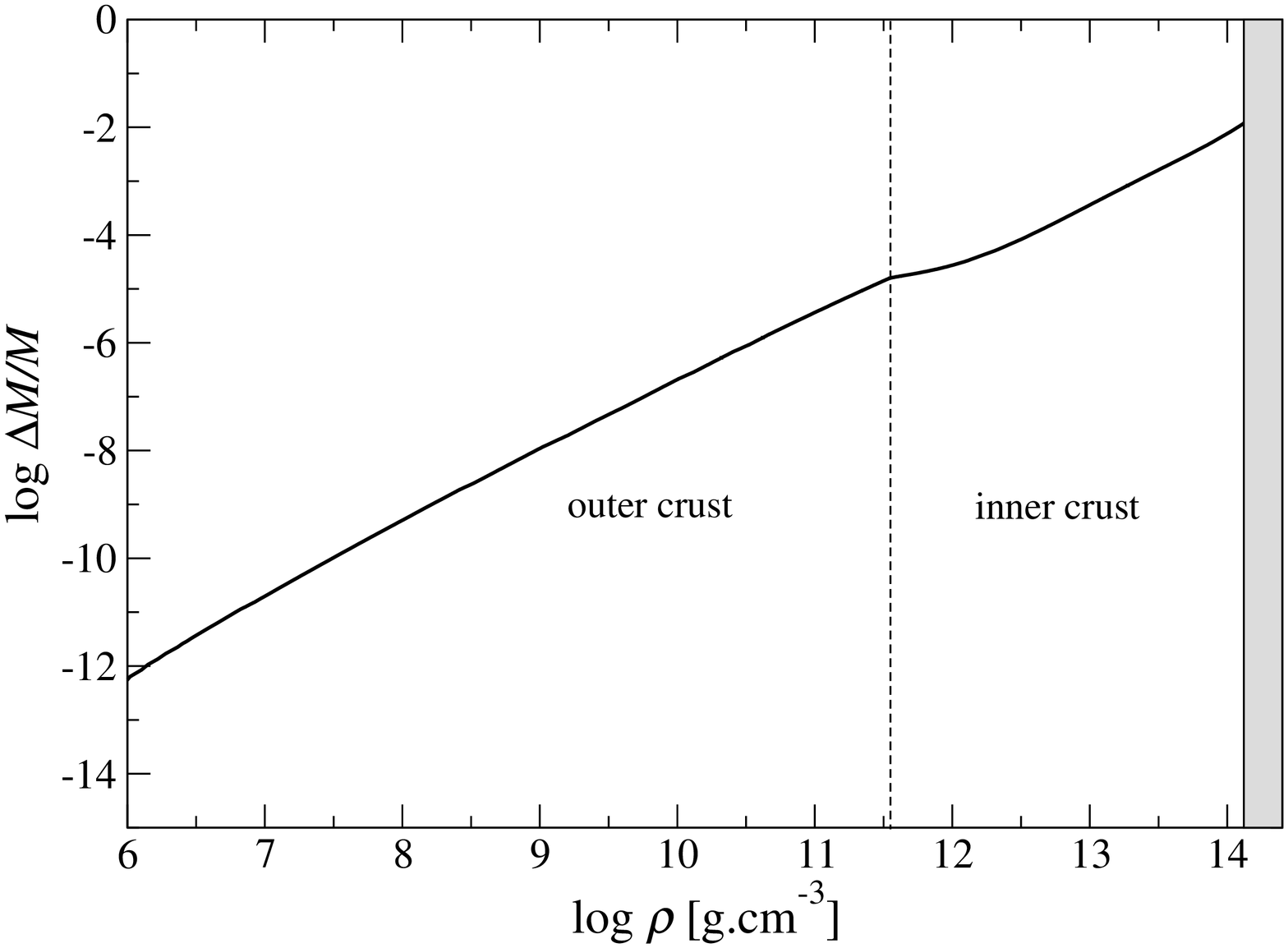}\includegraphics[scale=0.3]{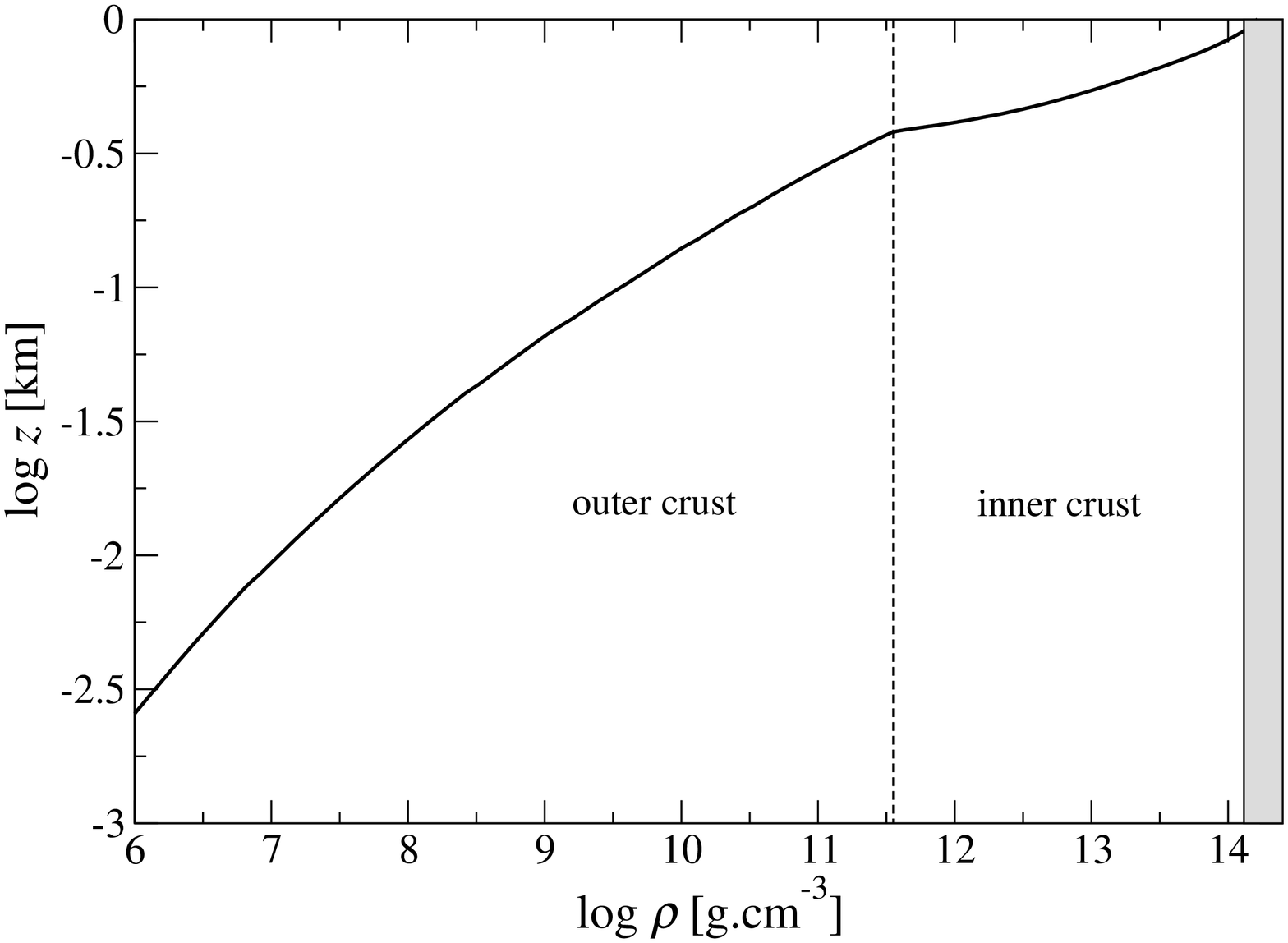}}
    \caption{Mass of the crust for the SLy EoS~\cite{douchin-01}. The
      neutron star mass is $M=1.4\,M_\odot$. For this EoS, spherical
      nuclei persist down to the crust-core interface. Left panel:
      fractional mass of the crust shell, $\Delta M/M$, vs.\ its
      bottom density $\rho$. Right panel: proper depth below the star
      surface, $z$ vs.\ mass density $\rho$.}
    \label{fig.sect.structure.Mcr-rho-z-SLy}
\end{figure}}

\epubtkImage{Mcr-rho-FPS-rho-z-FPS.png}{%
  \begin{figure}[htbp]
    \centerline{\includegraphics[scale=0.3]{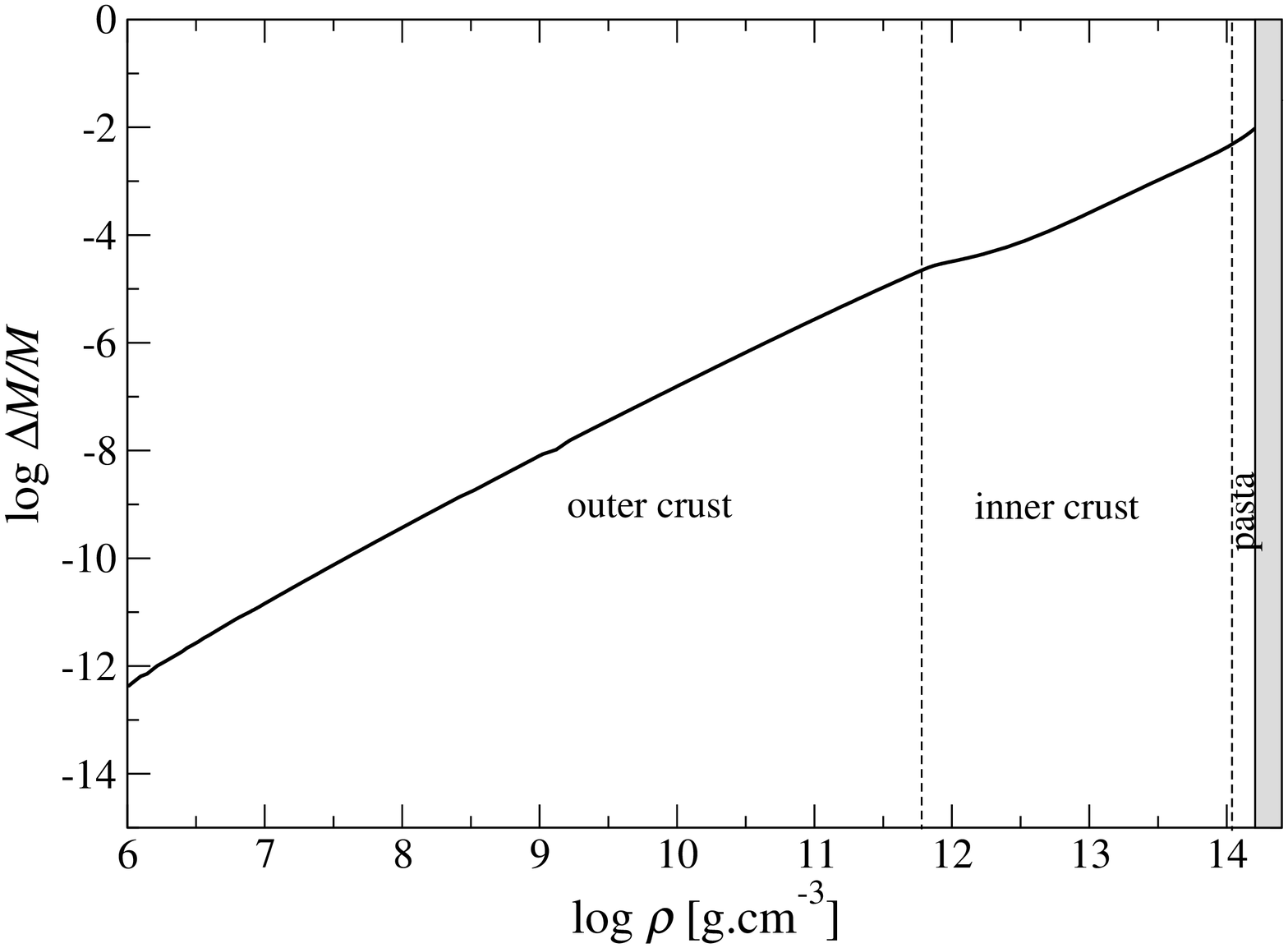}\includegraphics[scale=0.3]{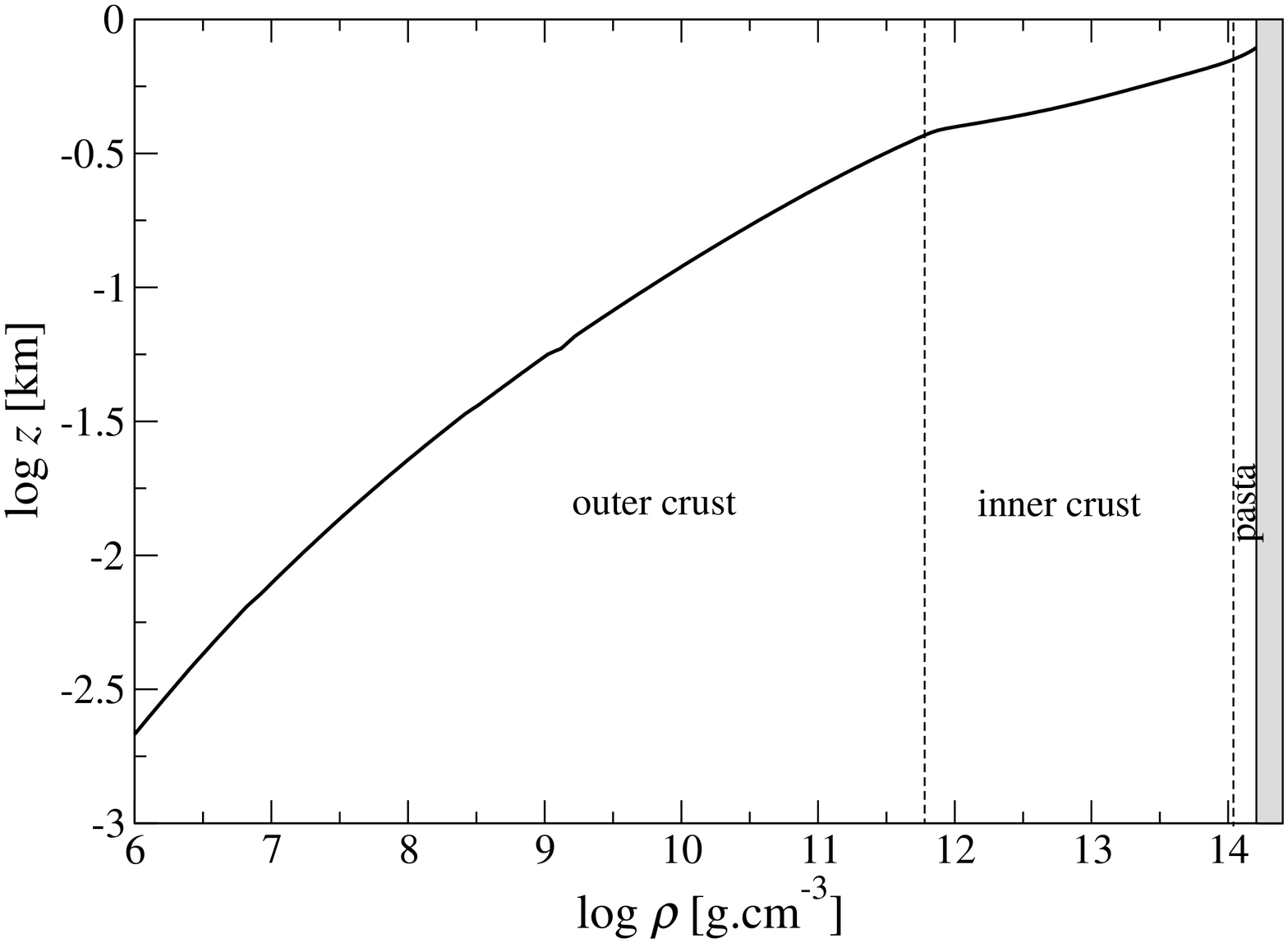}}
    \caption{Mass of the crust for the FPS EoS~\cite{lorenz-93}. The
      neutron star mass is $M=1.4\,M_\odot$. Notice the presence of
      the pasta layer, which are absent for the SLy EoS. The pasta
      phases occupy a thin density layer, but contain about 48\% of
      the crust mass. Left panel: fractional mass of the crust shell,
      $\Delta M/M$, vs.\ its bottom density $\rho$. Right panel:
      proper depth below the star surface, $z$ vs.\ mass density
      $\rho$.}
    \label{fig.sect.structure.Mcr-rho-z-FPS}
\end{figure}}

\epubtkImage{Mcr-z-GSC-ACC.png}{%
  \begin{figure}[htbp]
    \centerline{\includegraphics[scale=0.3]{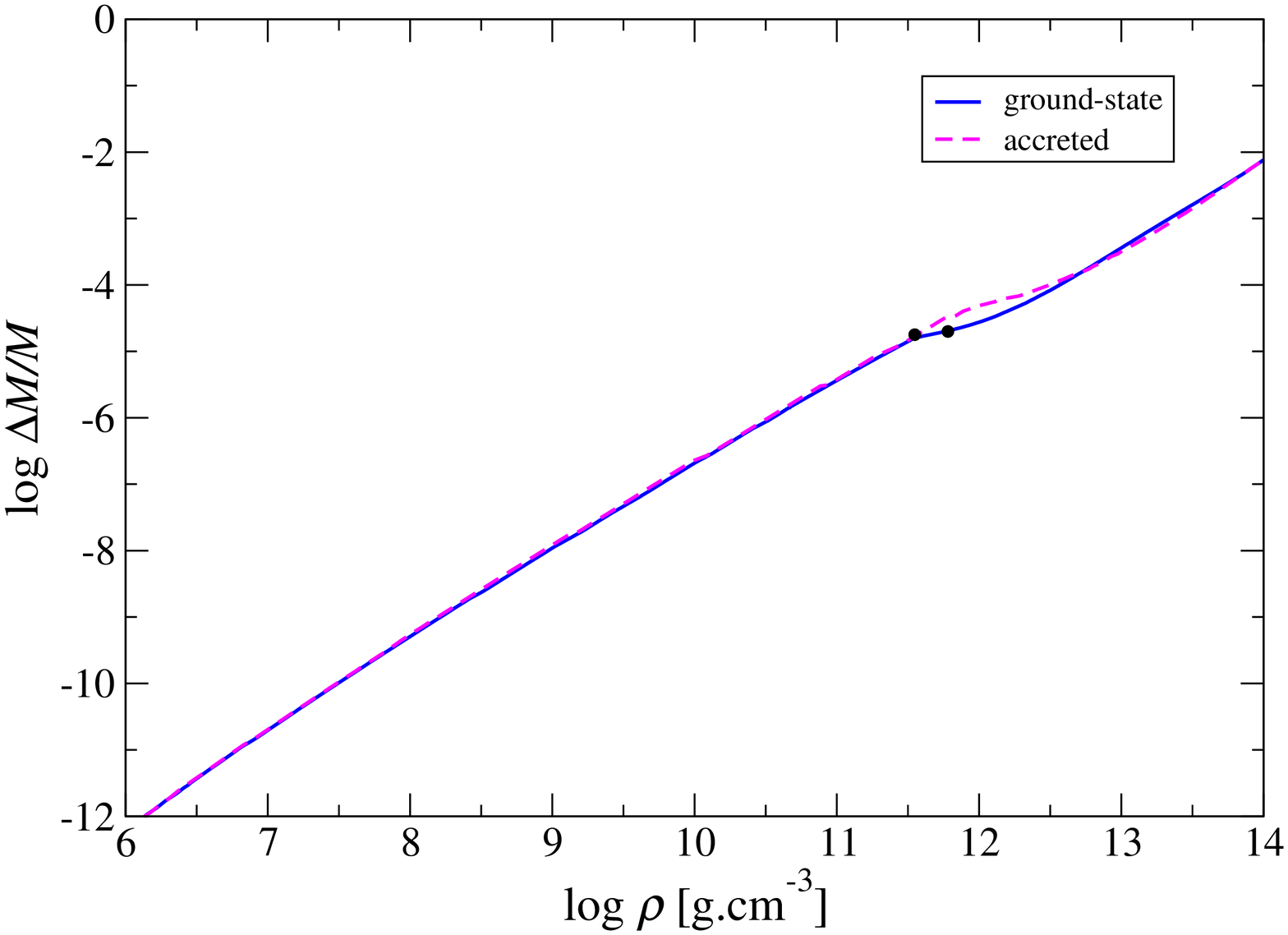}\includegraphics[scale=0.3]{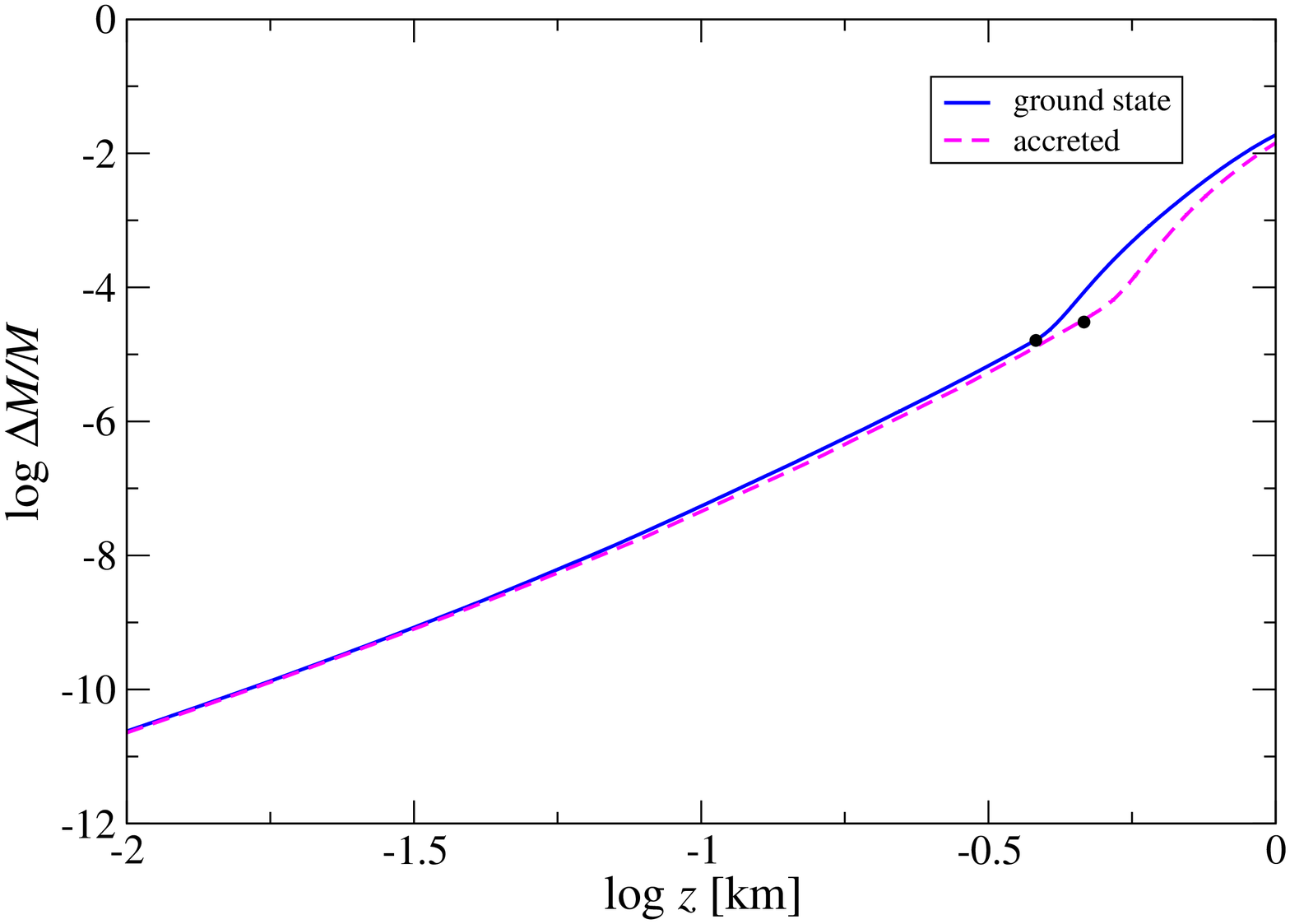}}
    \caption{Mass of the crust shell for the ground-state crust and
    for the accreted crust. The total stellar mass is
    $M=1.4\,M_\odot$. Accreted crust: EoS of Haensel \&
    Zdunik~\cite{haensel-90b}. Ground-state crust: same compressible
    liquid drop model of atomic nuclei of Mackie \&
    Baym~\cite{mackie-77}, but full thermodynamic equilibrium (cold
    catalyzed matter). The black dots indicate the neutron drip
    transition. Left panel: fractional mass of the crust shell,
    $\Delta M/M$, vs. its bottom density $\rho$. Right panel: $\Delta
    M/M$ versus depth below the star surface, $z$.}
    \label{fig.sect.structure.Mcr-z-GSC-ACC}
\end{figure}}

Consider a slow rigid rotation, meaning that the angular frequency of
rotation, $\Omega$, as measured by an observer at infinity,
 is small compared to the mass
shedding frequency, $\Omega_{\mathrm{ms}}$.  It can be shown that for slow
rigid rotation the moment of
inertia, $I$, then involves only the structure of the nonrotating neutron
star~\cite{hartle-67}
(see also \cite{glendenning-00, haensel-06}).
Using our notation,
\begin{equation}
I= {8\pi\over 3}\int_0^R\mathrm{d}r\, r^4 \left( \rho +{P\over
c^2}\right){\bar{\omega}\over \Omega} \mathrm{e}^{-\lambda
-\Phi} \, ,
\label{eq.sect.structure.Itot}
\end{equation}
where $\bar{\omega}$ is the local rotation frequency, as
measured in the local inertial frame. It can be calculated
from an ordinary differential equation derived by Hartle~\cite{hartle-67},
with boundary conditions,
\begin{equation}
\bar{
\omega}(R)=\Omega-{2GJ\over R^3 c^2} \, ,
\label{eq.sect.structure.Omega.R}
\end{equation}
where $J$ is the stellar angular momentum.

The contribution of the crust comes from the spherical outer
shell, $r_{\mathrm{cc}}<r<R$,
\begin{equation}
I_{\mathrm{cr}}= {8\pi\over 3}\int_{r_{\mathrm{cc}}}^R
\mathrm{d}r\, r^4 \rho {\bar{\omega}\over \Omega}
 \mathrm{e}^{-\lambda
-\Phi} \, ,
\label{eq.sect.structure.Icr-int}
\end{equation}
where $P/\rho c^2$ was neglected compared to one. Approximating
the integrand by its value at $r=R$, and using
$\mathrm{e}^{\Phi(R)}=\mathrm{e}^{-\lambda(R)}=(1-r_{\mathrm{g}}/R)^{1/2}$
and Equations~(\ref{eq.sect.structure.Omega.R})
and~(\ref{eq.sect.structure.Mcr}), we get
\begin{equation}
I_{\mathrm{cr}}={2\over 3} M_{\mathrm{cr}} R^2 \left(1- {r_{\mathrm{g}}\over
R} {I\over MR^2}\right) \, .
\label{eq.sect.structure.Icr-app}
\end{equation}
%

\subsection{Crust in rotating neutron stars}
\label{sect.structure.crust-rot}

We consider stationary neutron-star rotation and assume, for
the time being, perfect axial symmetry. Consequently, the spacetime metric
is axially symmetric too. Using coordinates $t$, $r$, $\theta$, and $\phi$,
we can write the spacetime metric in the form (we use the
notation of Haensel, Potekhin and Yakovlev~\cite{haensel-06})
\begin{equation}
\mathrm{d}s^2=c^2\mathrm{d}t^2\mathrm{e}^{2\Phi}-\mathrm{e}^{2\lambda}r^2
\sin^2\theta\;(\mathrm{d}\phi-\omega \mathrm{d}t)^2 - \mathrm{e}^{2\alpha}
(\mathrm{d}r^2 + r^2 \mathrm{d}\theta^2) \, ,
\label{eq.sect.structure.metric.rot}
\end{equation}
where the metric
functions $\Phi$, $\lambda$, $\omega$, and $\alpha$ depend on
$r$ and $\theta$. The metric function $\omega(r,\theta)$ has
the asymptotic behavior $\omega(r\longrightarrow \infty)= 0$.
It is usually referred to as the ``angular velocity of the
local inertial frames''. Einstein's equations for a stationary
rigid rotation of perfect fluid stars can be reduced to a set
of 2-D coupled partial differential
equations~\cite{bonazzola-93, stergioulas-03}.
 Their numerical solution can now be obtained using
 publicly available domain codes, for example, the code {\tt rotstar} from the
LORENE library~\cite{lorene} and the code {\tt RNS}~\cite{rns}.

\epubtkImage{rot-crust-NS.png}{%
\begin{figure}[htbp]
  \centerline{\includegraphics[scale=0.5]{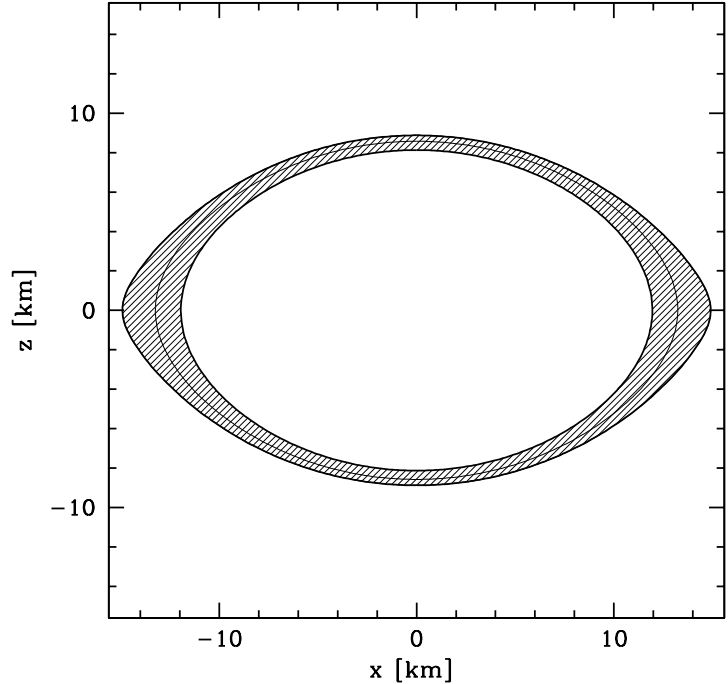}}
  \caption{Cross section in the plane passing through the rotation
 axis of a neutron star of $M=1.82\,M_\odot$, rotating at 1200~Hz.
 The SLy EoS for crust and core~\cite{douchin-01} is used. The coordinates are
 defined by: $x=r \sin \theta \cos \phi$ and $z=r \cos \theta$, while $r$, $\theta$, and 
$\phi$ are metric coordinates,
 Equation~(\ref{eq.sect.structure.metric.rot}). The contours are lines
 of constant density. Inner contour: crust-core
 interface. Intermediate contour: outer-inner crust interface. Outer
 contour: stellar surface. Figure made by J.L.\ Zdunik.}
\label{fig.sect.structure.rot.crust.NS}
\end{figure}}

\epubtkImage{B-crust.png}{%
\begin{figure}[htbp]
  \centerline{\includegraphics[scale=0.7]{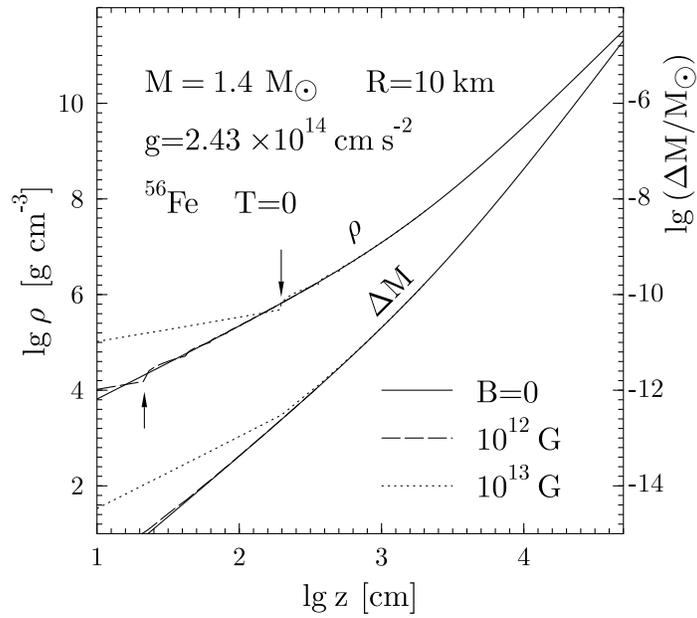}}
  \caption{Nonmagnetized and magnetized pure $^{56}$Fe
    crust in a neutron star with $M=1.4\,M_\odot$ and
    $R=10$~km and at  $T=0$.
    Matter density (left vertical scale) and the mass
    of the outer shell $\Delta M$ (right vertical axis) versus depth
    below the surface $z$, for $B=0$, $B=10^{12}$~G, and
    $B=10^{13}$~G. Arrows indicate densities (kinks) at which the
    $n=1$  Landau level starts to be populated  with increasing
    depth. Figure~6.15 from~\cite{haensel-06}.}
  \label{fig.sect.structure.B-crust}
\end{figure}}

As can be seen on
Figure~\ref{fig.sect.structure.rot.crust.NS}, the outer crust
is the most strongly deformed by the centrifugal force. The
spin frequency of 1200~Hz, significantly larger than the
highest detected up to now, 716~Hz, was chosen to make the
effect more spectacular. Let us nevertheless point out that
some evidence  suggesting the existence of  a rapidly-rotating
neutron star has recently been found in XTE~J1739$-$285~\cite{kaaret-07} with a spin frequency of about 1122~Hz.
Let us consider, for example, the difference between the
equatorial (maximal) and polar (minimal) thickness of the
crust, $\Delta R_{\mathrm{eq-pol}}= \Delta R_{\mathrm{eq}}-\Delta
R_{\mathrm{pol}}$. For a $\Omega$, which is not too close to the
mass-shedding limit, $\Delta R_{\mathrm{eq-pol}}\propto
\Omega^2$. Therefore, $\Delta R_{\mathrm{eq-pol}}|_{716 \mathrm{\ Hz}}
\approx 0.4\Delta R_{\mathrm{eq-pol}}|_{1200 \mathrm{\ Hz}}$.

The crustal \emph{baryon} mass (not to be confused with the
gravitational mass) $M_{\mathrm{b, cr}}(\Omega)$ of a neutron star
rotating at angular frequency $\Omega$, is larger than the
crustal baryon mass $M_{\mathrm{b, cr}}(0)$ of the static star (with
the same total baryon mass). The baryon mass (also called the rest
mass) of a star is equal to the number $A$ of baryons it
contains times an assumed baryon mass $m_{\mathrm{b}}$. One may
take $m_{\mathrm{b}}=m_n$ or $m_{\mathrm{b}}=m_{\mathrm{u}}$. We take
$M_{\mathrm{b, cr}}=A m_{\mathrm{u}}$. For $\Omega$ not too close to
$\Omega_{\mathrm{ms}}$, we have $M_{\mathrm{b, cr}}(\Omega)-M_{\mathrm{b, cr}}(0) \propto
\Omega^2$. Due to the radiation of electromagnetic waves and
particles, a pulsar spins down, so that $\dot{\Omega}<0$.
Consequently, the baryon mass of the pulsar crust decreases in
time, $\dot{M}_{\mathrm{b, cr}}\propto \dot{\Omega}{\Omega}<0$.
Nucleons pass from the crust to the liquid core, releasing
some heat. As shown in
Figure~\ref{fig.sect.structure.rot.crust.NS}, the crust is
decompressed near the equator and  compressed near the pole.
These deformations trigger various nuclear reactions involving
electrons, neutrons, and nuclei. These reactions tend to drive
the deformed crust towards its equilibrium shape and release
heat, which influences the cooling of a spinning down
pulsar~\cite{iida-97}. Additional heating results from crust
cracking when local shear strain exceeds the maximal one.

\subsection{Effects of magnetic fields on the crust structure}
\label{sect.structure.B}

The structure of the outer layers of the crust (neutron star
``envelope'') can be affected by the presence of a magnetic
field. Effects of magnetic fields on the EoS were briefly
mentioned in Section~\ref{sect.plasma.B}. Here we consider
examples showing effects of $\pmb{B}$ on the crust structure.

Typical values of the surface magnetic fields of radio pulsars
are $B\sim 10^{12}$~G. In
Figure~\ref{fig.sect.structure.B-crust} we compare plots of
$\rho(z)$ and $\Delta M(z)$ for the pure iron envelope with
$\rho\lesssim 10^5~\mdens$ of a neutron star with
$M=1.4\;M_\odot$ and $R=10~$km with and without a magnetic field.
The $T=0$ approximation for the crust is used. Typical values of
the surface magnetic field of radio pulsars are $B\sim
10^{12}$~G. At such magnetic fields, the effect of $B$ is seen
only in the outer envelope, which is $\sim 30~$cm thick. We can see
there quantum oscillations of the density as a function of
depth. They are associated with the filling of the lowest  Landau
levels  $n=0,1,...$ by the electrons
(Section~\ref{sect.plasma.B}). Increasing $B$ to $10^{13}\;$G,
associated with the most magnetized radio pulsars, leads to
much deeper magnetization of the crust, down to the depth of 30~m,
where the prevailing density reaches $10^{6}~\mdens$. The
effect of the magnetic field in the outer 30 cm of the crust is dramatic; in spite
of a gravitational acceleration $g=GM/R^2=2.43\times
10^{14} \mathrm{\ cm\ s}^{-2}$, the density is only slowly
decreasing, and is  still $\sim 10^5~\mdens$ at 10~cm depth,
ten times higher than in a nonmagnetized envelope at the same
depth. So, magnetized iron plasma is ``condensed'', and much
less compressible than nonmagnetized (Section~\ref{sect.plasma.B}). At $10^{13}$~G,  the $n=1$
Landau level begins to be populated only at $4\times
10^5~\mdens$ (depth 3~m), to be compared with $2\times
10^4~\mdens$ (depth 30~cm) at $10^{12}$~G. It should be
stressed that, as the surface temperature at pulsar age
$10^3\mbox{\,--\,}10^4$~y is $T_{\mathrm{s}}\sim 10^6$~K, inclusion of $T$
will weaken the  magnetization effects on the structure of the
crust~\cite{haensel-06}.

Typical surface magnetic fields of magnetars are $\sim
10^{15}$~G. In Figure~\ref{fig.sect.structure.B15G-crust} we
see that for $\rho<10^9~\mdens$, the effect of such a huge
$\pmb{B}$ on the crust EoS is strong and becomes dramatic at
lower $\rho$. For example, at $10^7~\mdens$, the matter
pressure decreases by two orders of magnitude, compared to the
$\pmb{B}=0$ case. The $n=1$ Landau level begins to be
populated only at $10^{8.7}~\mdens$, at a depth of about 50~m.

\epubtkImage{EOS-B15-crust.png}{%
\begin{figure}[htbp]
  \centerline{\includegraphics[scale=0.7]{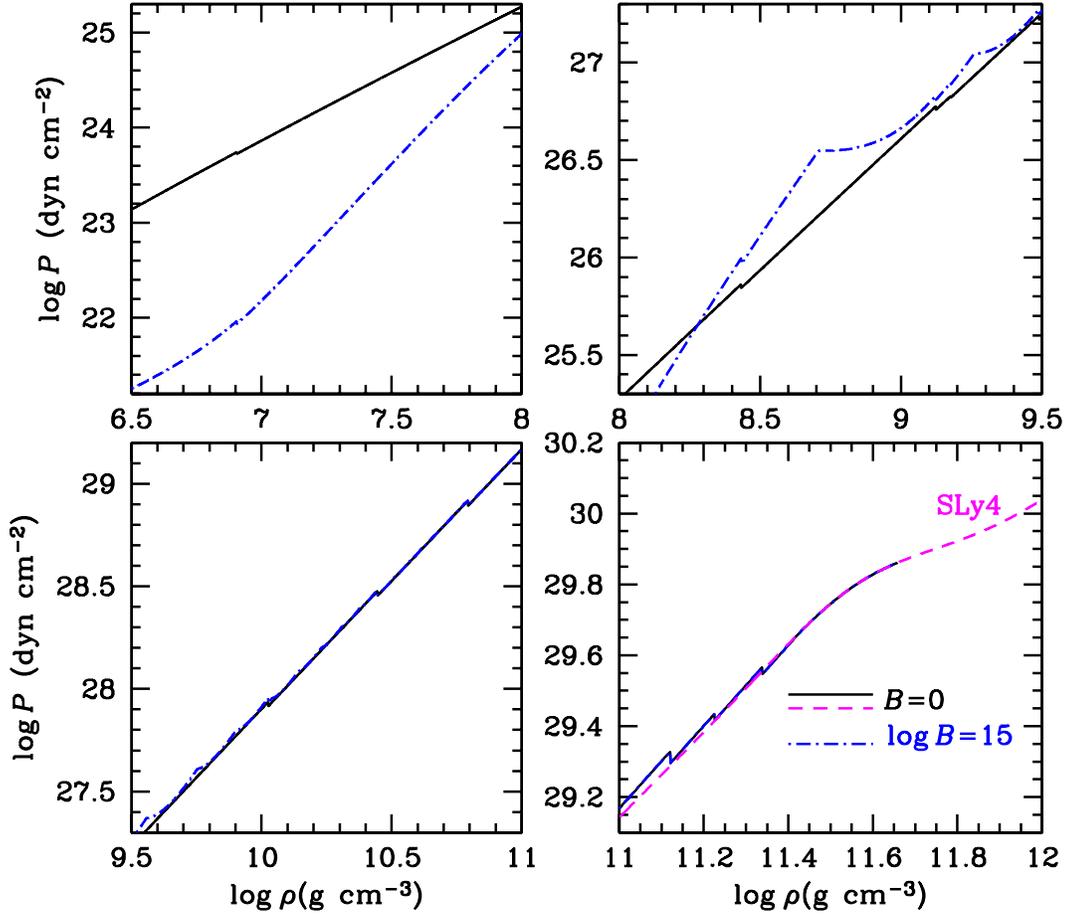}}
  \caption{Nonmagnetized (solid line),
  and magnetized  crust (dash-dotted line) calculated  with
  ground state composition calculated  at $\pmb{B}=0$.
  The dashed magenta line in the bottom-right panel corresponds to the
  $\pmb{B}=0$ compressible liquid drop model of \cite{douchin-01},
  and is smooth due to its quasiclassical nature, while the curve
  with discontinuous drops in pressure as obtained using
  the ground-state model of \cite{haensel-94}.
  Figure made by A.Y.\ Potekhin.}
  \label{fig.sect.structure.B15G-crust}
\end{figure}}

\newpage


\section{Elastic Properties}
\label{sect.elast}

A solid crust can sustain an {\it elastic strain} up to a critical
level, the breaking strain. Neutron stars are relativistic objects,
and therefore a relativistic theory of elastic media in a curved
spacetime should be used to describe elastic effects in neutron star
structures and dynamics. Such a theory of elasticity has been developed
by Carter \& Quintana~\cite{carter-72}, who applied it to rotating
neutron stars in ~\cite{carter-75a, carter-75b} (see also
Beig~\cite{beig-03} and references therein). Recently, Carter and
collaborators have extended this theory to include the effects of the
magnetic field~\cite{carter-06}, as well as the presence of the
neutron superfluid, which permeates the inner
crust~\cite{carterchachoua-06, cartersamuelsson-06}. For the time
being, for the sake of simplicity, we ignore  magnetic fields
 and free neutrons. However, in  Section~\ref{sect.elast.pasta}
 the effect of free neutrons on the elastic moduli of  the pasta phases is
 included,  within  the compressible liquid drop model.
 Since relativistic effects are not very large in the
crust, we shall restrict ourselves to the Newtonian approximation
(see, e.g., \cite{landau-86}).

The thermodynamic equilibrium of an element of neutron-star crust
corresponds to \emph{equilibrium positions} of nuclei, which will be
denoted by a set of vectors $\lbrace{\pmb r}\rbrace$, which are 
associated with the lattice sites.  Neutron star evolution, driven by
 spin-down, accretion of matter or some external forces,
 like tidal forces produced by   a close
massive body, or internal electromagnetic strains associated with strong
magnetic fields,  may lead to \emph{deformation} of this
crust element as compared to the equilibrium state.

For simplicity, we will neglect thermal contributions  to
thermodynamic quantities and restrict ourselves to the $T=0$
approximation. Deformation of a crust element with respect to the
equilibrium configuration implies a {\it displacement}
of nuclei into their new positions ${\pmb r}^\prime = {\pmb r} + {\pmb
  u}$, where ${\pmb u}={\pmb u}({\pmb r})$ is the displacement
vector. In the continuum limit, valid for  macroscopic phenomena, both
${\pmb r}$ and ${\pmb u}$ are treated as continuous fields. Nonzero
${\pmb u}$ is associated with \emph{elastic strain} (i.e., forces
which tend to return the matter element to the equilibrium state of
minimum energy density $\varepsilon_0$), and  with the \emph{deformation
  energy} density  $\varepsilon_{\mathrm{def}}=\varepsilon({\pmb
  u})-\varepsilon_0$~\epubtkFootnote{In this section, by ``energy'' we
  will usually mean energy of a unit volume (i.e.\ energy density).}.

A uniform translation does not contribute to $\varepsilon_{\mathrm{def}}$,
and the true  deformation is described by the (symmetric)
\emph{strain tensor}
\begin{equation}
{\tt u}_{ik}={\tt u}_{ki}={1\over 2}
\left(
{\partial u_i\over \partial x_k} +
{\partial u_k\over \partial x_i}
\right) \, ,
\end{equation}
where $i,j=1,2,3$, and $x_1=x,~x_2=y,~x_3=z$. This formula for ${\tt
  u}_{ik}$ is valid if the displacement vector  ${\pmb u}$ is  small,
  so that the terms quadratic in the components of ${\bf u}$ can be
  neglected compared to the linear ones~\cite{landau-86}.

Any  deformation can be decomposed into \emph{compression} and
\emph{shear} parts,
\begin{eqnarray}
{\tt u}_{ik}={\tt u}^{\mathrm{comp}}_{ik}+{\tt u}^{\mathrm{shear}}_{ik} \, , \mathrm{where}~~~
{\tt u}^{\mathrm{comp}}_{ik}={1\over 3}({\pmb \nabla}\cdot{\pmb u})\delta_{ik} \, ,
~{\tt u}^{\mathrm{shear}}_{ik}=
{\tt u}_{ik}-{1\over 3}({\pmb \nabla}\cdot{\pmb u})\delta_{ik} \, .
\end{eqnarray}
Under deformation, matter element volume  changes according to
$\mathrm{d}V^\prime =[1+({\pmb \nabla}\cdot{\pmb u})]\mathrm{d}V$. A pure
compression (no shear) of the matter element is described by  ${\tt
  u}_{ik}=\alpha\delta_{ik}$. For volume preserving  deformations
  $({\pmb \nabla}\cdot{\pmb u})=0$.

To lowest order, the deformation energy is quadratic in the deformation
tensor,
\begin{equation}
\varepsilon_{\mathrm{def}}=\sum_{iklm}{1\over 2}\lambda_{iklm}{\tt u}_{ik}{\tt
u}_{lm} \, .
\end{equation}
Since $\varepsilon_{\mathrm{def}}$ is a scalar,  $\lambda_{iklm}$ is a fourth rank
tensor. The total number of $\lambda_{iklm}$ components is
eighty one; general symmetry relations reduce the maximum number of
linearly-independent components (elastic moduli) to twenty one. The
number of independent elastic moduli decreases with increasing
symmetry of the elastic medium. It is  three in the case of a cubic
crystal, and two for an isotropic solid (see, e.g., \cite{landau-86}).

The elastic contribution to the  stress tensor
$\Pi_{ik}^{\mathrm{elast}}\equiv \sigma_{ik}$ is $\sigma_{ik}=\partial
\varepsilon_{\mathrm{def}}/\partial {\tt u}_{ik}$.

\subsection{Isotropic solid (polycrystal)}
\label{sect.elast.isotr.solid}

Microscopically, the ground state corresponds to a body-centered cubic
(bcc) crystal lattice. However, one usually assumes that
macroscopically, the neutron star crust is an  isotropic bcc
polycrystal. Elastic properties of an isotropic solid are described by
two elastic moduli. The deformation energy  can be  expressed as
\begin{equation}
\varepsilon_{\mathrm{def}}= {1\over 2} K ({\pmb \nabla}\cdot {\pmb u})^2
+\mu \left(u_{ik} -
{1\over 3}\delta_{ik}\;({\pmb \nabla}\cdot {\pmb u})
\right)^2 \, .
\end{equation}
Here, $\mu$ is the {\it shear modulus} and $K$ is the {\it compression
  modulus}. The elastic stress tensor is, therefore,
\begin{equation}
\sigma_{ik}={\partial \varepsilon_{\mathrm{def}}\over
\partial u_{ik}} = K ({\pmb \nabla}\cdot {\bf u})\delta_{ik} +
2\mu \left(u_{ik} - {1\over 3}({\pmb \nabla}\cdot {\bf u})
\delta_{ik}\right) \, .
\end{equation}
Considering a small pure uniform compression, one finds
\begin{equation}
K=n_{\mathrm{b}}{\partial P\over \partial n_{\mathrm{b}}}=\gamma P \, ,
\end{equation}
where $P$ is the total pressure given by
Equation~(\ref{sect.groundstate.outer.P}) and $\gamma$ is the
adiabatic index defined by Equation~(\ref{eq.sect.eos.gamma}).

Monte Carlo calculations of the effective shear modulus of a
polycrystalline bcc Coulomb solid were performed by Ogata \&
Ichimaru~\cite{ogata-90}. The deformation energy, resulting from the
application of a specific strain ${\tt u}_{ik}$, was evaluated through
Monte Carlo sampling.

As we have already mentioned, for an ideal cubic crystal lattice there
are only three independent elastic moduli, denoted traditionally as
$c_{11}$, $c_{12}$ and $c_{44}$ (see Chapter~3, pp.~80\,--\,87 of the
book  by Kittel~\cite{kittel-96}). For pure shear deformation,
only two independent elastic moduli are relevant,
\begin{equation}
\varepsilon_{\mathrm{def}}=b_{11}
\left(u_{xx}^2 + u_{yy}^2 + u_{zz}^2\right)
 + 2c_{44}\left(u_{xy}^2 + u_{xz}^2 + u_{yz}^2\right)
 ~~~\mathrm{for}~~
({\pmb \nabla}\cdot {\pmb u})=0 \, ,
\end{equation}
because  $b_{11}={1\over 2}(c_{11}-c_{12})$.
At $T=0$,  Ogata \& Ichimaru~\cite{ogata-90} find
\begin{equation}
b_{11}=0.0245\; n_{\mathrm{N}}\;\frac{(Ze)^2}{R_{\mathrm{cell}}} \, ,
\end{equation}
\begin{equation}
c_{44}=0.1827\; n_{\mathrm{N}}\;\frac{(Ze)^2}{R_{\mathrm{cell}}}\, .
\end{equation}
These values agree with the classical result of Fuchs~\cite{fuchs-36}.

\epubtkImage{crust18.png}{%
\begin{figure}[htbp]
  \centerline{\includegraphics[scale=0.6]{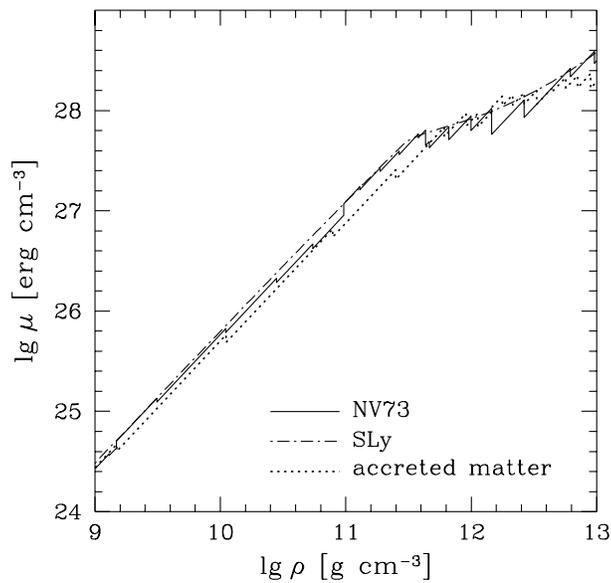}}
  \caption{Effective shear modulus $\mu$ versus density,
    for a  bcc  lattice. Solid line -- cold
    catalyzed matter (Haensel and Pichon 1994 model~\cite{haensel-94}
    for the outer crust (Section~\ref{sect.groundstate.outer}), and
    that of Negele and Vautherin 1973~\cite{nv-73} for the inner crust
    (Section~\ref{sect.groundstate.inner.quantum})). Dash-dotted line
    -- cold catalyzed matter calculated by Douchin and Haensel
    2000~\cite{douchin-00} (compressible liquid drop model, based on
    SLy effective N-N interaction, Section ~\ref{sect.groundstate.inner.LDM}).
    Dotted line --
    accreted crust model of Haensel and Zdunik 1990~\cite{haensel-90b}
    (Section~\ref{sect.accretion}). Figure made by A.Y.\ Potekhin.}
  \label{fig.sect.elast.mu}
\end{figure}}

The definition  of an ``effective'' shear modulus of a bcc polycrystal
deserves a comment. In numerous papers, a standard preferred choice
was $\mu=c_{44}$ (\cite{baym-71, pandharipande-76, mcdermott-88}, and
references therein). However, replacing $\mu$ by a single {\it
  maximal} elastic modulus of a strongly anisotropic bcc lattice is not
correct. An effective value of $\mu$ was calculated by Ogata \&
Ichimaru~\cite{ogata-90}. They  performed directional averages over
rotations of the Cartesian axes. At $T=0$, they obtained
\begin{equation}
\mu={1\over 5} \left(2b_{11}+3c_{44}\right)=
0.1194\; {n_{\mathrm{N}}\left(Ze\right)^2\over R_{\mathrm{cell}}} \, ,
\label{sect.elast.mueff}
\end{equation}
some 30\% smaller than $\mu =c_{44}$ used in previous
papers. Dependence of $\mu$ on temperature was studied, using the
Monte Carlo method, by Strohmayer et al.~\cite{strohmayer-91}. As
expected, the effective shear modulus decreases with increasing
temperature.

The formula for $\mu$, Equation~(\ref{sect.elast.mueff}), can be
rewritten as
\begin{equation}
\label{sect.elast.muP}
\mu=0.0159~\left(\frac{Z}{26}\right)^{2/3}~P_e \, ,
\end{equation}
where $P_e$ is the pressure of ultrarelativistic degenerate
electrons. Therefore,
\begin{equation}
\frac{\mu}{K} = 0.016~\left(\frac{Z}{26}\right)^{2/3}~\frac{P_e}{\gamma P}~ \ll 1 \, .
\end{equation}
The Poisson coefficient $\sigma\simeq 1/2$, while the
Young modulus $E\simeq 3\mu$.

Let us remember that the formulae given above hold for the outer crust,
where the size of the nuclei is very small compared to the lattice
spacing and $P\simeq P_e$. For the inner crust these formulae are only
approximate.

\subsection{Nuclear pasta}
\label{sect.elast.pasta}

Some theories of dense matter predict the existence of ``nuclear pasta''
--  rods, plates, tubes, bubbles -- in the bottom layer of the crust
with $\rho\gtrsim 10^{14}~\mdens$ (see
Section~\ref{sect.groundstate.pasta}). In what follows we will
concentrate on rods (\emph{spaghetti}) and plates
(\emph{lasagna}). They are expected to fill most of the bottom crust
layers. The matter phases containing rods and plates have  properties
intermediate between solids and liquids. The displacement of an element
of matter parallel to the plane containing  rods or plates is not
opposed by restoring forces: this lack of a shear strain is typical
for a liquid. On the contrary, an elastic strain opposes any bending
of planes or rods: this is a property of a solid. Being intermediate
between solids and liquids, these kinds of matter are usually called
\emph{mesomorphic} phases, or \emph{liquid crystals} (see,
e.g., \cite{landau-86}).

The elastic properties of rod and plate phases of neutron star matter were
studied by Pethick \& Potekhin~\cite{pethick-98}. As they stressed,
the physical reasons for the forming of mesomorphic phases in neutron star crusts
are very different from those relevant to liquid crystals in
laboratory. For terrestrial liquid crystals it is the interaction
between very nonspherical molecules, which drives them to form rods
and plates. In a neutron star crust the mechanism consists
in  {\it spontaneous symmetry breaking},  resulting from competition between the Coulomb energy and nuclear surface energy. We
will follow closely Pethick \& Potekhin~\cite{pethick-98}.
They calculated the energies of mesomorphic phases using the
generalized liquid drop model.  The plate phase has rotational
symmetry about any axis perpendicular to the plates, and is therefore
similar to the \emph{smectics A} phase of liquid
crystals~\cite{degennes-93}.  Let the z-axis coincide with the
symmetry axis of the equilibrium configuration. Only a displacement in the
z-direction is opposed by a restoring force. Therefore,  we consider
only ${\pmb u}=(0,0,u)$. The deformation energy of a unit volume is
then~\cite{degennes-93}
\begin{equation}
\varepsilon_{\mathrm{def}}= {1\over 2}B~\left[{\partial u\over
\partial z} - {1\over 2} \left({\pmb \nabla}_\perp
u\right)^2\right]^2 +{1\over 2} K_1 \left({\pmb  \nabla}_\perp^2
u\right)^2 \, ,
\end{equation}
where ${\pmb \nabla}_\perp \equiv (\partial/\partial
x,~\partial/\partial y,~0)$. Using  the generalized liquid drop model,
Pethick \& Potekhin~\cite{pethick-98} obtain
\begin{equation}
B=6\varepsilon_{\mathrm{Coul}}~~~~~~~~~~K_1={2\over 15}R_{\mathrm{cell}}^2~
\varepsilon_{\mathrm{Coul}}\left(1+2w-2w^2\right) \, ,
\end{equation}
where $\varepsilon_{\mathrm{Coul}}$ is the Coulomb energy density (in
equilibrium)
\begin{equation}
\varepsilon_{\mathrm{Coul}}={2\pi\over 3}\;
(e n_{p\mathrm{i}} R_{\mathrm{cell}})^2 {(1-w)^2\over w} \, ,
\end{equation}
 and where $R_{\mathrm{cell}}$ is the half-distance between the plates,
$n_{p\mathrm{i}}$ is the proton density in nuclear matter and  $w$ is the
volume fraction occupied by nuclear matter (all values calculated for
the relaxed system).

In the case of the rod phase, also called the \emph{columnar phase}~\cite{degennes-93}, the number of elastic moduli is larger. They
describe the increase in energy density due to compression,
dilatation, transverse shearing, and bending of the rod lattice. Elastic
moduli were calculated within the liquid drop model by Pethick \&
Potekhin~\cite{pethick-98} and  by Watanabe, Iida \&
Sato~\cite{watanabe-01, watanabe_err-03}.

At the microscopic scale (fermis), the elastic properties of the nuclear pastas
are very different from those of a body-centered--cubic crystal of spherical nuclei. 
Nevertheless, the effects of pasta phases on the elastic properties of neutron star crusts
 may not be so dramatic at large scales (let's say meters). Indeed these nuclear pastas are 
necessarily of finite extent since one and two-dimensional long-range crystalline orders cannot 
exist in infinite systems (see, for instance, \cite{gelfert-01} and references therein).
How the nuclear pastas arrange themselves remains to be studied, but it is likely that the resulting
configurations look more-or-less isotropic at macroscopic scales.

\newpage


\section{Superfluidity and Superconductivity}
\label{sect.super}

Except for a brief period after their birth, neutron stars are
expected to contain various superfluid and superconducting
phases~\cite{sauls-89, dean-03, baldo-05, sedrakian-06}. In this
section, after a brief discussion of superconductivity and its
possible occurrence in neutron star crusts, we will review our current
theoretical understanding of the static and dynamic properties of
neutron superfluid in the inner crust of neutron stars. For a general
introduction and a recent overview on superfluidity and
superconductivity, see, for instance, the book by
Annett~\cite{annett-05}.

\subsection{Superconductivity in neutron star crusts}
\label{sect.super.electron}

In the Bardeen--Cooper--Schrieffer (BCS) theory of
superconductivity~\cite{bcs-57} the coupling of electrons with
lattice vibrations leads to an effective attractive interaction
between electrons despite the repulsive Coulomb force. As a
result, the
electrons of opposite spins form pairs with zero total angular
momentum. These {\it Cooper pairs} behave like bosons. Unlike
fermions, bosons do not obey the Pauli exclusion principle, which
forbids multiple occupancy of single particle quantum states.
Consequently,  below some critical temperature, bosons condense into the
lowest-energy single-particle state, giving rise to superfluidity, as
in liquid helium-4. Loosely speaking, superconductivity can thus be
seen as Bose--Einstein condensation of bound electron pairs. However
the analogy with Bose--Einstein condensation should not be taken too
far. Indeed,  Cooper pairs do not exist above the superconducting
transition, while in  Bose--Einstein condensation bosons always
exist above and below the critical temperature. Besides the electrons
in a pair do not form well separated ``molecules'' like atoms in
liquid helium, but instead strongly overlap. The spatial extent of a Cooper
pair in a conventional superconductor is typically
several orders of magnitude larger than the mean inter-electron spacing. The
Bose--Einstein condensation (BEC) and the BCS regime are now understood
as two different limits of the same phenomenon as illustrated in
Figure~\ref{fig.sect.super.static.bcsvsbec}. The transition between
these two limits has recently been studied in ultra-cold atomic 
Fermi gases, for which the interaction can be adjusted experimentally~\cite{combescot-06}.

\epubtkImage{bcsvsbec.png}{%
\begin{figure}[htbp]
  \centerline{\includegraphics[scale=0.5]{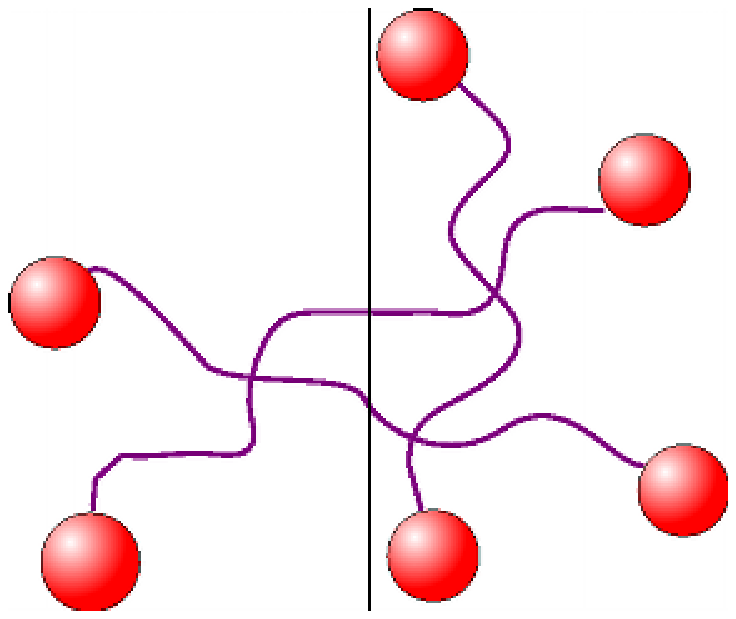}\hskip1cm\includegraphics[scale=0.5]{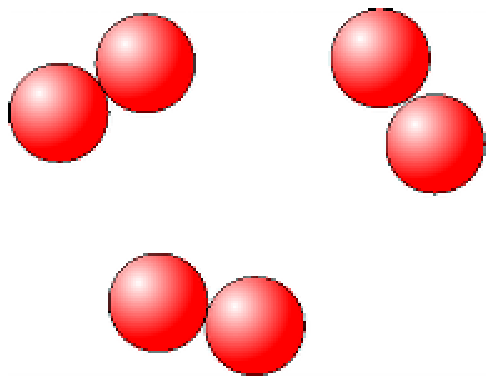}}
  \caption{Schematic picture illustrating the difference between
  the BCS regime (left) of overlapping loosely bound fermion pairs and
  the BEC regime (right) of strongly bound pairs.}
  \label{fig.sect.super.static.bcsvsbec}
\end{figure}}

In the outermost envelope of neutron stars, where the density is
similar to that of ordinary solids,  the critical temperature
for the onset of electron superconductivity is,
at most, on the order of a few K,
which is many orders of magnitude smaller than the expected
and observed surface temperature (see
Section~\ref{sect.obs.cooling}). Besides, it is well known that iron,
the most probable constituent of the outer layers of the crust
(Section~\ref{sect.groundstate.outer}), is not superconducting under
normal pressure. It was discovered in 2001~\cite{shimizu-01} that iron
becomes superconducting at ``high'' pressures\epubtkFootnote{Such
  pressures are very small in the context of neutron stars. For iron
  at room temperature they correspond to a density of about
  $8.2~\mdens$, in comparison to $7.86~\mdens$ under normal
  atmospheric pressure~\cite{bps-71}.} between $1.5\times 10^{11}$ and
$3\times 10^{11} \mathrm{\ dyn\ cm}^{-2}$, when the
temperature is below about 2~K. At higher densities, assuming that the
electrons are degenerate, we can estimate the critical temperature
from the BCS weak coupling approximation~\cite{bcs-57} (see also the
discussion by Ginzburg~\cite{ginzburg-69}).
\begin{equation}
T_{\mathrm{c}e}\sim T_{\mathrm{pi}} \exp(-2/v_{\mathrm{eff}}g_{\mathrm{F}e})\, ,
\end{equation}
where $T_{\mathrm{pi}}$ is the ion plasma frequency, $v_{\mathrm{eff}}$ is the
effective attractive electron-electron interaction and $g_{\mathrm{F}e}$
is the density of electron states at the Fermi level (per unit energy
and per unit volume). Neglecting band structure effects (which is a
very good approximation in dense matter; see, for instance, the
discussion of Pethick \& Ravenhall~\cite{pethick-95}), the density of
electron states is given by
\begin{equation}
g_{\mathrm{F}e}= \frac{k_{\mathrm{F}e}^2}{\pi^2\hbar v_{\mathrm{F}e} } \, ,
\end{equation}
so that the critical temperature takes the form
\begin{equation}
\label{eq.sect.super.static.electron.Tc}
T_{\mathrm{c}e}\sim T_{\mathrm{pi}} \exp(-\zeta \hbar v_{\mathrm{F}e}/e^2  )\, ,
\end{equation}
where $\zeta$ is a numerical positive coefficient of order
unity~\cite{ginzburg-69}. At densities much below $\sim
10^6~\mdens$, the electrons are nonrelativistic and their Fermi
velocity is given by  $v_{\mathrm{F}e}=\hbar k_{\mathrm{F}e}/m_e$. As a
result, the critical temperature decreases exponentially with the
average mass density $\rho$ as
\begin{equation}
\label{eq.sect.super.static.electron.Tc.nr}
T_{\mathrm{c}e}\sim T_{\mathrm{pi}} \exp(-\zeta^\prime (\rho/\rho_{\mathrm{ord}})^{1/3} )\, ,
\end{equation}
where $\zeta^\prime\equiv\zeta (9\pi Z/4 A)^{1/3} $ and $\rho_{\mathrm{ord}}=m_{\mathrm{u}}/(4\pi a_0^3/3)$ is a typical density
of ordinary matter ($a_0$ is the Bohr radius and $m_u$ the atomic mass
unit). Considering, for instance, a plasma of iron and electrons, and
adopting the value $\zeta=8/\pi$ calculated for the ``jellium'' model
by Kirzhnits~\cite{kirzhnits-69}, the critical temperature is
approximately given by
\begin{equation}
T_{\mathrm{c}e}\sim 3.6\times 10^3 \left(\frac{\rho}{1 \mathrm{\ g\
    cm}^{-3}}\right)^{1/2} \exp\biggl(-2.7 \left(\frac{\rho}{1
    \mathrm{\ g\ cm}^{-3}}\right)^{1/3} \biggr) \mathrm{\ K}.
\end{equation}
This rough estimate shows that the critical temperature decreases very
rapidly with increasing density, dropping from $\sim 30$~K at
$\rho=10~\mdens$ to $10^{-1}$~K at $\rho=10^2~\mdens$ and to
$10^{-7}$~K at $\rho=10^3~\mdens$! At densities above $\sim
10^6~\mdens$, electrons become relativistic, and  $v_{\mathrm{F}e}\sim
c$. According to Equation~(\ref{eq.sect.super.static.electron.Tc}),
the critical temperature of relativistic electrons is given by
\begin{equation}
\label{eq.sect.super.static.electron.Tc.r}
T_{\mathrm{c}e}\sim T_{\mathrm{pi}} \exp(-\zeta/\alpha )\, ,
\end{equation}
where $\alpha=e^2/\hbar c\simeq 1/137$ is the fine structure
constant. Due to the exponential factor, the critical temperature is
virtually zero.

We can, thus, firmly conclude that electrons in neutron star crusts (and,
{\it a fortiori}, in neutron star cores) are not
superconducting. Nevertheless, superconductivity in the crust is not
completely ruled out. Indeed, at the crust-core interface some protons
could be free in the ``pasta'' mantle
(Section~\ref{sect.groundstate.pasta}), and could be superconducting
due to pairing via strong nuclear interactions with a critical
temperature far higher than that of electron
superconductivity. Microscopic calculations in uniform nuclear matter
predict transition temperatures on the order of $T_{\mathrm{c}p}\sim
10^9\mbox{\,--\,}10^{10}$~K, which are much larger than typical temperatures in
mature neutron stars. Some properties of superconductors are discussed
in Sections~\ref{sect.super.dyn.fluxtubes}
and~\ref{sect.super.dyn.ns}.

\subsection{Static properties of neutron superfluidity}
\label{sect.super.static}

Soon after its formulation, the Bardeen--Cooper--Schrieffer (BCS) theory
of electron superconductivity~\cite{bcs-57} was successfully applied
to nuclei by Bohr, Mottelson and Pines~\cite{bohr-58} and
Belyaev~\cite{belyaev-59}. In a paper devoted to the moment of inertia
of nuclei, Migdal~\cite{migdal-59} speculated about the possibility
that  superfluidity could occur in the ``neutron core'' of stars (an idea
which was raised by Gamow and Landau in 1937 as a possible source of
stellar energy; see, for instance, \cite{haensel-06}). The
superfluidity inside neutron stars was first  studied by Ginzburg and
Kirzhnits in 1964~\cite{ginzburg-64a, ginzburg-64b}. Soon after,
Wolf~\cite{wolf-66} showed that the free neutrons in the crust are
very likely to be superfluid. It is quite remarkable that the
possibility of superfluidity inside neutron stars was raised
\emph{before} the discovery of pulsars by Jocelyn Bell and
Anthony Hewish in 1967. Later, this prediction seemed to be
 confirmed by the observation of the long relaxation time,
  on the order of months, following the first glitch in
  the Vela pulsar~\cite{baym-69}. The
neutron superfluid in the crust is believed to play a key role in the
glitch mechanism itself. Pulsar glitches are still
considered to be the strongest observational evidence
of superfluidity in neutron stars
(see Section~\ref{sect.obs.glitches}).

At the heart of BCS theory is the existence of an attractive
interaction needed for pair formation. In conventional
superconductors, this pairing interaction is indirect and weak. In the
nuclear case the occurrence of superfluidity is a much less subtle
phenomenon since the bare strong interaction between nucleons is
naturally attractive at not too small distances in many $JLS$ channels
($J$-total angular momentum, $L$-orbital angular momentum, $S$-spin of
nucleon pair). Apart from a proton superconductor similar to
conventional electron superconductors, two different kinds of neutron
superfluids are expected to be found in the interior of a neutron star
(for a review, see, for instance, \cite{sauls-89, lombardo-01, dean-03,
  baldo-05, sedrakian-06}). In the crust and in the outer core, the
neutrons are expected to form an isotropic superfluid like helium-4,
while in denser regions they are expected to form a more exotic kind
of (anisotropic) superfluid with each member of a pair having parallel
spins, as in superfluid helium-3. Neutron-proton pairs could also
exist in principle;  however,  their formation is not strongly favored in the 
asymmetric nuclear matter of neutron stars.

\subsubsection{Neutron pairing gap in uniform neutron matter at zero
  temperature}
\label{sect.super.static.uniform}

A central quantity in BCS theory is the gap function, which is
related to the binding energy of a pair. Neglecting for the time being
 nuclear clusters in the inner crust and considering pure neutron
matter, the gap equations at a given number density $n$ and at zero
temperature read, in the simplest approximation~\cite{ringschuck-80},
\begin{equation}
\label{eq.sect.super.static.BCS.gap}
\Delta(\pmb{k}) = - \frac{1}{2} \int \frac{d^3 \pmb{k^\prime}}{(2\pi)^3} \widetilde{V}_{\pmb{k},\pmb{k^\prime}} \frac{\Delta(\pmb{k^\prime})}{\sqrt{(\epsilon(\pmb{k^\prime})-\mu)^2+\Delta(\pmb{k^\prime})^2}  }\, ,
\end{equation}
together with
\begin{equation}
\label{eq.sect.super.static.BCS.dens}
n\equiv\frac{k_{\mathrm{F}}^3}{3 \pi^2}=\int \frac{d^3 \pmb{k}}{(2\pi)^3} \left(1-\frac{\epsilon(\pmb{k})-\mu}{\sqrt{(\epsilon(\pmb{k})-\mu)^2+\Delta(\pmb{k})^2}}\right)
\end{equation}
where $\widetilde{V}_{\pmb{k},\pmb{k^\prime}}$ is the matrix element
of the pairing interaction between time-reversed plane-wave states
with wave vectors $\pmb{k}$ and $\pmb{k^\prime}$, $\mu$ is the chemical
potential and $\epsilon(\pmb{k})$ is the single particle
energy. Whenever the ratio $\Delta({\pmb{k}})/\mu$ is small, 
the concept of a Fermi surface remains well defined and the two
equations can be decoupled by setting
$\mu=\epsilon(k_{\mathrm{F}})\equiv\epsilon_{\mathrm{F}}$.

Since the kernel in the gap integral peaks around the chemical potential
$\mu\simeq\epsilon_{\mathrm{F}}$, let us suppose that the pairing matrix elements
$\widetilde{V}_{\pmb{k},\pmb{k^\prime}}\simeq
\widetilde{V}_{k_{\mathrm{F}},k_{\mathrm{F}}}$ remain constant within
$|\epsilon(k)-\mu|< \epsilon_{\mathrm{C}}$ and zero elsewhere;
$\epsilon_{\mathrm{C}}$ is a \emph{cutoff energy}. With this schematic
interaction, the gap function becomes independent of $\pmb{k}$. In
conventional superconductors, the electron pairing is conveyed by
vibrations in the ion lattice. The ion plasma frequency thus provides
a natural cutoff $\epsilon_{\mathrm{C}}=\hbar \omega_{\mathrm{p i}}$ (see
Section~\ref{sect.super.electron}). In the nuclear case however, there
is no {\it a priori} choice of $\epsilon_{\mathrm{C}}$. A cutoff can still
be introduced in the BCS equations, provided the pairing interaction
is suitably renormalized, as shown by Anderson \&
Morel~\cite{anderson-61}. The BCS gap equations~(\ref{eq.sect.super.static.BCS.gap}) 
can be solved analytically in the weak coupling approximation, assuming that 
the pairing interaction is small, $g(\mu)\widetilde{V}_{k_{\mathrm{F}},k_{\mathrm{F}}}\ll 1$, 
where $g(\epsilon)$ is the density of single particle states at the energy $\epsilon$. 
Considering that $g(\epsilon)$ remains constant in the energy range 
$|\epsilon-\mu|< \epsilon_{\mathrm{C}}$, the gap
$\Delta(k_{\mathrm{F}})\equiv \Delta_{\mathrm{F}}$ at the Fermi
momentum $k_{\mathrm{F}}$ is given by
\begin{equation}
\label{eq.sect.super.static.BCS.weak}
\Delta_{\mathrm{F}} \simeq 2 \epsilon_{\mathrm{C}}
\exp\left( \frac{2}{g(\mu)
\widetilde{V}^{\mathrm{C}}_{k_{\mathrm{F}},k_{\mathrm{F}}}}\right) \, ,
\end{equation}
where the superscript C is to remind us that the strength of the pairing
interaction depends on the cutoff. This expression is usually not very
good for nuclear matter because the pairing interaction is
strong. Nevertheless it illustrates the highly nonperturbative nature
of the pairing gap. It also shows that the lower the density of
states, the lower the gap. Consequently the neutron pairing gap in
neutron star crust is expected to be smaller inside the nuclear
clusters (discrete energy levels) than outside (continuous energy
spectrum) as confirmed  by detailed calculations (see
Section~\ref{sect.super.static.crust}).

The pairing gap obtained by solving
Equations~(\ref{eq.sect.super.static.BCS.gap})
and~(\ref{eq.sect.super.static.BCS.dens}) for neutron matter using a
bare nucleon-nucleon potential and assuming a free Fermi gas single
particle spectrum
\begin{equation}
\label{eq.sect.super.static.freespec}
\epsilon(k)=\frac{\hbar^2 k^2}{2 m_n} \, ,
\end{equation}
where $m_n$ is the neutron mass, is almost independent of the
nucleon-nucleon potential (provided it fits scattering data) and is
shown in Figure~\ref{fig.sect.super.static.bcsfreegap}.

\epubtkImage{bcsfreegap.png}{%
\begin{figure}[htbp]
  \centerline{\includegraphics[scale=0.4]{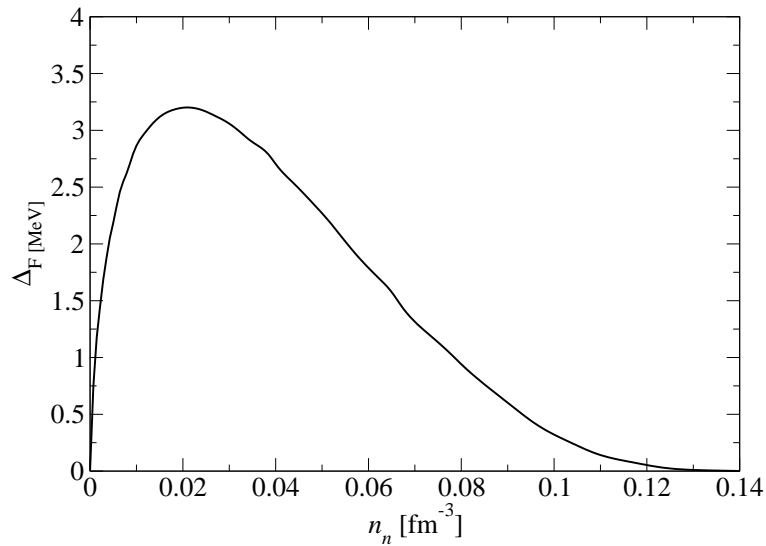}}
  \caption{Typical $^1S_0$ pairing gap in pure neutron matter as a
  function of the neutron number density, obtained in the BCS approximation
  with a bare nucleon-nucleon potential and the free energy
  spectrum (taken from Figure~7 of~\cite{lombardo-01}).}
  \label{fig.sect.super.static.bcsfreegap}
\end{figure}}

As can be seen, the neutron pairs are most strongly bound at
neutron densities around $n_n\simeq 0.02 \mathrm{\ fm}^{-3}$. At higher
densities, the pairing gap decreases due to the short range repulsive
part of the nucleon-nucleon interaction. The pairing gap is almost independent
of the nucleon-nucleon potential. The reason is that 
nucleon-nucleon potentials are constrained to reproduce the experimental phase shifts up to
scattering energies of order $E_{\mathrm{lab}}\sim 350$~MeV, which
corresponds to neutron densities of order $n_n\simeq 0.3 \mathrm{\ fm}^{-3}$. 
In fact, it can be shown that the
pairing gap is completely determined by the experimental $^1S_0$
nucleon-nucleon phase shifts~\cite{elgaroy-98}.
At small relative momenta $k$, the neutron-neutron $^1S_0$ phase shifts
$\delta(k)$ are well approximated by the expansion
\begin{equation}
\label{eq.sect.super.static.effrange}
k  \cot \delta(k) = -\frac{1}{a_{nn}}+\frac{1}{2} r_{nn} k^2  \, ,
\end{equation}
where $a_{nn}=-18.5\pm0.3$ fm and $r_{nn}=2.75\pm0.11$ fm are the
neutron-neutron scattering length and effective range,
respectively~\cite{slaus-89}. Large negative values of the scattering length are
associated with attractive interactions, which nearly lead to the existence
of a di-neutron. In magnetically-trapped Bose atomic gases, the scattering length 
can be varied experimentally by tuning the magnetic field~\cite{duine-04}. 

It can be shown that for Fermi wave vectors $k_{\mathrm{F}}\lesssim
0.5 \mathrm{\ fm}^{-1}$ ($n_{\mathrm{b}}\lesssim 4\times 10^{-3}
\mathrm{\ fm}^{-3}$), the pairing gap $\Delta_{\mathrm{F}}$ is
completely determined by these two parameters
only~\cite{elgaroy-98}. In the zero density limit $k_{\mathrm{F}}
|a_{nn}| \ll 1$ (i.e., $n_n \ll 5\times 10^{-6} \mathrm{\ fm}^{-3}$),
the gap equations are exactly solvable and the pairing gap can be
expressed analytically by the exact formula~\cite{papenbrock-99}
\begin{equation}
\label{eq.sect.super.static.BCS.exact}
\Delta_{\mathrm{F}} = \frac{8}{\exp(2)}\mu \exp\left(\frac{\pi}{2 k_{\mathrm{F}}
  a_{nn}}\right) \, .
\end{equation}
Let us emphasize that this formula is universal and valid for any
fermion system with attractive interactions
($a_{nn}<0$). Equation~(\ref{eq.sect.super.static.BCS.exact}) shows
that the pairing gap strongly depends on the density and the
nucleon-nucleon interaction. It can, thus, be foreseen that
modifications of the bare nucleon-nucleon interaction due to medium
polarization, which have been neglected, have a dramatic
effect. Indeed, it can be rigorously shown~\cite{gorkov-61} that in
the low density limit, polarization effects reduce the gap value
(\ref{eq.sect.super.static.BCS.exact}) by a factor of $4^{1/3}\exp(1/3)\sim 2$,
independent of the strength of the interaction!

The gap Equations~(\ref{eq.sect.super.static.BCS.gap})
and~(\ref{eq.sect.super.static.BCS.dens}) solved for the bare
interaction with the free single particle energy spectrum,
Equation~(\ref{eq.sect.super.static.freespec}), represent the simplest possible 
approximation to the pairing problem. A more consistent approach from the point of 
view of the many-body theory, is to calculate the single particle energies in the 
Hartree--Fock approximation (after regularizing the hard core of the bare nucleon-nucleon interaction). 
The next step is to ``dress'' the pairing interaction by medium polarization effects. 
Calculations have been carried out with phenomenological nucleon-nucleon interactions 
such as the Gogny force~\cite{gogny-80,farine-99}, that are constructed so as to reproduce some properties
of finite nuclei and nuclear matter.
Another approach is to derive this effective interaction
from a bare nucleon-nucleon potential (two-body and/or three-body
forces) using many-body techniques. 
Still the gap equations of form~(\ref{eq.sect.super.static.BCS.gap}) neglect 
important many-body aspects.

In many-body theory, the general equations describing a superfluid
Fermi system are the Nambu-Gorkov equations~\cite{abrikosov-75}, in
which the gap function $\Delta(\pmb{k},\omega)$ depends not only on the wave vector 
$\pmb{k}$ but also on the frequency $\omega$. This frequency dependence arises from 
dynamic effects. In this framework, it can be shown that BCS theory is
a mean field approximation to the many-body pairing problem. The
Gorkov equations cannot be solved exactly and some approximations have
to be made. Over the past years, this problem has been tackled using
different microscopic treatments and approximation
schemes. Qualitatively these calculations lead to the conclusion
that medium effects reduce the \emph{maximum} neutron pairing gap compared to the BCS
value (note that this includes the possibility that medium effects actually increase 
the pairing gap for some range of densities).
However, these calculations predict very different density
dependence of the pairing gap as illustrated in
Figure~\ref{fig.sect.super.static.BCS.realistic}.

\epubtkImage{realistic_gaps.png}{%
\begin{figure}[htbp]
  \centerline{\includegraphics[scale=0.4]{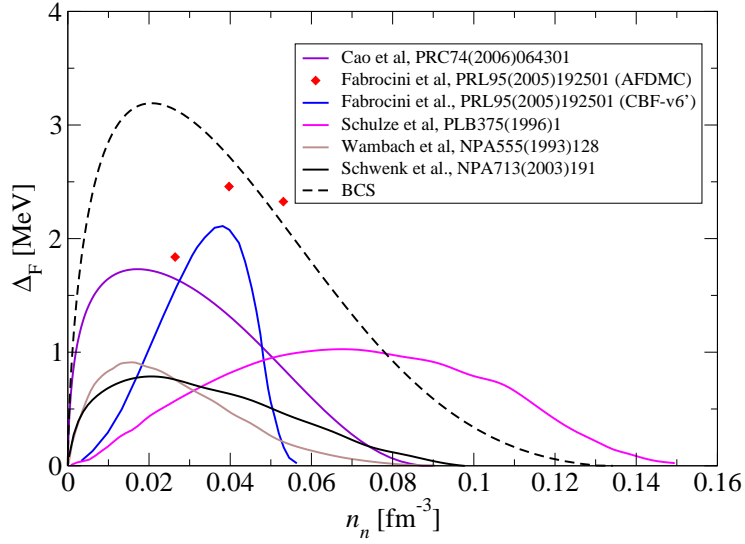}}
  \caption{$^1S_0$ pairing gap in pure neutron matter as a
  function of the neutron number density obtained from microscopic
  calculations with different approximations to account for medium
  effects.}
  \label{fig.sect.super.static.BCS.realistic}
\end{figure}}

Before concluding this section, we provide an analytic formula for a few
representative neutron pairing gaps, using the following expression
proposed by Kaminker et al.~\cite{kaminker-01}.
\begin{equation}
\label{eq.sect.super.static.fit}
\Delta_{\mathrm{F}}=\theta(k_{\max}-k_{\mathrm{F}})\, \Delta_0 \frac{k_{\mathrm{F}}^2}{k_{\mathrm{F}}^2+k_1^2}\frac{(k_{\mathrm{F}}-k_2)^2}{(k_{\mathrm{F}}-k_2)^2+k_3^2}\, ,
\end{equation}
where $k_{\mathrm{F}}=(3\pi^3 n_n)^{1/3}$ and $\theta$ is the Heaviside
step function $\theta(x)=1$ if $x>0$ and $\theta(x)=0$ otherwise. The
parameters $\Delta_0$, $k_1$, $k_2$, $k_3$ and $k_{\max}$ are given in
Table~\ref{table.sect.super.static.fit}. At low density, the pairing
gap $\Delta_{\mathrm{F}} \propto k_{\mathrm{F}}^2$ varies roughly as expected,
Equation~(\ref{eq.sect.super.static.BCS.exact}), remembering that $\mu
\sim \epsilon_{\mathrm{F}} \propto k_{\mathrm{F}}^2$.

\begin{table}[htbp]
  \caption[Parameters for the analytic formula of a few representative
  $^1S_0$ pairing gaps in pure neutron matter.]{Parameters for the
  analytic formula Equation~(\ref{eq.sect.super.static.fit}) of a few
  representative $^1S_0$ pairing gaps in pure neutron matter: BCS-BCS
  pairing gap shown in Figure~\ref{fig.sect.super.static.bcsfreegap},
  Brueckner -- pairing gap of Cao et al.~\cite{cao-06} based on
  diagrammatic calculations (shown in
  Figure~\ref{fig.sect.super.static.BCS.realistic}) and RG -- pairing
  gap of Schwenk et al.~\cite{schwenk-03} based on the Renormalization
  Group approach (shown in
  Figure~\ref{fig.sect.super.static.BCS.realistic}). $\Delta_0$ is
  given in MeV. $k_1$, $k_2$, $k_3$ and $k_{\max}$ are given in
  fm$^{-1}$.}
  \label{table.sect.super.static.fit}
  \vskip 4mm

  \centering
    \begin{tabular}{c c c c c c}
      \toprule
      model & $\Delta_0$ & $k_1$ & $k_2$ & $k_3$ & $k_{\max}$ \\
      \midrule
      BCS   & 910.603 & 1.38297 & 1.57068 & 0.905237 & 1.57 \\
      Brueckner & 11.4222 & 0.556092 & 1.38236 & 0.327517 & 1.37 \\
      RG & 16.5709 & 1.13084 & 1.47001 & 0.582515 & 1.5 \\
      \bottomrule
    \end{tabular}
\end{table}

\subsubsection{Critical temperature for neutron superfluidity}
\label{sect.super.static.Tc}

The BCS gap Equations~(\ref{eq.sect.super.static.BCS.gap}) at zero
temperature can be generalized to finite temperature $T$ (adopting the
standard notation $\beta\equiv 1/k_{\mathrm{B}} T$ where $k_{\mathrm{B}}$ is the
Boltzmann constant)
\begin{equation}
\label{eq.sect.super.static.BCS.T}
\Delta(\pmb{k}) = - \frac{1}{2} \int\frac{d^3\pmb{k^\prime}}{(2\pi)^3}\widetilde{V}_{\pmb{k},\pmb{k^\prime}}\frac{\Delta(\pmb{k^\prime})}{\sqrt{(\epsilon(\pmb{k^\prime})-\mu)^2+\Delta(\pmb{k^\prime})^2}}\tanh\frac{\beta}{2}\sqrt{(\epsilon(\pmb{k^\prime})-\mu)^2+\Delta(\pmb{k^\prime})^2}\, .
\end{equation}

Superfluidity disappears whenever the temperature exceeds some
critical threshold. Let us remark that  isotropic neutron
superfluidity can also be destroyed by a sufficiently strong
magnetic field, since it would force each spin of a neutron pair to be
aligned (as pointed out by Kirszshnits~\cite{kiszhnits-70}). It can be
shown on general grounds that the isotropic pairing gap
$\Delta_{\mathrm{F}}(T=0)$ at zero temperature (at Fermi momentum
$k_{\mathrm{F}}$) and the critical temperature $T_{\mathrm{c}}$ are
approximately related by~\cite{bcs-57}
\begin{equation}
\label{eq.sect.super.static.BCS.Tc}
\Delta_{\mathrm{F}}(T=0)=\pi \exp(-\gamma)\, k_{\mathrm{B}} T_{\mathrm{c}} \simeq 1.76\, k_{\mathrm{B}} T_{\mathrm{c}} \, ,
\end{equation}
where $\gamma$ is the Euler constant. This well-known result of
conventional electron superconductivity applies rather well to nucleon
superfluidity, especially for densities at which $\Delta_{\mathrm{F}}(0)$
takes its maximum value~\cite{lombardo-01}.

The temperature dependence of the pairing gap, for $T\leq T_{\mathrm{c}}$,
can be approximately written as~\cite{yakovlev-01}
\begin{equation}
\label{eq.sect.super.static.BCS.approx.T}
\Delta_{\mathrm{F}}(T)\simeq k_{\mathrm{B}} T \sqrt{1-\frac{T}{T_{\mathrm{c}}}}\left(1.456-0.157\left(\frac{T}{T_{\mathrm{c}}}\right)^{-1/2} +1.764\left(\frac{T}{T_{\mathrm{c}}}\right)^{-1} \right)\, .
\end{equation}

Zero temperature pairing gaps on the order of 1~MeV are therefore
associated with critical temperatures of the order $10^{10}$~K,
considerably larger than typical temperatures inside neutron stars
except for the very early stage of their formation. The existence of a
neutron superfluid in the inner crust of a neutron star is therefore well
established theoretically. Nevertheless the density dependence of the
critical temperature predicted by different microscopic calculations
differ considerably due to different approximations of the many-body
problem. An interesting issue concerns the cooling of neutron stars
and the crystallization of the crust: do the neutrons condense into a
superfluid phase before the formation of the crust or after?

Figure~\ref{fig.sect.super.static.NV.Tm-Tc} shows the melting
temperature $T_{\mathrm{m}}$ of the inner crust of neutron stars compared to
the critical temperature $T_{\mathrm{c}}$ for the onset of neutron
superfluidity. The structure of the crust is that calculated by Negele
\& Vautherin~\cite{nv-73}. The melting temperature has been calculated
from Equation~(\ref{eq.sect.plasma.noB.Tm}) with $\Gamma_{\mathrm{m}}=175$. The temperature $T_{\mathrm{c}}$ has been obtained from
Equation~(\ref{eq.sect.super.static.BCS.Tc}), considering a uniform
neutron superfluid, with the density $\tilde{\rho}_{\mathrm{G}}$ of unbound
neutrons given by Negele \& Vautherin~\cite{nv-73}. Several critical
temperatures are shown for comparison. As discussed in
Section~\ref{sect.super.static.uniform}, the BCS value represents the simplest
approximation to the true critical temperature. The other two
critical temperatures have been obtained from more realistic pairing-gap calculations, which include medium effects  using different
many-body approximations. The calculation of Cao et al.~\cite{cao-06}
is based on diagrammatic calculations, while that of
Schwenk et al.~\cite{schwenk-03} relies on the renormalization group.

For the BCS and Brueckner calculations of the pairing gap, in the
density range of $\sim 10^{12}\mbox{\,--\,}10^{14}~\mdens$, the neutrons may
become superfluid \emph{before} the matter crystallizes into a solid
crust. As discussed in Section~\ref{sect.super.dyn.rot}, as a result
of the rotation of the star, the neutron superfluid would be threaded
by an array of quantized vortices. These vortices might affect the
crystallization of the crust by favoring nuclear clusters along the
vortex lines, as suggested by
Mochizuki et al.~\cite{mochizuki-99}. On the contrary, the
calculations of Schwenk et al.~\cite{schwenk-03} indicate that, at any
density, the solid crust would form before the neutrons become
superfluid. Recently, it has also been  shown,  by taking into account the
effects of the inhomogeneities on the neutron superfluid,  that in the
shallow layers of the inner crust, the neutrons might remain in the
normal phase even long after the formation of the crust, when the
temperature has dropped below 10$^9$~K~\cite{monrozeau-07}.

\epubtkImage{NV_Tm-Tc.png}{%
\begin{figure}[htbp]
  \centerline{\includegraphics[scale=0.4]{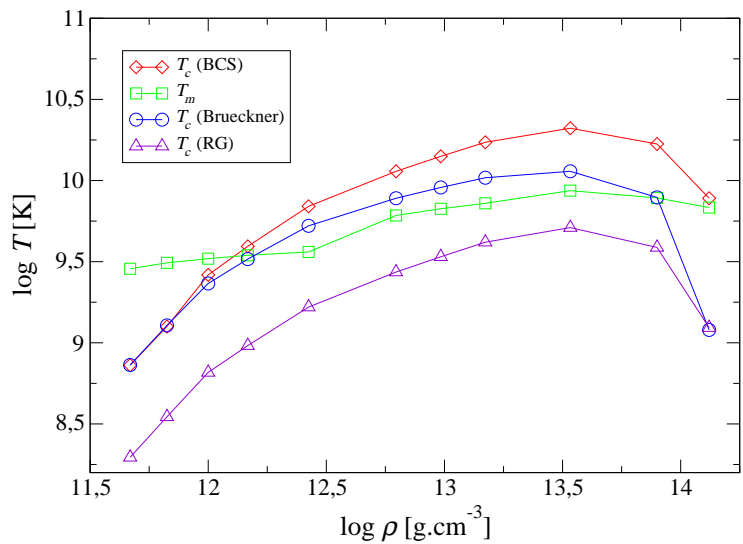}}
  \caption{Melting temperature $T_{\mathrm{m}}$ of the crust
  and critical temperature $T_{\mathrm{c}}$ for the onset of neutron
  superfluidity as a function of the density $\rho$. The model
  of the inner crust is based on Negele \& Vautherin\cite{nv-73}.
  Three representative cases (as shown in Figure~\ref{fig.sect.super.static.BCS.realistic}) are: the BCS pairing gap $\Delta_{\mathrm{F}}$ and the
  more realistic pairing gaps of Cao et al.~\cite{cao-06} and
  Schwenk et al.~\cite{schwenk-03}.}
  \label{fig.sect.super.static.NV.Tm-Tc}
\end{figure}}

\subsubsection{Pairing gap in neutron star crusts}
\label{sect.super.static.crust}

In this section, we will discuss the effects of the nuclear clusters
on the pairing properties of the neutron superfluid in neutron star
crusts. The relative importance of these effects is determined by the
coherence length, defined as the root mean square radius of the pair
wave function. Broadly speaking, the coherence length represents the
size of a neutron pair. This is an important length scale, which
determines many properties of the superfluid. For instance, the
coherence length is of the order of the size of the superfluid vortex
cores. According to Anderson's theorem~\cite{degennes-66}, the effects
of the inhomogeneities (here - nuclear clusters)
on the neutron superfluid are negligible
whenever the coherence length is much larger than the characteristic
size of the inhomogeneities. Assuming weak coupling, the coherence
length can be roughly estimated from Pippard's expression
\begin{equation}
\label{eq.sect.super.static.BCS.xi}
\xi = \frac{\hbar^2 k_{\mathrm{F}}}{\pi m_n \Delta_{\mathrm{F}}} \, ,
\end{equation}
where $\Delta_{\mathrm{F}}$ is the neutron pairing gap at the Fermi
momentum $k_{\mathrm{F}}$ (see
Section~\ref{sect.super.static.uniform}). This expression gives only a
lower bound for the coherence length, since medium effects tend to
reduce the pairing correlations as discussed in
Section~\ref{sect.super.static.uniform}. Nevertheless, this estimate
is rather close to the value obtained in more detailed
calculations~\cite{blasio-97, matsuo-06}. As  can be seen in
Figure~\ref{fig.sect.super.static.NV.xi-Rc}, the coherence length is
smaller than the lattice spacing\epubtkFootnote{For a body-centered--cubic lattice, the lattice spacing $a$ is related to the
  Wigner--Seitz radius $R_{\mathrm{cell}}$ by $a=2(\pi/3)^{1/3} R_{\mathrm{cell}}$.} except for the densest layers of the crust. Consequently
the effects of the solid crust on neutron superfluidity cannot be
neglected. This situation is in sharp contrast to that encountered in
ordinary type I superconductors, where the electron Cooper pairs are
spatially extended over mesoscopic distances of $\sim 10^3\mbox{\,--\,}10^4$~\AA\,
and as a result the pairing gap is nearly insensitive to the details
of the atomic crystal structure, since the typical lattice spacing is
of order a few \AA. In Figure~\ref{fig.sect.super.static.NV.xi-Rc} we
also displayed the mean inter-neutron spacing defined by
\begin{equation}
\label{eq.sect.super.static.dn}
d_n = 2 \left(\frac{3}{4\pi n_n}\right)^{1/3} \, .
\end{equation}

In the denser layers of the crust, the coherence length is smaller
than the mean inter-neutron spacing, suggesting that the neutron
superfluid is a Bose--Einstein condensate of strongly-bound neutron
pairs, while in the shallower layers of the inner crust the neutron
superfluid is in a BCS regime of overlapping loosely-bound
pairs. Quite remarkably, for screened pairing gaps like those of
Schwenk et al.\cite{schwenk-03}, the coherence length is larger than
the mean inter-neutron spacing in the entire inner crust, so that in
this case, at any depth, neutron superfluid is in the BCS regime.

\epubtkImage{NV_xi-Rc.png}{%
\begin{figure}[htbp]
  \centerline{\includegraphics[scale=0.4]{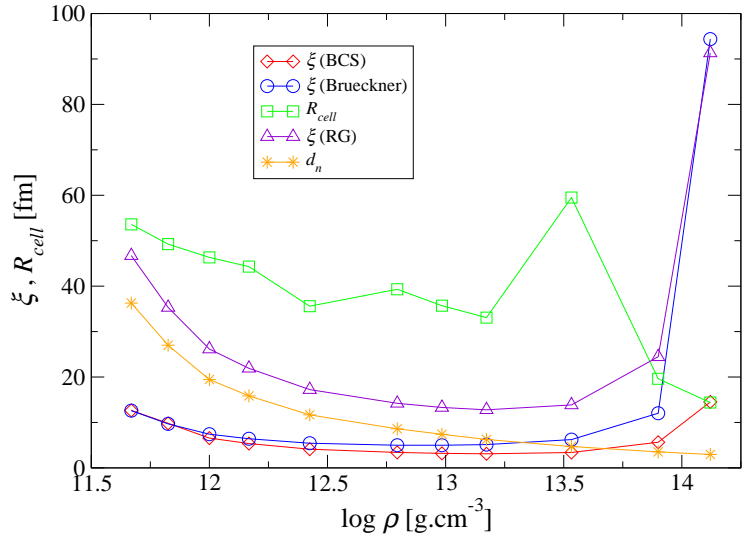}}
  \caption{Pippard's coherence length for the neutron star crust
  model  of Negele \& Vautherin\cite{nv-73}. The coherence length
  has been  calculated from
  Equation~(\ref{eq.sect.super.static.BCS.xi}), assuming that the
  neutron superfluid is uniform with the density of unbound neutrons
  denoted by $\tilde{\rho}_{\mathrm{G}}$ in \cite{nv-73}. Three
  representative cases have been considered: the BCS pairing gap
  $\Delta_{\mathrm{F}}$,  and the more
  realistic pairing gaps of Cao et al.~\cite{cao-06} and
  Schwenk et al.~\cite{schwenk-03}. The gaps are shown in 
  Figure~\ref{fig.sect.super.static.BCS.realistic}. For comparison, we
  also show the radius $R_{\mathrm{cell}}$ of the Wigner--Seitz sphere and
  the  mean inter-neutron spacing $d_n$.}
  \label{fig.sect.super.static.NV.xi-Rc}
\end{figure}}

Since the formulation of the BCS theory, considerable theoretical
efforts have been devoted to the microscopic calculation of pairing
gaps in uniform nuclear matter using the many body theory. On the
other hand, until recently the superfluidity in neutron star crusts has
not attracted much attention despite its importance in many
observational phenomena like pulsar glitches (see
Section~{\ref{sect.obs}). The pairing correlations in an inhomogeneous
  superfluid system can be described in terms of a pairing field
  $\Delta(r)$. In early studies~\cite{broglia-94, blasio-96,
    elgaroy-96} the pairing field has been calculated assuming that
  the matter is locally homogeneous (local density
  approximation). Such calculations predict, in particular, that the
  value of the pairing field inside the nuclear clusters is almost the
  same for different layers of the crust. The reason lies in the
  nuclear saturation: the density inside heavy nuclei is essentially
  constant, independent of the number of bound nucleons. In some
  cases, the pairing field was found to vanish inside the
  clusters~\cite{elgaroy-96}. The local density approximation is valid
  if the coherence length is smaller than the characteristic scale of
  density variations. However, this condition is never satisfied in the
  crust. As a result, the local density approximation overestimates
  the spatial variation of the pairing field. Due to ``proximity
  effects'', the free superfluid neutrons induce pairing correlations
  of the bound neutrons inside clusters and {\it vice versa} leading
  to a smooth spatial variation of the neutron pairing
  field~\cite{barranco-98}. As a remarkable consequence, the value of
  the neutron pairing field outside (resp.\ inside) the nuclear
  clusters is generally \emph{smaller} (resp.\ \emph{larger}) than that
  obtained in uniform neutron matter for the same
  density~\cite{baldo-07}. In particular, the neutrons inside the
  clusters are also superfluid. The neutron superfluid in the crust
  should, therefore, be thought of as an inhomogeneous superfluid rather
  than a superfluid flowing past clusters like obstacles. The effects
  of nuclear clusters on neutron superfluidity have been
  investigated in the Wigner--Seitz approximation by several
  groups. These calculations have been carried out at the mean field
  level with realistic nucleon-nucleon potentials~\cite{barranco-97},
  effective nucleon-nucleon interactions~\cite{barranco-98,
    montani-04, sandulescu-04, sandulescu-04b, khan-05} and with
  semi-microscopic energy functionals~\cite{baldo-05, baldo-06,
    baldo-07}. Examples are shown in
  Figure~\ref{fig.sect.super.static.pairing-field}. The effects of
  medium polarization have been considered by the Milano
  group~\cite{gori-04, vigezzi-05}, who found that these effects lead to
  a reduction of the pairing gap, as in uniform neutron
  matter. However, this quenching is less pronounced than in uniform matter due
  to the presence of nuclear clusters. Apart from uncertainties in the pairing interaction, it has recently been
  shown~\cite{baldo-06, baldo-07} that the pairing field is very
  sensitive to the choice of boundary conditions, especially in the
  bottom layers of the crust (as also found by the other
  groups). Consequently, the results obtained in the Wigner--Seitz
  approximation should be interpreted with caution, especially when
  calculating thermodynamic quantities like the neutron specific heat,
  which depends exponentially on the gap.

\epubtkImage{pairing_field.png}{%
\begin{figure}[htbp]
  \centerline{\includegraphics[scale=0.4]{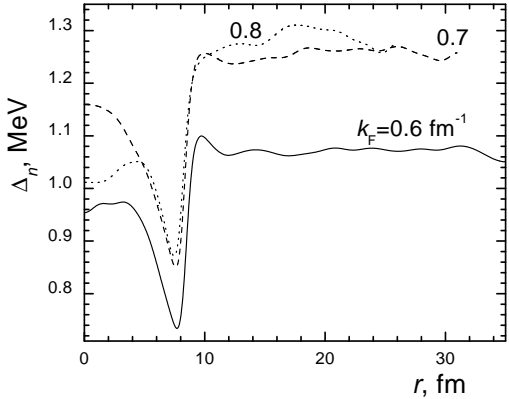}}
  \caption{Neutron pairing fields in the inner crust, calculated by
  Baldo et al.~\cite{baldo-07b}. Results are shown inside the
  Wigner--Seitz sphere. $k_{\mathrm{F}}$ is the average Fermi momentum
  defined by $k_{\mathrm{F}}=(3\pi^2 n_{\mathrm{b}})^{1/3}$, where
  $n_{\mathrm{b}}$ is the  baryon density.}
  \label{fig.sect.super.static.pairing-field}
\end{figure}}

\subsection{Superfluid hydrodynamics}
\label{sect.super.dyn}

\subsubsection{Superflow and critical velocity}
\label{sect.super.dyn.Vc}

The basic property of superfluid is that it can flow without
dissipation. In a normal fluid, friction and viscosity arise because
particles are randomly scattered. Such scattering events are forbidden
in  superfluid because energy and momentum cannot be simultaneously
conserved. The key argument of Landau is that in a superfluid like
helium-4, the particles are strongly correlated so that the concept of
single particles becomes meaningless. However,  at \emph{low enough}
temperatures, the system is still assumed to be described in terms of
noninteracting ``quasiparticles'', which do not correspond to material
particles but to many-body motions (excitations). The
energy spectrum of these quasiparticles can be very different from
that of single particles. Using this idea, Landau~\cite{landau-41} was
able to explain the origin of nondissipative superflow and the
existence of a critical velocity beyond which superfluidity
disappears. The argument is the following. Let us consider a
macroscopic body of mass $M$ flowing through the superfluid. At  low
temperatures, its velocity $\pmb{V}$ can only be changed in scattering
processes where one or more quasiparticles are created, assuming that
the flow is not turbulent. For a quasiparticle of energy $E(p)$ and
momentum $p$ to be created, energy conservation implies that
\begin{equation}
\frac{1}{2}M V^2 > \frac{1}{2}M V^{\prime 2}+E(p) \, ,
\end{equation}
where $\pmb{V^\prime}$ is the body's velocity after the event. However,
 momentum must also be conserved,
\begin{equation}
M \pmb{V}=M \pmb{V^\prime} + \pmb{p} \, .
\end{equation}

These two conditions can only be satisfied if
\begin{equation}
E(p) < \pmb{V}\cdot \pmb{p} - \frac{p^2}{2 M} \, .
\end{equation}

Since the perturbing body contains a macroscopic collection of particles,
the mass $M$ is very large so that the second term can be
neglected. The resulting inequality cannot be satisfied unless the
velocity exceeds some critical value
\begin{equation}
\label{eq.sect.super.dyn.Landau.Vc}
V_c = \min \biggl \lbrace \frac{E(p)}{p} \biggr\rbrace \, ,
\end{equation}
where $\min \lbrace x \rbrace$ is the smallest value of
$x$ in a set  $\lbrace x \rbrace$.
This means that for velocities smaller than $V_c$, the creation of
quasiparticles is forbidden and therefore the fluid flows without
dissipation. In a normal liquid, the single particle energy is given
by an expression of the form
\begin{equation}
\label{eq.sect.super.dyn.sp.energy}
E(p)=\frac{p^2}{2 m} \, ,
\end{equation}
where $m$ is an effective mass suitably renormalized to include
many-body effects. Consequently, the critical velocity, according to
Landau's criterion~(\ref{eq.sect.super.dyn.Landau.Vc}), is
zero,  $V_c=0$. Since liquid helium-4 is superfluid,
Landau~\cite{landau-41, landau-41b} postulated a different energy
spectrum. At low momenta, the quasiparticle excitations are sound
waves (phonons) as illustrated in
Figure~\ref{fig.sect.super.dyn.phonons}. The dispersion relation is
thus given by
\begin{equation}
\label{eq.sect.super.dyn.qp.phonons}
E(p)=c_{\mathrm{s}} p \, ,
\end{equation}
where $c_{\mathrm{s}}$ is a sound speed.

\epubtkImage{phonons.png}{%
\begin{figure}[htbp]
  \centerline{\includegraphics[scale=0.8]{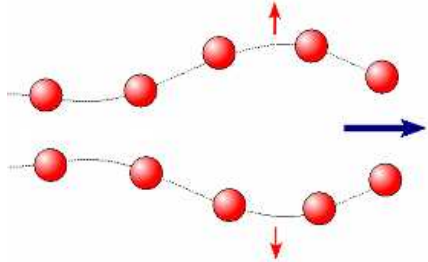}}
  \caption{Schematic picture illustrating  collective motions
  of particles associated with a low momentum quasiparticle (phonon).}
  \label{fig.sect.super.dyn.phonons}
\end{figure}}

At very high momenta, the dispersion relation coincides with that of a
normal liquid, Equation~(\ref{eq.sect.super.dyn.sp.energy}). In
between, the dispersion relation exhibits a local minimum and is
approximately given by
\begin{equation}
\label{eq.sect.super.dyn.qp.rotons}
E(p)=\Delta_{\mathrm{r}}+\frac{(p-p_0)^2}{2 m_{\mathrm{r}}} \, .
\end{equation}

The quasiparticles associated with this minimum were dubbed ``rotons''
by I.E.\ Tamm as reported by Landau~\cite{landau-41}. Landau postulated
that these rotons are connected with a rotational velocity flow,  hence
the name. These rotons arise due to the interactions between
the particles. Feynman~\cite{feynman-56} argued that a roton can be
associated with the motion of a single atom. As the atom moves through the
fluid, it pushes neighboring atoms out of its way forming a ring
of particles rotating backwards as illustrated in
Figure~\ref{fig.sect.super.dyn.rotons}. The net result is a vortex ring
of an atomic size.

\epubtkImage{rotons.png}{%
\begin{figure}[htbp]
  \centerline{\includegraphics[scale=0.8]{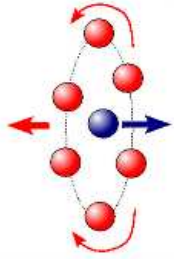}}
  \caption{Schematic picture illustrating collective motions
  of particles associated with a roton quasiparticle according to
  Feynman's interpretation.}
\label{fig.sect.super.dyn.rotons}
\end{figure}}

The roton local minimum has also been interpreted as a characteristic
feature of density fluctuations marking the onset of
crystallization~\cite{horner-72, pomeau-94, nozieres-04}. According to
Nozi\`eres~\cite{nozieres-04}, rotons are ``ghosts of Bragg
spots''. Landau's theory has been very successful in explaining the
observed properties of superfluid helium-4 from the postulated energy
spectrum of quasiparticles.

In weakly interacting dilute Bose gases, as in ultra-cold Bose atomic
gases, the energy of the quasiparticles are given by (see for instance
Section~21 of Fetter \& Walecka~\cite{fetter-03})
\begin{equation}
\label{eq.sect.super.dyn.qp.bec}
E(p)=\sqrt{ \left(\frac{p^2}{2 m}\right)^2 + p^2 c_{\mathrm{s}}^2} \, ,
\end{equation}
where $c_{\mathrm{s}}$ is the speed of sound. At low momentum $p$, it reduces
to Equation~(\ref{eq.sect.super.dyn.qp.phonons}), while at high momentum it tends to
Equation~(\ref{eq.sect.super.dyn.sp.energy}). Dilute
Bose gases, thus, have only phonon excitations. Quite remarkably, the
ideal Bose gas (which is characterized by the dispersion relation~(\ref{eq.sect.super.dyn.qp.bec})
with $c_{\mathrm{s}}=0$) exhibits a Bose-Einstein condensation at low enough temperatures, but is not superfluid
since its critical velocity is equal to zero.

Owing to the specific energy spectrum of quasiparticles in both atomic
Bose gases and helium-4, the critical velocity does not vanish, thus
explaining their superfluid properties. Landau's critical velocity,
Equation~(\ref{eq.sect.super.dyn.Landau.Vc}), of superfluid helium-4
due to the emission of rotons, is given by
\begin{equation}
\label{eq.sect.super.dyn.helium.Vc}
V_c=\frac{\Delta_{\mathrm{r}}}{p_0}\simeq 60 \mathrm{\ m\ s}^{-1}\, .
\end{equation}
This value has been confirmed by ion propagation
experiments~\cite{ellis-85}.  However, in most experiments, much smaller
critical velocities are measured due to the existence of other kinds of
excitations. The critical velocity of atomic Bose gases has also been
measured, using laser beams instead of a macroscopic
object~\cite{ketterle-00}. Again, velocities smaller than Landau's
velocity (which is equal to the velocity of sound in this case) have been
found.

The previous discussion of the critical velocity of Bose liquids
can be easily extended to fermionic superfluids. In the BCS theory,
fermions form bound pairs, which undergo Bose condensation when the temperature
falls below a critical temperature
(Section~\ref{sect.super.static}). The quasiparticle energies for a
uniform Fermi system are given by
\begin{equation}
\label{eq.sect.super.dyn.qp.BCS}
E(p)=\sqrt{(\epsilon(p)-\mu)^2+\Delta(p)^2} \, ,
\end{equation}
where $p=\hbar k$, $\epsilon(p)$ is the single particle energy, $\mu$
the chemical potential and $\Delta(p)$ the pairing gap at momentum
$p$. According to Landau's argument, the critical
velocity~(\ref{eq.sect.super.dyn.Landau.Vc}) is equal 
to
\begin{equation}
\label{eq.sect.super.dyn.Vc.BCS}
V_c = \frac{\Delta_{\mathrm{F}}}{\hbar k_{\mathrm{F}}} \, .
\end{equation}

This expression can be derived more rigorously from the microscopic
BCS theory~\cite{bardeen-62}. It shows that a system of fermions is
superfluid (i.e.\ the critical velocity is not zero) whenever the
interactions are attractive, so that the formation of pairs becomes
possible. It is also interesting to note that the BCS spectrum can be 
interpreted in terms of rotons. Indeed, expanding
Equation~(\ref{eq.sect.super.dyn.qp.BCS}) around the minimum 
leads, to lowest order, to an expression similar to
Equation~(\ref{eq.sect.super.dyn.qp.rotons}). In this case, $p_0$ is obtained by 
solving $\epsilon(p)=\mu$. The other parameters are given by 
$\Delta_{\mathrm{r}}=\Delta(p_0)$ and $m_{\mathrm{r}}=\Delta(p_0)/v_0^2$
where $v_0=d\epsilon/dp$ is the group velocity evaluated at $p_0$.

The presence of an ``external'' potential affects 
superfluidity. This issue has recently attracted a lot of theoretical, as
well as experimental, interest in the field of optically-trapped
ultra-cold atomic Bose gases~\cite{morsch-06}. It is also relevant in the context of
neutron stars, where the solid crust is immersed in a neutron
superfluid (and possibly a proton superconductor in the liquid crystal
mantle, where nuclear ``pastas'' could be present; see
Section~\ref{sect.groundstate.pasta}). In the BCS approximation~(\ref{eq.sect.super.static.BCS.gap}),
considering a periodic potential (induced by the solid crust
in neutron stars), the quasiparticle energies still take a form
similar to Equation~(\ref{eq.sect.super.dyn.qp.BCS}). However, the
dependence on the momentum is no more isotropic. As shown by Carter,
Chamel \& Haensel~\cite{cch-05},
Equation~(\ref{eq.sect.super.dyn.Vc.BCS}) for the critical velocity
should then be replaced by
\begin{equation}
\label{eq.sect.super.dyn.Vc.BCS.crust}
V_c =\min_{_{\mathrm{FS}}} \biggl\lbrace  \frac{\Delta(\pmb{k})}{m_\star v(\pmb{k})} \biggr\rbrace \, ,
\end{equation}
where $\pmb{k}$ is the Bloch wave vector
(Section~\ref{sect.groundstate.inner.beyond}),
$\pmb{v}(\pmb{k})=\hbar^{-1}\pmb{\nabla_k}\, \epsilon(\pmb{k})$ is the
group velocity of the fermions and $m_\star$ is an effective mass,
which arises from the interactions between the particles and the
lattice as discussed in Section~\ref{sect.super.dyn.entr}. The
subscript FS means that the minimum is to be searched on the Fermi
surface\epubtkFootnote{The Fermi surface is the surface in $k$-space
  defined by $\epsilon(\pmb{k})=\mu$. Note that, in general, it is not
  spherical.}. In the neutron star crust, the effective mass
of the unbound neutrons could be very large $m_\star\gg m_n$, as shown
by Chamel~\cite{chamel-05, chamel-06}.  In the limit of homogeneous
matter, the single particle energies are given by an expression of the
form~(\ref{eq.sect.super.dyn.sp.energy}) with $m=m_n$ and $\pmb{p}=\hbar \pmb{k}$. 
Consequently, $m_\star=m_n$ and $\pmb{v}=\hbar\pmb{k}/m_n$, so that 
Equation~(\ref{eq.sect.super.dyn.Vc.BCS.crust}) reduces to
Equation~(\ref{eq.sect.super.dyn.Vc.BCS}) using the extended Brillouin zone scheme.
Equation~(\ref{eq.sect.super.dyn.Vc.BCS.crust}) shows that superfluidity disappears 
whenever the pairing gap vanishes at some point on the Fermi surface.

The real critical velocity is expected to be smaller than
that given by Equation~(\ref{eq.sect.super.dyn.Vc.BCS.crust}) due to
finite temperature and many-body effects beyond the mean
field. Likewise, the critical velocity of superfluid helium-4 obtained
in Landau's quasiparticle model is only an upper bound because in
this model, the quasiparticles are assumed to be noninteracting. The
experimentally-measured critical velocities are usually much smaller, in
particular, due to the nucleation of vortices. Indeed, relative motions
between superfluid and the vortices lead to mutual friction forces and, hence, to dissipative effects. In general, any \emph{curved} vortex
line does not remain at rest in the superfluid reference frame and, 
therefore, induces dissipation. Feynman~\cite{feynman-55} derived the
critical velocity associated with the formation of a vortex ring in a
channel of radius $R$,
\begin{equation}
\label{eq.sect.super.dyn.Feynman.Vc}
V_c\simeq \frac{\hbar}{m R} \log\left(\frac{R}{r_\upsilon}\right) \, ,
\end{equation}
where $r_\upsilon$ is the radius of the vortex core. This result shows
that the critical velocity depends, in general, on experimental
set-up. More generally, the critical velocity scales like $V_c\propto
\hbar/m L$, where $L$ is a characteristic length scale in the
experiment. The theoretical determination of the breakdown of
superfluidity is still an open issue, which requires a detailed
understanding of superfluid dynamics and, in particular, the dynamics 
of vortices.

\subsubsection{Rotating superfluid and vortices}
\label{sect.super.dyn.rot}

Superfluidity is closely related to the phenomenon of
Bose--Einstein condensation as first envisioned by Fritz
London~\cite{london-38}. In the superfluid phase, a
macroscopic collection of particles condense into the lowest
quantum single particle state, which (for a uniform system) is a
plane wave state  with zero momentum (therefore a constant). Soon
after the discovery of the superfluidity of liquid helium,
Fritz London introduced the idea of a macroscopic wave function
$\Psi(\pmb{r})$, whose squared modulus is proportional to the
density $n_\Psi$ of particles in the condensate. This density
$n_\Psi$, which should not be confused with the superfluid
density $n_{\cal S}$ introduced in the two-fluid model of
superfluids (Section~\ref{sect.super.dyn.entr}), can be
rigorously defined from the one-particle density
matrix~\cite{penrose-56}. The wave function is defined up to a
global phase factor. The key distinguishing feature of a 
superfluid is the symmetry breaking of this gauge invariance
by imposing that the phase $\phi(\pmb{r})$ be \emph{local}.
The macroscopic wave function thus takes the form
\begin{equation}
\label{eq.sect.super.dyn.macro.psi}
\Psi(\pmb{r})=\sqrt{n_\Psi}\, e^{\mathrm{i}\, \phi(\pmb{r})} \, .
\end{equation}

Applying the momentum operator $-\mathrm{i}\hbar \nabla$ to this
wave function shows that  superfluid carries a net momentum (per superfluid particle)
\begin{equation}
\label{eq.sect.super.dyn.momentum}
\pmb{p}=\hbar \pmb{\nabla} \phi \, .
\end{equation}

This implies that the superflow is characterized by the condition
\begin{equation}\label{eq.sect.super.dyn.irrot}
\pmb{\nabla} \times \pmb{p} = 0\, .
\end{equation}

In the absence of any entrainment effects (as discussed in
Section~\ref{sect.super.dyn.entr}),  the
momentum is given by $\pmb{p}=m\, \pmb{v}$, where $m$ is the mass of
the superfluid ``particles''\epubtkFootnote{In the case of fermionic
  superfluids, the superfluid particles are fermion pairs.} and
$\pmb{v}$ is the velocity of
superfluid. Equation~(\ref{eq.sect.super.dyn.irrot}) thus implies that
the flow is irrotational. This means, in particular, that a superfluid
in a rotating bucket remains at rest with respect to the laboratory
reference frame. However, this {\it Landau state} is destroyed whenever the rotation
rate exceeds the critical threshold for the formation of vortices
given approximately by
Equation~(\ref{eq.sect.super.dyn.Feynman.Vc}). Experiments show that
the whole superfluid then rotates like an ordinary fluid. The
condition~(\ref{eq.sect.super.dyn.irrot}) can therefore be locally
violated as first suggested by Onsager~\cite{onsager-49} and discussed
by Feynman~\cite{feynman-55}. Indeed, since the phase of the
macroscopic wave function is defined modulo $2 \pi$, the momentum
circulation over any closed path is quantized
\begin{equation}
\label{eq.sect.super.dyn.vortices}
\oint \pmb{p} \cdot \pmb{d \ell} = N h \, ,
\end{equation}
where $N$ is any integer and $h$ is Planck's constant. 
Equation~(\ref{eq.sect.super.dyn.vortices}) is simply the
Bohr-Sommerfeld quantization rule. This rule can be equivalently
deduced from the following argument. Considering  superfluid as a
macroscopic quantum state, its momentum $p$ is given by $p=h/\lambda$,
where $\lambda$ is the de Broglie wavelength. The quantization
rule~(\ref{eq.sect.super.dyn.vortices}) thus follows from the
requirement that the length of any closed path must be an integral
multiple of the de Broglie wavelength.

In a rotating superfluid the flow quantization implies the formation
of vortex lines, each carrying  a quantum $\hbar$ of angular
momentum, the quantum number $N$ being the number of vortices (the
formation of a single vortex carrying all the angular momentum is not
energetically favored). The size of the core of a vortex line is
roughly on the order of the superfluid coherence length (see
Section~\ref{sect.super.static.crust} for estimates of the coherence
length). In some cases, however, it may be much
smaller~\cite{deblasio-99, elgaroy-01}, so that the coherence length is
only an upper bound on the vortex core size. In the presence of
vortices, Equation~(\ref{eq.sect.super.dyn.irrot}) must therefore be
replaced by
\begin{equation}
\label{eq.sect.super.dyn.irrot.vortices}
\pmb{\nabla} \times \pmb{p} = m\, \pmb{\kappa} \, ,
\end{equation}
$\pmb{\kappa}$ being the circulation of the $N$ vortex lines. The
existence of quantized vortices was demonstrated by
Vinen~\cite{vinen-61} in 1961, however, they were not observed until
much later, in 1974 at Berkeley~\cite{williams-74, yarmchuk-79}. At
length scales much larger than the superfluid coherence length, the
finite size of the vortex core can be neglected and the circulation is,
thus, given by
\begin{equation}
\label{eq.sect.super.dyn.tot.circ}
\pmb{\kappa} = \sum_{\alpha=1}^N \delta^{(2)}(\pmb{r}-\pmb{r_\alpha})\, \pmb{\kappa_\alpha} \, ,
\end{equation}
where $\delta^{(2)}$ is the two-dimensional Dirac distribution, and
$\pmb{\kappa_\alpha}$ is the circulation of a given vortex $\alpha$
\begin{equation}
\label{eq.sect.super.dyn.circ}
\pmb{\kappa_\alpha} = \pmb{\hat\kappa_\alpha}\, \frac{1}{m}
\oint \pmb{p} \cdot \pmb{d\ell} = \frac{h}{m} \pmb{\hat\kappa_\alpha} \, ,
\end{equation}
$\pmb{\hat\kappa_\alpha}$ being a unit vector directed along the
vortex line. Equation~(\ref{eq.sect.super.dyn.irrot.vortices}) is
formally similar to Ampere's law in magnetostatics. The momentum
$\pmb{p}$ induced by the vortex lines is, thus, given by the Biot-Savart
equation, substituting $1/c$ with the mass $m$ divided by $4\pi$ in
Gaussian cgs units (or replacing the magnetic permeability $\mu_0$ by
the mass $m$ in SI units) and the electric current $I$ by $\kappa$,
\begin{equation}
\label{eq.sect.super.dyn.biot-savart}
\pmb{p} (\pmb{r}) = \frac{m}{4\pi} \int \mathrm{d}r^\prime \, \pmb{\kappa}(\pmb{r^\prime}) \times \frac{ \pmb{r}-\pmb{r^\prime}}{|r-r^\prime|^3} \, ,
\end{equation}
where the integral is taken along the vortex lines as shown in
Figure~\ref{fig.sect.super.dyn.biotsavart}. The analogy between
hydrodynamics and magnetostatics shows, in particular, that a vortex
ring should  move along its symmetry axis with a velocity inversely
proportional to its radius.

\epubtkImage{biotsavart.png}{%
\begin{figure}[htbp]
  \centerline{\includegraphics[scale=0.3]{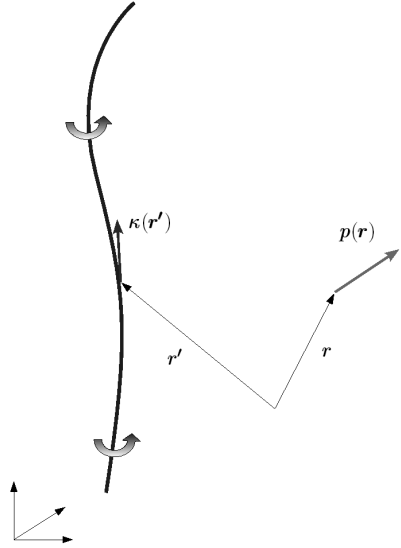}}
  \caption{Momentum $\pmb{p}(\pmb{r})$ induced at a position $\pmb{r}$
  by the vortex line with circulation $\pmb{\kappa}$.}
\label{fig.sect.super.dyn.biotsavart}
\end{figure}}

As shown by Tkachenko~\cite{tkachenko-65, tkachenko-66}, quantized
vortices tend to arrange themselves on a regular triangular array. Such
patterns of vortices have been observed in superfluid helium and more
recently in atomic Bose--Einstein condensates. The intervortex spacing
$d_\upsilon$ is given by
\begin{equation}
\label{eq.sect.super.dyn.intervortex}
d_\upsilon=\sqrt{\frac{h}{\sqrt{3} m \Omega}}\, ,
\end{equation}
where $\Omega$ is the angular frequency. At length scales much larger
than the intervortex spacing $d_\upsilon$, as a result of the
superposition of the flow pattern of all the vortex lines,
superfluid flow mimics rigid body rotation. Since at this scale a fluid
element is threaded by many vortex lines, it is relevant to smoothly
average the hydrodynamic equations governing the flow of the
superfluid. In particular,
Equation~(\ref{eq.sect.super.dyn.irrot.vortices}) now reads
\begin{equation}
\label{eq.sect.super.dyn.irrot.vortices2}
\pmb{\nabla} \times \pmb{p} = m n_\upsilon \pmb{\kappa} \, ,
\end{equation}
where $n_\upsilon$ is the surface density of vortices given (in the 
absence of entrainment effects) by
\begin{equation}
\label{eq.sect.super.dyn.vortex.density}
n_\upsilon = \frac{m \Omega}{\pi \hbar} \, ,
\end{equation}
and the vector $\pmb{\kappa}$, whose norm is equal to $h/m$, is
aligned with the average angular velocity (generalization of 
Equation~(\ref{eq.sect.super.dyn.vortex.density}) to account for 
entrainment effects is discussed in Section~\ref{sect.hydro.super}).

\subsubsection{Type~II superconductors and magnetic flux tubes}
\label{sect.super.dyn.fluxtubes}

Let us remark that the condition~(\ref{eq.sect.super.dyn.irrot}) for
superfluids also applies to superconductors, like the proton
superconductor in the liquid core and possibly in the ``pasta'' mantle
of neutron stars (Section~\ref{sect.groundstate.pasta}). The momentum
of a  superconductor is given by $\pmb{p}\equiv m\, \pmb{v}+q
\pmb{A}$ (in this section, we use SI units), where $m$, $q$, and $v$ are the mass, electric charge and velocity of ``superconducting'' particles
respectively\epubtkFootnote{For a proton superconductor, $m=2m_p$ and
  $q=2e$, where $m_p$ and $e$ are the proton mass and proton electric
  charge, respectively.}, and $\pmb{A}$ the electromagnetic potential
vector. Introducing the density $n$ of superconducting particles and
their electric current density (referred to simply as ``supercurrent'')
$\pmb{\cal J}=nq\pmb{v}$, Equation~(\ref{eq.sect.super.dyn.irrot})
leads to the London equation
\begin{equation}
\label{eq.sect.super.dyn.London}
\pmb{\nabla} \times \pmb{\cal J} = - \frac{n q^2}{m} \pmb{B} \, ,
\end{equation}
where $\pmb{B} =\pmb{\nabla}\times \pmb{A}$ is the magnetic field
induction. According to the Bohr-Sommerfeld quantization
rule~(\ref{eq.sect.super.dyn.vortices}), the London
Equation~(\ref{eq.sect.super.dyn.London}) corresponds to
$N=0$. Situations with $N>0$ are encountered in type~II
superconductors. Considering a closed contour outside a sample of
such a superconductor for which ${\cal J}=0$ and integrating the
momentum $\pmb{p}$ along this contour, leads to the quantization of
the total magnetic flux into $N$ flux tubes 
\begin{equation}
\oint \pmb{A} \cdot \pmb{d\ell} = \frac{h N}{q}\, .
\end{equation}

These flux tubes tend to arrange themselves into a triangular lattice,
the Abrikosov lattice, with a spacing given by
\begin{equation}
d_\upsilon=\sqrt{\frac{2 h}{\sqrt{3} q B}}\, .
\end{equation}
Averaging at length scales much larger than $d_\upsilon$, the surface
density of flux tubes is given by
\begin{equation}
n_\upsilon = \frac{q B}{h} \, .
\end{equation}

\subsubsection{Superfluid vortices and magnetic flux tubes in neutron
  stars}
\label{sect.super.dyn.ns}

For neutron superfluid in neutron stars, superfluid particles
are neutron pairs, so that $m=2m_n$. As early as in 1964, Ginzburg \&
Kirzhnits~\cite{ginzburg-64a, ginzburg-64b} suggested the existence of
quantized vortex lines inside neutron stars. The critical velocity for
the nucleation of vortices can be roughly estimated from $V_c\sim
\hbar/m_n R$ where $R$ is the radius of a neutron star. For $R=10$ km,
$V_c\sim 10^3$ fm/s, which is, by several orders of
magnitude, smaller  than  characteristic velocities of
matter flows within the  star. The
interior of neutron stars is thus threaded by a huge number of
vortices. The intervortex spacing is
\begin{equation}
\label{eq.sect.super.dyn.intervortex.NS}
d_\upsilon\simeq 3.4\times 10^{-3} \sqrt{\frac{10^{2} \mathrm{\
      s}^{-1}}{\Omega}} \mathrm{\ cm} \, ,
\end{equation}
which is much larger than the coherence length, so that the assumption
of infinitely-thin vortex lines in
Equation~(\ref{eq.sect.super.dyn.tot.circ}) is justified. Assuming
that  neutron superfluid is uniformly co-rotating with  the star,
the density of vortices per square kilometer is
then given by $n_\upsilon \simeq 10^{14}/P$, where $P$ is the
rotation period in seconds. With a rotation period of 33
milliseconds, a pulsar like the Crab is threaded by an array of about
$10^{18}$ vortex lines (assuming a radius of 10~km)! Likewise,
proton superconductors (assumed to be of type~II~\cite{baym-69}) in the
core of neutron stars (and possibly in the pasta layers; see
Section~\ref{sect.groundstate.pasta}) is threaded by quantized
magnetic flux tubes each carrying a magnetic field on the order of
$10^{15}$~G. The surface density of flux tubes is about $10^{13}$
times that of neutron vortices. The interactions between the neutron
vortices and the flux tubes are likely to affect the dynamic
evolution of the star~\cite{ruderman-98, ruderman-98err}.

\subsubsection{Dynamics of superfluid vortices}
\label{sect.super.dyn.vortices}

In this section, we will discuss the nonrelativistic dynamics of
superfluid vortices. The generalization to relativistic dynamics
has been discussed in detail by Carter~\cite{carter-99}. According to
the Helmholtz theorem, the vortex lines are frozen in  superfluid
and  move with the same velocity unless some force acts on
them. The dynamics of a vortex line through the crust is governed by
different types of forces, which depend on the velocities
$\pmb{v_{\mathrm{f}}}$, $\pmb{v_\upsilon}$ and $\pmb{v_{\mathrm{c}}}$
of the bulk neutron superfluid, the vortex and the solid crust,
respectively.

\begin{itemize}

\item A viscous drag force (not to be confused with entrainment,
which is a nondissipative effect; see Section~\ref{sect.super.dyn.entr})
opposes  relative motion between a vortex line and the crust, inducing
dissipation. At sufficiently small relative velocities,
the force per unit length of the vortex line can be written as
\begin{equation}
\pmb{{\cal F}_{\mathrm{d}}} = -{\cal R} (\pmb{v_\upsilon} - \pmb{v_{\mathrm{c}}}) \, ,
\end{equation}
where ${\cal R}$ is a positive resistivity coefficient, which is
determined by the interactions of the neutron vortex line with the
nuclear lattice and the electron gas. The pinning of the vortex line
to the crust is the limit of very strong drag entailing  that
$\pmb{v_\upsilon}= \pmb{v_{\mathrm{c}}}$.

\item Relative motion of a vortex line with respect to  bulk
  superfluid (caused by drag or pinning) gives rise to a Magnus or
  lift force (analog to the Lorentz force), given by
\begin{equation}
\pmb{{\cal F}_{\mathrm{m}}} = \rho_{\mathrm{f}} \pmb{\kappa} \times (\pmb{v_{\mathrm{f}}}-\pmb{v_\upsilon}) \, ,
\end{equation}
where $\rho_{\mathrm{f}}$ is the mass density of the free superfluid
neutrons and $\pmb{\kappa}$ is a vector oriented along the superfluid
angular velocity and whose norm is given by $h/2m_n$  (see Carter \&
  Chamel~\cite{CCII-05} for the generalization to multi-fluid systems).

\item A tension force resists the bending of the vortex line and is
  given by
\begin{equation}
\pmb{{\cal F}_{\mathrm{t}}} = - \rho_{\mathrm{f}} \kappa
{\cal C}_{\mathrm{t}} \frac{\partial^2 \pmb{u}}{\partial z^2} \, ,
\end{equation}
where $\pmb{u}$ is the two-dimensional displacement vector of the
vortex line directed along the $z$-axis. ${\cal C}_{\mathrm{t}}$ is a
rigidity coefficient of order
\begin{equation}
{\cal C}_{\mathrm{t}} \sim \frac{\kappa}{4\pi} \ln
\frac{d_\upsilon}{r_\upsilon} \, ,
\end{equation}
where $\kappa=h/2 m_n$, $d_\upsilon$ is the intervortex spacing and
$r_\upsilon$ the size of the vortex core~\cite{sonin-87}.

\end{itemize}

All forces considered above are given per unit length of the vortex
line. Let us remark that even in the fastest millisecond pulsars, the
intervortex spacing (assuming a regular array) of order
$d_\upsilon\sim 10^{-3}\mbox{\,--\,}10^{-4}$~cm is much larger than
the size of the vortex core $r_\upsilon\sim
10\mbox{\,--\,}100$~fermis. Consequently the vortex-vortex
interactions can be neglected.

The dynamic evolution of a vortex line is governed by
\begin{equation}
\label{eq.sect.super.dyn.vortex.evol}
\frac{m_\upsilon}{L_\upsilon} \frac{\mathrm{d}\pmb{v_\upsilon}}{\mathrm{d}t} = \pmb{{\cal F}_{\mathrm{d}}} + \pmb{{\cal F}_{\mathrm{t}}} + \pmb{{\cal F}_{\mathrm{m}}} \, ,
\end{equation}
where $m_\upsilon$ is the inertial mass of the vortex line and
$L_\upsilon$ its length. Since the free neutron density inside the
vortex core is typically much smaller than outside (unless the line is
pinned to nuclei)~\cite{elgaroy-01, yu-03, avogadro-07}, the motion of
the vortex line is accompanied by a rearrangement of the free
neutrons. The inertial mass of the vortex line is approximately equal
to the mass density of the neutron superfluid $\rho_{\mathrm{f}}$ times the
volume of the line $\pi r_\upsilon^2 L_\upsilon$ so that
$m_\upsilon/L_\upsilon\sim \rho_{\mathrm{f}}\pi r_\upsilon^2$.

On a scale much larger than the intervortex spacing, the drag force
$\pmb{{\cal F}_{\mathrm{d}}}$ acting on every vortex line leads to a mutual
friction force between the neutron superfluid and the normal
constituents. Assuming that the vortex lines are rigid and form a
regular array, the mutual friction force, given by $\pmb{f_{\mathrm{d}}}=n_\upsilon\, \pmb{{\cal F}_{\mathrm{d}}}$, can be obtained from
Equation~(\ref{eq.sect.super.dyn.vortex.evol}) after multiplying by
the surface density $n_\upsilon$. Since $n_\upsilon\sim 1/\pi
d_{\upsilon}^2$, the inertial term on the
left-hand side of Equation~(\ref{eq.sect.super.dyn.vortex.evol}) is
proportional to $(r_\upsilon/d_\upsilon)^2\ll 1$ and can be
neglected. Solving the force balance equation yields the mutual
friction force (per unit volume)~\cite{andersson-06}
\begin{equation}
\label{eq.sect.super.dyn.mutual.friction}
\pmb{f_{\mathrm{d}}}=n_\upsilon\, \pmb{{\cal F}_{\mathrm{d}}} = \frac{{\cal B}^2}{1+{\cal B}^2}\rho_{\mathrm{f}} n_\upsilon \pmb{\kappa} \times (\pmb{v_{\mathrm{N}}} - \pmb{v_{\mathrm{f}}})
+\frac{{\cal B}}{1+{\cal B}^2} \rho_{\mathrm{f}} n_\upsilon \pmb{\hat\kappa} \times \pmb{\kappa} \times (\pmb{v_{\mathrm{c}}} - \pmb{v_{\mathrm{f}}}) \, ,
\end{equation}
where $n_\upsilon$ is the surface density of vortices in a plane
perpendicular to the axis of rotation  and $\cal B$ is a dimensionless
parameter defined by
\begin{equation}
{\cal B} = \frac{\cal R}{\rho_{\mathrm{f}} \kappa} \, .
\end{equation}

Different dissipative mechanisms giving rise to a mutual friction
force have been invoked: scattering of electrons/lattice vibrations
(phonons)/impurities/lattice defects by thermally excited neutrons in
vortex cores~\cite{feibelman-71, harding-78, jones-90a}, electron
scattering off the electric field around a vortex
line~\cite{bildsten-89}, and coupling between phonons and vortex line
oscillations (Kelvin modes)~\cite{epstein-92, jones-92}. In the weak
coupling limit ${\cal B}\rightarrow 0$, the vortices co-rotate with
the bulk superfluid (Helmholtz theorem), while in the opposite limit
${\cal B}\rightarrow +\infty$, they are ``pinned'' to the
crust. In between these two limits,  in a frame co-rotating with the
crust, the vortices move radially outward at angle $\mathrm{atan} ({\cal
  R})$ with respect to the azimuthal direction. The radial component
of the vortex velocity reaches a maximum at ${\cal R}=1$.

Vortex pinning plays a central role in theories of pulsar
glitches. The strength of the interaction between a \emph{small
  segment} of the vortex line and a nucleus remains a controversial
issue~\cite{alpar-77, epstein-88, pizzochero-97, elgaroy-01,
  donati-03, donati-04, donati-06, avogadro-07}. The actual
``pinning'' of the vortex line (i.e., $\pmb{v_\upsilon}=
\pmb{v_{\mathrm{c}}}$) depends not only on the vortex-nucleus interaction,
but also on the structure of the crust, on the rigidity of
lines and on the vortex dynamics. For instance, assuming that the crust is a
polycrystal, a rigid vortex line would not pin to the crust simply
because the line cannot bend in order to pass through the nuclei,
independent of the strength of the vortex-nucleus interaction!
Recent observations of long-period precession in
PSR~1828$-$11~\cite{stairs-00}, PSR~B1642$-$03~\cite{shabanova-01} and
RX~J0720.4$-$3125~\cite{haberl-06} suggest that, at least in those neutron
stars, the neutron vortices cannot be pinned to the crust and must be
very weakly dragged~\cite{shaham-77, link-02}.

Let us stress that the different forces acting on a vortex
vary along the vortex line. As a consequence, the vortex lines may not
be straight~\cite{hirasawa-01}. The extent to which the lines are
distorted depends on the vortex dynamics. In particular,
Greenstein~\cite{greenstein-70} suggested a long time ago that vortex
lines may twist and wrap about the rotation axis giving rise to a
turbulent flow. This issue has been more recently addressed by several
groups~\cite{peralta-05, peralta-06, melatos-07, andersson-07}. In
such a turbulent regime the mutual friction force takes the
form~\cite{gorter-49}
\begin{equation}
\label{eq.sect.super.dyn.mutual.friction.GM}
\pmb{f_{\mathrm{d}}}= \frac{{\cal C}_{\mathrm{d}}}{\kappa} (\pmb{v_{\mathrm{c}}} - \pmb{v_{\mathrm{f}}})^2 (\pmb{v_{\mathrm{c}}} - \pmb{v_{\mathrm{f}}}) \, ,
\end{equation}
where ${\cal C}_{\mathrm{d}}$ is a dimensionless temperature-dependent
coefficient, assuming a dense random tangle of vortex lines.

\subsubsection{Superfluid hydrodynamics and entrainment}
\label{sect.super.dyn.entr}

One of the striking consequences of superfluidity is the allowance for several 
distinct dynamic components. 
In 1938, Tisza~\cite{tisza-38} introduced a two-fluid model in order
to explain the properties of the newly discovered superfluid phase of
liquid helium-4, which behaves either like a fluid with no viscosity
in some experiments or like a classical fluid in other
experiments. Guided by the Fritz London's idea that superfluidity is
intimately related to Bose--Einstein condensation (which is now
widely accepted), Tisza proposed that liquid helium is a mixture of
two components, a superfluid component, which has no viscosity, and a
normal component, which is viscous and conducts heat, thus, carrying all
the entropy of the liquid. These two fluids are allowed to flow with
different velocities. This model was subsequently developed by
Landau\cite{landau-41, landau-41b} and justified on a microscopic
basis by several authors,  especially Feynman~\cite{feynman-98}.
Quite surprisingly,
 Landau never mentioned Bose--Einstein condensation in his work on
superfluidity. According to Pitaevskii (as recently cited by
Balibar~\cite{balibar-07}), Landau might have reasoned that
superfluidity and superconductivity were similar phenomena (which is
indeed true), incorrectly concluding that they could not depend on
the Bose or Fermi statistics (see also the discussion by 
Feynman in Section 11.2 of his book~\cite{feynman-98}). 

In Landau's two-fluid model, the normal part with particle density
$n_{_{\cal N}}$ and velocity $\pmb{v_{_{\cal N}}}$ is identified with
the collective motions of the system or ``quasiparticles'' (see
Section~\ref{sect.super.dyn.Vc}). The viscosity of the normal fluid is
accounted for in terms of the interactions between those
quasiparticles (see, for instance, the book by
Khalatnikov~\cite{khalatnikov-89} published in 1989 as a reprint of an
original 1965 edition). Following the traditional notations, the
superfluid component, with a particle density $n_{_{\cal S}}$ and a
``velocity'' $\pmb{v_{_{\cal S}}}$, is locally irrotational except at
singular points (see the discussion in
Section~\ref{sect.super.dyn.rot})
\begin{equation}
\pmb{\nabla} \times \pmb{v_{_{\cal S}}} = 0\, .
\end{equation}
As pointed out many times by Brandon Carter, unlike $\pmb{v_{_{\cal N}}}$  
the ``superfluid velocity'' is not a true velocity but is \emph{defined} through
\begin{equation}
\label{eq.super.dyn.Vs}
\pmb{p}\equiv m\, \pmb{v_{_{\cal S}}}\, ,
\end{equation}
where $\pmb{p}$ is the true momentum per particle of the superfluid
and $m$ is the mass of a helium atom. Although deeply anchored in the
Lagrangian and Hamiltonian formulation of classical mechanics, the
fundamental distinction between velocities and their canonical
conjugates, namely momenta, has been traditionally obscured in the
context of superfluidity. Note also that the superfluid density $n_{_{\cal S}}$
coincides neither with the density $n$ of helium atoms (except at $T=0$) 
nor with the density $n_\Psi$ of atoms in the condensate\epubtkFootnote{For instance at $T=0$, 
$n_{_{\cal S}}=n$ while $n_\Psi \simeq 0.1 n$.}.
The density of helium atoms is given by
\begin{equation}
\label{eq.super.dyn.tot.density}
n=n_{_{\cal S}}+n_{_{\cal N}}\, .
\end{equation}

The confusion between velocity and momentum is very misleading and makes generalizations of the two-fluid model to multi-fluid systems (like the interior of neutron stars) unnecessarily difficult. 
Following the approach of Carter (see Section~\ref{sect.hydro}), the two-fluid model
can be reformulated in terms of the real velocity $\pmb{v}$ of the helium atoms instead of 
the superfluid ``velocity'' $\pmb{v_{_{\cal S}}}$. The normal fluid with velocity $\pmb{v_{_{\cal N}}}$
is then associated with the flow of entropy and the corresponding number density is given by 
the entropy density. At low temperatures, heat dissipation occurs via the emission of 
phonons and rotons. As discussed in Section~\ref{sect.super.dyn.Vc}, these quasiparticle excitations represent 
collective motions of atoms with no net mass transport (see, in particular, Figures~\ref{fig.sect.super.dyn.phonons} 
and~\ref{fig.sect.super.dyn.rotons}). Therefore, the normal fluid does not carry any mass, i.e., its associated mass 
is equal to zero. 

Following the general principles reviewed in Section~\ref{sect.hydro}, the momentum $\pmb{p}$ of 
the superfluid helium atoms can be written as
\begin{equation}
\label{eq.super.dyn.entr}
\pmb{p} = m_\star \pmb{v} + (m-m_\star) \pmb{v_{_{\cal N}}} \, .
\end{equation}
Note that the momentum $\pmb{p}$ of the helium atoms is not simply equal to 
$m\, \pmb{v}$ due to the scattering of atoms by quasiparticles. 
In the rest frame of the normal fluid, in which  $v_{_{\cal N}}=0$, the
momentum and the velocity of the superfluid are aligned. However, the
proportionality coefficient is not the (bare) atomic mass $m$ of helium but an effective mass
$m_\star$. This effective mass $m_\star$ is related to the
hydrodynamics of superfluid and should not be confused with the
definitions employed in microscopic many-body theories. Before
going further, let us remark that in the \emph{momentum} rest frame of
the normal component, the relation $\pmb{p}=m\, \pmb{v}$ holds! This
can easily be shown from Equation~(\ref{eq.super.dyn.tot.momentum}), by
using the identity~(\ref{eq.super.dyn.noether}) and remembering that
the normal fluid is massless.

Comparing Equations~(\ref{eq.super.dyn.entr}) and~(\ref{eq.super.dyn.Vs})
shows that the ``superfluid velocity'' in the original two-fluid model of Landau
is not equal to the velocity of the helium atoms but is a linear combination of both
velocities $\pmb{v}$ and $\pmb{v_{_{\cal N}}}$
\begin{equation}
\label{eq.super.dyn.Vs2}
\pmb{v_{_{\cal S}}} = \frac{m_\star}{m} \pmb{v} + (1-\frac{m_\star}{m}) \pmb{v_{_{\cal N}}} \, .
\end{equation}
The current of helium atoms is given by the sum of the normal and superfluid
currents
\begin{equation}
\label{eq.super.dyn.tot.current}
n\,\pmb{v}=n_{_{\cal S}}\pmb{v_{_{\cal S}}}+n_{_{\cal
    N}}\pmb{v_{_{\cal N}}} \, .
\end{equation}
Substituting Equation~(\ref{eq.super.dyn.Vs2}) into Equation (\ref{eq.super.dyn.tot.current})
yields the relations
\begin{equation}
n_{_{\cal S}} = n \frac{m}{m_\star} \, , \hskip 0.5cm n_{_{\cal N}} = n \left(1-\frac{m}{m_\star}\right) \, ,
\end{equation}
which clearly satisfies $n=n_{_{\cal S}}+n_{_{\cal N}}$.
The superfluid and normal densities can be directly measured in the
experiment devised by Andronikashvili~\cite{andronikashvili-46}. A
stack of disks, immersed in superfluid, can undergo torsional
oscillations about its axis. Due to viscosity, the normal component is
dragged by  motion of the disks, while the superfluid part remains
at rest. The normal and superfluid densities can, thus, be obtained at
any temperature by measuring the oscillation frequency of the
disks. Since $n_{_{\cal S}}(T)/n=m/m_\star(T)$, the dynamic effective mass $m_\star(T)$ can be
determined experimentally. In particular, it is equal
to the bare mass at $T=0$, $m_\star(T=0)=m$, and
goes up  as the temperature is raised, diverging at the critical point when superfluidity 
disappears. At any temperature $T>0$ the dynamic effective mass of an helium atom is therefore
\emph{larger} than the bare mass.

Entrainment effects, whereby momentum and velocity are not aligned, exist in any 
fluid mixtures owing to the microscopic interactions between the particles. 
But they are usually not observed in ordinary fluids due to the viscosity, which tends to 
equalize velocities. 
Even in superfluids like liquid Helium~II, 
entrainment effects may be hindered at finite temperature\epubtkFootnote{Entrainment 
effects disappear as $T$ goes to zero since $m_\star(T=0)=m$ so that $\pmb{p}=m\,\pmb{v}$ 
according to Equation~(\ref{eq.super.dyn.entr}).} by dissipative processes. 
For instance, when a superfluid is put into a
rotating container, the presence of quantized vortices induces a mutual
friction force between the normal and superfluid components
(as discussed in Section~\ref{sect.super.dyn.vortices}). As a consequence, 
in the stationary limit the velocities of the two fluids become equal. 
Substituting $\pmb{v}=\pmb{v_{_{\cal N}}}$ in Equation~(\ref{eq.super.dyn.entr}) 
implies that $\pmb{p}=m\,\pmb{v}$, as in the absence of entrainment.

\subsubsection{Entrainment effects in neutron stars}
\label{sect.super.dyn.entr.ns}

A few years after the seminal work of Andreev \& Bashkin~\cite{andreev-76}
on superfluid $^3$He\,--\,$^4$He mixtures, it was realized that entrainment effects 
could play an important role in the dynamic evolution of neutron 
stars~(see, for instance,~\cite{sauls-89} and references therein). 
For instance, these effects are very important for studying 
the oscillations of neutron star cores, composed of superfluid neutrons 
and superconducting protons~\cite{andersson-01c}. Mutual entrainment not only 
affects the frequencies of the modes but, more surprisingly, (remembering that 
entrainment is a nondissipative effect) also affects their damping. 
Indeed, entrainment effects induce a flow of protons around each neutron superfluid 
vortex line. The outcome is that each vortex line carries a huge
magnetic field $\sim 10^{14}$~G~\cite{alpar-84}. The electron
scattering off these magnetic fields leads to a mutual friction force between the neutron
superfluid and the charged particles (see~\cite{andersson-06} and
references therein). This mechanism, which is believed to be the main
source of dissipation in the core of a neutron star, could also be at
work in the bottom layers of the crust, where some protons might be unbound
and superconducting (as discussed in Section~\ref{sect.groundstate.pasta}). 

It has recently been pointed out that entrainment effects 
are also important in the inner crust of neutron stars, where free neutrons 
coexist with a lattice of nuclear clusters~\cite{cchI-06, cchII-05}. 
It is well known in solid state physics that free electrons in ordinary 
metals move as if their mass were replaced by a dynamic effective mass  $m_\star^e$  
(usually referred to as an optical mass in the literature) due to Bragg scattering 
by the crystal lattice (see, for instance, 
the book by Kittel~\cite{kittel-96}).  
The end result is that, in the rest frame of the solid, 
the electron momentum is given by
\begin{equation}
  \pmb{p^e} = m_\star^e \pmb{v_e},
\end{equation} 
where $\pmb{v_e}$ is the electron velocity. This implies that in an arbitrary frame, 
where the solid (ion lattice) is moving with velocity $\pmb{v_{_{\mathrm{I}}}}$, the electron momentum
is not aligned with the electron velocity but is given by
\begin{equation}
  \pmb{p^e} = m_\star^e \pmb{v_e} + (m_e-m^e_\star) \pmb{v_{_{\mathrm{I}}}} \, ,
\end{equation}
which is similar to Equation~(\ref{eq.super.dyn.entr}) for the momentum of superfluid Helium~II. 
The concept of dynamic effective mass was introduced in the context of neutron
diffraction experiments ten years ago~\cite{zeilinger-86} and has only recently been extended to the inner crust of neutron stars by
Carter, Chamel \& Haensel~\cite{cchI-06, cchII-05}. While the dynamic effective electron mass 
in ordinary metallic elements differs moderately from the bare mass $m_\star^e\sim 1\mbox{\,--\,}2\,m_e$ 
(see, for instance, \cite{huttner-96}), Chamel~\cite{chamel-05} has shown that the dynamic 
effective mass of free neutrons in neutron star crust could be very large, $m_\star^{\mathrm{f}}\sim 10\mbox{\,--\,}15\,m_n$. 
The dynamics of the free neutrons is deeply affected by these entrainment effects, which have to be properly 
taken into account (see Section~\ref{sect.hydro}). Such effects are important for modeling 
various observed neutron star phenomena, like pulsar glitches (see Section~\ref{sect.obs.glitches}) 
or neutron star oscillations (see Sections~\ref{sect.obs.gw} and~\ref{sect.obs.sgr}).

\newpage


\section{Conductivity and Viscosity}
\label{sect.cond}

\subsection{Introduction}
\label{sect.cond.introd}

In this section, we will consider the transport of heat, electric
charge, and momentum in neutron star crusts.
In the absence of magnetic fields and assuming that the solid
crust is isotropic\epubtkFootnote{This assumption may not remain valid
  in the ``nuclear pasta'' layers  at the bottom of the crust discussed in
  Section~\ref{sect.groundstate.pasta}.}, the transport of heat and electric
charge is described by the thermal conductivity $\kappa$ and the 
electrical conductivity $\sigma$ (Section~\ref{sect.cond.cond}) respectively. 
Under the same conditions, the transport of momentum is characterized by the shear viscosity $\eta$ and 
the bulk viscosity $\zeta$ (Section~\ref{sect.cond.vis}).

\epubtkImage{kappa-envel.png}{%
\begin{figure}[htbp]
  \centerline{\includegraphics[scale=0.4]{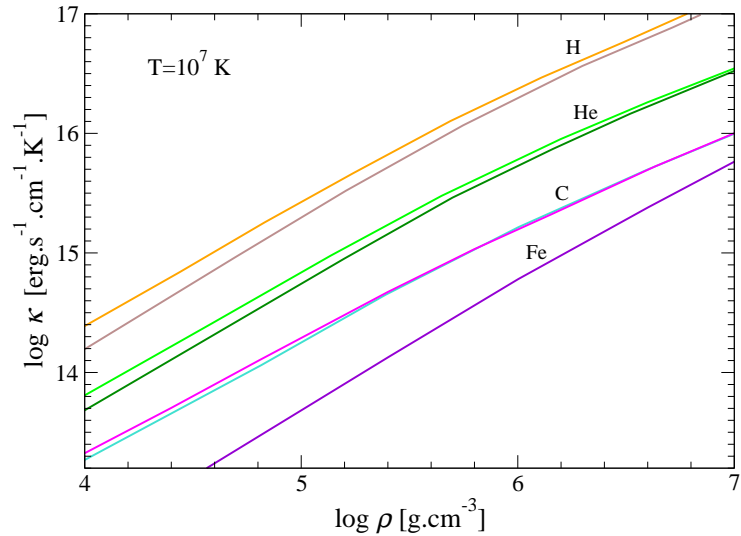}}
  \caption{Thermal conductivity vs. mass density at
    $T=10^7$~K for four types of ions in the neutron star
    envelope. Lower curves: for each composition, electron-ion and
    electron-electron collisions included. Upper curves: electron-ion
    collisions only. Based on Figure~6 from~\cite{potekhin-99}.}
  \label{fig.sect.cond.kappa-envel}
\end{figure}}

Except for the very outer envelope, the main carriers in the
transport processes in the outer crust are electrons,
and they scatter mainly off
ions (exceptions will be mentioned at the end
of the corresponding sections). Theoretical techniques for
the calculation of the transport coefficients in neutron star
crusts are to a large extent borrowed from solid state
physics, the classical reference still remaining the book of
Ziman~\cite{ziman-01}. However, one has to remember that the
density/temperature conditions within neutron star crusts are
tremendously different from those in terrestrial solids, so
that special care concerning the approximations used should
be taken.

\subsection{Boltzmann equation for electrons and its solutions}
\label{sect.cond.BE.electrons}

The electron distribution function is $f({\pmb p},{\pmb r},t,s)$,
where ${\pmb p}$ and ${\pmb r}$ are electron momentum and position
vectors, respectively, $t$ is time, and $s$ is the electron spin
projection on the spin quantization axis.  The distribution function
$f({\pmb p},{\pmb r},t,s)$ satisfies the Boltzmann equation (BE) for
electrons. At first glance, the validity of the BE (originally derived for a
gas of particles) for a super dense plasma of electrons may seem
paradoxical. However, electrons are strongly degenerate, so that only
electron states in a thin shell around the chemical potential $\mu_e$
with energies $|\epsilon - \mu_e|\lesssim k_{\mathrm{B}}T$ are involved in
the transport phenomena. In other words, the gas of ``electron
excitations'' is dilute. Also, the infinite range of Coulomb
interactions in vacuum is no longer a problem in dense
electron-nuclear plasma because of screening. Moreover, as the kinetic
Fermi energy of electrons is much larger than the Coulomb energy per
electron, the Coulomb energy  can be treated as a small perturbation and the
electrons  can be considered as a quasi-ideal Fermi gas.

We consider spin-unpolarized plasma. Then $f$ does not depend on $s$ and the BE for electrons reads
\begin{equation}
{\partial f\over \partial t} + {\pmb v}\cdot {\partial
f\over \partial {\pmb r}} +
{\pmb F}\cdot {\partial f\over \partial {\pmb p}}=I_e[f] \, ,
\label{eq.sect.cond.BoltzEq-e}
\end{equation}
where ${\pmb v}$ is the electron velocity, and ${\pmb F}$ is the external
force acting on the electrons (e.g., electrostatic force ${\pmb F}=-e{\pmb
  E}$, where electron electric charge is $-e$ ). The total collision
  integral $I_e[f]$, which is a functional of $f$, is a sum of
  collision integrals corresponding to
collisions of electrons with nuclei, electrons, and impurities in the
crystal lattice, $I_e[f]=I_{e\mathrm{N}}[f]+I_{ee}[f] + I_{\mathrm{imp}}[f]$.

The additivity of partial collision integrals is valid when the
scatterers are uncorrelated. This assumption may seem surprising
for a crystal. The electron scatters off a lattice  by exciting it,
i.e., transferring  energy and momentum to the lattice.
This process of electron-lattice interactions corresponds
to the creation and absorption of
phonons, which are the elementary excitations of the crystal lattice.
In this way the  electron-lattice interaction is equivalent to the
scattering of electrons by phonons. At temperatures well below
the Debye temperature, $T<{1\over 4}\Theta$, the gas of phonons is dilute,
and e-N scattering, represented by $I_{e\mathrm{N}}[f]$, is actually the
electron scattering by single phonons. These
phonons form a Bose gas, and their
number density  and mean energy depend on $T$.

In the absence of external forces, the solution of
Equation~(\ref{eq.sect.cond.BoltzEq-e}) is the Fermi--Dirac
distribution function, $f^{(0)}$, corresponding to full thermodynamic
equilibrium. The collision integrals then vanish, $I_{ej}[f^{(0)}]=0$,
with $j=\mathrm{N},e,\mathrm{imp}$. We shall now show how to calculate the
conductivities $\kappa$ and $\sigma$. Let us consider small stationary 
perturbations characterized by gradients of temperature ${\pmb \nabla}T$, 
of the electron chemical potential ${\pmb \nabla}\mu_e$, and let us apply a weak constant
electric field ${\pmb E}$. The plasma will  become slightly
nonuniform, with weak charge and heat currents flowing through it. We
assume that the length scale of this nonuniformity is much larger
than the electron mean free path. Therefore, any  plasma element
will be close to a local thermodynamic equilibrium. However,
gradients of $T$ and $\mu_e$, as well as ${\pmb E}$, will induce a
deviation of $f$ from $f^{(0)}$ and will produce
heat and charge currents.

The next step consists in writing  $f=f^{(0)}+\delta f$, where
$\delta f$ is a small correction to $f^{(0)}$, linear in ${\pmb
  \nabla}T$, ${\pmb \nabla}\mu_e$, and ${\pmb E}$. We introduce the
enthalpy per electron $h=\mu_e+(S_e/n_e)T$, where $S_e$ is the electron
entropy density. The linearized  left-hand side
of Equation~(\ref{eq.sect.cond.BoltzEq-e}) is then
\begin{equation}
 -\left[{\epsilon - h\over T}{\pmb \nabla T} + {\pmb E} +
 {{\nabla \mu_e}\over e} + {S_e\over n_e}{{\pmb \nabla}T\over
 e}\right]\cdot {\pmb v}{\partial f^{(0)}\over \partial
 \epsilon} \, .
\label{eq:LHS_BE}
\end{equation}
The  general form of $\delta f$ is~\cite{ziman-01}
\begin{equation}
\delta f = \Phi {\partial f^{(0)}\over \partial  \epsilon}=
-{\Phi\over k_{\mathrm{B}}T}f^{(0)}(1-f^{(0)}) \, ,
\label{eq:delta.f}
\end{equation}
where $\Phi$ is a slowly varying function of electron energy
$\epsilon_e$.
This specific form of $\delta f$ results from the BE, and deserves a
comment. In the limit of strong electron degeneracy $T\ll
T_{\mathrm{F}e}$ where $T_{\mathrm{F}e}$ is given by
Equation~(\ref{sect.plasma.noB.TFe}), the derivative $\partial
f^{(0)}/\partial \epsilon$ is strongly peaked at $\epsilon \approx
\mu_e$. Actually, for $T/T_{\mathrm{F}e}\longrightarrow 0$ we get
$\partial f^{(0)}/\partial \epsilon \longrightarrow -\delta(\epsilon -
\mu_e)$. On the contrary, $\Phi$ is a slowly varying function of
$\epsilon$.

In our case, the general
form of $\Phi$ linear in $\pmb{\nabla}T$, $\pmb{\nabla}\mu_e$
and $\pmb{E}$ can be written as~\cite{ziman-01} 
\begin{equation}
\Phi={\epsilon - h\over T}A_T(\epsilon)\; {\pmb v}\cdot {\pmb
\nabla T} + e A_e(\epsilon)\; {\pmb v}\cdot {\pmb E}^* \, ,
\label{eq:Phi}
\end{equation}
where
\begin{equation}
{\pmb E}^* ={\pmb E} + {{\nabla \mu_e}\over e} + {S_e\over n_e}{{\pmb \nabla}T\over e} \, ,
\label{eq:def.Estar}
\end{equation}
and the coefficients $A_e(\epsilon)$ and $A_T(\epsilon)$ are functions of
electron energy $\epsilon$. They fully determine $\kappa$ and
$\sigma$. However, to determine them, we have to linearize the
collision integrals with respect to $\Phi$, and then solve the
linearized BE.

As the nuclei are very heavy compared to the electrons,
the typical electron energy transferred during a collision is much smaller
than $k_{\mathrm{B}}T$. The collision integral then takes the simple form
(see, e.g., Ziman~\cite{ziman-01})
\begin{equation}
I_{e\mathrm{N}}[f]\approx -{\delta f \over \tau_0(\epsilon)} \, .
\label{eq:I_eN-relax}
\end{equation}
This is the \emph{relaxation time approximation}, and
$\tau_0(\epsilon)$ is an \emph{effective relaxation time} for
the electron distribution function at an energy $\epsilon$. For
strongly degenerate electrons we should put $\epsilon=\mu_e$ in the
argument of $\tau_0(\epsilon)$.

This simple relaxation time approximation breaks down at $T\lesssim
T_{\mathrm{pi}}$, when the quantum effects in the phonon gas become
pronounced so that the typical energies transferred become $\sim
k_{\mathrm{B}}T$ (and the number of phonons becomes exponentially small).
The dominance of electron-phonon scattering breaks down at very low
 $T$. Simultaneously, $I_{ee}$ has a characteristic
low-$T\ll T_{\mathrm{F}e}$ behavior $I_{ee}\propto (T/T_{\mathrm{F}e})^2$.
All this implies the dominance of the e-impurity
 scattering in the low-$T$ limit, $I_{\mathrm{imp}}\gg I_{ee},I_{e\mathrm{N}}$.

The scattering of electrons on ions (nuclei) can be
 calculated from the Coulomb interaction, including medium effects
 (screening). An effective scattering frequency of an electron of energy
 $\epsilon$ , denoted $\nu_{e\mathrm{N}}(\epsilon)$, is  related to the
 corresponding transport scattering cross section $\sigma_{\mathrm{tr}}(\epsilon)$ by
\begin{equation}
 \nu_{e\mathrm{N}}(\epsilon)={1\over \tau_0(\epsilon)}=
 n_{\mathrm{N}}\;v\;\sigma_{\mathrm{tr}}(\epsilon) \, ,
\label{eq:nu_ei.epsilon}
\end{equation}
where $v=\partial \epsilon/\partial p$ is electron velocity.
The transport scattering cross section is calculated by the
integration of the differential scattering cross section,
\begin{equation}
 \sigma_{\mathrm{tr}}(\epsilon)=
 2\pi \int_0^\pi\mathrm{d}\vartheta \sin\vartheta\;
 \sigma\;(\epsilon,\vartheta)\;
 (1-\cos\vartheta) \, ,
\label{eq:sigma.tr}
\end{equation}
where $\vartheta$ is the electron scattering angle.

The electron-nucleus scattering is
quasi-elastic at $T>T_{\mathrm{pi}}$, with electron energy change $\delta\epsilon\ll
k_{\mathrm{B}}T$. The function $\sigma_{\mathrm{tr}}(\epsilon)$ can be
calculated, including screening and relativistic effects. After
scattering, the electron momentum changes by $\hbar q$ within
$q_{\min}\le q \le q_{\max}$. Therefore, the formula for
$\sigma_{\mathrm{tr}}(\epsilon)$ can be rewritten as
\begin{equation}
 \sigma_{\mathrm{tr}}(\epsilon)=
 4\pi {Z^2e^4\over p^2v^2}\Lambda_{e\mathrm{N}}(\epsilon) \, ,
\label{eq:sigma.tr.Lambda}
\end{equation}
where $\Lambda_{e\mathrm{N}}(\epsilon)$ is the famous {\it Coulomb
logarithm} of the plasma transport theory. The Coulomb
logarithm is directly related to the Fourier transform of the
complete electron-nucleus interaction, $\phi_q$, by
 \begin{equation}
  \Lambda_{e\mathrm{N}}(\epsilon)=
  \int_{q_{\min}}^{q_{\max}}
  \mathrm{d}q\, q^3 \mid \phi_q \mid ^2 \, ,
 \label{eq:Lambda.phi_q}
 \end{equation}
where $q_{\min}$ is related to the screening and
$q_{\max}=2p_{\mathrm{F}}/\hbar$ for strongly degenerate electrons.
In a concise numerical form
\begin{equation}
  \tau_{e\mathrm{N}}={1\over \nu_{e\mathrm{N}}(\mu_e)}=
  {5.70\times 10^{-17} \mathrm{\ s}\over \gamma_{\mathrm{r}}Z \Lambda_{e\mathrm{N}}} \, ,
\label{eq:tau_ei.Lambda}
\end{equation}
where $\gamma_{\mathrm{r}}$ is given by
Equation~(\ref{eq.sect.plasma.noB.gamma.r.e}). In the relaxation time
approximation the calculation of transport coefficients
reduces to the calculation of the Coulomb logarithm.

A second important approximation (after the relaxation time one) is
expressed as the {\it Matthiessen rule}. In reality, the electrons
scatter not only off nuclei ($e\mathrm{N}$), but also by themselves
($ee$), and off randomly distributed impurities (imp), if there are
any. The Matthiessen rule (valid under strong degeneracy of electrons)
states that the total effective scattering frequency is the sum of
frequencies on each of the scatterers.

For heat conduction the Matthiessen rule gives, for the total
effective scattering frequency of electrons,
\begin{equation}
\nu_\kappa=\nu^\kappa_{e\mathrm{N}}
+\nu^\kappa_{ee}+\nu^\kappa_{\mathrm{imp}} \, .
\label{eq:cond.Mathiessen.kappa}
\end{equation}
Notice that, as $ee$ scattering does not change the electric
current, it will not contribute to electrical conductivity and
to $\nu_\sigma$, so that the Matthiessen rule gives
\begin{equation}
\nu_\sigma=\nu^\sigma_{e\mathrm{N}}+\nu^\sigma_{\mathrm{imp}} \, .
\label{eq:cond.Mathiessen.sigma}
\end{equation}

Electron scattering on randomly distributed
impurities in some lattice sites is similar to the
scattering by ions with charge $Z-Z_{\mathrm{imp}}$. The scattering frequency
of electrons by impurities is
\begin{equation}
\nu_{\mathrm{imp}}={4\pi e^4 \over p_{\mathrm{F}e}^2 v_{\mathrm{F}e}}\;
\sum_{\mathrm{imp}}\left(Z-Z_{\mathrm{imp}}\right)^2
n_{\mathrm{imp}}\;\Lambda_{e\mathrm{imp}} \, ,
\label{eq:nu.imp}
\end{equation}
where $n_{\mathrm{imp}}$ is the number density of impurities of a given
type ``imp'' and the sum is over all types ``imp''.
Detailed calculations of $\nu^{\kappa}_{e\mathrm{N}}$ and
$\nu^{\sigma}_{e\mathrm{N}}$ in liquid and solid plasma of neutron star envelopes, taking into account additional
effects, such as electron-band structures and multi-phonon
processes, are presented in \cite{potekhin-99}. There one can find
analytic fitting formulae, which are useful for applications.

Recently, the calculation of  $\nu^{\kappa}_{ee}$ has been
revised, taking into account the Landau damping of transverse
plasmons \cite{shternin-06}. This effect strongly
reduces $\nu^{\kappa}_{ee}$ for ultrarelativistic electrons
at $T<T_{\mathrm{p}e}$.

In the presence of a magnetic field ${\pmb B}$, transport
properties become anisotropic,
as briefly described in
Section~\ref{sect.cond.mag}.

\subsection{Thermal and electrical conductivities}
\label{sect.cond.cond}

A small temperature gradient $\pmb{\nabla}T$ and a weak
constant electric field $\pmb{E}$ induce a heat current
${\pmb j}_T$ and an electric current $\pmb{j}_e$,
\begin{equation}
\pmb{j}_T=-Q_T T\pmb{j}_e -\kappa \pmb{\nabla}T \, ,
\label{eq:j.conduct.1}
\end{equation}
\begin{equation}
\pmb{j}_e=\sigma\pmb{E}^* + \sigma Q_T\pmb{\nabla}T \, ,
\label{eq:j.conduct.2}
\end{equation}
where $Q_T$ is the thermopower, $\sigma$ is electrical
conductivity, and $\kappa$ is thermal conductivity.
Here, $\pmb{E}^*$ is an ``effective electric field'',
defined in Equation~(\ref{eq:def.Estar}).

Both conductivities can be expressed in terms of the
corresponding effective scattering
frequency, calculated in the preceding section as
\begin{equation}
\sigma = {e^2 n_e \over m^*_e \nu_\sigma} \, ,
\label{eq:sigma.gen}
\end{equation}
\begin{equation}
\kappa = {\pi^2 k_{\mathrm{B}}^2 T n_e \over 3m^*_e \nu_\kappa} \, ,
\label{eq:kappa.gen}
\end{equation}
where the electron effective mass is given by
Equation~(\ref{eq.sect.plasma.noB.mstar.e}).

In the relaxation time
approximation and for strongly degenerate electrons we get
\begin{equation}
\nu_\kappa=\nu_\sigma\simeq \nu_{e\mathrm{N}}(\mu_e) \, ,
\label{eq:tau_kappa.tau_sigma}
\end{equation}
so that the \emph{Wiedemann--Franz law} is satisfied:
\begin{equation}
\kappa={\pi^2 k_{\mathrm{B}}^2 T \over 3 e^2}\sigma \, .
\label{eq:Wied-Franz_law}
\end{equation}
Let us remind ourselves, that the
equality~(\ref{eq:tau_kappa.tau_sigma}) is violated when the $ee$
scattering is not negligible compared to electron scattering by nuclei
and by impurities. This and other effects leading to violation of the
Wiedemann--Franz law are discussed in \cite{potekhin-99}.

\epubtkImage{condGSC.png}{%
\begin{figure}[htbp]
  \centerline{\includegraphics[scale=0.3]{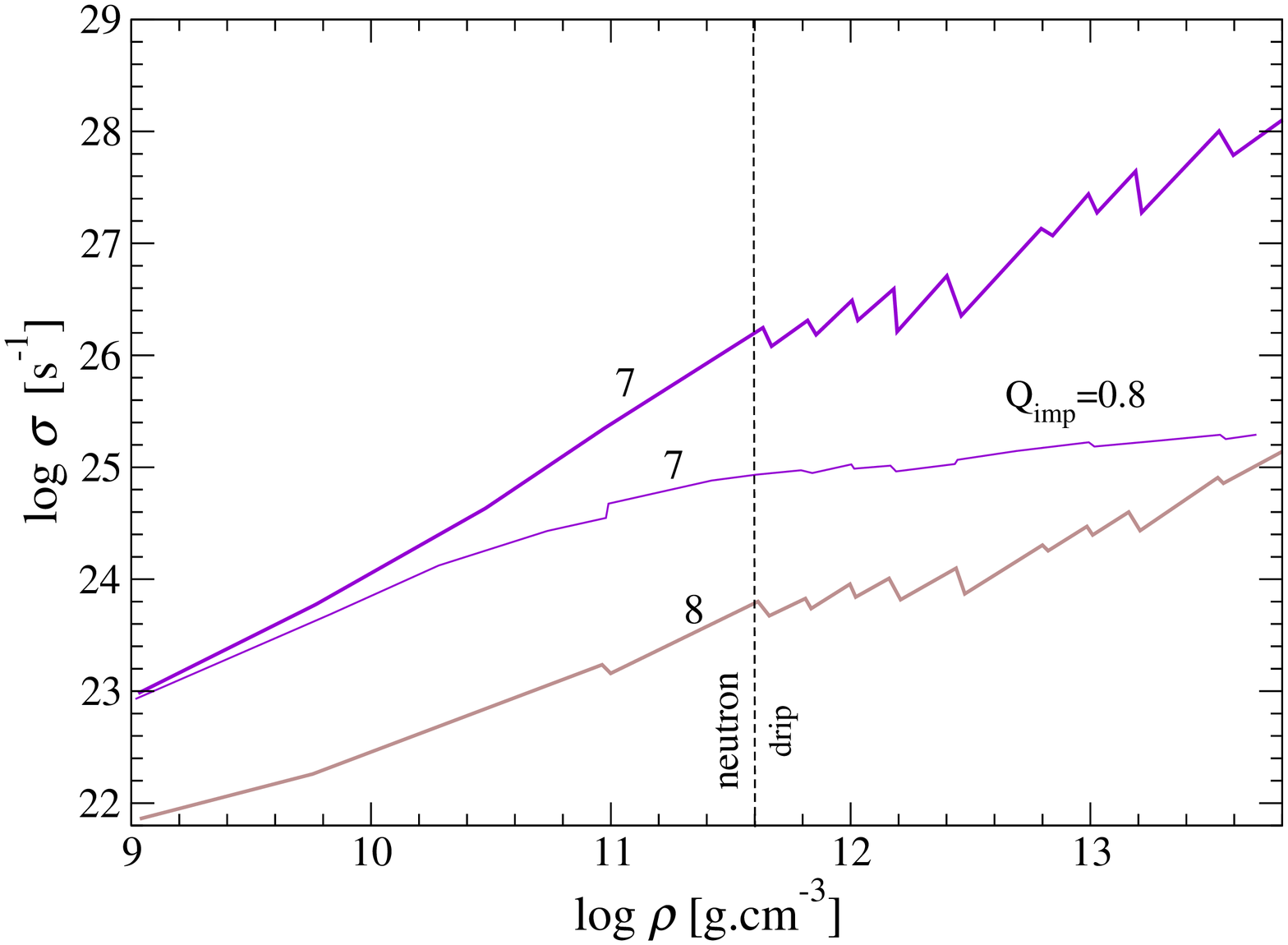}\includegraphics[scale=0.3]{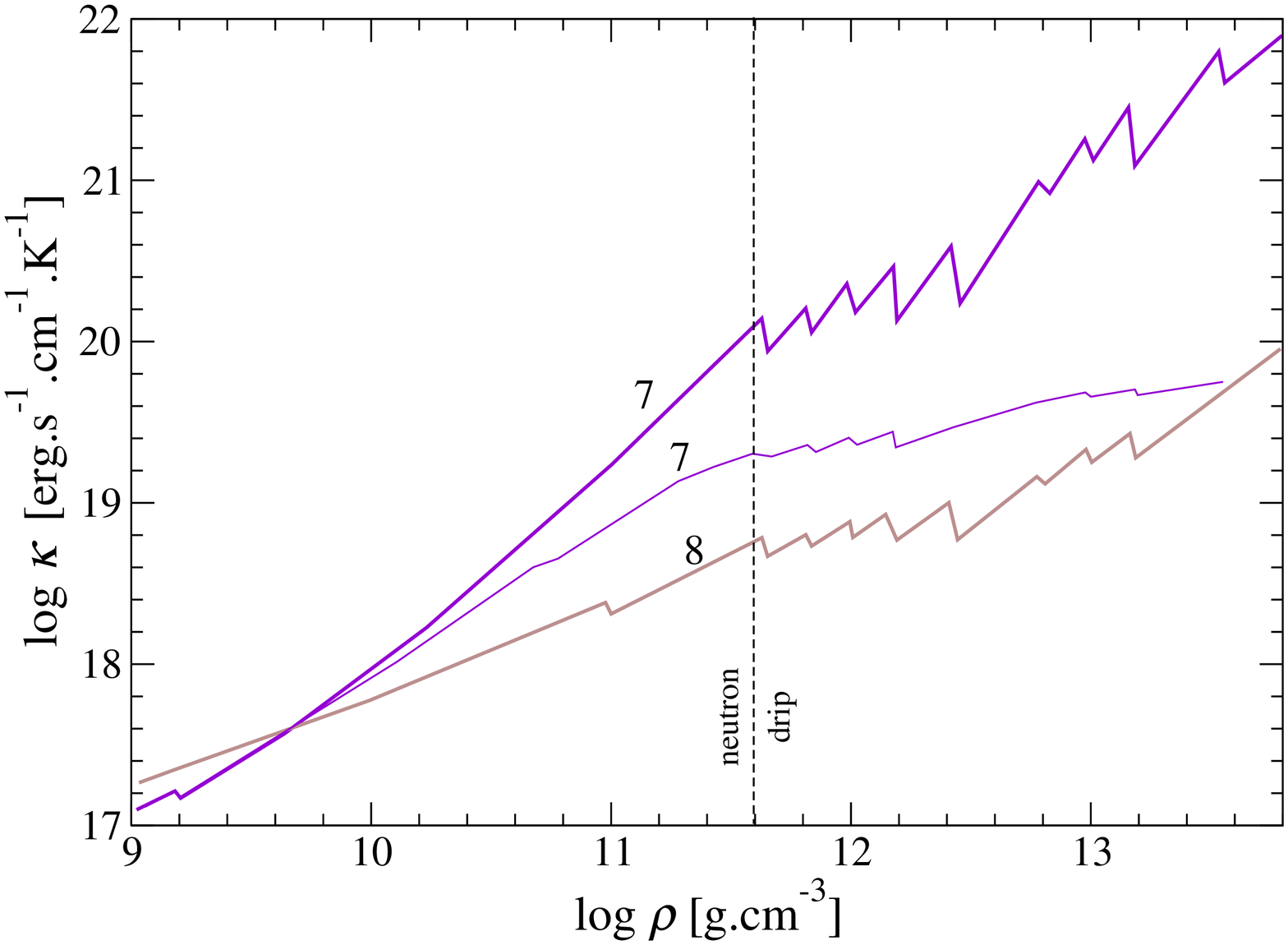}}
  \caption{Electrical conductivity $\sigma$ and thermal conductivity
  $\kappa$ of the outer and inner crust, calculated for the
  ground-state model of Negele \& Vautherin~\cite{nv-73}. Labels 7 and
  8 refer to $\log_{10}T[\mathrm{K}]=7$ and 8, respectively. The thin
  line with label 7 corresponds to an impure crust, which contains  in
  the lattice sites 5\% impurities --  nuclei with
  $|Z_{\mathrm{imp}}-Z|= 4$. Based on a figure made by A.Y.\
  Potekhin.}
  \label{fig.sect.cond.cond.gsc}
\end{figure}}

\epubtkImage{condACC.png}{%
\begin{figure}[htbp]
  \centerline{\includegraphics[scale=0.3]{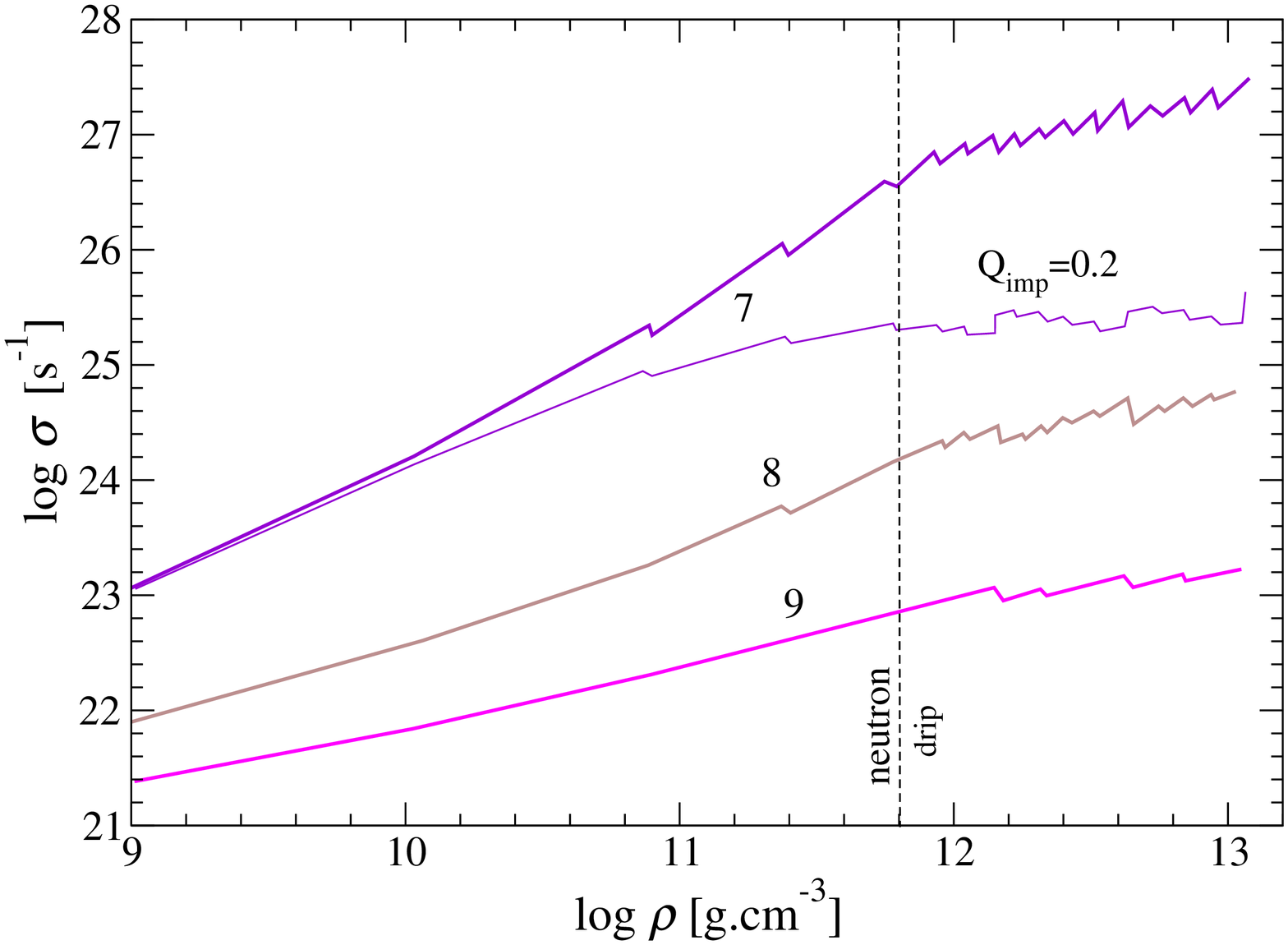}\includegraphics[scale=0.3]{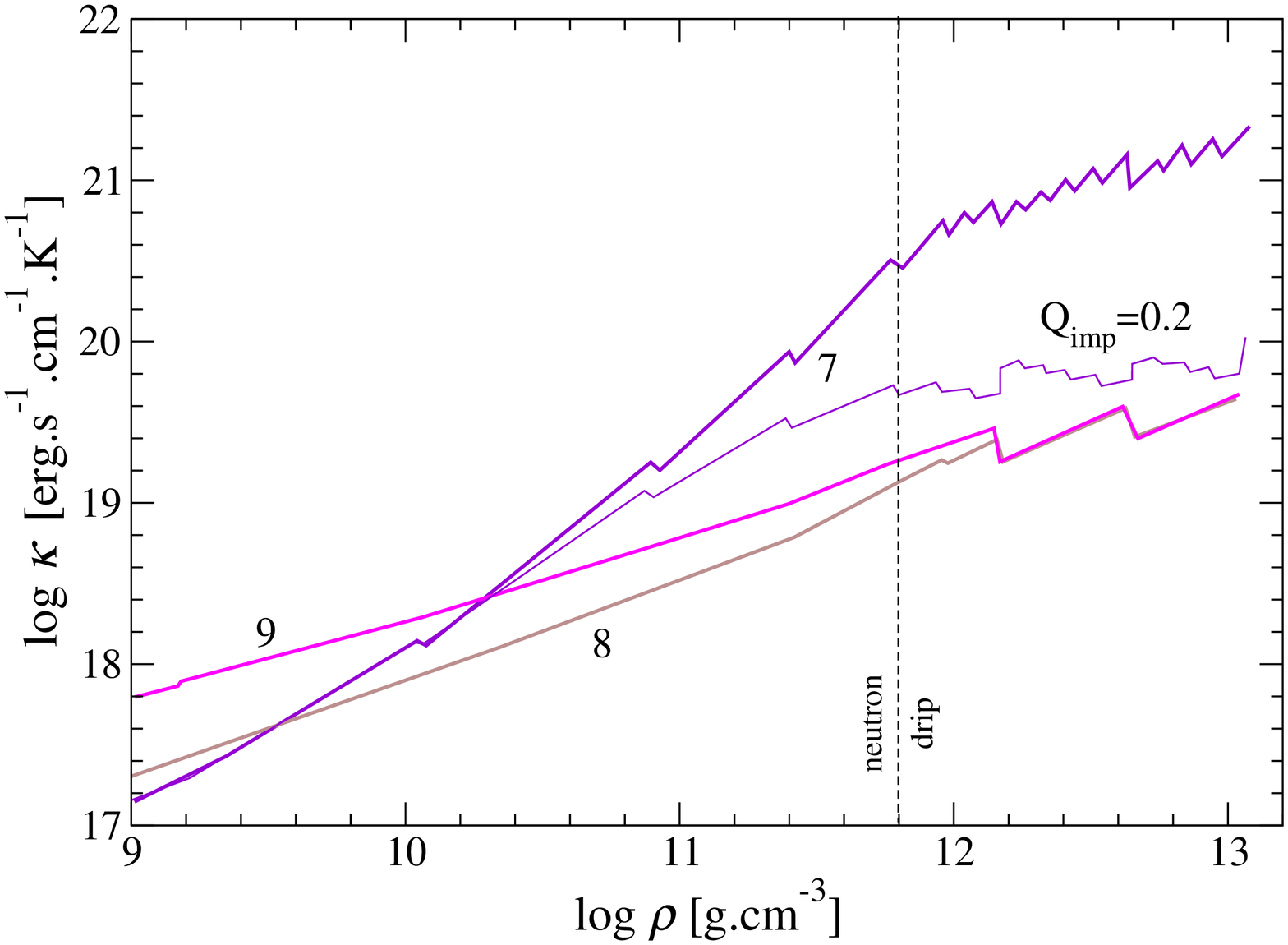}}
  \caption{The same as in Figure~\ref{fig.sect.cond.cond.gsc} but
    calculated for the accreted-crust model of Haensel \&
    Zdunik~\cite{haensel-90b}. Results  obtained
    assuming 5\% of nuclei -- impurities with $|Z_{\mathrm{imp}}-Z|=2$. Based on a
    figure made by A.Y.\ Potekhin.}
  \label{fig.sect.cond.cond.acc}
\end{figure}}

The role of neutron gas in the inner crust deserves comment. Its normal component contributes to $\kappa$, so that
\begin{equation}
\kappa=\kappa_e+\kappa_n \, .
\label{eq:kappa.e.n}
\end{equation}
However, there are actually two contributions to $\kappa_n$. The first
one results from the scattering and is therefore of a standard
``diffusive'' nature. This contribution to $\kappa_n$ is
\begin{equation}
\kappa^{\mathrm{diff}}_n =
{\pi^2 k_{\mathrm{B}}^2 T n_n \over 3 m^*_n \nu_n} \, ,
\label{eq:kappa_n}
\end{equation}
where $n_n$ is the number density of the gas of ``neutron
excitations'' (neutrons are strongly degenerate and moreover
superfluid), $m^*_n$ is their effective mass, and $\nu_n$ is their
scattering frequency. Neutron excitations scatter by  nuclear clusters
and  by  themselves via strong interactions,
\begin{equation}
\nu_n=\nu_{n\mathrm{N}}+\nu_{nn} \, .
\label{eq:nu.n-sum}
\end{equation}
However, due to a much smaller neutron-neutron cross section as compared to
the neutron-cluster one,  and due to the low density of neutron
excitations, we get $\nu_{n\mathrm{N}}\gg \nu_{nn}$, so that
$\nu_n\approx \nu_{n\mathrm{N}}$. The second contribution to $\kappa_n$
is characteristic of superfluids and has a convective character
(``convective counterflow'', see, e.g., Tilley \&
Tilley~\cite{tilley-74}); we denote it by $\kappa_n^{\mathrm{conv}}$.

It can be noted that for $T<T_{\mathrm{m}}$  ``neutron
excitations'' scatter by the lattice phonons.
Complete calculation of $\kappa_n$, taking
due account of the effect of the crystal lattice on neutron scattering and
neutron superfluidity remains to be done.

The presence of impurities considerably decreases electrical and
thermal conductivities at low temperature and high density; see
Figure~\ref{fig.sect.cond.cond.gsc}. At $T=10^7$~K, 5\% of impurities
with $|Z_{\mathrm{imp}}-Z|=4$ reduces $\sigma$ and $\kappa$ at
$\rho=10^{13}~\mdens$ by two orders of magnitude. Accreted crusts
are characterized by nuclides with lower values of $A$ and $Z$ than
those in the ground-state crust. Accordingly, accreted crusts have
higher electrical and thermal conductivities than the ground-state crust
of the same $\rho$ and  $T$. This is
illustrated in Figure~\ref{fig.sect.cond.cond.acc}. Notice the
differences between the $\sigma$ and $\kappa$ plots at $10^8$~K and
$10^9$~K. They are due to an additional factor $T$ in $\kappa$,
reflected in the Wiedermann--Franz law~(\ref{eq:Wied-Franz_law}).

Recent calculations of $\kappa_{ee}$, taking into account the Landau
damping of transverse plasmons, give a much larger contribution from
$ee$ scattering than the previous ones, using the static screening, on
which Figures~\ref{fig.sect.cond.cond.gsc}
and~\ref{fig.sect.cond.cond.acc} are based. As shown by Shternin and
Yakovlev~\cite{shternin-06}, the Landau damping of transverse plasmons
strongly reduces $\nu^{\kappa}_{e}$ in the inner crust at $T\lesssim
10^7$~K.

The contribution of ions to $\kappa$ was recently calculated by
Chugunov and Haensel~\cite{chugunov-07}, who also quote older
papers on this subject. As a rule, $\kappa_{\mathrm{N}}$ can be
neglected compared to $\kappa_e$. A notable exception,
relevant for magnetized neutron stars, is discussed in
Section~\ref{sect.cond.mag}.

\subsection{Viscosity}
\label{sect.cond.vis}

In this section we consider the viscosity of the
crust, which can be in a liquid or a solid phase. For strongly
nonideal ($\Gamma\gg 1$) and solid plasma the transport is
mediated mainly by electrons. We will limit ourselves to this case
only. For solid crust, we will assume a polycrystal structure,
so that on a macroscopic scale the crust will behave as an
isotropic medium.

Let us denote a stationary macroscopic  hydrodynamic
velocity field, imposed on the plasma, by  ${\pmb U}({\pmb r})$. In an
isotropic plasma the viscous part of the stress tensor can be 
written as
\begin{equation}
\Pi_{ij}^{\mathrm{vis}}=\eta\left({\partial U_i\over \partial x_j}
+ {\partial U_j\over \partial x_i} -{2\over 3} \delta_{ij}
{\pmb \nabla}\cdot{\pmb U}\right)
+\zeta\delta_{ij}{\pmb \nabla}\cdot {\pmb U} \, ,
\label{eq:cond_stress_vis}
\end{equation}
where $\eta$ is the shear viscosity and $\zeta$ is the bulk
viscosity. The viscous component of the stress tensor enters
the equations of neutron star hydrodynamics and is
relevant for neutron star pulsations.

\epubtkImage{vis-e-Chugunov.png}{%
\begin{figure}[htbp]
  \centerline{\includegraphics[width=10cm]{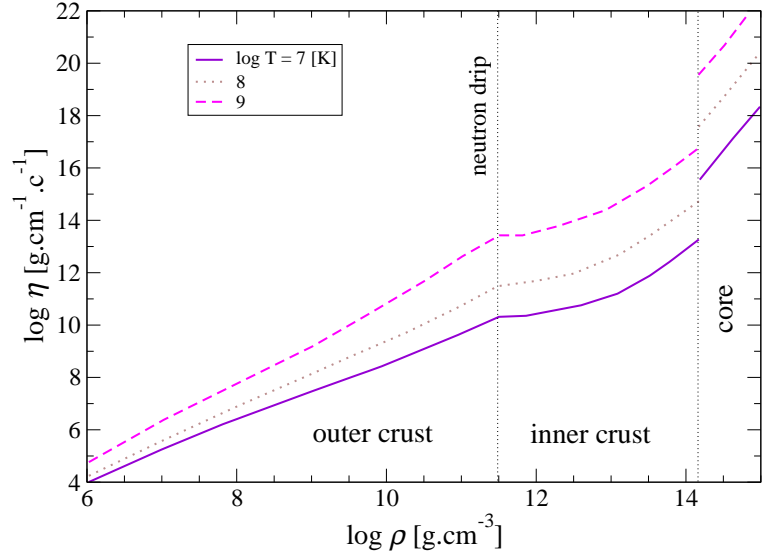}}
  \caption{Electron shear viscosity of the crust and the upper
    layer of the core for $\log_{10}T[\mathrm{K}]=7,8,9$. Calculated by Chugunov \&
    Yakovlev~\cite{chugunov-05} with a smooth composition model of the
    ground-state (Appendix~B of Haensel, Potekhin, and
    Yakovlev~\cite{haensel-06}).}
  \label{sect.cond.vis.elec-chug}
\end{figure}}

\epubtkImage{viscos-e.png}{%
\begin{figure}[htbp]
  \centerline{\includegraphics[width=10cm]{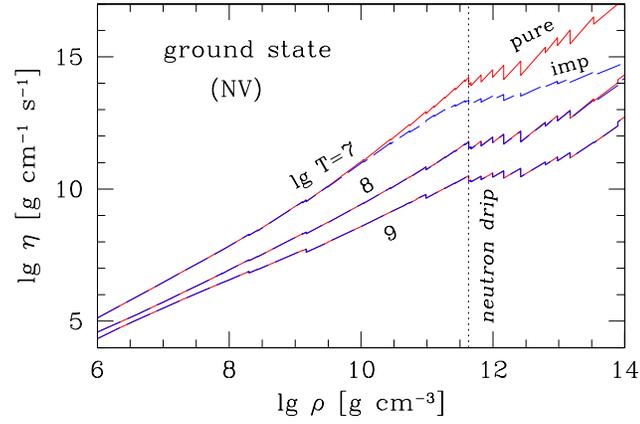}}
  \caption{Electron contribution to crust viscosity and effect of
    impurities. Solid lines -- perfect one-component plasma.
    Dashed  line -- admixture of impurities with
    $n_{\mathrm{imp}}=0.05 n_{\mathrm{i}}$ and $|Z-Z_{\mathrm{imp}}|=2$. Curves
    are labeled by $\log_{10}T[\mathrm{K}]$. Impurity contribution to
    $\eta_e$ becomes dominant for $T\ll T_{\mathrm{pi}}$: this is visible
    for a $\log_{10}T[\mathrm{K}]=7$ curve. Ground-state crust model
    of Negele \& Vautherin~\cite{nv-73} is used. Figure made by A.Y.\
    Potekhin.}
  \label{sect.cond.vis.elec}
\end{figure}}

First, let us consider volume preserving flows, characterized by ${\pmb
  \nabla}\cdot {\pmb U}=0$. A schematic view of such a flow in the
solid crust, characteristic of torsional  oscillations of the crust, is
  shown in Figure~\ref{fig.sect.cond.vis.shear}. The dissipation
  resulting in the entropy production is determined by
  the shear viscosity $\eta$. In the outer crust, $\eta$ is
  a sum of the electron and nuclei contributions, $\eta=\eta_e +
  \eta_{\mathrm{N}}$, but for $\rho>10^5~\mdens$  $\eta_e\gg \eta_{\mathrm{N}}$
  and   $\eta\approx \eta_e$.\epubtkFootnote{Ion contribution to $\eta$ can be important
in the very outer layers with  $\rho<10^4~\mdens$,
where $\eta\approx \eta_{\mathrm{N}}$}.  In the inner crust, an additional
  contribution from the normal component of the gas of dripped
  neutrons should be added.

\epubtkImage{shear.png}{%
\begin{figure}[htbp]
  \centerline{\includegraphics[scale=0.8]{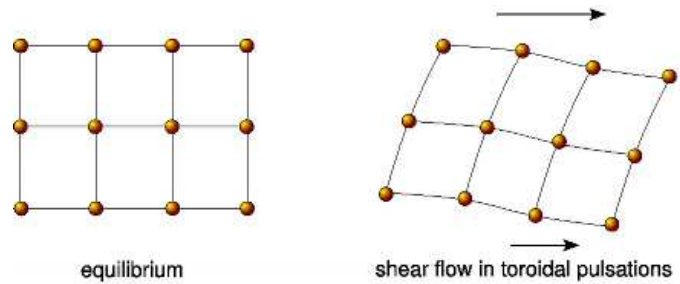}}
  \caption{Schematic picture of torsional oscillations in neutron
  star crust. Left: equilibrium structure of the solid crust,
  represented as a two-dimensional square lattice. Right: shear flow
  in the crust (the shear velocity is indicated by arrows).}
  \label{fig.sect.cond.vis.shear}
\end{figure}}

The electrons are scattered on nuclei, on impurity nuclei, and on
themselves, so that the effective frequency of their scattering is
given by the sum $\nu_e^\eta=\nu_{e\mathrm{N}}^\eta+\nu_{\mathrm{imp}}^\eta+\nu_{ee}^\eta$. However, as long as the temperature is not
too low, the approximation $\nu_e \approx \nu_{e\mathrm{N}}$ can be
used.

To calculate $\eta_e$ from the BE for
electrons, we have to determine $\delta f=f-f^{(0)}$ due to the presence
of a weak plasma velocity field, ${\pmb U}$. The solution of
the BE, linearized in $\pmb{U}$ and in $\delta f$, has the form
\begin{equation}
\delta f= A_\eta(\epsilon)[{\pmb v}\cdot{\pmb \nabla}({\pmb
p}\cdot {\pmb U})]\;{\partial f^{(0)}\over \partial \epsilon} \, ,
\label{eq:delta_f.viscos}
\end{equation}
where $A_\eta(\epsilon)$ is a function to be determined from
the BE. In the relaxation time approximation and for
strongly degenerate electrons $A_\eta=
\tau^\eta_{e\mathrm{N}}=1/\nu_{e\mathrm{N}}$ is the effective
relaxation time due to $e\mathrm{N}$ collisions, calculated at
the electron Fermi surface.

The scattering frequency $\nu_{e\mathrm{N}}$, in turn, can be expressed
in terms of the effective Coulomb logarithm $\Lambda_{e\mathrm{N}}$ by
\begin{equation}
\nu_{e\mathrm{N}}^\eta =12\pi {Z^2e^4n_{\mathrm{N}}\over
  p_{\mathrm{F}e}^2v_{\mathrm{F}e}}\Lambda_{e\mathrm{N}} \, .
\label{eq:nu_ei}
\end{equation}
Having calculated the Coulomb logarithm, we get
the electron viscosity using  a standard formula
\begin{equation}
\eta_e ={n_e p_{\mathrm{F}e}v_{\mathrm{F}e}\over 5\nu_{e\mathrm{N}}},
\label{eq:cond-nu_e}
\end{equation}
At low temperature, impurities can become the main scatterers
of electrons, with
\begin{equation}
\nu_{\mathrm{imp}}={12\pi e^4\over p_{\mathrm{F}e}^2v_{\mathrm{F}e}}
\sum_{\mathrm{imp}}(Z-Z_{\mathrm{imp}})^2n_{\mathrm{imp}}\Lambda_{e\mathrm{imp}} \, .
\label{eq:nu_imp}
\end{equation}
In the inner crust, one must also consider the gas of dripped
neutrons.

Calculations of the shear viscosity for the liquid phase were done by
Flowers \& Itoh~\cite{flowers-76b, flowers-79} and Nandkumar \&
Pethick~\cite{nandkumar-84}. Recently, calculations of the shear viscosity
of the neutron star crust  were done for both the liquid and the crystal
phases, by Chugunov \& Yakovlev~\cite{chugunov-05}; their results are
displayed in Figure~\ref{sect.cond.vis.elec-chug}. These authors also give
analytic fitting formulae for the effective Coulomb logarithms,
which can  be used for different models of the crust. The
electron-impurity scattering becomes dominant at low $T$, when
electron-lattice scattering (via phonons) is suppressed by quantum
effects. This is visualized in Figure~\ref{sect.cond.vis.elec}.
Recently, the contribution to $\eta_e$ resulting from the $ee$
scattering, was recalculated by Shternin~\cite{shternin-08}, 
who took into account the Landau
damping of transverse plasmons. The Landau damping of
transverse plasmons leads to a significant suppression of
$\eta_{ee}$ for ultra-relativistic electrons, and modifies
the temperature dependence of $\eta_{ee}$.

We are not aware of any calculations of the bulk viscosity of the
crust, $\zeta$. We just mention that it is generally assumed that
the bulk viscosity of the crust is much smaller than the shear
one, $\zeta\ll \eta$.

\subsection{Transport in the presence of strong magnetic fields} 
\label{sect.cond.mag}

\epubtkImage{cond_mag_Fe.png}{%
  \begin{figure}[htbp]
    \centerline{\includegraphics[scale=0.5]{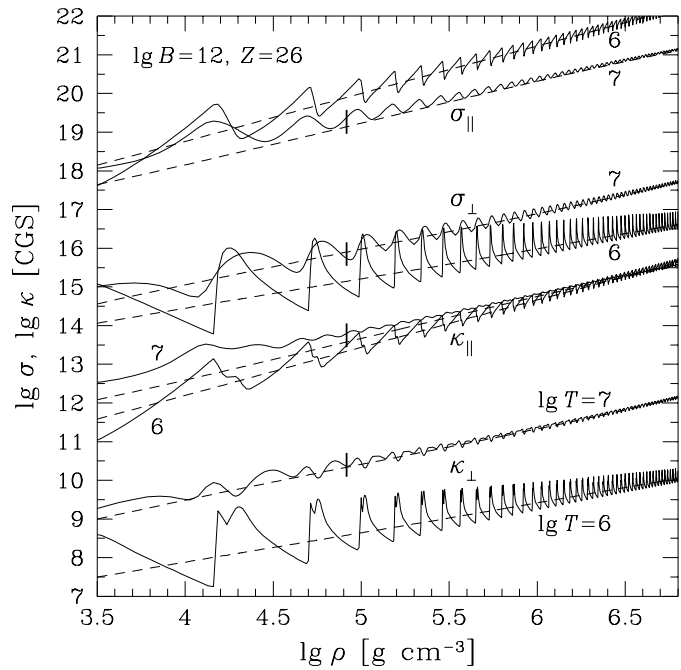}}
    \caption{Longitudinal ($\parallel$) and transverse ($\perp$)
    electrical and thermal conductivities in the outer envelope
    composed of $^{56}$Fe for $B=10^{12}$~G and
    $\log_{10}T[\mathrm{K}]=6,7$. Quantum calculations (solid lines)
    are compared with classical ones (dash lines). Vertical bars:
    liquid-solid transition at $T=10^7$~K. Based on Figure~5
    from~\cite{potekhin-99a}.}
    \label{sect.cond-mag.Fe}
\end{figure}}

We consider a surface magnetic field to be {\it strong}  if $B\gg
10^9$~G. Such magnetic fields affect the accretion of plasma onto
the neutron star and modify the properties of atoms in the
atmosphere. On the contrary, a magnetic field $B\lesssim
10^9$~G, such as associated with millisecond pulsars or with
most of the X-ray bursters, is considered to be {\it weak}.
Typical pulsars are magnetized neutron stars, with the value of
${\pmb B}$ near the magnetic pole $\sim 10^{12}$~G. Much
stronger magnetic fields are associated with magnetars, $\sim
10^{14}\mbox{\,--\,}10^{15}$~G; such magnetic fields with $B\gtrsim 10^{14}$~G are often called
``super-strong''.  These magnetic fields can strongly affect transport
processes within neutron star envelopes. Electron transport
processes in magnetized neutron star envelopes and crusts are
reviewed in~\cite{potekhin-99a, ventura-01}. In the
present section we limit ourselves to a very brief overview.

Locally, a magnetic field can be considered uniform. We will
choose the $z$ axis of a coordinate system along ${\pmb B}$, so
that ${\pmb
  B}=[0,0,B]$. We will limit ourselves to the case of strongly degenerate
electrons and we will assume the validity of the relaxation time
approximation.  Let relaxation time for $B=0$ be $\tau_0$. An
important timescale associated with magnetic fields is the
electron gyromagnetic frequency\epubtkFootnote{The gyromagnetic frequency
of electrons
  should not be confused with the electron cyclotron frequency
  $\omega_{\mathrm{c}e}=eB/m_e c$ entering the formula for the energies
  of the Landau levels; see Section~\ref{sect.plasma}.}
\begin{equation}
  \omega_B={eB\over m^*_e c} \, .
  \label{eq:omegaB}
\end{equation}
The magnetic field bends electron trajectories in the
$(x,y)$ plane, and suppresses the electron transport 
 across ${\pmb B}$. Therefore, the transport
properties become anisotropic, and we must consider tensors
$\kappa_{ij}$ and $\sigma_{ij}$, which  can be written as
\begin{equation}
\kappa_{ij}={\pi^2 k_{\mathrm{B}}^2 T n_e\over 3 m^*_e}\xi^{\kappa}_{ij} \, ,
\label{eq:kappa_ij}
\end{equation}
\begin{equation}
\sigma_{ij}={e^2 n_e\over m^*_e}\xi^{\sigma}_{ij} \, .
\label{eq:sigma_ij}
\end{equation}
In what follows we will consider three basic regimes of
transport in magnetic fields.

\subsubsection{Nonquantizing magnetic fields}
\label{sect:transport-B-nonquant}

Many Landau orbitals are populated and quantum effects are smeared by
thermal effects because $k_{\mathrm{B}}T>\hbar \omega_{\mathrm{c}e}$. Transport along the magnetic field is not affected by ${\pmb B}$,
while transport across ${\pmb B}$ is fully described by the {\it
  Hall magnetization parameters}  $\omega_B\tau_0^{\kappa,\sigma}$,
\begin{equation}
\omega_B\tau_0^{\kappa,\sigma}\approx 1760 {B_{12}\over \gamma_{\mathrm{r}}}
{\tau_0^{\kappa,\sigma}\over 10^{-16} \mathrm{\ s}} \, ,
 \label{eq:magnet-param}
\end{equation}
 where $\tau_0^{\kappa,\sigma}$ are the effective relaxation times at $\pmb{B}=0$ 
for the thermal and electric conductivities, respectively.
The nonzero components of the $\xi^{\kappa,\sigma}_{ij}$ tensors are
\begin{equation}
\xi_{zz}=\tau_0^{\kappa,\sigma} \, , \
\xi^{\kappa,\sigma}_{xx}=\xi^{\kappa,\sigma}_{yy}=
   {\tau_0^{\kappa,\sigma}\over 1+(\omega_B\tau_0^{\kappa,\sigma})^2}
   \, , \
   \xi^{\kappa,\sigma}_{xy}=\xi^{\kappa,\sigma}_{yx}={\omega_B(\tau_0^{\kappa,\sigma})^2\over 1+(\omega_B\tau_0^{\kappa,\sigma})^2} \, .
\label{eq:tau_ij-nonQ}
\end{equation}
As an example, consider strongly degenerate electrons and
dominating $e\mathrm{N}$ scattering, $\tau_0\approx \tau_{e\mathrm{N}}$. Then, Equation~(\ref{eq:tau_ei.Lambda}) yields
$\omega_B\tau_0^{\kappa,\sigma}\approx 1003\; B_{12}/(\gamma_{\mathrm{r}}^2 Z
\Lambda_{e\mathrm{N}})$.

\subsubsection{Weakly-quantizing magnetic fields}
\label{sect:transport-B-weak-quant}

Many Landau levels are populated by electrons, but
quantization effects are well pronounced because
$k_{\mathrm{B}}T<\hbar \omega_{\mathrm{c}e}$. There are two relaxation
times, $\tau_{_{\parallel}}^{\kappa,\sigma}$ and $\tau_{_{\perp}}^{\kappa,\sigma}$, which
oscillate with density (see below). As shown by Potekhin~\cite{potekhin-99a}, the formulae
for the nonzero components of the $\xi^{\kappa,\sigma}_{ij}$ tensors
can be written in a form similar to
Equation~(\ref{eq:tau_ij-nonQ}):
\begin{equation}
\xi^{\kappa,\sigma}_{zz}=\tau_{_{\parallel}}^{\kappa,\sigma} \, , \
  \xi^{\kappa,\sigma}_{xx}=\xi^{\kappa,\sigma}_{yy}=
  {\tau_{_\perp}^{\kappa,\sigma}\over
  1+(\omega_B\tau_{_\perp}^{\kappa,\sigma})^2} \, , \
  \xi^{\kappa,\sigma}_{xy}=\xi^{\kappa,\sigma}_{yx}=
  {\omega_B(\tau_{_\perp}^{\kappa,\sigma})^2\over
  1+(\omega_B\tau_{_\perp}^{\kappa,\sigma})^2} \, .
\label{eq:tau_ij-strongQ}
\end{equation}
At a fixed temperature and in the presence of a weakly
quantizing ${\pmb B}$, the density dependence of the
components of the $\sigma$ and $\kappa$ tensors exhibits
characteristic oscillations around the nonquantized (classical)
values. Each oscillation corresponds to the filling of a new Landau level. The amplitude of these oscillations decreases
with decreasing density. An example of the density
dependence of $\sigma_{_{\parallel}}\equiv \sigma_{zz}$
and $\sigma_{_{\perp}}\equiv \sigma_{xx}=\sigma_{yy}$, and of
the same components of the $\kappa_{ij}$ tensor, is presented
in Figure~\ref{sect.cond-mag.Fe}. As we see in
Figure~\ref{sect.cond-mag.Fe}, the ``density period''of
oscillation decreases with increasing $\rho$,
and the oscillation amplitude decreases with increasing $T$. At $T=10^7$~K,
a magnetic field of $10^{12}$~G is weakly quantizing at
$\rho>10^{4.2}~\mdens$.

\subsubsection{Strongly-quantizing magnetic fields}
\label{sect:transport-B-strong-quant}

Not only is $k_{\mathrm{B}}T<\hbar \omega_{\mathrm{c}e}$, but also
most of the electrons are populating the ground Landau level. Both the
values of $\sigma_{ij}$ and $\kappa_{ij}$  and their density
dependence are dramatically different from those of the nonquantizing
(classical) case.  As shown by Potekhin~\cite{potekhin-99a}, 
the formulae for $\sigma_{ij}$ and
$\kappa_{ij}$ are still given by Equations~(\ref{eq:tau_ij-strongQ}).
Analytical fitting formulae for $\tau_{_{\parallel}}^{\kappa,\sigma}$ and 
$\tau_{_{\perp}}^{\kappa,\sigma}$ are given in \cite{potekhin-99a}. As seen 
in Figure~\ref{sect.cond-mag.Fe}, at
$T=10^6$~K  a field of  $10^{12}$~G is strongly quantizing for
$\rho<10^{4.2}~\mdens$

\subsubsection{Possible dominance of ion conduction}
\label{sect:transport-ion-cond}

Thermal conduction by ions  is much smaller than that
by electrons {\it along} $\pmb{B}$. However, the electron
conduction {\it across} $\pmb{B}$ is strongly suppressed. In
outer neutron star crust, heat flow across $\pmb{B}$  can
be dominated by ion/phonon conduction\cite{chugunov-07}.
This is important for the heat conduction across $\pmb{B}$ in
cooling magnetized neutron stars. Correct inclusion of the ion
heat conductivity then leads to a significant reduction of the
thermal  anisotropy in the envelopes of magnetized neutron
stars\cite{chugunov-07}.

\newpage



\section{Macroscopic Model of Neutron Star Crusts}
\label{sect.hydro}

The understanding of many observed phenomena occurring in neutron
stars (and briefly reviewed in Section~\ref{sect.obs}, for instance, pulsar glitches 
or torsional oscillations in Soft Gamma Repeaters) requires
modeling the dynamic evolution of the crust. 
So far theoretical efforts have been mainly devoted to
modeling the dynamic evolution of the liquid core
by considering a mixture of superfluid neutrons and superconducting
protons (see, for instance, the recent review by Andersson \&
Comer~\cite{anderssoncomer-07}). 

Macroscopic models of neutron star crusts, taking into account the presence of the neutron superfluid
at $\rho>\rho_{\mathrm{ND}}$ (see Section~\ref{sect.super}),
have been developed by Carter and collaborators. They have shown how to extend the two-fluid picture of 
neutron star cores~\cite{chamel-08b} to the inner crust layers in the Newtonian framework~\cite{cchI-06,chamelcarter-06}. 
They have also discussed how to calculate the microscopic coefficients of this model~\cite{cchII-05,cch-05}. 
More elaborate models treating the crust as a neutron superfluid in an elastic medium and taking into 
account the effects of a frozen-in magnetic field have been very recently developed both 
in general relativity~\cite{carter-06, cartersamuelsson-06} and in the Newtonian 
limit~\cite{carter-06, carterchachoua-06}. All these models are based on an action principle that will be briefly
reviewed in Section~\ref{sect.hydro.action}. We will consider a simple nonrelativistic two-fluid model of 
neutron star crusts in Section~\ref{sect.hydro.twofluid} using the fully-4D covariant formulation of 
Carter \& Chamel~\cite{CCI-04, CCII-05, CCIII-05}. 
Entrainment effects and superfluidity will be discussed in Sections~\ref{sect.hydro.entrainment} 
and~\ref{sect.hydro.super}, respectively.

\subsection{Variational formulation of multi-fluid hydrodynamics}
\label{sect.hydro.action}
 
In the convective variational approach of hydrodynamics developed by
Carter~\cite{carter-89, carter-98}, and recently reviewed by
Gourgoulhon~\cite{gourgoulhon-06} and Andersson \&
Comer~\cite{anderssoncomer-07}, the dynamic equations 
are obtained from an action principle by considering variations of the
fluid particle trajectories. First developed in the context of general
relativity, this formalism has been adapted to the Newtonian
framework in the usual 3+1 spacetime decomposition by
Prix~\cite{prix-04, prix-05}. As shown by Carter \& Chamel in a series of papers~\cite{CCI-04,
  CCII-05, CCIII-05}, the Newtonian hydrodynamic equations can be written in a very
concise and elegant form in a fully-4D covariant framework. Apart from facilitating 
the comparison between relativistic and nonrelativistic fluids, this approach 
sheds a new light on Newtonian hydrodynamics following the steps of 
Elie Cartan, who demonstrated in the 1920's that the effects of gravitation in Newtonian theory 
can be expressed in geometric terms as in general relativity. 

The variational formalism of Carter provides a very general framework for deriving the 
dynamic equations of any fluid mixture and for obtaining conservation laws, using 
exterior calculus. In particular, this formalism is very well suited to describing 
superfluid systems, like laboratory superfluids or neutron star interiors, by making a 
clear distinction between particle velocities and the corresponding momenta 
(see the discussion in Section~\ref{sect.super.dyn.entr}).

The dynamics of the system (either in relativity or in the Newtonian limit) 
is thus governed by a Lagrangian density $\Lambda$, which depends on the particle
4-currents $n_{_{\mathrm{X}}}^\mu=n_{_{\mathrm{X}}}u_{_{\mathrm{X}}}^\mu$, where
$n_{_{\mathrm{X}}}$ and $u_{_{\mathrm{X}}}^\mu$ are the particle number density and the 4-velocity 
of the constituent X, respectively. We will use Greek letters 
for spacetime indices with the Einstein summation convention for repeated indices. 
The index X runs over the different constituents in the system. Note that repeated chemical
indices X will not mean summation unless explicitly specified. 

The dynamic equations for a mixture of several interacting fluids (coupled by 
entrainment effects) can be obtained by requiring that the action integral
\begin{equation}
{\cal S} = \int  \Lambda \mathrm{d}{\cal V} \,
\end{equation}
(where $\cal V$ is the 4-volume element)
be stationary under variations of the 4-currents $n_{_{\mathrm{X}}}^\mu$. These variations
are not arbitrary because they have to conserve the number of particles. 
In classical mechanics of point-like particles, the equations of motion can be 
deduced from an action integral by considering displacements of the particle 
trajectories. Likewise, considering variations of the 4-currents induced by 
displacements of the fluid-particle worldlines yield
\begin{equation}
\label{eq.sect.hydro.Euler}
n_{_{\mathrm{X}}}^\nu \varpi^{_{\mathrm{X}}}_{\!\nu\mu}+\pi^{_{\mathrm{X}}}_\mu\nabla_\nu n_{_{\mathrm{X}}}^\nu=f^{_{\mathrm{X}}}_\mu \, ,
\end{equation}
where $\nabla_\mu$ denotes the covariant derivative. 
$\pi^{_{\mathrm{X}}}_\mu$, defined by
\begin{equation}
\label{eq.sect.hydro.4momentum}
\pi^{_{\mathrm{X}}}_\mu = \frac{\partial \Lambda}{\partial n_{_{\mathrm{X}}}^\mu} \, ,
\end{equation}
is the 4-momentum per particle associated with the 4-current $n_{_{\mathrm{X}}}^\mu$, 
$\varpi^{_{\mathrm{X}}}_{\!\nu\mu}$ is the vorticity 2-form defined by the 
exterior derivative of the corresponding 4-momentum
\begin{equation}
\varpi^{_{\mathrm{X}}}_{\!\mu\nu}=2\nabla_{\![\mu}\pi^{_{\mathrm{X}}}_{\,\nu]} =\nabla_\mu \pi^{_{\mathrm{X}}}_\nu-\nabla_\nu \pi^{_{\mathrm{X}}}_\mu \, ,
\end{equation}
and $f^{_{\mathrm{X}}}_\mu$ is the (nongravitational) 4-force density
acting on the constituent X.
Equation~(\ref{eq.sect.hydro.4momentum}) is the generalization to fluids of 
the definition introduced in classical Lagrangian mechanics for the momentum of 
point-like particles. This equation shows that momentum and velocity are intrinsically 
different mathematical objects since the former is a co-vector while the latter is a vector. 
The vorticity 2-form is closely analogous to the electromagnetic 2-form $F_{\mu\nu}$. 
Equation~(\ref{eq.sect.hydro.Euler}) is the covariant generalization of Euler's equation 
to multi-fluid systems. 
The stress-energy-momentum tensor of this multi-fluid system is given by
\begin{equation}
T^{\mu}_{\ \nu} = \Psi \delta_\nu^\mu + \sum_{_{\mathrm{X}}} n_{_{\mathrm{X}}}^\mu \pi^{_{\mathrm{X}}}_\nu    \, ,
\label{eq.sect.hydro.Tmunu}
\end{equation}
where the generalized pressure $\Psi$ is defined by
\begin{equation}
\label{eq.sect.hydro.Psi}
\Psi=\Lambda-\sum_{_{\mathrm{X}}} n_{_{\mathrm{X}}}^\mu \pi^{_{\mathrm{X}}}_\mu \, .
\end{equation}
Note that so far we have made no assumptions regarding the spacetime geometry so that 
Equations~(\ref{eq.sect.hydro.Euler}), (\ref{eq.sect.hydro.Tmunu}) and (\ref{eq.sect.hydro.Psi}) 
are valid for both relativistic and nonrelativistic fluids.
The presence of a frozen-in magnetic field and the elasticity of the solid crust can be taken into 
account within the same variational framework both in (special and general) relativity~\cite{carter-06, cartersamuelsson-06} and in the Newtonian limit~\cite{carter-06, carterchachoua-06}.

\subsection{Two-fluid model of neutron star crust}
\label{sect.hydro.twofluid}

In this section, we will review the simple model for neutron star 
crust developed by Carter, Chamel \& Haensel~\cite{cchI-06} (see also Chamel \&
Carter~\cite{chamelcarter-06}). The crust is treated 
as a two-fluid mixture containing a superfluid of {\bf f}ree neutrons
(index f) and a fluid of nucleons {\bf c}onfined  inside nuclear
clusters (index c), in a uniform background of degenerate relativistic electrons. This model 
includes the effects of stratification (variation of the crust
structure and composition with depth; see
Section~\ref{sect.groundstate}) and allows for
entrainment effects (Section~\ref{sect.super.dyn.entr.ns}) that have
been shown to be very strong~\cite{chamel-05}. However, this model does not take into account
either the elasticity of the crust or magnetic fields. 
For simplicity, we will restrict ourselves to the case of zero temperature and we will 
not consider dissipative processes (see, for instance, Carter \& Chamel~\cite{CCIII-05}, 
who have discussed this issue in detail and have proposed a three-fluid model 
for hot neutron star crust). 

This model has been developed in the Newtonian framework,
since relativistic effects are expected to be small in crust layers, but using a
4D fully-covariant formulation in order to facilitate the link 
with relativistic models of neutron star cores~\cite{chamel-08b}. 
In Newtonian theory, the 4-velocities are defined by
\begin{equation}
u_{_{\mathrm{X}}}^\mu=\frac{\mathrm{d}x^\mu}{\mathrm{d}t} \, ,
\end{equation}
$t$ being the ``universal'' time. The components of the 4-velocity vectors have the form 
$u_{_{\mathrm{X}}}^{\,_0}=1$, $u_{_{\mathrm{X}}}^{\, i}=v_{_{\mathrm{X}}}^{\, i}$ in
``Aristotelian'' coordinates (representing the usual kind of 3+1 spacetime decomposition with respect to the rest frame of the star). This
means that the time components of the 4-currents are simply equal
to the corresponding particle number densities $n_{_{\mathrm{X}}}^{_0} =
n_{_{\mathrm{X}}}$,  while the space components are those of the current
 3-vector $n_{_{\mathrm{X}}}^i =n_{_{\mathrm{X}}}v_{_{\mathrm{X}}}^i$ (using Latin letters $i,j,...$ 
for the space coordinate indices). 

The basic variables of the two-fluid model considered here are the particle 4-current vectors
$n_{\mathrm{c}}^\mu$, $n_{\mathrm{f}}^\mu$ and the number density $n_{\mathrm{N}}$ of 
nuclear clusters, which accounts for stratification effects. 
For clusters with mass number $A$, we have the relation
$n_{\mathrm{c}}=A n_{\mathrm{N}}$. 
In the following we will neglect the small
neutron-proton mass difference and we will write simply $m$ for the
nucleon mass (which can be taken as the atomic mass unit, for example). 
The total mass density is thus given by $\rho=m (n_{\mathrm{c}}+n_{\mathrm{f}})$. 
The Lagrangian density $\Lambda$, which contains the microphysics of the system, 
has been derived by Carter, Chamel \& Haensel~\cite{cchI-06, cchII-05}.

The general dynamic equations~(\ref{eq.sect.hydro.Euler}) are given, in this case, by
\begin{equation}
\label{eq.sect.hydro.Euler1}
n_{\mathrm{f}}^\nu
\varpi^{\mathrm{f}}_{\!\nu\mu}+\pi^{\mathrm{f}}_\mu\nabla_\nu
n_{\mathrm{f}}^\nu=f^{\mathrm{f}}_\mu \, ,
\end{equation}
\begin{equation}
\label{eq.sect.hydro.Euler2}
n_{\mathrm{c}}^\nu \varpi^{\mathrm{c}}_{\!\nu\mu}+\pi^{\mathrm{c}}_\mu\nabla_\nu n_{\mathrm{c}}^\nu=f^{\mathrm{c}}_\mu \, .
\end{equation}
The time components of the 4-momentum co-vectors $\pi^{_{\mathrm{X}}}_{_{\mu}}$ are interpretable as the opposite of 
the chemical potentials of the corresponding species in the Aristotelian frame, while the space components 
coincide with those of the usual 3-momentum co-vectors $p^{_{\mathrm{X}}}_i$, defined by
\begin{equation}
\label{eq.super.dyn.def.momentum}
p^{_{\mathrm{X}}}_i = \frac{\partial \Lambda}{\partial n_{_{\mathrm{X}}}^i} \, .
\end{equation}
In general, as a result of entrainment effects~\cite{andreev-76}, 
the momentum co-vector $p^{_{\mathrm{X}}}_i$ can be
decomposed into a purely kinetic part and a chemical part,
\begin{equation}
\label{eq.super.dyn.tot.momentum}
p^{_{\mathrm{X}}}_i = m_{_{\mathrm{X}}} v_{_{\mathrm{X}} i} + \chi^{_{\mathrm{X}}}_i \, .
\end{equation}
The chemical momentum $\chi^{_{\mathrm{X}}}_i$ arises from interactions
between the particles constituting the fluids, and is defined by
\begin{equation}
\label{eq.super.dyn.chem.momentum}
\chi^{_{\mathrm{X}}}_i = \frac{\partial \Lambda_{\mathrm{int}}}{\partial
  n_{_{\mathrm{X}}}^i} \, .
\end{equation}
In this case, $\Lambda_{\mathrm{int}}$ is the internal contribution to the Lagrangian
density defined by
\begin{equation}
\Lambda_{\mathrm{int}} = \Lambda-\Lambda_{\mathrm{kin}} \, ,
\end{equation}
where
\begin{equation}
\Lambda_{\mathrm{kin}} = \frac{1}{2} \sum_{_{\mathrm{X}}} m_{_{\mathrm{X}}} n_{_{\mathrm{X}}}
v_{_{\mathrm{X}}}^2 \, .
\end{equation}
According to the Galilean invariance, $\Lambda_{\mathrm{int}}$ can only
depend on the relative velocities between the fluids, which implies the
following Noether identity
\begin{equation}
\label{eq.super.dyn.noether}
\sum_{_{\mathrm{X}}} n_{_{\mathrm{X}}} \chi^{_{\mathrm{X}}}_j \gamma^{ij} = 0 \, ,
\end{equation}
where $\gamma^{ij}$ is the Euclidean space metric. Consequently, unlike the
individual momenta~(\ref{eq.super.dyn.tot.momentum}), the \emph{total}
momentum density is simply given by the sum of the kinetic momenta
\begin{equation}
\sum_{_{\mathrm{X}}} n_{_{\mathrm{X}}} p^{_{\mathrm{X}}}_j \gamma^{ij}  = \sum_{_{\mathrm{X}}} n_{_{\mathrm{X}}} m_{_{\mathrm{X}}} v_{_{\mathrm{X}}}^i \, .
\end{equation}
Let us stress that entrainment is a nondissipative effect and is different from drag. 

The cluster 4-momentum co-vector is purely timelike since the Lagrangian density depends only on $n_{\mathrm{N}}$.
It can thus be written as $\pi^\mathrm{N}_\mu = -\mu_{\mathrm{N}} t_\mu$, where $t_\mu=\nabla_\mu t$ 
is the gradient of the universal time $t$ and $\mu_{\mathrm{N}}$ is a cluster potential, whose gradient 
leads to stratification effects. The dynamic equation of the nuclear clusters therefore reduces to
\begin{equation}
\label{eq.sect.hydro.fN}
n_{\mathrm{N}}\nabla_\mu\, \mu_{\mathrm{N}}-t_\mu \nabla_\nu(n_{\mathrm{N}}^\nu \mu_{\mathrm{N}})=f^\mathrm{N}_\mu\, ,
\end{equation}
where $n_{\mathrm{N}}^\mu=n_{\mathrm{N}} u_{\mathrm{c}}^\mu$. 
The space components of the 4-force density co-vectors $f^{\mathrm{f}}_\mu$, $f^{\mathrm{c}}_\mu$ and $f^\mathrm{N}_\mu$ 
coincide with those of the usual 3-force density co-vectors while the time components are related to the rate of energy
loss as discussed in more detail by Carter \& Chamel~\cite{CCIII-05}. 
In the nondissipative model considered here, the total
external force density co-vector vanishes:
\begin{equation}
\label{eq.sect.hydro.force.balance}
f^{\mathrm{f}}_\nu+f^{\mathrm{c}}_\nu+f^\mathrm{N}_\nu=0 \, .
\end{equation}
At this point, let us remark that, in general, the total force density co-vector
may not vanish due to elastic crustal stresses, as shown by Chamel \&
Carter~\cite{chamelcarter-06}. Moreover for a secular evolution of
pulsars, it would also be necessary to account for the external
electromagnetic torque. 

Both the cluster number and baryon number have to be conserved:
\begin{equation}
\label{eq.sect.hydro.cluster}
\nabla_\mu n_{\mathrm{N}}^\mu = 0 \, ,
\end{equation}
\begin{equation}
\label{eq.sect.hydro.baryon}
\nabla_\mu n_{\mathrm{f}}^\mu + \nabla_\mu n_{\mathrm{c}}^\mu = 0 \, .
\end{equation}
On a short time scale, relevant for pulsar glitches or high frequency oscillations, 
it can be assumed that the composition of the crust remains frozen, i.e., the 
constituents are separately conserved so that we have the additional
condition
\begin{equation}
\label{eq.sect.hydro.neutron}
\nabla_\mu n_{\mathrm{f}}^\mu  = 0 \, .
\end{equation}
However, on a longer time scale, the free and confined nucleon currents
may not be separately conserved owing to electroweak interaction processes,
which transform neutrons into protons and {\it vice versa}. The relaxation times 
are strongly dependent on temperature~\cite{yakovlev-01} and on superfluidity~\cite{villain-05}. 
A more realistic assumption in such cases is therefore to suppose that the system is in
equilibrium, which can be expressed by 
\begin{equation}
\label{eq.sect.hydro.equilibrium}
{\cal A}=0 \, ,
\end{equation}
where the chemical affinity ${\cal A}$~\cite{CCIII-05} is defined by the chemical potential
difference in the crust rest frame
\begin{equation}
{\cal A}=u_{\mathrm{c}}^\nu (\pi^{\mathrm{f}}_\nu - \pi^{\mathrm{c}}_\nu) \, .
\end{equation}
%

\subsection{Entrainment and effective masses}
\label{sect.hydro.entrainment}

The momenta of the free superfluid neutrons and of the confined nucleons
can be expressed in terms of the velocities as
\begin{equation}
\label{eq.super.dyn.entr_crust1}
\pmb{p^{\mathrm{f}}} = m_{\mathrm{f}}^\star \pmb{v_{\mathrm{f}}} + (m-m_{\mathrm{f}}^\star) \pmb{v_{\mathrm{c}}} \, ,
\end{equation}
\begin{equation}
\label{eq.super.dyn.entr_crust2}
\pmb{p^{\mathrm{c}}} = m_{\mathrm{c}}^\star \pmb{v_{\mathrm{c}}} + (m-m_{\mathrm{c}}^\star) \pmb{v_{\mathrm{f}}},
\end{equation}
where $m_{\mathrm{f}}^\star$ and $m_{\mathrm{c}}^\star$ are dynamic effective masses (for 
a generalization to relativistic fluids, see Chamel~\cite{chamel-08b}) so
that in the crust (resp.\ neutron superfluid) rest frame we have
$\pmb{p^{\mathrm{f}}} = m_{\mathrm{f}}^\star \pmb{v_{\mathrm{f}}}$
(resp.\ $\pmb{p^{\mathrm{c}}} = m_{\mathrm{c}}^\star \pmb{v_{\mathrm{c}}}$). These
effective masses arise because of the momentum transfer between the
free neutrons and the nuclear lattice (Bragg scattering). Due to
Galilean invariance, these effective masses are not independent but
are related by
\begin{equation}
m_{\mathrm{f}}^\star - m = \frac{\rho_{\mathrm{c}}}{\rho_{\mathrm{f}}} (m_{\mathrm{c}}^\star-m) \, ,
\end{equation}
where $\rho_{\mathrm{f}}$ and $\rho_{\mathrm{c}}$ are the mass densities of the
neutron superfluid and confined nucleons, respectively. This entails
that the \emph{total} momentum density is simply $\rho_{\mathrm{f}}\pmb{v_{\mathrm{f}}}+\rho_{\mathrm{c}}\pmb{v_{\mathrm{c}}}$. These dynamic effective
masses can be expressed directly from the internal Lagrangian density
$\Lambda_{\mathrm{int}}$ of the system by (X=f,c)
\begin{equation}
\frac{m^{_{\mathrm{X}}}_\star}{m} = 1 + \frac{2}{\rho_{_{\mathrm{X}}}}\frac{\partial \Lambda_{\mathrm{int}}}{\partial \omega^2} \, ,
\end{equation}
where $\rho_{_{\mathrm{X}}}=n_{_{\mathrm{X}}} m$ and
$\pmb{\omega}=\pmb{v_{\mathrm{f}}}-\pmb{v_{\mathrm{c}}}$ is the relative velocity between the neutron
superfluid and the crust. 

Alternatively, we could introduce different kinds of effective masses
$m_{\mathrm{f}}^\sharp$ and $m_{\mathrm{c}}^\sharp$, so that the momentum and the
velocity of one fluid are aligned in the \emph{momentum} rest frame of
the other fluid,
\begin{equation}
m_{\mathrm{f}}^\sharp - m = \frac{m}{m_{\mathrm{c}}^\star} (m_{\mathrm{f}}^\star-m) \, ,
\end{equation}
\begin{equation}
m_{\mathrm{c}}^\sharp - m = \frac{m}{m_{\mathrm{f}}^\star} (m_{\mathrm{c}}^\star-m) \, .
\end{equation}
This shows, in particular, that these effective masses are either
\emph{all} greater or \emph{all} smaller than the nucleon mass
$m$. From a stability analysis, Carter, Chamel and
Haensel~\cite{cchI-06} proved that these effective masses obey the
following inequalities (X=f,c)
\begin{equation}
\frac{m_{_{\mathrm{X}}}^\star}{m}>\frac{\rho_{_{\mathrm{X}}}}{\rho} \, ,
\end{equation}
\begin{equation}
\frac{m_{_{\mathrm{X}}}^\sharp}{m}<\frac{\rho}{\rho_{_{\mathrm{X}}}} \, ,
\end{equation}
where $\rho=\rho_{\mathrm{f}} + \rho_{\mathrm{c}}$ is the total mass
density. Microscopic calculations carried out by
Chamel~\cite{chamel-05, chamel-06}, using for the first time the band
theory of solids (see Section~\ref{sect.groundstate.inner.beyond}),
show that these effective masses can be much larger than the bare
nucleon mass $m$. This is in sharp contrast to the situation
encountered in the liquid core, where effective masses are slightly
smaller than $m$ (see, for instance, Chamel \&
Haensel~\cite{chamelhaensel-06} and references therein). 
Note, however, that the \emph{relativistic} effective neutron mass can be 
slightly larger than the bare mass in the liquid core, as recently shown 
by Chamel~\cite{chamel-08b}.

\subsection{Neutron superfluidity}
\label{sect.hydro.super}

As already discussed in Section~\ref{sect.super.dyn.rot}, a superfluid is locally characterized by condition~(\ref{eq.sect.super.dyn.irrot}), which, in the 4D covariant framework~\cite{CCI-04,
  CCII-05}, reads
\begin{equation}
\label{eq.sect.hydro.super.irrot}
\varpi^{\mathrm{f}}_{\!\mu\nu}=0\, .
\end{equation}
However, at length scales that are large compared to the intervortex spacing, the
neutron vorticity 2-form will not vanish. Since the vorticity of 
superfluid is carried by quantized vortex lines (as illustrated in
Figure~\ref{fig.sect.super.dyn.vortex}), mathematically this means that
the Lie derivative of the vorticity 2-form along the 4-velocity vector
$u_\upsilon^\mu$ of the vortex lines, vanishes:
\begin{equation}
\label{eq.sect.hydro.super.vortices1}
\vec u_\upsilon \pounds \varpi^{\mathrm{f}}_{\!\mu\nu}=0 \, .
\end{equation}
This condition is equivalent to
\begin{equation}
\label{eq.sect.hydro.super.vortices2}
 u_\upsilon^\mu\varpi^{\mathrm{f}}_{\!\mu\nu}=0 \, .
\end{equation}
\epubtkImage{vortex.png}{%
\begin{figure}[htbp]
  \centerline{\includegraphics[scale=0.4]{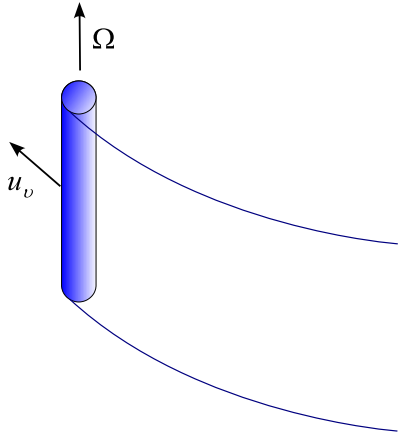}}
  \caption{Sketch of the 2-surface swept out by a quantized vortex
  line moving with the 4-velocity $u_\upsilon$ ; $\Omega$ is the
  superfluid angular velocity.}
  \label{fig.sect.super.dyn.vortex}
\end{figure}}

Let us remark that the definition of neutrons that have to be counted 
as ``free'' is not unique and there is
some arbitrariness in the above model. Nonetheless, it can be shown
that the 4-momentum co-vector $\pi^{\mathrm{f}}_\mu$ of the neutron superfluid is
invariant under such ``chemical'' readjustments and the
above superfluidity conditions are well defined~\cite{cchI-06}.
Note also that these conditions are valid for both relativistic and nonrelativistic 
superfluids. 

As discussed by Chamel \& Carter~\cite{chamelcarter-06}, there are two
cases, which are consistent with the nondissipative models considered
here. The first possibility is that the neutron vortices are free and
are co-moving with the superfluid, i.e.\ $u_\upsilon^\mu=u_{\mathrm{f}}^\mu$. On a sufficiently short time scale, it may be further assumed
that the free neutron current is conserved, which implies from
Equations~(\ref{eq.sect.hydro.Euler}) and
(\ref{eq.sect.hydro.super.vortices2}) that the force $f^{\mathrm{f}}_\mu$
vanishes. Since no external force is supposed to be exerted on the
system, the force acting on the confined nucleons also
vanishes. However, on longer time scales, as discussed in
Section~\ref{sect.hydro.twofluid}, it would be more appropriate to replace
Equation~(\ref{eq.sect.hydro.neutron}) by the equilibrium condition
Equation~(\ref{eq.sect.hydro.equilibrium}). In this case, there will
still be a force acting on the superfluid (hence, also a force, acting
on the confined nucleons) owing to the conversion of free neutrons into
confined protons and {\it vice versa}. The other possibility is that
the vortices are pinned to the crust, so that $u_\upsilon^\mu=u_{\mathrm{c}}^\mu$. As shown by Chamel \& Carter~\cite{chamelcarter-06}, the
pinning condition $u_{\mathrm{c}}^\mu\varpi^{\mathrm{f}}_{\!\mu\nu}=0$ is
equivalent to imposing that the individual vortices be subject to the
corresponding Magnus force. 

The dynamics of the neutron superfluid vortices in the crust play a
major role in the understanding of pulsar glitches and 
post-glitch relaxation. Due to entrainment effects, the distribution of
vortices is not simply given by
Equation~(\ref{eq.sect.super.dyn.vortex.density}), but also depends on
the angular velocity of the crust according to the following formula
derived by Chamel \& Carter~\cite{chamelcarter-06}:
\begin{equation}
n_\upsilon=\frac{2 m \Omega_{\mathrm{f}}}{\pi \hbar} + \frac{2 m}{\pi \hbar}\left(\frac{m_{\mathrm{f}}^\star}{m}-1\right) (\Omega_{\mathrm{f}}-\Omega_{\mathrm{c}})\, ,
\end{equation}
assuming that the neutron superfluid and the crust are uniformly rotating 
around the same axis with angular
velocities $\Omega_{\mathrm{f}}$ and $\Omega_{\mathrm{c}}$, respectively.
In addition, spatial variations of the effective masses are neglected. Since
the superfluid rotates faster than the crust and since the dynamic 
effective neutron mass is larger than the bare mass, the
entrainment effects increase the surface density of the vortices (whereas
the entrainment effects \emph{decrease} the surface density of neutron
vortices in the liquid core).

\newpage


\section{Neutrino Emission}
\label{sect.neutrino}

\subsection{Neutrino emission processes -- an overview}

\begin{table}[htbp]
  \caption[Main neutrino emission processes in neutron star
  crusts.]{Main neutrino emission processes in neutron star
  crusts. Symbols: $\gamma$ stands for a photon or a plasmon; $(A,Z)$
  stands for a nucleus with charge number $Z$ and mass number $A$;
  lepton symbol $x=e,\mu,\tau$; neutron quasiparticle (neutron-like
  elementary excitation) in superfluid neutron gas is denoted by
  $\widetilde{n}$.}
  \label{table.sect.neutrino.proc}
  \vskip 4mm

  \centering
    \begin{tabular}{c|c}
      \toprule
      Process name &   Reaction \\
      \midrule
      &\\
      $e^-e^+$ pair annihilation & $e^-+ e^+
      \longrightarrow \nu_x +\bar{\nu}_x$\\
      &\\
      Plasmon decay & $\gamma \longrightarrow \nu_x +\bar{\nu}_x$\\
      &\\
      Photoneutrino emission  & $\gamma +e^-\longrightarrow
      e^-+\nu_x +\bar{\nu}_x$\\
      &\\
      Electron synchrotron radiation&
      $ e^-\longrightarrow e^- +\nu_x +\bar{\nu}_x $\\
      &\\
      Electron-nucleus Bremsstrahlung &
      $e^-  + (A,Z)
      \longrightarrow e^- + (A,Z) +\nu_x +\bar{\nu}_x$\\
      &\\
      Cooper pair formation &
      $\widetilde{n}+\widetilde{n}\longrightarrow \nu_x +\bar{\nu}_x$\\
      &\\
      \bottomrule
    \end{tabular}
\end{table}

There is a great wealth of neutrino emission processes in hot neutron
star crusts. These processes are associated with weak interaction
involving electrons, positrons, nuclei, and free nucleons. As soon as
the crust becomes neutrino-transparent, which occurs in about a minute
after neutron star birth, these neutrinos freely leave the crust,
taking away thermal energy, and contributing in this way to crust
cooling. Let us notice that, because of a finite thermal equilibration
timescale, during the first few decades of a neutron star's life the thermal
evolution of the crust is decoupled from that of the liquid core.

In this section we do not pretend to give a complete review of
neutrino emission processes. We will limit ourselves to a
brief presentation of six main neutrino emission mechanisms,
listed in Table~\ref{table.sect.neutrino.proc}. Some other
neutrino emission processes are briefly mentioned in
Section~\ref{sect.neutrino.other}. We will dwell on the qualitative
features of main contributions to the neutrino emissivity of
the crust, $Q_\nu$, which is defined as the energy radiated in
neutrinos (in erg) from $1 \mathrm{\ cm}^3$ during 1~s. In
particular, we will discuss, at the qualitative level, the
 density and temperature dependences of the contributions from
a given process. A reader interested in a more complete and
detailed description of neutrino emission processes from the
crust is referred to the review  of
Yakovlev et al.~\cite{yakovlev-01}.

Processes reviewed in
Sections~\ref{sect.nu.pairs}\,--\,\ref{sect.neutrino.CP} have been
studied mainly at $\pmb{B}=0$. The effects of ${\pmb B}$ on $Q_\nu$ were
calculated only for some processes, and will be briefly
mentioned at the end of the corresponding sections. The
presence of ${\pmb B}$ makes possible the synchrotron radiation of
electrons, considered in Section~\ref{sect.neutrino.synchrotron}.

A summary of the $\rho-T$ dependence of different contributions to
$Q_\nu$, and a discussion of their relative importance in
different regions of the $\rho-T$ plane, are presented in
Section~\ref{sect.nu.summary}.

\subsection{Electron-positron pair annihilation}
\label{sect.nu.pairs}

This process was proposed by Chiu \& Morrison~\cite{chiu-60}. It
requires the presence of positrons and is, therefore, important at high
temperatures and low densities. Such conditions prevail in the outer
layer of a newly-born neutron star. The matter there is opaque to
photons. Electrons, positrons and photons are in thermodynamic
equilibrium, with number densities $n_{e^-}$, $n_{e^+}$, $n_\gamma$,
and the corresponding chemical potentials $\mu_{e^-}$, $\mu_{e^+}$,
$\mu_\gamma$, respectively. As the number of photons is not fixed,
their chemical potential $\mu_\gamma=0$. Therefore, equilibrium with
respect to reactions
\begin{equation}
e^+  + e^- \leftrightarrow 2\gamma
\label{eq:pair.gamma}
\end{equation}
implies $\mu_{e^-}=-\mu_{e^+}$. Electrons and positrons can be
treated as ideal relativistic Fermi gases. An electron or a positron of
momentum $\pmb{p}$ has energy
\begin{equation}
\epsilon({\pmb{p}})=\epsilon(p)=\sqrt{(m_e c^2)^2 + p^2 c^2} \, .
\label{eq:e_p}
\end{equation}
The electron and positron number densities are given by (writing $\mu_{e^-}=\mu_e$)
\begin{equation}
n_{e^-}={1\over \pi^2
\hbar^3}\int_0^\infty\mathrm{d}p\, {p^2\over 1
  +\mathrm{exp}[(\epsilon(p) -\mu_e)/k_{\mathrm{B}}T]} \, ,
\label{eq:n.e-}
\end{equation}
and
\begin{equation}
n_{e^+}={1\over
\pi^2 \hbar^3}\int_0^\infty\mathrm{d}p\, { p^2\over 1
  +\mathrm{exp}[(\epsilon(p) +\mu_e)/ k_{\mathrm{B}}T]} \, .
\label{eq:n.e+}
\end{equation}
Charge neutrality implies
\begin{equation}
n_{e^+}+n_{e^-}=n_{\mathrm{N}} Z \, ,
\label{eq:e+e-.neutral}
\end{equation}
where $n_{\mathrm{N}}$ is the density of nuclei.
The calculation of the neutrino emissivity $Q_{\mathrm{pair}}$ from
reactions
\begin{equation}
e^+ + e^-\longrightarrow \nu_x +\bar{\nu}_\ell \, ,~~\ell=e,\mu,\tau \, ,
\label{eq:pair.reactions}
\end{equation}
is described in detail, e.g., in
Yakovlev et al.~\cite{yakovlev-01}. Here we limit ourselves to a
qualitative discussion of two limiting cases. Let us first
consider the case of nondegenerate electrons
and positrons, $k_{\mathrm{B}}T> \mu_e$; such conditions prevail
in the supernova shock  and in the shocked envelope of a newly 
born proto-neutron star.   Then, $n_{e^+}\approx
n_{e^-}\gg n_{\mathrm{N}}Z$. The mean
energies of electrons, positrons, photons and neutrinos  are then
``thermal'', $\langle\epsilon\rangle\sim k_{\mathrm{B}}T$. The
cross section $\sigma_{e^+e^-\rightarrow \nu\bar{\nu}}$ for
process~(\ref{eq:pair.reactions}) is quadratic in the center-of-mass
energy. Therefore, the temperature dependence of $Q_\nu^{\mathrm{pair}}$
can be evaluated as
\begin{equation}
Q_\nu^{\mathrm{pair}}\propto \langle\epsilon_\nu\rangle n_{e^-}
n_{e^+} \sigma_{e^+e^-\rightarrow \nu\bar{\nu}}
\propto T\times T^3\times T^3\times T^2\propto T^9 \, .
\label{eq:Q.pair.nondeg}
\end{equation}
Let us now consider the opposite limit of degenerate ultra-relativistic
electrons, $k_{\mathrm{B}}T\ll \mu_e$ and $\mu_e \gg m_e c^2$. The
positron density is then exponentially
small. This is because  $\mu_{e^+}=-\mu_{e}$ is large and negative, so
that $n_{e^+}\propto \mathrm{exp}(-\mu_e/k_{\mathrm{B}}T)\ll n_{e^-}$. Therefore, 
the pair annihilation process is
strongly suppressed for degenerate electrons, with decreasing temperature or increasing
density. Detailed formulae for $Q_\nu^{\mathrm{pair}}$, valid in different
density-temperature domains, are given in~\cite{yakovlev-01}.

The pair annihilation process can be affected by a strong magnetic field
$\pmb{B}$. General expressions for $Q_\nu^{\mathrm{pair}}$ for
arbitrary $\pmb{B}$ were derived by Kaminker et al.~\cite{kaminker-92,kaminker-92b}. In these
papers one can also find practical expressions for a hot,
nondegenerate plasma in arbitrary $\pmb{B}$, as well as
interpolating expressions for $Q_\nu^{\mathrm{pair}}$ in a plasma
of any degeneracy and in any $\pmb{B}$. In a hot,
nondegenerate plasma  with $T\gtrsim 10^{10}$~K, $B\gg 10^{15}$~G must be huge
to affect $Q_\nu^{\mathrm{pair}}$. However,
at $T\lesssim  10^9$~K, even $B\sim 10^{14}$~G may quantize
the motion of positrons  and increase substantially their
number density. Consequently, $B\sim 10^{14}$~G strongly
increases $Q_\nu^{\mathrm{pair}}$ al low densities. This is
visualized in Figure \ref{fig.sect.neutrino.nucrust2}.

\subsection{Plasmon decay}
\label{sect.neutrino.plasmon}

Quanta of electromagnetic waves in a plasma
have different properties than in vacuum. They are called
(electron) plasmons, and appear in two basic modes.
We first consider the case $\pmb{B}=0$. Those modes, which consist
of transverse oscillations, are called transverse plasmons.
The relation between the frequency of transverse plasmons
$\omega$ and their wave-number $k=2\pi/\lambda$ (where $\lambda$ is
the wavelength) for the simplest case of nonrelativistic electrons is
\begin{equation}
\omega=\sqrt{\omega_{\mathrm{p}e}^2 + k^2 c^2} \, ,
\label{eq:omega.k.t-plasmon}
\end{equation}
where $\omega_{\mathrm{p}e}$ is the electron plasma
frequency~(\ref{eq.sect.plasma.noB.omega_pe}). Only plasmons with
$\omega>\omega_{\mathrm{p}e}$ can propagate in the crust. At a given
temperature $T$, the number density of plasmons is given by the
Bose--Einstein formula
\begin{equation}
n_\gamma\propto  \int\mathrm{d}p\, {p^2\over
  \mathrm{e}^{\hbar\omega/k_{\mathrm{B}}T}-1} \, .
\label{eq:n_gamma-plas}
\end{equation}
While a photon in vacuum cannot decay into a
neutrino-antineutrino pair, a plasmon in a plasma can,
\begin{equation}
\gamma \rightarrow  \nu_x + \bar{\nu}_\ell \, ,
~~ \ell=e,\mu,\tau \, .
\label{eq:plasm.pair}
\end{equation}
 This process of neutrino emission was first considered  in detail
 by Inman \& Ruderman~\cite{inman-64}. For $T\ll  T_{\mathrm{p}e}$,
the value of $Q_\nu^{\mathrm{plas}}$ is strongly temperature dependent,
\begin{equation}
Q_\nu^{\mathrm{plas}}\propto \exp(-T_{\mathrm{p}e}/T) \, ,
\label{eq:Q.plas}
\end{equation}
where $T_{\mathrm{p}e}$ is the electron plasma temperature defined by
Equations~(\ref{eq.sect.plasma.noB.omega_pe})
and~(\ref{eq.sect.plasma.noB.T_pe}). For $T\ll T_{\mathrm{p}e}$, the plasmon
decay process is therefore negligible. It is also strongly density dependent
through $T_{\mathrm{p}e}$ in Equation~(\ref{eq:Q.plas}). Generally,
$Q_\nu^{\mathrm{plas}}$ is switched-off by decreasing temperature and
increasing density. Detailed formulae for $Q^{\mathrm{plas}}_\nu$ can be
found in~\cite{yakovlev-01}.

The plasmon decay is influenced by a strong magnetic field,
because $\pmb{B}$ modifies the plasma dispersion relation
(relation between plasmon frequency $\omega$ and its
wavenumber $k$). In particular, new plasma modes may appear.
The effects of magnetic fields are important if $\omega_B \equiv
e B/(m_e^*c)>\omega_{\mathrm{p}e}$. At $\rho\sim 10^{11}~\mdens$  this
requires $B\gtrsim 3\times 10^{15}$~G. The magnitude of $B$
required to modify the plasmon dispersion relation grows as
$\rho^{2/3}$.

\subsection{Photoneutrino emission}
\label{sect.nu.photo}

This process differs from the plasmon decay described in
Section~\ref{sect.neutrino.plasmon}, by the participation of an
electron,
\begin{equation}
\gamma + e^- \longrightarrow \nu_\ell + \bar{\nu}_\ell +
e^- \, ,~~\ell=e,\mu,\tau \, .
\label{eq:photo.pair}
\end{equation}
 The process was proposed by Ritus~\cite{ritus-61a,
  ritus-61b} and Chiu \& Stabler~\cite{chiu-61}; the most recent
formulae for $Q^{\mathrm{phot}}_\nu$ were derived by
Itoh et al.~\cite{itoh-89, itoh-89err,
  itoh-96}. Formula~(\ref{eq:n_gamma-plas}) implies
  that if $T\ll T_{\mathrm{p}e}$ then $n_\gamma \propto
  \mathrm{e}^{-T_{\mathrm{p}e}/T}$. Therefore, $Q^{\mathrm{phot}}_\nu$
  is exponentially suppressed with  decreasing temperature or increasing
density. Additional damping of $Q^{\mathrm{phot}}_\nu$ results from
electron degeneracy. However, if electrons are nondegenerate,
  $Q^{\mathrm{phot}}_\nu$ can be stronger than
  $Q_\nu^{\mathrm{plas}}$.

\subsection{Neutrino Bremsstrahlung from electron-nucleus collisions}
\label{sect.nu.Bremss}

Proposed by Pontecorvo~\cite{pontecorvo-59a, pontecorvo-59b} and
Gandelman~\cite{gandelman-59a, gandelman-59b},
neutrino Bremsstrahlung from electron-nucleus collisions is one of the major neutrino emission mechanisms in the
crust. The process can be written as
\begin{equation}
e^- + (A,Z) \longrightarrow  e^-  + (A,Z) + \nu_\ell +
\bar{\nu}_\ell \, , ~~\ell=e,\mu,\tau \, .
\label{eq:proc.Bremss}
\end{equation}
With obvious notations, the total momentum of an initial state  is
$\pmb{P}=\pmb{p}_e+\pmb{p}_A$, and that of a final state $\pmb{P}^\prime
=\pmb{p}^\prime_e+\pmb{p}^\prime_A
+\pmb{p}_{\nu_\ell}+\pmb{p}^\prime_{\bar{\nu}_\ell}$. The corresponding
total energies will be denoted by $E$ and $E^\prime$. We first
consider the case of  $T>T_{\mathrm{m}}$, when nuclei form a Coulomb liquid,
$T_{\mathrm{m}}$ being the melting temperature of the crust defined by
Equation~(\ref{eq.sect.plasma.noB.Tm}).
Let us also neglect, for the sake of simplicity, the Coulomb
correlations between ions.
The neutrino emissivity from process~(\ref{eq:proc.Bremss}) is then
calculated by integrating the energy
emission rate over initial and final
momenta,
\begin{eqnarray}
Q_\nu^{\mathrm{Brem}}&~&\propto \int \mathrm{d}^3p_e\mathrm{d}^3p^\prime_e
\mathrm{d}^3p_\nu\mathrm{d}^3p_{\bar{\nu}}
\mathrm{d}^3p_{A}\mathrm{d}^3p^\prime_{A}\cr\cr
&~&\times W_{\mathrm{Brem}}\delta(E^\prime-E)\delta({\pmb P}^\prime-{\pmb P})
f_{A}f_e\left[1-f_{e^\prime}\right]
(\epsilon_\nu +\epsilon_{\bar{\nu}}) \, ,
\end{eqnarray}
where $f_{A}$ is the Boltzmann distribution of ions (nuclei),
 $f_e$ is the Fermi--Dirac distribution for electrons,
\begin{equation}
f_e(\pmb{p})={1\over
\mathrm{e}^{[(\epsilon(p)-\mu_e)/k_{\mathrm{B}}T]}+1} \, ,
\label{eq:f_e}
\end{equation}
and $W_{\mathrm{Brem}}$ is the square of the transition amplitude corresponding to 
elementary process~(\ref{eq:proc.Bremss}).
The factor $\left[1-f_{e^\prime}\right]$ takes care of the
Pauli exclusion principle for the electron in the final state.
Let us consider the integration over the neutrino momenta.
We can rewrite $\mathrm{d}^3 p_\nu =
\mathrm{d}^2 \hat{\pmb p}_\nu p_\nu^2 \mathrm{d} p_\nu $
($\hat{\pmb p}_\nu\equiv \pmb{p}_\nu/p_\nu$) and replace
$p_\nu$ with $\epsilon_\nu /c$. Relevant neutrino energies are
$\epsilon_\nu\sim k_{\mathrm{B}}T $.
Electrons are degenerate, and the momenta contributing
to the integral are close to the electron Fermi surface.
 Consequently, the integration over energies
is restricted to  a thin shell around $\mu_e$ of a thickness
$\sim k_{\mathrm{B}}T$. Integration over neutrinos and
antineutrinos yields a factor of $T^3$ each,
neutrino energies give $T$, and the energy delta function removes
one $T$ factor. Moreover, $W_{\mathrm{Brem}}\propto Z^2$ and $n_{\mathrm{N}}
\propto \rho/A$. To account for the nonideality of the plasma of nuclei, one
introduces an additional dimensionless factor $L$
\cite{haensel-96}. All in all,
\begin{equation}
Q_\nu^{\mathrm{Brem}}\propto T^6{Z^2\rho L /A} \, .
\label{eq:Q.Bremss}
\end{equation}
For $\Gamma \ll 1$ we have $L\approx 1$, but for a strongly
coupled plasma ($\Gamma\gg 1$) $L$ can be significantly
smaller than one (see~\cite{haensel-96}).

Things become more complicated at low temperatures.
Then the  electron states are no longer described by plane waves.
Instead, the Bloch functions consistent with crystal symmetry should be used
(see the review of band theory in Section~\ref{sect.groundstate.inner.beyond}).
The electron energy spectrum is no longer continuous,
but is formed of bands.
At high $T$, the thermal motion of electrons ``smears out'' this band structure.
However,  the gaps between energy bands strongly
suppress $Q_\nu^{\mathrm{Brem}}$ \cite{kaminker-99}.
Detailed formulae valid for different domains of the density-temperature plane can be
found in~\cite{yakovlev-01}.

Strong magnetic fields affect the motion of electrons
scattered off nuclei. However, the effect of $\pmb{B}$ on
$Q_\nu^{\mathrm{Brem}}$ has not been calculated.

\subsection{Cooper pairing of neutrons}
\label{sect.neutrino.CP}

For temperatures $T<T_{\mathrm{c}n}$, the free neutrons in the inner crust are
superfluid (see Section~\ref{sect.super}). The elementary excitations
in the superfluid neutron gas, denoted by $\widetilde{n}$, are fermions.
They have a specific energy spectrum, as discussed in Section~\ref{sect.super.dyn.Vc}.
The elementary excitations are called (Bogoliubov) quasiparticles. The simplest excitation 
in a superfluid is a quasiparticle pair. Annihilation of a quasiparticle
pair can go via an emission of a
neutrino-antineutrino pair,
\begin{equation}
\widetilde{n}+\widetilde{n}\longrightarrow \nu_\ell +
\bar{\nu}_\ell \, , ~\ell=e,\mu,\tau \, .
\end{equation}
Such a process was first considered by
Flowers et al.~\cite{flowers-76}. Neglecting for simplicity
the presence of the nuclear clusters and considering a uniform superfluid 
(see Section~\ref{sect.super.static}), the temperature dependence of the 
neutrino emissivity $Q_\nu^{\mathrm{CP}}$ from these processes is given 
by
\begin{equation}
Q_\nu^{\mathrm{CP}}\propto T^7\;F(T/T_{\mathrm{c}n}) \, ,
\label{eq:Q.CP}
\end{equation}
where $F(x\ge 1)=0$ and $F(x\longrightarrow 0)=0$. 
Detailed derivations and formulae can be found in~\cite{yakovlev-01}.

The actual value of the numerical prefactor in Equation~(\ref{eq:Q.CP}) 
has recently become a topic of lively
discussion. Recent calculations have shown that previous results may
severely overestimate $Q_\nu^{\mathrm{CP}}$ (see,
e.g., \cite{sedrakian-07,kolomeitsev-08}, and references
therein). However, the  actual reduction of $Q_\nu^{\mathrm{CP}}$ is still
a matter of debate.

\subsection{Synchrotron radiation from electrons}
\label{sect.neutrino.synchrotron}

The magnetic field not only modifies the rates of neutrino emission
processes, but  also  opens  new channels of neutrino
emission. Let us consider a coordinate system such that $\pmb{B}=[0,0,B]$. 
The electron energy levels are then given by 
$\epsilon_{n}(p_z)= c(m_e^2c^2+2\hbar\omega_{\mathrm{c}}m_e n
+p_z^2)^{1/2}$, where $n$ is the Landau quantum number. The transverse
component of the electron momentum is not conserved, and this allows
for the electron synchrotron radiation process,
\begin{equation}
e^-(\pmb{B})\longrightarrow
e^-(\pmb{B})+\nu_\ell+\bar{\nu}_x \, ,~~\ell=e,\mu,\tau \, ,
\label{eq:syn.proc}
\end{equation}
where the notation $\pmb{B}$ is to remind us of the necessary presence of the magnetic
field. For nonquantizing $\pmb{B}$ ($T\gg T_B$) and nondegenerate
nonrelativistic electrons, the neutrino emission rate is
temperature independent $Q^{\mathrm{syn}}_\nu\propto B^6$. But for a hot
($T\gg T_B$) ultra-relativistic ($k_{\mathrm{B}}T\gg m_e c^2$)
electron-positron gas, we get a 
density independent $Q^{\mathrm{syn}}_\nu\propto B^2 T^5$
(see~\cite{yakovlev-01}).

\subsection{Other neutrino emission mechanisms}
\label{sect.neutrino.other}

There are many other mechanisms of neutrino emission. For example,
there is the possibility of $\nu\bar{\nu}$ pair Bremsstrahlung
emission accompanying $nn\longrightarrow nn \nu\bar{\nu}$
scattering of dripped neutrons, and scattering of neutrons on nuclear
clusters, $n(A,Z)\longrightarrow n(A,Z)\nu\bar{\nu}$. Moreover,
in a newly-born neutron star beta processes involving electrons,
positrons and nuclei, e.g., $e^-(A,Z)\longrightarrow (A,Z-1)\nu_e$,
$(A,Z-1)\longrightarrow (A,Z)e^-\bar{\nu}_e$, etc., are a source
of neutrino emission. These are the famous  Urca processes, proposed
in the early 1940s;  their intriguing history is described, e.g., in
Section~3.3.5 of Yakovlev et al.~\cite{yakovlev-01}. One can also
contemplate a photo-emission from nuclei, $\gamma(A,Z)\longrightarrow
(A,Z)\nu\bar{\nu}$. Finally, we should also mention the
interesting possibility of a very efficient neutrino emission by the
direct Urca process in some ``pasta layers'' (see
Section~\ref{sect.groundstate.pasta}) near the bottom of the
crust~\cite{lorenz-93, leinson-93, leinson-95, gusakov-04}. As this
mechanism, restricted to a bottom layer of the neutron star crust, could
be a very efficient neutrino emission channel, we will describe it in
more detail.

\subsubsection{Direct Urca process in the pasta phase of the crust}

It is well known that the direct Urca process is the  most efficient
mechanism of neutrino emission~\cite{lattimer-91a}. The direct Urca
reactions  in a dense degenerate plasma composed mainly of neutrons,
with an admixture of protons and electrons, are the  neutron beta
decay and the inverse reaction of electron capture on a proton,
\begin{equation}
n\longrightarrow p+e^-+\bar{\nu}_e \, ,
~~ p + e^- \longrightarrow n +\nu_e \, .
\label{eq:dUrca.proc}
\end{equation}
These reactions are  allowed by momentum conservation, if the
Fermi momenta of neutrons, protons, and electrons satisfy the
triangle condition $p_{\mathrm{F}n}<p_{\mathrm{F}p}+p_{\mathrm{F}e}$.
The triangle condition implies that the proton fraction in the $npe$
plasma should be greater than $1/9\approx 11\%$. For the time
being, we ignore whether this
condition is satisfied in the cores of the most massive
neutron stars. If the direct Urca (dUrca) process  is forbidden, then the
main neutrino emission mechanism from the nonsuperfluid
neutron star core is the modified Urca (mUrca)  process,
\begin{equation}
n+X\longrightarrow p+X+e^-+\bar{\nu}_e \, ,
~~ p+X + e^- \longrightarrow n+X +\nu_e \, ,
\label{eq:mUrca.proc}
\end{equation}
where $X=n$ or $X=p$ is an additional ``spectator'' nucleon needed for
momentum conservation. By strong (nuclear) interaction with $n$ or
$p$, $X$ absorbs (supplies)  the excessive  (missing) momentum of the nucleons
participating in the processes~(\ref{eq:mUrca.proc}), without changing
its own nucleon state. The difference in emissivities from the dUrca
and mUrca processes is huge.  For nonsuperfluid neutron star cores, 
$Q^{\mathrm{dUrca}}_\nu/Q^{\mathrm{mUrca}}_\nu\sim 10^6\;T_9^{-2}$, where
$T_9\equiv T/(10^9~\mathrm{K})$ \cite{lattimer-91}.

If the ``pasta mantle'' of the crust exists (see
Section~\ref{sect.groundstate.pasta}), it allows for a partial
opening of the dUrca process in those phases, in which the $npe$
matter component fills most of the space. This happens in the phases with 
tubes or bubbles filled with a neutron gas. However, because of the periodicity of the
lattice of tubes or bubbles, neutrons and protons in the $npe$ plasma move in a periodic nuclear
single-particle potential. This means that the nucleon single particle
wave functions are no longer the eigenfunctions of momentum (plane waves), but have to be replaced by
Bloch wave functions (see Section~\ref{sect.groundstate.inner.beyond}). All in all, the dUrca
process becomes ``slightly open''  in the relevant  pasta layers of
the crust~\cite{leinson-93, leinson-95, gusakov-04}. The
emissivities calculated by  Gusakov et al.~\cite{gusakov-04} can be presented as
\begin{equation}
Q^{\mathrm{dUrca(m)}}_\nu = {\cal R}(\rho)\;
Q^{\mathrm{dUrca(0)}}_\nu(\rho,T) \, ,
\label{eq:mUrca.cr}
\end{equation}
where $Q^{\mathrm{dUrca(0)}}_\nu$ is the dUrca emissivity for a homogeneous
$npe$ plasma, calculated using plane waves and ignoring momentum
conservation, and ${\cal R}$ is the reduction factor, resulting from
momentum and energy  constraints in the presence of a periodic
lattice of tubes or bubbles filled with a neutron gas. The calculation
performed for the bubble phase shows that ${\cal R}\sim 10^{-5}$,
but even this strong reduction still allows  $Q^{\mathrm{dUrca(m)}}_\nu$
to be much larger than that from any other process of neutrino emission in the neutron
star crust~\cite{gusakov-04}. However, it should be stressed that
$Q^{\mathrm{dUrca(m)}}_\nu$ acts in a rather thin bottom layer of the
crust. In the model developed by Gusakov et al.~\cite{gusakov-04}, it
was localized in the ``Swiss-cheese'' layer in the density range
$(10^{14.14}\mbox{\,--\,}10^{14.16})~\mdens$. Within this layer, and at
temperature $T=3\times 10^8$~K, $Q^{\mathrm{dUrca(m)}}_\nu$
exceeds all other crust emissivities  by a factor of at least $10^4$.

\subsection{Neutrinos from the crust -- summary in the $T-\rho$ plane}
\label{sect.nu.summary}

In this section we will summarize results for neutrino
emission from a neutron star crust. We will limit ourselves to densities $\rho
\gtrsim 10^7~\mdens$, so that electrons will always
be ultra-relativistic (see Section~\ref{sect.groundstate}).

Contributions to $Q_\nu$ from various neutrino emission mechanisms
(except for the  Cooper-pair mechanism, which will be considered
later in this section)  versus $\rho$  are plotted in
Figures~\ref{fig.sect.neutrino.nucrust1},
\ref{fig.sect.neutrino.nucrust2} and~\ref{fig.sect.neutrino.nucrust3}.

\epubtkImage{nucrust1.png}{%
  \begin{figure}[htbp]
    \centerline{\includegraphics[scale=0.6]{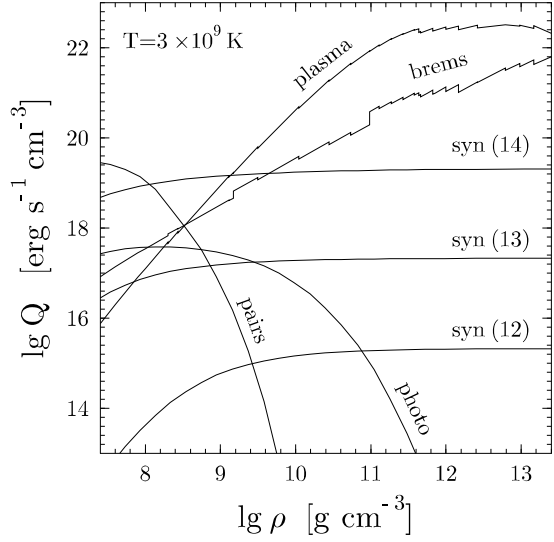}}
    \caption{Neutrino emissivities associated with different
    mechanisms of neutrino emission acting in a neutron star crust,
    versus $\rho$, at temperature $T=3\times 10^9$~K. Numbers in
    parentheses indicate $\log_{10}B$. Effect of $B=10^{14}$~G
    (and {\it a fortiori} -- effect of a lower $B$) on $Q_\nu^{\mathrm{pair}}$ is
    insignificant. $Q_\nu^{\mathrm{plas}}$, $Q_\nu^{\mathrm{Brem}}$, and
    $Q_\nu^{\mathrm{phot}}$ were calculated at $B=0$. Label ``syn (14)'' --
    synchrotron radiation by electrons in constant magnetic field
    $B=10^{14}$~G, etc. Ground-state composition of the crust is
    assumed: Haensel \& Pichon~\cite{haensel-94} model for the outer
    crust, and Negele \& Vautherin~\cite{nv-73} model for the inner
    crust. For further explanations see the
    text. From~\cite{yakovlev-01}.}
    \label{fig.sect.neutrino.nucrust1}
\end{figure}}

We start with the crust of a very young neutron star, with a
temperature $T=3\times 10^9$~K (age $\lesssim$~1~year),
Figure~\ref{fig.sect.neutrino.nucrust1}. For density $\rho \lesssim
10^8~\mdens$, the contribution $Q^{\mathrm{pair}}_\nu$ is
dominant. However, with increasing density, electrons become
degenerate, and positrons disappear in the matter, so that
$Q^{\mathrm{pair}}_\nu$ is strongly suppressed at
$\rho>10^9~\mdens$. We also notice that $Q^{\mathrm{phot}}_\nu$  is
never important in the inner crust, because of the strong electron
degeneracy. $Q^{\mathrm{plas}}_\nu$  from the plasmon decay gives the
dominant contribution to $Q_\nu$ from  $\rho\approx 10^9 \mathrm{\ g\
  cm}^{-3}$ down to the bottom of the crust. We notice also that
$Q^{\mathrm{syn}}_\nu$ behaves differently than the other
contributions. Namely, at $\rho\gtrsim 10^9~\mdens$ its density
dependence is very weak, and $Q^{\mathrm{syn}}_\nu$ scales
approximately with the magnetic field $B$ as $\propto B^2$. Finally,
one notices jumps  of $Q^{\mathrm{Brem}}_\nu$ and
$Q^{\mathrm{plas}}_\nu$, which result from jumps in $Z$ and $A$ in the
ground-state matter. As we will see, this feature is even more
pronounced at lower temperatures $T$.

\epubtkImage{nucrust2.png}{%
  \begin{figure}[htbp]
    \centerline{\includegraphics[scale=0.6]{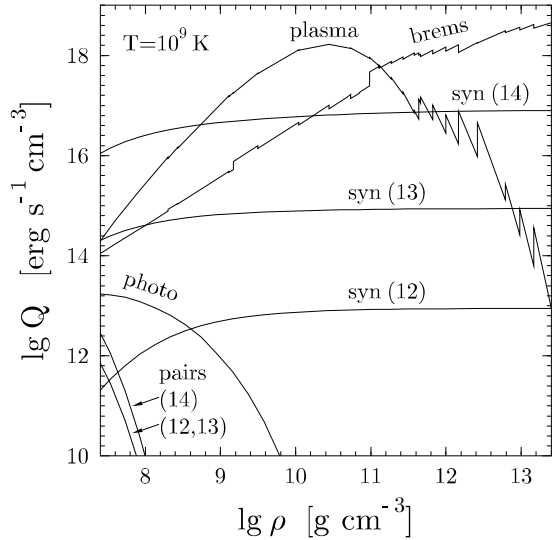}}
    \caption{Same as for Figure~\ref{fig.sect.neutrino.nucrust1}, but
    at $T=10^9$~K. $Q^{\mathrm{pair}}_\nu$ is increased by $B$, as shown by
    the labels (14), (13), (12). Notice, that $Q^{\mathrm{pair}}_\nu$ at
    $B=0$ is too low to be seen. From~\cite{yakovlev-01}.}
    \label{fig.sect.neutrino.nucrust2}
\end{figure}}

Let us now consider the case of a colder crust at $T=10^{9}$~K,
Figure~\ref{fig.sect.neutrino.nucrust2}. Except for $Q^{\mathrm{syn}}_\nu$,
which is just scaled down due to the decrease of temperature, there is
a dramatic change in the overall landscape. For a magnetic field
$B=10^{14}$~G, $Q^{\mathrm{syn}}_\nu$ dominates in the lowest-density
region. On the contrary, $Q^{\mathrm{pair}}_\nu$ is of marginal importance,
and is influenced by $\pmb{B}$ (increases with $B$). Moreover,
contribution of $Q^{\mathrm{phot}}_\nu$ is  negligible. Neglecting the
effect of magnetic fields, one concludes that $Q^{\mathrm{plas}}_\nu$
dominates in the outer crust, while $Q^{\mathrm{Brem}}_\nu$
dominates in the inner crust.  Let us notice that $Q^{\mathrm{plas}}_\nu$
reaches its maximum near $10^{10.5}~\mdens$ and then
decreases by four orders of magnitude when the density falls below $\rho
\sim 10^{13}~\mdens$; this characteristic behavior is due to
the $\exp(-T_{\mathrm{p}e}/T)$ factor, Equation~(\ref{eq:Q.plas}).
On the contrary, $Q^{\mathrm{Brem}}_\nu$ rises steadily with increasing density.

\epubtkImage{nucrust3.png}{%
  \begin{figure}[htbp]
    \centerline{\includegraphics[scale=0.6]{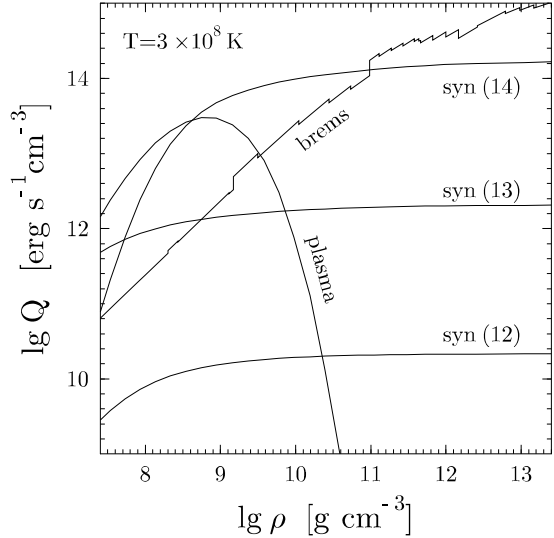}}
    \caption{Same as in Figure~\ref{fig.sect.neutrino.nucrust1}, but
    calculated at  $T=3\times 10^8$~K. $Q^{\mathrm{pair}}_\nu$ and
    $Q^{\mathrm{phot}}_\nu$ are too small to be
    seen. From~\cite{yakovlev-01}.}
    \label{fig.sect.neutrino.nucrust3}
\end{figure}}

\epubtkImage{nucrust9.png}{%
  \begin{figure}[htbp]
    \centerline{\includegraphics[scale=0.6]{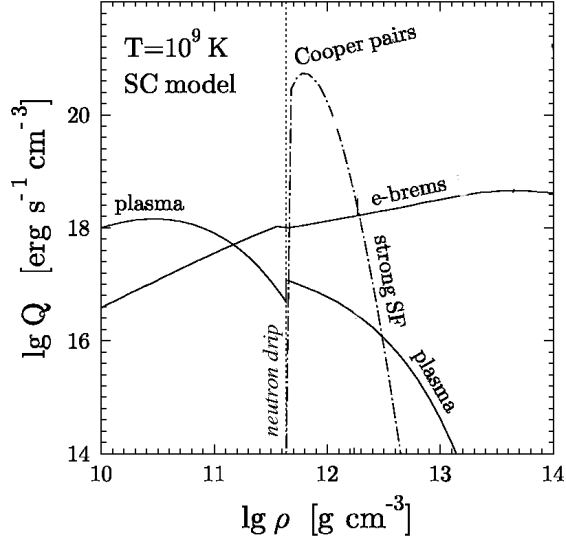}}
    \caption{Neutrino emissivity from the Cooper pair formation mechanism,
    calculated for strong  (uniform) neutron superfluidity with the
    maximum critical temperature $T_{\mathrm{c}n}^{\max}=1.6\times
    10^{10}$~K (a model from \cite{kaminker-99}). For comparison,
    $Q_\nu$ from two most efficient other mechanisms, plasmon decay
    and electron Bremsstrahlung, are also plotted. The smooth
    composition ground-state crust model of
    Kaminker et al.~\cite{kaminker-99} is used; it predicts a specific
    $\rho_n(\rho)$ in the inner crust. This crust model is  described
    in detail in Appendix~B.2 of
    Haensel et al.~\cite{haensel-06}). From~\cite{yakovlev-01}.}
    \label{fig.sect.neutrino.CP}
\end{figure}}

Finally, in Figure~\ref{fig.sect.neutrino.nucrust3} we consider an
even colder  crust  at $T=3\times 10^8$~K. Pair and photoneutrino
constributions have disappeared completely, while
$Q^{\mathrm{plas}}_\nu$ dominates  for $\rho<10^{9}~\mdens$, whereas
at higher densities, $Q^{\mathrm{Brem}}_\nu$ is the main source of
neutrino emission. At the magnetic field $B=10^{14}$~G, characteristic
of magnetars, synchrotron radiation dominates in the density interval
near $\sim 10^{10} \mathrm{\ g\ cm}^{-3}$, but then at  $\rho\gtrsim
10^{11}~\mdens$, $Q^{\mathrm{Brem}}_\nu$ becomes the strongest
neutrino radiation mechanism.

Two general remarks are in order. First, as we have already mentioned,
 jumps in $Q^{\mathrm{Brem}}_\nu$ and $Q^{\mathrm{plas}}_\nu$ are due to
 specific factors involving $Z^2$ and $A$ and reflect the  jumps in
 $Z$ and $A$ in the layered crust. For the other mechanisms, the
 electron chemical potential $\mu_e$ with its  smooth dependence on
 $\rho$ plays the role of the crucial plasma parameter, and
 therefore no jumps are seen. Secondly, were the magnetic field $B\ge
 10^{15}$~G, $Q^{\mathrm{syn}}_\nu$ would be overall dominant for
 $T<10^9$~K and $\rho>10^9~\mdens$.

The Cooper-pair mechanism of neutrino radiation differs fundamentally from
the other  mechanisms of neutrino cooling, discussed above, and therefore
we consider it separately.  $Q^{\mathrm{CP}}_\nu$ depends sensitively on
the interplay between temperature $T$ and the $^{1}\mathrm{S}_0$ pairing
gap $\Delta_{\mathrm{F}}$ of the dripped neutrons.
 The gap itself depends on $T$, rising from zero at $T=T_{\mathrm{c}n}$ to the
asymptotic value $\Delta_0\equiv \Delta_{\mathrm{F}}(T=0)$ for $T\ll T_{\mathrm{c}n}$ (see
Section~\ref{sect.super.static.Tc}). As we already discussed in
Section~\ref{sect.super.static.uniform}, the dependence of $\Delta_0$
on the free neutron density, $\rho_n$,  is very poorly understood, and this introduces
a large uncertainty in the calculated values of
$Q^{\mathrm{CP}}_\nu$. Notice that the relation $\rho_n(\rho)$, needed to get
$Q^{\mathrm{CP}}_\nu(\rho)$,  depends on the model of the inner crust.

Figure~\ref{fig.sect.neutrino.CP} refers to  $T=10^{9}$~K,
a selected model of neutron superfluidity, and a selected model
of the inner crust.  
In the BCS theory
(Section~\ref{sect.super.static.uniform}), the maximum of $\Delta_0(\rho)$, 
denoted by $\Delta_0^{\max}$, corresponds to the maximum of 
$T_{\mathrm{c}n}(\rho)$, given by  $T_{\mathrm{c}n}^{\max}=0.5669\Delta_0^{\max}/k_{\mathrm{B}}$
(Section~\ref{sect.super.static.Tc}). In the case presented in
Figure~\ref{fig.sect.neutrino.CP}, $T_{\mathrm{c}n}^{\max}=1.6\times
10^{10}$~K is significantly larger than $T$. The Cooper-pair
mechanism is efficient only within a narrow range of temperature below
$T_{\mathrm{c}n}$, namely for $0.7\lesssim T/T_{\mathrm{c}n}<1$, and is
strongly damped outside this region. In view of this, $Q^{\mathrm{CP}}_\nu$
usually has two maxima, around $\rho_1$ and $\rho_2$,
which are the two solutions of $T_{\mathrm{c}n}(\rho)=T$ (remembering the bell shape 
of the pairing gap as a function of density). Only
the lower-density maximum can be seen in
Figure~\ref{fig.sect.neutrino.CP}. Because $T_{\mathrm{c}n}^{\max}\gg
T$, the localization of the peaks (at $\sim 10^{12}~\mdens$
and at $\sim 10^{14}~\mdens$) does not change much with
decreasing temperature. However, the heights of the peaks decrease
very fast. For the selected superfluidity model, and at $T=10^{9}$~K,
$Q^{\mathrm{CP}}_\nu$ in the peak region dominates over all other neutrino
emission mechanisms. Let us notice that a proper inclusion of the
in-medium modification of the weak interactions could significantly
decrease the maximum  value of  $Q^{\mathrm{CP}}_\nu$  by about two orders of
magnitude~\cite{sedrakian-07,kolomeitsev-08}.

\newpage


\section{Observational Constraints on Neutron Star Crusts}
\label{sect.obs}

\subsection{Supernovae and the physics of hot dense inhomogeneous matter}
\label{sect.obs.supernova}

The stellar evolution of massive stars with a mass $M\sim 10\mbox{\,--\,}20\,M_\odot$ ends with the catastrophic gravitational collapse of the
degenerate iron core (for a recent review, see, for instance, \cite{janka-07} and
references therein). Photodissociation of iron nuclei and electron captures 
lead to the neutronization of matter. As a result, the internal pressure resisting the
gravitational pull drops, thus accelerating the collapse, which
proceeds on a time scale of $\sim 0.1$~s. When the matter density
inside the core reaches $\sim 10^{12}~\mdens$, neutrinos become
temporarily trapped, thus hindering electron captures and providing
additional pressure to resist gravity. However, this is not sufficient
to halt the collapse and the core contraction  proceeds until the
central density reaches about twice the saturation density $\rho\simeq
2.8 \times 10^{14}~\mdens$ inside atomic nuclei. After that, due
to the stiffness (incompressibility) of nuclear matter, the collapse
halts and the core bounces, generating a shock wave. The shock wave
propagates outwards against the infalling material and eventually
ejects the outer layers of the star, thus spreading heavy elements into
the interstellar medium. A huge amount of energy, $\sim 10^{53}$~erg, is
released, almost entirely (99\%) in the form of neutrinos and
antineutrinos of all flavors. The remaining energy is lost into
electromagnetic and gravitational radiation. This scenario of
core-collapse supernova explosion proved to be consistent with dense
matter theory and various observations of the supernova 1987A in the
Large Magellanic Cloud (discovered on February~23, 1987). In particular, the
observation of the neutrino outburst provided the first direct
estimate of the binding energy of the newly-born neutron star. With
the considerable improvement of neutrino detectors and the development
of gravitational wave interferometers, future observations of galactic
supernova explosions would bring much more restrictive constraints onto
theoretical models of dense matter. Supernova observations would
indirectly improve our knowledge of neutron star crusts despite very
different conditions, since in collapsing stellar cores and neutron star crusts 
the constituents are the same and are therefore described by the same 
microscopic Hamiltonian.

In spite of intense theoretical efforts, numerical simulations of
supernovae still fail to reproduce the stellar explosion, which
probably means that some physics is missing and more realistic physics
input is required~\cite{martinez-06}.  One of the basic ingredients
required by supernova simulations is the equation of state of hot
dense matter for both the inhomogeneous and homogeneous phases, up to
a few times nuclear saturation densities
(Section~\ref{sect.eos.supernova}). The equation of states (EoS) plays
an important role in core collapse, the formation of the shock
and its propagation~\cite{sumiyoshi-05, janka-07}. The key parameter
for the stability of the star is the adiabatic index defined by
Equation~(\ref{eq.sect.eos.supernova.gamma}). The stellar core becomes
unstable to collapse when the pressure-averaged value of the adiabatic
index inside the core falls below some critical threshold
$\gamma_c$. A stability analysis in Newtonian gravitation shows that
$\gamma_c=4/3$. The effects of general relativity increase the
critical value above 4/3. The precise value of the adiabatic index in
the collapsing core depends on the structure and composition of
the hot dense matter and, in particular, on the presence of nuclear
pastas, as can be seen in
Figure~\ref{fig.sect.eos.supernova.gamma.pasta}. The composition of
the collapsing core and its evolution into a proto-neutron star depend
significantly on the EoS. The mass fractions of the various components
present inside the stellar core during the collapse are shown in
Figure~\ref{fig.sect.obs.supernova.composition} for two different EoS,
the standard Lattimer \& Swesty~\cite{lattimer-91} EoS (L\&S) based
on a compressible liquid drop model and the recent relativistic mean
field EoS of Shen et al.~\cite{shen-98, shen-98b} (note however that the 
treatment of the inhomogeneous phases is not quantal but is based on 
the semi-classical Thomas--Fermi approximation, discussed in 
Section~\ref{sect.groundstate.inner.TF}). As seen in
Figure~\ref{fig.sect.obs.supernova.composition}, the L\&S EoS
predicts a larger abundance of free protons than the Shen EoS. As a
consequence, the L\&S EoS enhances electron captures compared to the
Shen EoS and leads to a stronger deleptonization of the core, thus
affecting the formation of a shock wave. The effects of the EoS are
more visible during the late period of the propagation of the shock wave as
shown in Figure~\ref{fig.sect.obs.supernova.shock}. The L\&S EoS
leads to a more compact proto-neutron star, which is therefore hotter
and has higher neutrino luminosity, as can be seen in
Figure~\ref{fig.sect.obs.supernova.luminosity}.

\epubtkImage{sumiyoshi2.png}{%
  \begin{figure}[htbp]
    \centerline{\includegraphics[scale=0.5]{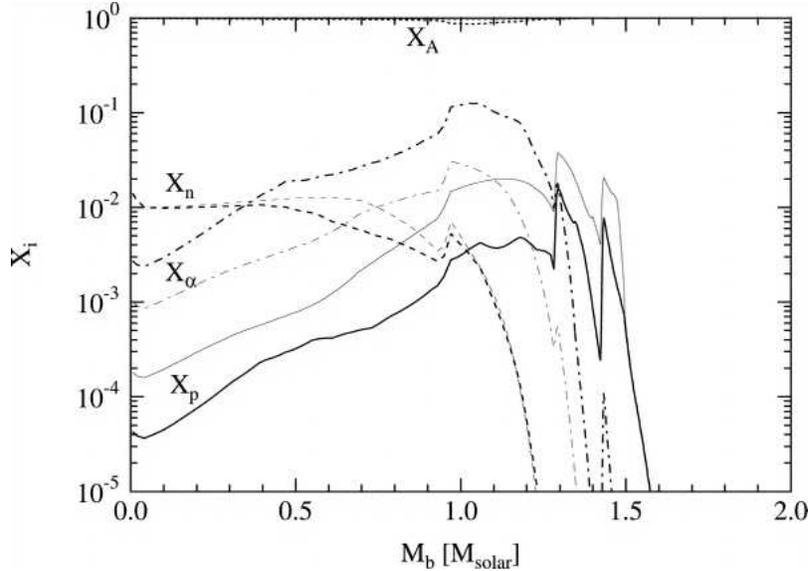}}
    \caption{Mass fractions of different particles in a supernova core
    as a function of baryon mass coordinate at the time when the
    central density reaches $10^{11}~\mdens$. Solid, dashed,
    dotted, and dot-dashed lines show mass fractions of protons,
    neutrons, nuclei, and alpha particles, respectively. The results
    are given for two equations of state: the compressible liquid drop
    model of Lattimer \& Swesty~\cite{lattimer-91} (thin lines) and
    the relativistic mean field theory in the local density
    approximation of Shen et al.~\cite{shen-98, shen-98b} (thick
    lines). See~\cite{sumiyoshi-05} for details.}
    \label{fig.sect.obs.supernova.composition}
\end{figure}}

\epubtkImage{sumiyoshi1.png}{%
  \begin{figure}[htbp]
    \centerline{\includegraphics[scale=0.5]{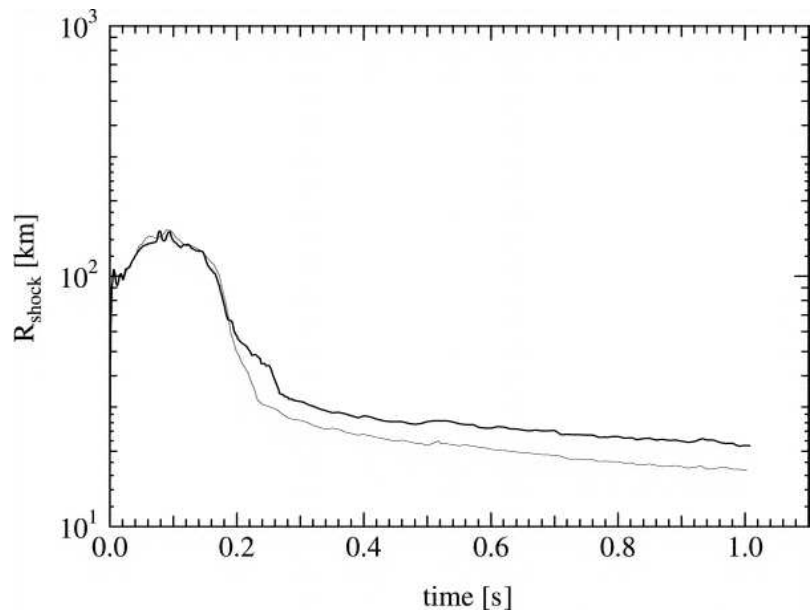}}
    \caption{Radial positions of shock waves as a function of time
    after bounce (the moment of greatest compression of the central
    core corresponding to a maximum central density) for two different
    equations of states: the compressible liquid drop model of
    Lattimer \& Swesty~\cite{lattimer-91} (thin line) and the
    relativistic mean field theory in the local density approximation
    of Shen et al.~\cite{shen-98, shen-98b} (thick
    line). See~\cite{sumiyoshi-05} for details. Notice that these
    particular models failed to produce a supernova explosion.}
    \label{fig.sect.obs.supernova.shock}
\end{figure}}

The collapse of the supernova core and the formation of the
proto-neutron star are governed by weak interaction processes and
neutrino transport~\cite{langanke-06}. Numerical simulations generally
show that as the shock wave propagates outwards, it loses energy due
to the dissociation of heavy elements and due to the pressure of the
infalling material so that it finally stalls around $\sim 10^2$~km, as
can be seen, for instance, in
Figure~\ref{fig.sect.obs.supernova.shock}. According to the delayed
neutrino-heating mechanism, it is believed that the stalled shock is
revived after $\sim$~100~ms by neutrinos, which deposit energy in the
layers behind the shock front. The interaction of neutrinos with
matter is therefore crucial for modeling supernova explosions. 
The microscopic structure of the supernova core  has a strong
influence on the neutrino opacity and, therefore, on the
neutrino diffusion timescale. In the relevant core layers, 
neutrinos form a nondegenerate gas, with a de Broglie
wavelength $\lambda_\nu=2\pi\hbar c/E_\nu$,
where $E_\nu\sim 3k_{\mathrm{B}}T\sim 5\mbox{\,--\,}10$~MeV. If
$\lambda_\nu>2R_A$, where $R_A$ is the radius of a spherical
cluster, then thermal neutrinos ``do not see'' the individual
nucleons inside the cluster and scatter coherently on the 
$A$ nucleons. Putting it differently, a neutrino couples to a
single weak current of the cluster of $A$ nucleons. If the
neutrino scattering amplitude on a single nucleon is $f$, then
the scattering amplitude on a cluster is $Af$, and the
scattering cross section is $\sigma_A^{\mathrm{coh}}=A^2|f|^2$
(\cite{freedman-74}, for a review, see~\cite{shapiro-83}). Consider now the opposite case of
$\lambda_\nu\ll 2R_A$. Neutrinos scatter on every nucleon
inside the cluster. As a result, the scattering amplitudes add \emph{incoherently}, 
and the neutrino-nucleus scattering cross section $\sigma_A^{\mathrm{incoh}}=A|f|^2$, similar to that for a gas of $A$ nucleons. In
this way, $\sigma_A^{\mathrm{coh}}/\sigma_A^{\mathrm{incoh}}\approx A
\sim 100$. One therefore concludes, that the presence of
clusters in hot matter can dramatically increase
the neutrino opacity. The neutrino transport in supernova cores 
depends not only on the characteristic size of the clusters, but also
on their geometrical shape and topology. 
In particular, the presence of an heterogeneous plasma
(due to thermal statistical distribution of $A$ and $Z$)  in
the supernova core~\cite{caballero-06} or the existence of
nuclear pastas instead of spherical clusters~\cite{horowitz-05, sonoda-07} 
have a sizeable effect on the neutrino propagation. The outcome is that the
neutrino opacity of inhomogeneous matter is considerably increased 
compared to that of uniform matter.

\epubtkImage{sumiyoshi4.png}{%
  \begin{figure}[htbp]
    \centerline{\includegraphics[width=\textwidth]{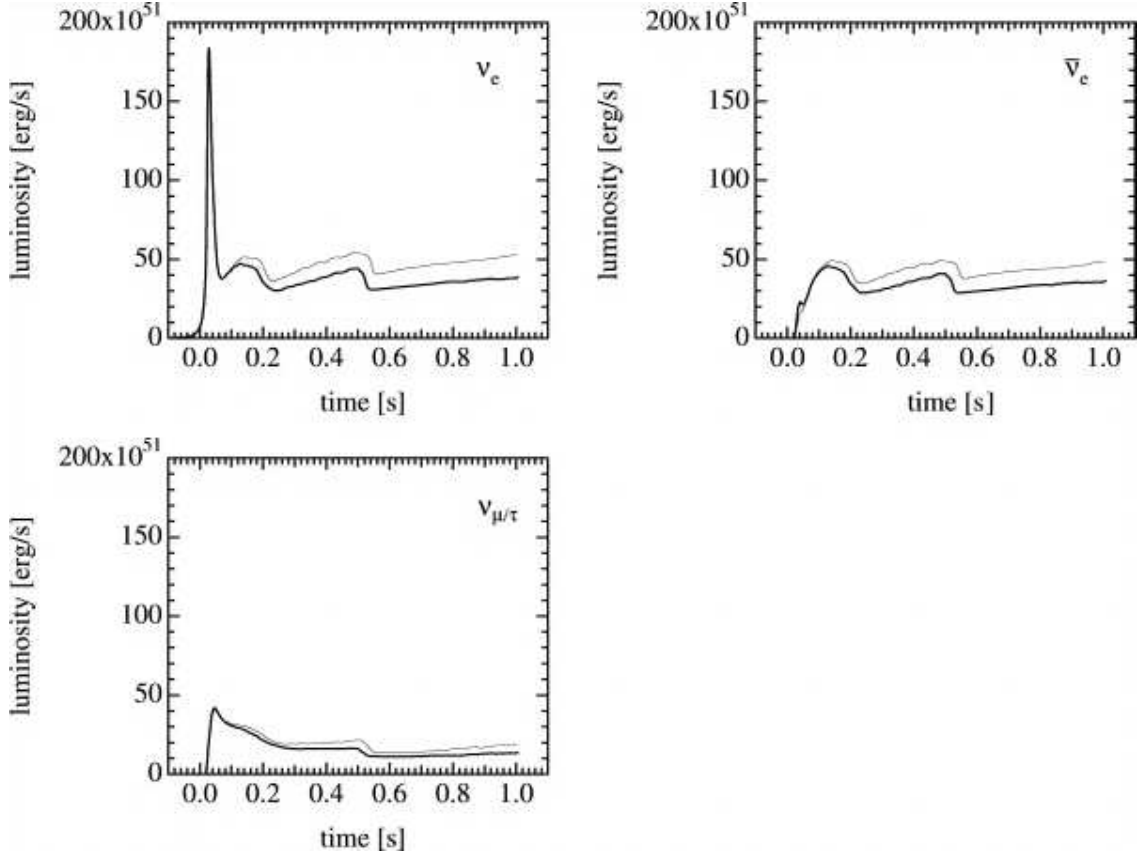}}
    \caption{Luminosities of $\nu_e$, $\bar\nu_e$, and
    $\nu_{\mu/\tau}$ as a function of time after bounce (the moment of
    greatest compression of the central core corresponding to a
    maximum central density) for two different equations of state:
    the compressible liquid drop model of Lattimer \&
    Swesty~\cite{lattimer-91} (thin lines) and the relativistic mean
    field theory in the local density approximation of
    Shen et al.~\cite{shen-98, shen-98b} (thick
    lines). See~\cite{sumiyoshi-05} for details.}
    \label{fig.sect.obs.supernova.luminosity}
\end{figure}}

\subsection{Cooling of isolated neutron stars}
\label{sect.obs.cooling}

Neutron stars are born in the core collapse supernova explosions of
massive stars, as briefly reviewed in
Section~\ref{sect.obs.supernova}. During the first tens of seconds,
the newly formed proto-neutron star with a radius of $\sim 50$~km
stays very hot with temperatures on the order of
$10^{11}\mbox{\,--\,}10^{12}$~K. In the following stage, the star becomes
transparent to neutrinos generated in its interior via various
processes (see Section~\ref{sect.neutrino}). Within $\sim 10\mbox{\,--\,}20$~s the
proto-neutron star thus rapidly cools down by powerful neutrino
emission and shrinks into an ordinary neutron star. The last cooling
stage, after about $10^4\mbox{\,--\,}10^5$~years, is governed by the emission of
thermal photons due to the diffusion of heat from the interior to the
surface (for a recent review of neutron star cooling, see, for
instance, \cite{yakovlev-04, page-06} and references therein). Neutron
stars in X-ray binaries may be heated as a result of the accretion of
matter from the companion star. Observational data and references have
been collected on the UNAM webpage~\cite{unam}.

The cooling of a young neutron star is very
sensitive to its crust physics including, for example, neutron
superfluidity, as shown in Figure~\ref{fig.sect.obs.cooling.curve}. Superfluidity
of free neutrons in the inner crust suppresses heat capacity. 
Moreover, superfluidity opens a new channel for neutrino emission. Indeed, the
formation of a bound neutron pair liberates energy, which can be
converted into a neutrino-antineutrino pair, as discussed in
Section~\ref{sect.neutrino.CP}.

\epubtkImage{cooling1.png}{%
\begin{figure}[htbp]
  \centerline{\includegraphics[scale=0.4]{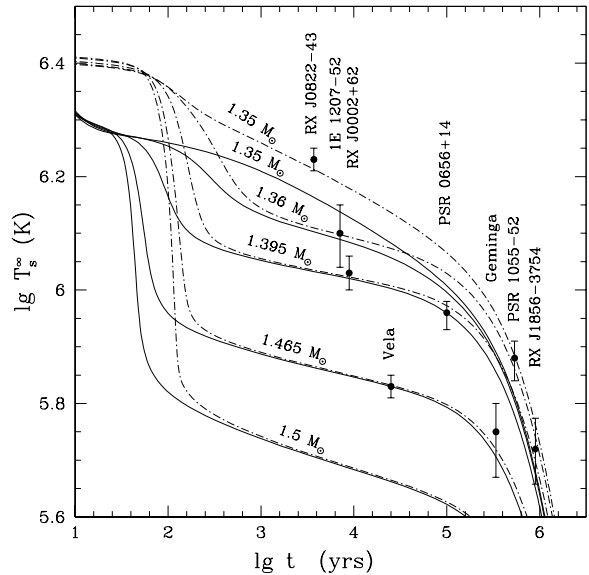}}
  \caption{Redshifted surface temperatures (as seen by an observer
  at infinity) vs.\ age of neutron stars
  with different masses as compared with observation. Dot-dashed curves
  are calculated with only proton superfluidity in the core. Solid
  curves also include neutron superfluidity in the crust and outer
  core~\cite{yakovlev-01}.}
  \label{fig.sect.obs.cooling.curve}
\end{figure}}

\subsubsection{Thermal relaxation of the crust}

Due to its relatively low neutrino emissivity, the crust of a newly-born neutron star cools less rapidly than the core and thus stays
hotter. As a result, the surface temperature decreases slowly during
the first ten to hundred years and then drops sharply when the cooling
wave from the core reaches the surface as illustrated in
Figure~\ref{fig.sect.obs.cooling.curve}. After time $t_w$, the star
becomes isothermal except for the very outer layers. The relaxation
time $t_w$ of reaching a quasi-isothermal state depends, in particular, on
the specific heat $C_v$ and on the thermal conductivity $\kappa$ of
the inner crust (see Section~\ref{sect.cond.cond}) and is
approximately given by~\cite{lattimer-94, gnedin-01}
\begin{equation}
\label{equation:relaxation_time}
t_w \sim (\Delta R)^2 \left(1-\frac{r_{\mathrm{g}}}{R}\right)^{-3/2} \frac{C_{\mathrm{v}}}{\kappa},
\end{equation}
where $\Delta R$ is the thickness of the crust, $R$ the circumferential radius of the
star and $r_{\mathrm{g}}=2G M/c^2$, the Schwarzschild radius. 
The ratio of specific heat $C_{\mathrm{v}}$ to thermal conductivity
$\kappa$ has to be taken at half nuclear saturation density,
slightly lower than  the crust bottom density $\rho_{\mathrm{cc}}$
(see \cite{yakovlev-01}; in general, the relaxation time is the most
sensitive to $\kappa$ and $C_{\mathrm{v}}$ in the density range
$0.1\rho_{\mathrm{cc}}<\rho<\rho_{\mathrm{cc}}$).
The thermal conductivity of the crust comes
mainly from electrons scattering off atomic nuclei and electrically
charged impurities. It is crucially dependent on the structure and 
composition of the crust (see Section~\ref{sect.cond}). The
crustal specific heat is dominated by free neutrons if
they are not superfluid. Otherwise the neutron specific heat is
strongly suppressed and its contribution to the total heat capacity is
negligible as can be seen in
Figure~\ref{fig.sect.obs.cooling.heatcap}. However, the density range
and the critical temperatures for neutron superfluidity in the crust
are still not very well known. The presence of nuclear inhomogeneities 
can have a significant effect on the specific heat by
reducing the neutron pairing correlations inside the nuclei 
especially in the shallow layers of the inner crust at densities
$\rho \sim 10^{11}\mbox{\,--\,}10^{12}~\mdens$~\cite{pizzochero-02,
  sandulescu-04, khan-05, monrozeau-07}. The cooling curves of a
$1.5\,M_\odot$ neutron star for different crust models are shown in
Figure~\ref{fig.sect.obs.cooling.crust}. Observations of young neutron
stars could thus put constraints on the thermal properties of the
crust, which in turn depend on its structure and 
composition. Such young neutron stars have not been observed yet. One
reason might be that neutron stars born in type~II supernova
explosions remain hidden by the expanding supernova envelopes for
many years.

\epubtkImage{heat_cap.png}{%
  \begin{figure}[htbp]
    \centerline{\includegraphics[scale=0.6]{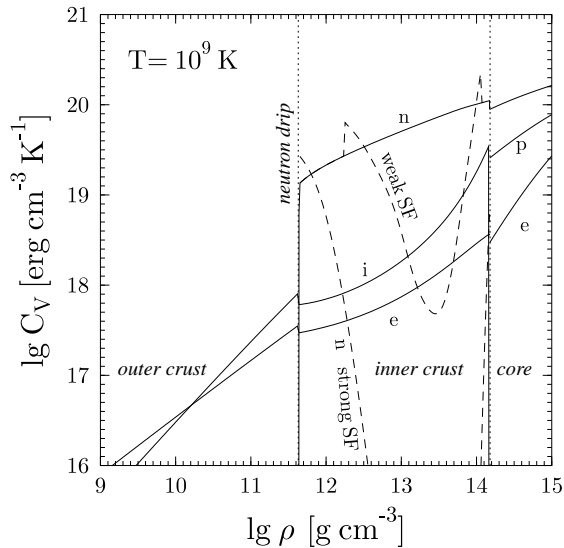}}
    \caption{Neutron star specific heat at $T = 10^9$~K. Solid lines:
    partial heat capacities of ions (i), electrons (e) and free
    neutrons (n) in nonsuperfluid crusts, as well as of neutrons,
    protons (p) and electrons in nonsuperfluid cores. Dashed lines:
    heat capacities of free neutrons in the crust modified by
    superfluidity. Two particular models of weak and strong
    superfluidity are considered. The effects of the nuclear
    inhomogeneities on the free neutrons are
    neglected. From~\cite{gnedin-01}.}
    \label{fig.sect.obs.cooling.heatcap}
\end{figure}}

\epubtkImage{crust_cool.png}{%
  \begin{figure}[htbp]
    \centerline{\includegraphics[scale=1.0]{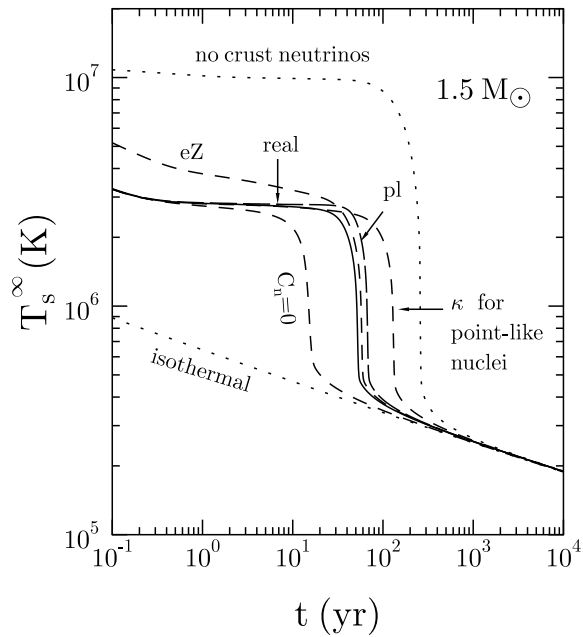}}
    \caption{Effective surface temperature (as seen by an observer at
      infinity) of a $1.5\,M_\odot$ neutron star during the first
      hundred years for different crust models. Dotted lines: cooling
      without neutrino emission from the crust (upper line), infinite
      $\kappa$ at $\rho>10^{10}~\mdens$. Solid line: cooling curve for
      the best values of $\kappa$, $C_{\mathrm{v}}$, and $Q_\nu$. Dashed
      line $C_n=0$: dripped neutrons heat capacity removed. Dashed
      curve $\kappa$: thermal conductivity calculated assuming
      point-like nuclei. Two other dashed lines: neutrino emission
      processes removed except for plasmon decay (pl) or
      electron-nucleus Bremsstrahlung ($eZ$). See also line
      $1.5\,M_\odot$ in Table~2 of~\cite{gnedin-01}.}
    \label{fig.sect.obs.cooling.crust}
\end{figure}}

\subsubsection{Observational constraints from thermal X-ray emission}

In cooling simulations, the neutron star is usually decomposed into
the stellar interior, which becomes isothermal after a few tens to
hundreds of years and the outer heat blanketing (insulating) envelope,
where temperature gradients persist due to low thermal
conductivity. The boundary between the interior and the envelope is
conventionally set at $\rho=10^{10}~\mdens$. The relationship between the surface temperature $T_{\mathrm{s}}$
and the temperature $T_{\mathrm{b}}$ at the bottom of the heat blanketing
envelope is very sensitive to the structure and the composition of the
crust and to the presence of a magnetic field. The outermost envelope of a
neutron star, composed mainly of iron
(Section~\ref{sect.groundstate.outer}), may be covered by a thin layer
of light elements due to accretion, which strongly enhances heat
transport and increases the surface temperature for a given $T_{\mathrm{b}}$ (let us remember that
the electron thermal conductivity in a Coulomb plasma of ions with
charge $Z$ varies as $\sim 1/Z$). Strong magnetic fields also affect
heat transport, leading to a nonuniform surface-temperature
distribution and, in particular, hot caps near the magnetic poles, as
illustrated in Figure~\ref{fig.sect.obs.cooling.Bcrust}.

\epubtkImage{b3e12_nsccru_1e6.png}{%
  \begin{figure}[htbp]
    \centerline{\includegraphics[scale=0.6]{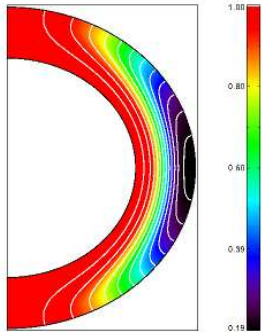}}
    \caption{Magnetic field lines and temperature distribution in a
    neutron star crust for an axisymmetric dipolar magnetic field
    $B=3\times 10^{12}$~G and an isothermal core with temperature
    $T_{\mathrm{core}}=10^6$~K. The temperature is measured in units of
    $T_{\mathrm{core}}$. The magnetic field is confined to the
    crust. From~\cite{geppert-04}.}
    \label{fig.sect.obs.cooling.Bcrust}
\end{figure}}

The effects of different crust models on the relationship between the
surface temperature $T_{\mathrm{s}}$ and the temperature $T_{\mathrm{b}}$ at the
bottom of the heat-blanketing envelope, are illustrated in
Figure~\ref{fig.sect.obs.cooling.Ts-Tb}. Grigorian~\cite{grigorian-06}
recently argued that cooling models predicting neutron stars with an
age between about $10^3\mbox{\,--\,}10^4$ years to be hotter than those already
observed, should be rejected since if such stars existed in our
galaxy, they would have already been detected. This {\it brightness
  constraint} puts restrictions on the
$T_{\mathrm{s}}\mbox{\,--\,}T_{\mathrm{b}}$ relationship hence on
crust models.

\epubtkImage{Ts_Tb.png}{%
\begin{figure}[htbp]
  \centerline{\includegraphics[scale=0.6]{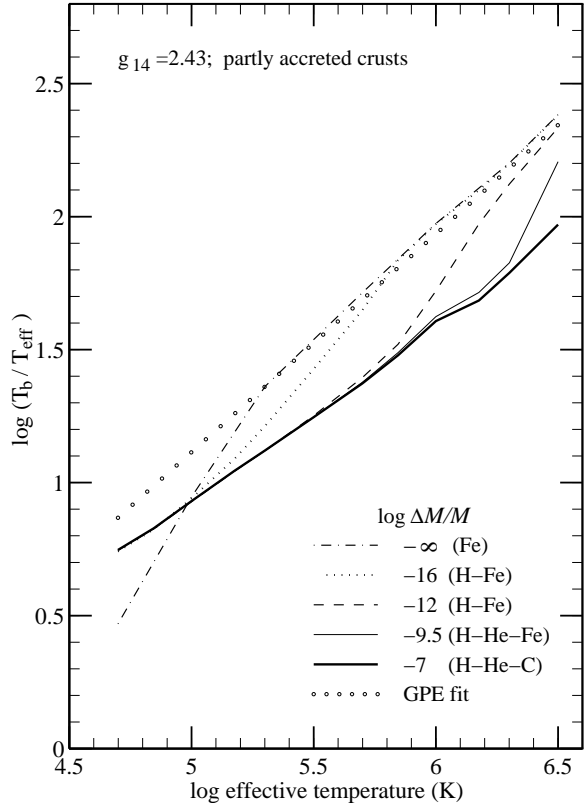}}
  \caption{Relationship between the effective surface temperature
  $T_{\mathrm{eff}}$, as measured by an observer at infinity, and the local
  temperature at the bottom of the heat-blanketing envelope,
  $T_{\mathrm{b}}$ (at
  $\rho_{\mathrm{b}}=10^{11}~\mdens$). Calculations performed for the
  ground-state (Fe) and partly-accreted envelopes of mass $\Delta
  M$. Numbers $-16,\ldots,-7$ indicate $\log_{10}(\Delta M/M)$ for a
  $1.4\,M_\odot$ star with surface gravity $g_{14}=2.43$ (in units of
  $10^{14} \mathrm{\ cm\ s}^{-2}$). Symbols in parentheses indicate
  chemical composition of accreted envelope.}
  \label{fig.sect.obs.cooling.Ts-Tb}
\end{figure}}

\subsection{r-process in the decompression of cold neutron star crusts}
\label{sect.obs.rproc}

The location of the astrophysical site for the rapid neutron
capture process (r-process), thought to be responsible for the
production of many heavy neutron-rich nuclei with $A>60$ in the
universe, still remains uncertain (for a recent review, see, for
instance, \cite{arnould-07}). Many possible sites have been considered,
but they all have serious problems. The most
studied scenarios are related to neutrino-driven wind during type~II
supernova explosions or $\gamma$-ray bursts. Nevertheless, apart from many
uncertainties in the explosion mechanism, the conditions for the
r-process to occur are difficult to reach and require a fine tuning of
 model parameters. Lattimer et al.~\cite{lattimer-77}
suggested a long time ago that the r-process could also occur during
the decompression of cold crustal matter ejected into the interstellar medium. 
This possibility has remained
largely unexplored until very recently
(see~\cite{goriely-05, arnould-07} and references therein). This
scenario is, however, promising because the presence of neutron-rich
nuclei, the large neutron-to-seed ratio and the low electron fraction
in the decompressing crustal matter are favorable conditions for the
r-process to occur. Various scenarios can be envisioned. Matter could be 
ejected into the interstellar medium by outflows from newly-born proto-neutron 
stars or jets such as those recently observed in Circinus X-1~\cite{heinz-07}. 
Neutron stars very rapidly spinning beyond the mass-shedding limit would also 
expel matter. In the early years of pulsar astronomy, Dyson~\cite{dyson-69} 
suggested that neutron stars might have volcanic activity. This idea of cataclysmic events has been more recently 
revived by the observations of giant flares in magnetars, thought to be the signature of
magnetic crustquakes. From observations of the radio afterglow~\cite{gelfand-07} it 
has been estimated that more than $10^{-9}\,M_\odot$ was ejected
during the December~27, 2004 event in SGR~1806$-$20. More exotic
astrophysical events have been proposed, such as the explosion of a
neutron star below the minimum mass~\cite{sumiyoshi-98} or the phase
transition into a strange quark star
(quark-novae)~\cite{jaikumar-07}. However the merging of a neutron
star and a black hole or of two neutron stars
(see~\cite{konstantin-06} for a recent review on compact binaries) is
probably the most likely scenario for the ejection of large amounts of
matter. This tidal disruption of two merging neutron stars has recently been
studied in detail~\cite{goriely-05, arnould-07}, motivated by the
results of hydrodynamic simulations, which show that up to
$10^{-2}\,M_\odot$ could be ejected in this manner. This study has proven that the
solar system abundance pattern can be qualitatively reproduced by
considering the decompression of clumps of neutron-star--crust matter
with different initial densities, as shown in Figure~\ref{fig.sect.obs.rproc}.

\epubtkImage{rproc.jpg}{%
  \begin{figure}[htbp]
    \centerline{\includegraphics[width=\textwidth]{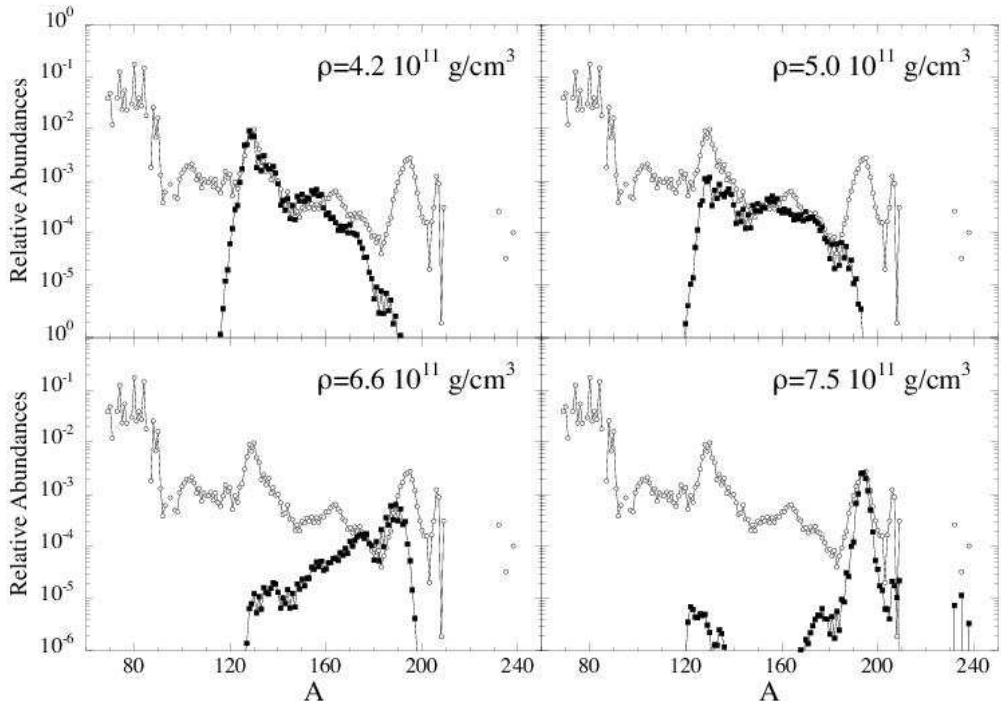}}
    \caption{Final composition of clumps of ejected neutron star crust
    with different initial densities (solid squares). The open circles
    correspond to the solar system abundance of
    r-elements. From~\cite{arnould-07}.}
    \label{fig.sect.obs.rproc}
\end{figure}}

\subsection{Pulsar glitches}
\label{sect.obs.glitches}

Since the discovery by Jocelyn Bell and Anthony Hewish
in 1967 of highly-periodic radio sources soon identified with rotating
neutron stars (Hewish was awarded the Nobel Prize in Physics in 
1974~\cite{hewish-nobel}), more than 1700 pulsars have been found at the time of
writing (pulsar timing data are available online
at~\cite{atnf}). Pulsars are the most precise clocks with
rotation periods ranging from about 1.396 milliseconds for the recently
discovered pulsar J1748$-$2446ad~\cite{hessels-06} up to several
seconds. The periodicity of arrival time of pulses is extremely
stable. The slight delays associated with the spin-down of the star
are at most of a few tens of microseconds per year. Nevertheless,
longterm monitoring of pulsars has revealed irregularities in their
rotational frequencies.

\epubtkImage{g11.png}{%
\begin{figure}[htbp]
  \centerline{\includegraphics[scale=0.6]{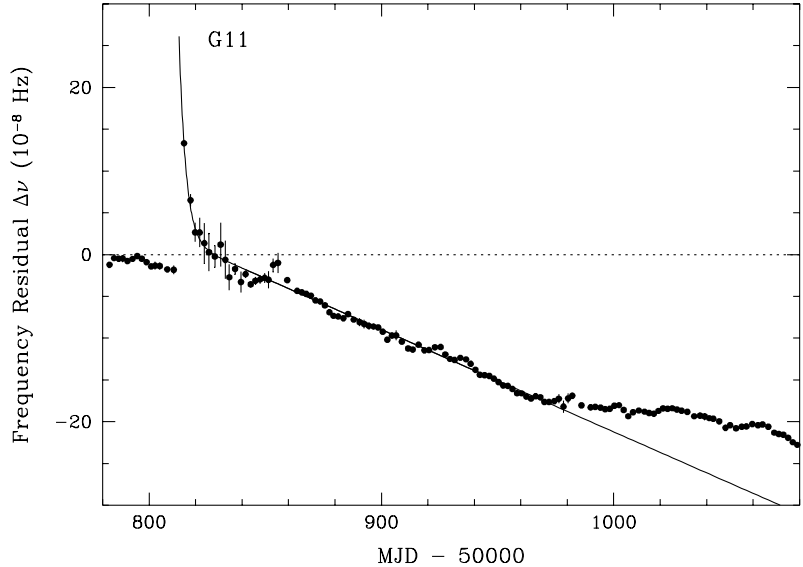}}
  \caption{Glitch $\Delta\Omega/\Omega \sim 9\times 10^{-9}$
  observed in the Crab pulsar by Wong et al.~\cite{wong-01}}.
  \label{fig.sect.obs.glitch.Crab}
\end{figure}}

The first kind of irregularity, called timing noise, is random
fluctuations of pulse arrival times and is present mainly in young
pulsars, such as the Crab, for which the slow down rate is larger than
for older pulsars. Indeed, correlations have been found between the
spin-down rate and the noise amplitude~\cite{lyne-98}. Timing noise might 
result from irregular transfers of angular momentum between the crust
and the liquid interior of neutron stars. A second kind of
irregularity is the sudden jumps or ``glitches'' of the rotational
frequency, which have been observed in radio pulsars and more recently
in anomalous X-ray pulsars~\cite{kaspi-00, kaspi-03, dallosso-03,
  kaspi-03b, morii-05}. An example of a glitch is shown in
Figure~\ref{fig.sect.obs.glitch.Crab}. Evidence of glitches have also been
reported in accreting neutron stars~\cite{galloway-04}. These
glitches, whose amplitude vary from $\Delta\Omega/\Omega \sim 10^{-9}$
up to $\Delta \Omega/\Omega \sim 16\times 10^{-6}$ for
PSR~J1806$-$2125~\cite{hobbs-02}, as shown in
Figure~\ref{fig.sect.obs.glitch.amp}, are followed by a relaxation
over days to years and are sometimes accompanied by a sudden change
of the spin-down rate from $|\Delta\dot{\Omega}/\dot{\Omega}|\sim
10^{-6}\mbox{\,--\,}10^{-5}$ to $|\Delta\dot{\Omega}/\dot{\Omega}|\sim
10^{-3}\mbox{\,--\,}10^{-2}$. By the time of this writing, 171 glitches have been
observed in 50 pulsars. Their characteristics and
the references can be found at~\cite{glitch-table}. The time between
two successive glitches is usually a few years. One of the most active
pulsars is PSR~J1341$-$6220, for which 12 glitches have been detected
during 8.2~years of observation~\cite{wang-00}.

\epubtkImage{glitch_amp.png}{%
\begin{figure}[htbp]
  \centerline{\includegraphics[scale=0.6,angle=-90]{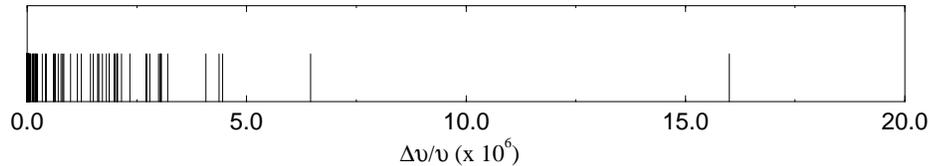}}
  \caption{Amplitudes of 97 pulsar glitches, including the very
  large glitch $\Delta \Omega/\Omega \sim 16\times 10^{-6}$ observed
  in PSR~J1806$-$2125~\cite{hobbs-02}.}
  \label{fig.sect.obs.glitch.amp}
\end{figure}}

Very soon after the observations of the first glitches in the Crab and
Vela pulsars, superfluidity in the interior of neutron stars was
invoked to explain the long relaxation times~\cite{baym-69}. The
possibility that dense nuclear matter becomes superfluid at low
temperatures was suggested theoretically much earlier, even
before the discovery of the first pulsars (see
Section~\ref{sect.super.static}). Following the first observations,
several scenarios were proposed to explain the origin of these glitches,
such as magnetospheric instabilities, pulsar disturbance by a planet,
hydrodynamic instabilities or collisions of infalling massive objects
(for a review of these early models, see
Ruderman~\cite{ruderman-72}). Most of these models had serious
problems. The most convincing interpretation was that of
starquakes, as briefly reviewed in
Section~\ref{sect.obs.glitches.starquake}. However, large amplitude
glitches remained difficult to explain. The possible role of superfluidity in 
pulsar glitches was first envisioned by Packard in 1972~\cite{packard-72}.
Soon after, Anderson and Itoh proposed a model of glitches based on the motion of neutron
superfluid vortices in the crust~\cite{anderson-75}. Laboratory experiments 
were carried out to study similar phenomena in 
superfluid helium~\cite{campbell-79, tsakadze-79, tsakadze-80}. 
It is now widely accepted that neutron superfluidity plays a major role
in pulsar glitches. As discussed
in~Sections~\ref{sect.obs.glitches.2comp}
and~\ref{sect.obs.glitches.recent}, the glitch phenomenon seems to
involve at least two components inside neutron stars: the crust and the
neutron superfluid. Section~\ref{sect.obs.glitches.constraint} shows
how the observations of pulsar glitches can put constraints on the
structure of neutron stars.

\subsubsection{Starquake model}
\label{sect.obs.glitches.starquake}

Soon after the observations of the first glitches in Vela and in the
Crab pulsars, Ruderman~\cite{ruderman-69} suggested that these events
could be the manifestations of starquakes (see also \cite{haensel-06}
and references therein). As a result of centrifugal forces, rotating
neutron stars are not spherical but are slightly deformed, as can be
seen in Figure~\ref{fig.sect.structure.rot.crust.NS}. If the star were
purely fluid, a deceleration of its rotation would entail
a readjustment of the stellar shape to a more spherical
configuration. However, a solid crust prevents such
readjustment and consequently the star remains more oblate. 
The spin-down of the star thus builds up stress
in the crust. When this stress reaches a critical level, the crust
cracks and the star readjusts its shape to reduce its
deformation. Assuming that the angular momentum is conserved during a
starquake, the decrease $\Delta I<0$ of the moment of inertia $I$ is
therefore accompanied by an increase $\Delta\Omega>0$ of the
rotational frequency $\Omega$ according to
\begin{equation}
\frac{\Delta \Omega}{\Omega} = -\frac{\Delta I}{I} \, .
\end{equation}
For a purely fluid rotating star, the moment of inertia can be written
as
\begin{equation}
I=I_0(1+\epsilon) \, ,
\end{equation}
where $I_0$ is the moment of inertia of a nonrotating spherical
star. The parameter $\epsilon$ is proportional to $\Omega^2$ and is
typically very small. For instance, for the Crab and Vela pulsars,
$\epsilon\sim 10^{-4}$ and $10^{-5}$, respectively. Since the decrease
of the moment of inertia of a pulsar can be at most equal to that of a
purely fluid star, this model predicts that the glitch amplitude is 
\begin{equation}
\frac{\Delta \Omega}{\Omega} < \epsilon \frac{\delta t}{\tau} \, ,
\end{equation}
where $\tau=\dot\Omega/2\Omega$ and $\delta t$ is the time between two
successive starquakes (on the order of years for the Vela and Crab
pulsars). This model is consistent with the glitches observed in the
Crab pulsar and explains the weak glitch activity of young pulsars by the
fact that the internal temperatures are still too high for the
crust to store a large stress. However, for the Vela pulsar, with
$\tau\sim 10^4$ years and $\epsilon\sim 10^{-5}$, this model predicts
glitch amplitude of $\Delta \Omega/\Omega\sim 10^{-9}$, about three
orders of magnitude smaller than those observed. The starquake model
fails to explain all the observations of pulsar glitches. Therefore, other
mechanisms have to be invoked.

\subsubsection{Two-component models}
\label{sect.obs.glitches.2comp}

Due to the interior magnetic field, the plasmas of electrically
charged particles inside neutron stars are strongly coupled and 
co-rotate with the crust on very long timescales on the order of the 
pulsar age~\cite{easson-79a}, thus following the long-term
spin-down of the star caused by the electromagnetic
radiation. Besides, the crust and charged particles are rotating
at the observed angular velocity of the pulsar due to coupling
with the magnetosphere. In contrast, neutrons are 
electrically neutral and superfluid. As a consequence, they can rotate 
at a different rate by forming
quantized vortex lines (Section~\ref{sect.super.dyn.rot}). This
naturally leads to the consideration of the stellar interior as a
two-fluid mixture. A model of this kind was first suggested by
Baym et al.~\cite{baym-69b} for interpreting pulsar glitches as a
transfer of angular momentum between the two components. Following a
sudden spin-up of the star after a glitch event, the plasma of charged
particles readjusts to a new rotational frequency within a few
seconds~\cite{easson-79b}. Moreover, as discussed in Section~\ref{sect.super.dyn.entr.ns}, 
neutron superfluid vortices carry magnetic flux giving rise to an effective mutual friction force
acting on the superfluid. As a result, the neutron superfluid in the core is
dynamically coupled to the crust and to the charged particles, on a
time scale much shorter than the post-glitch relaxation time of months
to years observed in pulsars like Vela, suggesting that glitches are
associated with the neutron superfluid in the crust. This conclusion
assumes that the distribution of proton flux tubes in the liquid core
is uniform. Nevertheless, one model predicts that every neutron vortex
line is surrounded by a cluster of proton flux
tubes~\cite{sedrakian-95I, sedrakian-95II}. In this vortex-cluster
model, the coupling time between the core superfluid and the crust
could be much longer than the previous estimates and could be
comparable to the postglitch relaxation times.

The origin of pulsar glitches relies on a sudden release of stresses
accumulated in the crust, similar to the starquake model. However,
the transfer of angular momentum from the rapidly-rotating neutron
superfluid to the magnetically-braked solid crust and charged
constituents during a glitch allows much larger spin-up than that due
solely to the readjustment of the stellar shape. Neutron
superfluid is weakly coupled to a normal charged component by mutual
friction forces and thus follows the spin-down of the crust via a
radial motion of the vortices away from the rotation axis unless the
vortices are pinned to the crust. In the latter case, the lag between
the superfluid and the crust induces a Magnus force, acting on the
vortices producing a crustal stress. When the lag exceeds a critical
threshold, the vortices are suddenly unpinned. Vortex motion could
also be initiated by a temperature perturbation, for instance the heat released
after a starquake~\cite{link-96}. As a result, the
superfluid spins down and, by the conservation of the total angular
momentum, the crust spins up leading to a
glitch~\cite{anderson-75}. If the pinning is strong enough, the crust
could crack before the vortices become unpinned, as suggested by
Ruderman~\cite{ruderman-76, ruderman-91a, ruderman-91b,
  ruderman-91c}. These two scenarios lead to different predictions for
the internal heat released after a glitch event. It has been argued
that observations of the thermal X-ray emission of glitching pulsars
could thus put constraints on the glitch mechanism~\cite{larson-02}.

In the vortex creep model~\cite{alpar-84b} a postglitch relaxation
is interpreted as a motion of vortices due to thermal
fluctuations. Even at zero temperature, vortices can become unpinned
by quantum tunneling~\cite{link-93}. The vortex current increases with
the growth of temperature and can prevent the accumulation of large crustal
stress in young pulsars, thus explaining the low glitch activity of
these pulsars. In the model of Alpar et al.~\cite{alpar-93, alpar-96}
a neutron star is analogous to an electric circuit with a capacitor
and a resistor, the vortices playing the role of the electric charge carriers. 
The star is, thus, assumed to be formed of resistive
regions, containing a continuous vortex current, and
capacitive regions devoid of vortices. A glitch can then be viewed as
a vortex ``discharge'' between resistive regions through capacitive
regions. The permanent change in the spin-down rate observed in some
pulsars is interpreted as a reduction of the moment of inertia due to
the formation of new capacitive regions. A major difficulty of this
model is to describe the unpinning and repinning of vortices.

Ruderman developed an alternative view based on the interactions
between neutron vortices and proton flux tubes in the core,
assuming that the protons form a type~II
superconductor~\cite{ruderman-98, ruderman-98err}. Unlike the vortex
lines, which are essentially parallel to the rotation axis, the
configuration of the flux tubes depends on the magnetic field and 
may be quite complicated. Recalling that the number of flux tubes per
vortex is about $10^{13}$ (see Sections~\ref{sect.super.dyn.fluxtubes}
and~\ref{sect.super.dyn.ns}), it is therefore likely that neutron
vortices and flux tubes are strongly entangled. As superfluid spins
down, the vortices move radially outward dragging along the flux
tubes. The motion of the flux tubes results in the build up of stress
in the crust. If vortices are strongly pinned to the crust, the stress
is released by starquakes fracturing the crust into plates like the
breaking of a concrete slab reinforced by steel rods when pulling on
the rods. These plates and the pinned vortices will move toward the
equator thus spinning down superfluid and causing a glitch. Since the
magnetic flux is frozen into the crust due to very high electrical
conductivity, the motion of the plates will affect the configuration
of the magnetic field. This mechanism naturally explains the increase
of the spin-down rate after a glitch observed in some pulsars like the
Crab, by an increase of the electromagnetic torque acting on the
pulsar due to the increase of the angle between the magnetic axis and
the rotation axis.

\subsubsection{Recent theoretical developments}
\label{sect.obs.glitches.recent}

Other scenarios have recently been proposed for explaining pulsar
glitches, such as transitions from a configuration of straight
neutron vortices to a vortex tangle~\cite{peralta-06}, and more exotic
mechanisms invoking  the possibility of crystalline color
superconductivity of quark matter in a neutron star
core~\cite{alford-01}. These models, and those briefly reviewed in
Section~\ref{sect.obs.glitches.2comp}, rely on rather poorly known
physics. The strength of the vortex pinning forces and the type of
superconductivity in the core are controversial issues (for a recent
review, see, for instance, \cite{sedrakian-06} and references
therein). Besides, it is usually implicitly assumed that superfluid
vortices extend throughout the star (or at least throughout the inner
crust). However, microscopic calculations show that the superfluidity
of nuclear matter strongly
depends on density (see Section~\ref{sect.super.static}). It
should be remarked that even in the inner crust, the outermost and
innermost layers may be nonsuperfluid, as discussed in
Section~\ref{sect.super.static.Tc}. It is not clear how superfluid
vortices arrange themselves if some regions of the star are nonsuperfluid. 
The same question also arises for magnetic flux tubes if
protons form a type~II superconductor.

Andersson and collaborators~\cite{andersson-04} have suggested that
pulsar glitches might be explained by a Kelvin--Helmholtz instability
between neutron superfluid and the conglomeration of charged
particles, provided the coupling through entrainment (see
Section~\ref{sect.super.dyn.entr.ns}) is sufficiently strong. It
remains to be confirmed whether such large entrainment effects can
occur. Carter and collaborators~\cite{carter-00} pointed out a few
years ago that a mere deviation from the mechanical and chemical
equilibrium induced by the lack of centrifugal buoyancy is a source of
crustal stress. This mechanism is always effective, independently of
the vortex motion and proton superconductivity. In particular, even if
the neutron vortices are not pinned to the crust, this model leads to
crustal stress of similar magnitudes than those obtained in the pinned
case. Chamel \& Carter~\cite{chamelcarter-06} have recently demonstrated that
the magnitude of the stress is independent of the interactions between 
neutron superfluid and normal crust giving rise to entrainment effects. 
But they have shown that stratification induces additional crustal stress. 
In this picture, the stress builds up
until the lag between neutron superfluid and the crust reaches a
critical value, at which point the crust cracks, triggering a
glitch. The increase of the spin-down rate observed in some pulsars
like the Crab can be explained by the crustal plate tectonics of
Ruderman~\cite{ruderman-76, ruderman-91a, ruderman-91b, ruderman-91c},
assuming that neutron superfluid vortices remain pinned to the
crust. Even in the absence of vortex pinning,
Franco et al.~\cite{franco-00} have shown that, as a result of
starquakes, the star will oscillate and precess before relaxing to a
new equilibrium state, followed by an increase of the angle between the
magnetic and rotation axis (thus increasing the spin-down rate).

\subsubsection{Pulsar glitch constraints on neutron star structure}
\label{sect.obs.glitches.constraint}

Basing their work on the two-component model of pulsar glitches,
Link et al.~\cite{link-99} derived a constraint on the ratio $I^{\mathrm{f}}/I$ of the moment of inertia $I^{\mathrm{f}}$ of the free 
superfluid neutrons in the crust to the total moment of inertia $I$ of the Vela pulsar, 
from which they inferred an inequality involving the mass and radius of the
pulsar. However, they neglected entrainment effects
(see Sections~\ref{sect.super.dyn.entr} and \ref{sect.super.dyn.entr.ns}), 
which can be very strong in the crust, as shown by
Chamel~\cite{chamel-05,chamel-06}. We will demonstrate here how the constraint 
is changed by including these effects, following the analysis of 
Chamel \& Carter~\cite{chamelcarter-06}.

The total angular momentum $J$ of a rotating neutron star is the sum of
the angular momentum $J^{\mathrm{f}}$ of free superfluid neutrons in the crust and of the
angular momentum  $J^{\mathrm{c}}$ of the ``crust'' (this includes
not only the solid crust but also the liquid core, as discussed in Section~\ref{sect.obs.glitches.2comp}). 
As reviewed in Sections~\ref{sect.hydro.twofluid} and \ref{sect.hydro.entrainment}, momentum and velocity 
of each component are not aligned due to (nondissipative) entrainment effects.
Likewise, it can be shown that the angular momentum of each component is a superposition of both angular 
velocities $\Omega_{\mathrm{f}}$ and $\Omega_{\mathrm{c}}$~\cite{chamelcarter-06};
\begin{equation}
J^{\mathrm{f}}=I^{\mathrm{ff}} \Omega_{\mathrm{f}} + I^{\mathrm{fc}} \Omega_{\mathrm{c}} \, ,
\end{equation}
\begin{equation}
J^{\mathrm{c}}=I^{\mathrm{cf}} \Omega_{\mathrm{f}} + I^{\mathrm{cc}} \Omega_{\mathrm{c}} \,,
\end{equation}
where $I^{\mathrm{ff}}$, $I^{\mathrm{fc}}=I^{\mathrm{cf}}$ and $I^{\mathrm{cc}}$ are 
partial moments of inertia, which determine $I^{\mathrm{f}}=I^{\mathrm{ff}}+I^{\mathrm{fc}}$ 
and $I^{\mathrm{c}}=I^{\mathrm{cf}}+I^{\mathrm{cc}}$. As discussed in~\cite{chamelcarter-06},
$I^{\mathrm{fc}}=I^{\mathrm{cf}}$ is expected to be positive in the core and negative in the crust.

\epubtkImage{glitch.png}{%
  \begin{figure}[htbp]
    \centerline{\includegraphics[scale=0.4,angle=0]{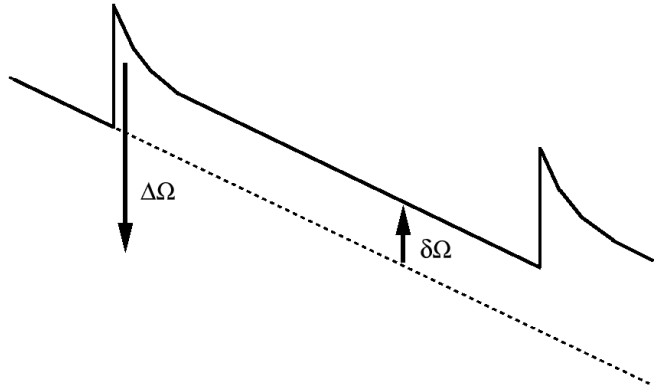}}
    \caption{Schematic picture showing the variations $\Delta \Omega$
    and $\delta \Omega$ of the pulsar angular frequency $\Omega$,
    during a glitch and in the interglitch period, respectively.}
    \label{fig.sect.obs.glitch}
\end{figure}}

Let us denote (discontinuous) variations of some quantity $Q$ during a glitch by $\Delta Q$ and (continuous) 
variations of this quantity during the interglitch period by $\delta Q$, as illustrated in 
Figure~\ref{fig.sect.obs.glitch}.
The total angular momentum $J=J^{\mathrm{f}}+J^{\mathrm{c}}$ can be assumed to be 
conserved during a glitch, therefore, 
\begin{equation}
\label{eq.sect.obs.glitches.constraint.1}
\Delta J^{\mathrm{f}}=-\Delta J^{\mathrm{c}}.
\end{equation}
If no torque were acting on the neutron superfluid in the interglitch period, 
its angular momentum $J^{\mathrm{f}}$ would be conserved and we would have $\delta J^{\mathrm{f}}=0$. 
However, neutron superfluid is weakly coupled to the magnetically-braked crust via 
friction forces induced by the dissipative motion of quantized vortex lines, as discussed in 
Section~\ref{sect.super.dyn.vortices}. Consequently, $J^{\mathrm{f}}$ does not remain exactly 
constant but decreases $\delta J^{\mathrm{f}}\leq 0$, 
\begin{equation}\label{eq.sect.obs.glitches.constraint.2}
\delta\Omega_{\mathrm{f}} \leq -\frac{I^{\mathrm{fc}}}{I^{\mathrm{ff}}} \delta
\Omega_{\mathrm{c}} \, .
\end{equation}
Friction effects prevent a long term build up of too large a deviation of the 
superfluid angular velocity $\Omega_{\mathrm{f}}$ from the externally observable
value $\Omega=\Omega_{\mathrm{c}}$. This means that the average over many glitches 
(denoted by $\langle ... \rangle$) of the change of relative angular velocity 
$\Omega_{\mathrm{f}}-\Omega$ should be approximately zero
\begin{equation}\label{eq.sect.obs.glitches.constraint.3}
\langle\Delta\Omega_{\mathrm{f}}+\delta\Omega_{\mathrm{f}}\rangle \simeq \langle \Delta\Omega
+\delta \Omega\rangle \, .
\end{equation}
Combining Equations~(\ref{eq.sect.obs.glitches.constraint.1}), (\ref{eq.sect.obs.glitches.constraint.2})
and~(\ref{eq.sect.obs.glitches.constraint.3}), it can be shown that the partial moments of inertia 
are constrained by the following relation obtained by Chamel \& Carter~\cite{chamelcarter-06}
\begin{equation}
\label{eq.sect.obs.glitch.constraint}
\frac{(I^{\mathrm{f}})^2}{I I^{\mathrm{ff}}}\geq {\cal G} \, .
\end{equation}
The dimensionless coupling parameter $\cal G$~\cite{link-99} depends
only on observable quantities and is defined by
\begin{equation}
{\cal G}=A_g \frac{\Omega}{|\dot\Omega_{\mathrm{av}}|} \, ,
\end{equation}
where $A_g$ is the glitch activity~\cite{mckenna-90}
\begin{equation}
A_g=\frac{1}{t}\sum_i \frac{\Delta \Omega_i} {\Omega} \, ,
\end{equation}
where the sum is over the glitches occurring during a time $t$ 
and $\dot\Omega_{\mathrm{av}}$ is the average spin-down rate. According to
the statistical analysis of 32 glitches in 15 pulsars by
Lyne et al.~\cite{lyneshemar-00}, the parameter $\cal G$ in Vela-like
pulsars is ${\cal G} \simeq 0.017$. The coupling parameters for
selected pulsars are shown in
Figure~\ref{fig.sect.obs.glitch.activity}. In particular, an analysis
of the Vela pulsar shows that ${\cal G} \simeq 0.014$.

\epubtkImage{glitch_activity.png}{%
  \begin{figure}[htbp]
    \centerline{\includegraphics[scale=0.5]{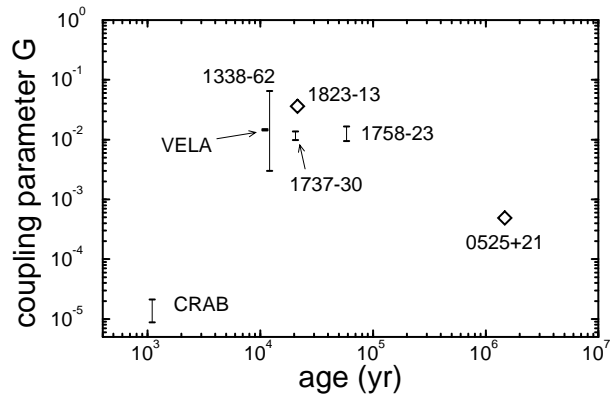}}
    \caption{Coupling parameter $\cal G$ for selected glitching
    pulsars. From~\cite{link-99}.}
    \label{fig.sect.obs.glitch.activity}
\end{figure}}

Microscopic calculations~\cite{chamel-05, chamel-06} show that the ratio $I^{\mathrm{f}}/I^{\mathrm{ff}}$ is 
smaller than unity~\cite{chamelcarter-06} (assuming that only
neutron superfluid in the crust participates in the glitch
phenomenon). We, thus, have
\begin{equation}
  \frac{I^{\mathrm{f}}}{I} \geq\frac{(I^{\mathrm{f}})^2}{I I^{\mathrm{ff}}} \geq 0  \, .
\end{equation} 
Adopting this upper bound and substituting in 
Equation~(\ref{eq.sect.obs.glitch.constraint}), Link et al.\cite{link-99} derived the 
following constraint on the mass $M$  and radius $R$ of the Vela pulsar
\begin{equation}
\label{eq.sect.obs.glitch.constraint.vela}
R\geq 3.6+3.9\frac{M}{M_\odot} \mathrm{\ km}\, .
\end{equation}
Let us emphasize that this constraint has been obtained by neglecting
entrainment effects between neutron superfluid and crust. However, these effects
 are known to be very strong~\cite{chamel-05}, so that the 
inequality~(\ref{eq.sect.obs.glitch.constraint.vela}) should be taken with a grain of salt.

\subsection{Gravitational wave asteroseismology}
\label{sect.obs.gw}

The development of gravitational wave detectors like LIGO~\cite{ligo},
VIRGO~\cite{virgo}, TAMA300~\cite{tama} and GEO600~\cite{geo} is opening up
a new window of astronomical observations. With a central density on
the order of $\sim 10^{15}~\mdens$, neutron stars are among the most
compact objects in the universe, and could be efficient sources 
of gravitational waves. The existence of such waves
predicted by Einstein's theory of general relativity was beautifully
confirmed by the observations of the binary pulsar PSR~1913+16 by
Russel Hulse and Joseph Taylor (who were awarded the Nobel Prize in
1993~\cite{hulse-taylor-nobel}). To compare 
the importance of general relativistic effects in binary pulsars with
those around ordinary stars like the Sun, let us remember that the
pulsar's periastron in PSR~1913+16 advances every day by the
same amount as Mercury's perihelion advances in a century!

\subsubsection{Mountains on neutron stars}

A spherical neutron star does not radiate gravitational waves, in
accordance with Birkhoff's theorem. This is still true for an axially-symmetric neutron star. However, a neutron star with nonaxial deformations,
rigidly rotating with the angular frequency $\Omega$, radiates 
gravitational waves and thus loses energy at a rate given by the formula
\begin{equation}
\dot{E}_{\mathrm{GR}}=-\frac{32}{5}\frac{G}{c^5}I^2\epsilon^2\Omega^6 \, ,
\end{equation}
where $\epsilon$ is a dimensionless parameter characterizing 
deformations of the star. Pulsar timing data can be used to
derive an upper bound on this parameter $\epsilon$. Since the energy
radiated away in gravitational waves can be at most equal to the loss
of kinetic energy due to the spinning down of the pulsar, this implies
\begin{equation}
\label{equation:triaxiality_spindown}
\epsilon < 3\times 10^{-9}\left(\frac{P}{1 \mathrm{\ ms}}\right)^{3/2}\left(\frac{\dot{P}}{10^{-19}}\right)^{1/2} \left(\frac{I}{10^{45} \mathrm{\ g\ cm}^2}\right)^{-1/2} \, ,
\end{equation}
where $P$, $\dot{P}$ and $I$ are, respectively, the pulsar's period,
period derivative and moment of inertia.

The parameter $\epsilon$ can be constrained, independent of the
pulsar timing data, by \emph{direct} observations with gravitational
wave detectors~\cite{jaranowski-98}
\begin{equation}
\label{equation:triaxiality_gw}
\epsilon < 0.237 \frac{h}{10^{-24}}\frac{d}{1 \mathrm{\ kpc}}\left(\frac{f}{1 \mathrm{\ Hz}}\right)^{-2}\left(\frac{I}{10^{45} \mathrm{\ g\ cm}^2}\right)^{-1} \, ,
\end{equation}
where $h$ is the amplitude of a gravitational wave signal, $f$ the
neutron star's spin frequency and $d$ is the distance to the source. 
78 radio pulsars have recently been observed using the LIGO
detector~\cite{ligo-07}. The analysis of the data imposes upper limits
on the maximum amplitude $h$ of the gravitational waves emitted by
these pulsars. The two inequalities
(\ref{equation:triaxiality_spindown}) and
(\ref{equation:triaxiality_gw}) can be combined to put constraints on
the moment of inertia $I$ and on the parameter $\epsilon$ of a given
pulsar. The example of the Crab pulsar~\cite{ligo-08} is shown in
Figure~\ref{fig.sect.obs.gw.crab}. Note that the constraint
from the gravitational wave data has already reached the level of the
spin-down limit.

\epubtkImage{CrabIeplane.png}{%
  \begin{figure}[htbp]
    \centerline{\includegraphics[scale=0.5,angle=0]{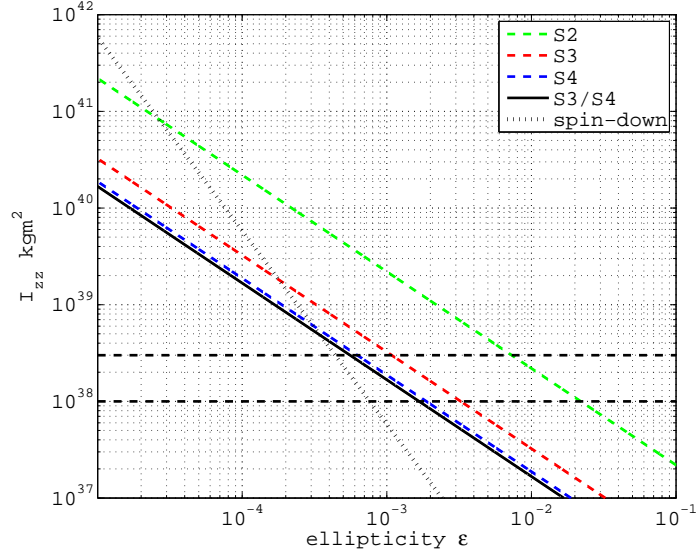}}
    \caption{The moment of inertia $I$ vs.\ the deformation parameter
    $\epsilon$ for the Crab pulsar over the S5 run of the LIGO
    detector. The areas to the right of the diagonal lines are the
    experimentally excluded regions. The horizontal lines represent
    theoretical upper and lower limits on the moment of inertia. The
    lines labelled ``uniform prior'' and ``restricted prior'' are the
    results obtained respectively without and with prior information
    on the gravitational wave signal parameters. From~\cite{ligo-08}.}
    \label{fig.sect.obs.gw.crab}
\end{figure}}

The crust of a neutron star contains only a very small percentage of the mass
of the star. Nevertheless, its elastic response to centrifugal,
magnetic and tidal forces determines the overall shape of the
star. The presence of mountains on the surface of the star leads to a
nonvanishing value of the parameter $\epsilon$. If the star is
rotating around one of the principal axes of the inertia tensor, 
$\epsilon$ is given by~\cite{shapiro-83}
\begin{equation}
\epsilon=\frac{I_{1}-I_{2}}{I}\, ,
\end{equation}
where $I_1$ and $I_2$ are the moments of inertia of the star with
respect to the principal axes orthogonal to the rotation axis. The
topography of the surface of a neutron star depends on the elastic
properties of the solid crust. In particular, the size of the highest
mountain depends on the breaking strain $\sigma_{\max}$, beyond
which the crust will crack. This parameter is believed to lie in the
range $10^{-5}\mbox{\,--\,}10^{-2}$, as argued by
Smoluchowski~\cite{smoluchowski-70}. Considering small perturbations in 
a Newtonian star composed of an incompressible liquid of density $\rho$,
surrounded by a thin crust with a constant shear modulus $\mu$, the
deformation parameter $\epsilon$ is given by the
formula~\cite{haskell-06}
\begin{equation}
\epsilon = \frac{9}{80}\frac{\mu\sigma_{\max} }{G \rho^2} \frac{\Delta R}{R^3},
\end{equation}
where $\Delta R$ the thickness of the solid crust. Adopting the value
$\mu=10^{30} \mathrm{\ g\ cm}^{-1}\mathrm{\ s}^{-2}$ (see
Section~\ref{sect.elast.isotr.solid}), $\epsilon$ can be written as
\begin{equation}
\label{equation:triaxility_crust}
\epsilon \simeq 5\times 10^{-5}
\left(\frac{\sigma_{\max}}{10^{-3}}\right)
\left(\frac{1.4\,M_\odot}{M}\right)^2 \left(\frac{R}{10 \mathrm{\ km}}\right)^3 \left(\frac{\Delta R}{1 \mathrm{\ km}}\right) \, .
\end{equation}
The size of the highest mountain of order $\sim \epsilon R$ can
be roughly estimated from the
inequality~(\ref{equation:triaxiality_spindown}). For instance, the
mountains on the surface of the Crab pulsar could be as high as a few
meters, while in PSR~1957+20 they cannot exceed
a few microns! A neutron star can also be deformed by its
 magnetic field (Section~\ref{sect.structure.B}, see also~\cite{haskell-07} 
and references therein). The deformation
depends on the configuration of the magnetic field and scales like
\begin{equation}
\epsilon  \propto \frac{B^2 R^4}{G M^2} \, .
\end{equation}
For a Newtonian star composed of an incompressible liquid, the deformation induced
by an internal magnetic field is approximately given by~\cite{haskell-07}
\begin{equation}
\label{equation:triaxility_B}
\epsilon \simeq 10^{-12} \left(\frac{R}{10 \mathrm{\ km}}\right)^4\left(\frac{M}{1.4\,M_\odot}\right)^{-2}\left(\frac{\widetilde{B}}{10^{12} \mathrm{\ G}}\right)^2\, ,
\end{equation}
where $\widetilde{B}$ is a suitably-averaged magnetic field. Comparing
Equations~(\ref{equation:triaxility_crust})
and~(\ref{equation:triaxility_B}) we conclude that for a canonical
neutron star ($M=1.4\,M_\odot$, $R=10$~km, $\Delta R=1$~km,
$\widetilde{B}=10^{12}$ G) the deformations induced by the magnetic
field are much smaller than those supportable by the elasticity of the
crust. However, for $\widetilde{B}\sim 10^{15}$~G, characteristic of
magnetars, both deformations have comparable magnitudes. Moreover, it
has been shown that neutron stars with large toroidal fields evolve into
configurations, where the angular momentum of the star is orthogonal to
the magnetic axis. Such configurations are associated with strong
gravitational wave emission (see~\cite{cutler-02} and references
therein).

\subsubsection{Oscillations and precession}

The presence of magnetic fields inside neutron stars or mountains on
the surface are not the only mechanisms for the emission of
gravitational waves. Time-dependent nonaxisymmetric deformations can
also be caused by oscillations. For instance, a neutron star with a
solid crust, rotating about some axis (i.e.\ not aligned with
any principal axis of the stellar moment of inertia tensor) will
precess. For a small wobble angle $\theta$, the deformation parameter is
given by~\cite{shapiro-83}
\begin{equation}
\epsilon=\frac{\theta}{4}\frac{I_{1}-I_{3}}{I} \, ,
\end{equation}
where $I_1$, $I_2$ and $I_3$ are the principal moments of inertia
(assuming $I_1=I_2$), $\theta$ is the angle between the direction of
the angular momentum and the stellar symmetry axis. Observational
evidence of long-period precessions have been reported in
PSR~1828$-$11~\cite{stairs-00}, PSR~B1642$-$03~\cite{shabanova-01} and
RXJ~0720.4$-$3125~\cite{haberl-06}.

A large number of different nonaxisymmetric neutron-star--oscillation
modes exists, for instance, in the liquid surface layers (``ocean''), in the
solid crust, in the liquid core and at the interfaces between the
different regions of the star. These oscillations can be excited by
thermonuclear explosions induced by the accretion of matter from a
companion star, by starquakes, by dynamic instabilities growing on a
timescale on the order of the oscillation period or by secular
instabilities driven by dissipative processes and growing on a much
longer timescale. Oscillation modes are also expected to be excited
during the formation of the neutron star in a supernova explosion. The
nature of these modes, their frequency, their growing and damping
timescales depend on the structure and composition of the star
(for a review, see, for
instance,~\cite{mcdermott-88, kostas-99, andersson-03}).

\subsubsection{Crust-core boundary and r-mode instability}

Of particular astrophysical interest are the inertial modes
or Rossby waves (simply referred to as r-modes) in 
neutron star cores. They can be made unstable by the radiation of
gravitational waves on short timescales of a few seconds in the most
rapidly-rotating neutron stars~\cite{andersson-01}. However, the growth
of these modes can be damped. One of the main damping
mechanisms is the formation of a viscous boundary Ekman layer at the
crust-core interface~\cite{bildsten-00} (see
also~\cite{glampedakis-06b} and references therein). It has been
argued that the heat dissipated in this way could even melt the
crust~\cite{lindblom-00}. The damping rate depends crucially on the
structure of bottom layers of the crust and scales like $(\delta
v/v)^2$, where $\delta v$ is the slippage velocity at the crust-core
interface~\cite{levin-01}. Let us suppose that the liquid in the core
does not penetrate inside the crust, like a liquid inside a
bucket. The slippage velocity in this case is very large $\delta v
\sim v$ and as a consequence the r-modes are strongly damped. However,
these assumptions are not realistic. First of all, the crust is not
perfectly rigid, as discussed in Section~\ref{sect.elast}. On the
contrary, the crust is quite ``soft'' to shear deformations because
$\mu/P\sim 10^{-2}$, where $\mu$ is the shear modulus and $P$ the
pressure. The oscillation modes of the liquid core are coupled to the
elastic modes in the crust, which results in much smaller damping
rates~\cite{levin-01, glampedakis-06a}. Besides, the transition between
the crust and the core might be quite smooth. Indeed, neutron
superfluid in the core permeates the inner crust and
the denser layers of the crust could be formed of nuclear ``pastas''
with elastic properties similar to those of liquid crystals (see
Section~\ref{sect.elast.pasta}). The slippage velocity at the bottom
of the crust could, therefore, be very small $\delta v \ll
v$. Consequently, the Ekman damping rate of the r-modes could be much
weaker than the available estimates. If the crust were purely
fluid, the damping rate would be vanishingly small. However the
presence of the magnetic field would also affect the damping time
scale and should be taken into
account~\cite{mendell-01, kinney-03}. Besides the character of the core
 oscillation modes is likely to be affected by coupling with
the crust. The role of the crust in the dissipation of the r-mode
instability is, thus, far from being fully understood. 
Finally, let us mention that  by far the strongest  damping
mechanism of r-modes, due to a huge  bulk viscosity,  may be located in
the inner neutron star core, provided it contains hyperons
(see, e.g., \cite{haensel-02,lindblom-02} and
references therein).

\subsection{Giant flares from Soft Gamma Repeaters}
\label{sect.obs.sgr}

The discovery of quasi-periodic oscillations (QPO) in  X-ray flux
following giant flares from Soft Gamma Repeaters (SGR) has recently
triggered a burst of intense theoretical research. Oscillations were
 detected at frequencies 18, 26, 29, 92.5, 150, 626.5 and 1837~Hz
during the spectacular  December 27, 2004 giant flare (the most intense
ever observed in our Galaxy) from SGR~1806$-$20~\cite{israel-05, watts-06, strohmayerw-06} 
and at 28, 54, 84 and
155~Hz during the August 27, 1998 giant flare from SGR~1900+14~\cite{strohmayerw-05}. 
Evidence has also been reported for
oscillations at 43.5~Hz during the March 5, 1979 event in SGR~0526$-$66~\cite{barat-83}. 
Soft Gamma Repeaters (SGR) are believed to be
strongly-magnetized neutron stars or magnetars endowed with magnetic
fields as high as $10^{14}\mbox{\,--\,}10^{15}$~G (for a recent review, see, for
instance, \cite{woods-06} and references therein; also see the home
page of Robert C.\ Duncan~\cite{duncan}). Giant flares are
interpreted as crustquakes induced by magnetic stresses. Such
catastrophic events are likely to be accompanied by global seismic
vibrations, as observed by terrestrial seismologists  after large
earthquakes. Among the large variety of possible oscillation modes,
torsional shear modes in the crust are the most likely~\cite{duncan-98}. 
Shear flow in the crust is illustrated
in Figure~\ref{fig.sect.cond.vis.shear}. If 
confirmed, this would be the first \emph{direct} detection of
oscillations in a neutron star crust.

Neglecting the effects of rotation and magnetic fields, and ignoring the
presence of neutron superfluid in the inner crust, but taking into
account the elasticity of the crust in general relativity, the
frequency of the fundamental toroidal crustal mode of multipolarity 
$\ell=2$ (the case $\ell = 1$ corresponding to the crust uniformly rotating 
around the static core is ignored), is approximately given
by~\cite{samuelsson-07}
\begin{equation}
\label{equation:tor_modes0}
f_{n=0,\ell=2} \simeq 2\pi v_t \sqrt{\frac{1-r_{\mathrm{g}}/R}{R r_{\mathrm{cc}}}}, \hskip 1cm f_{n=0,\ell}\simeq f_{n=0,\ell=2} \sqrt{(\ell-1)(\ell+2)},
\end{equation}
where $r_{\mathrm{g}}$ is the Schwarzschild radius of the star, 
$r_{\mathrm{cc}}$ the radius of the crust and $v_t$ is the speed of shear waves propagating in an angular
direction and polarized in the mutually orthogonal angular
direction. The frequencies of the higher fundamental ($n=0$) modes
scale like
\begin{equation}
\label{equation:tor_modes0l}
f_{n=0,\ell}\simeq f_{n=0,\ell=2} \sqrt{(\ell-1)(\ell+2)}\, .
\end{equation}
The frequencies of 
overtones $n>0$ are independent of $\ell$ to a good approximation
(provided $\ell$ is not too large compared to $n$), and can be roughly
estimated as
\begin{equation}
\label{equation:tor_modesn}
f_{n>0} \simeq (1-\frac{r_{\mathrm{g}}}{R})\frac{2\pi^2 n v_r}{\Delta R} \, ,
\end{equation}
where $v_r$ is the speed of shear waves propagating radially with
polarization in an angular direction. For a reasonable crustal
equation of state, the crust thickness $\Delta R\equiv R-r_{\mathrm{cc}}$
can be estimated as~\cite{samuelsson-07}
\begin{equation}
\frac{\Delta R}{R} \simeq \biggl[1+21.5 \frac{r_{\mathrm{g}}}{R} (1-\frac{r_{\mathrm{g}}}{R})\biggr]^{-1} \, .
\end{equation}
If the crust is isotropic, the velocities $v_r$ and $v_t$ are equal.

The analysis of QPOs in SGRs can potentially provide valuable
information on the properties of the crust and, more generally, on the
structure of neutron stars. The identification of both the 29~Hz
and 626.5~Hz QPOs in the 2004 giant flare from SGR~1806$-$20 as the
fundamental $n=0$, $\ell=2$ toroidal mode and the first overtone
$n=1$, $\ell=1$, respectively, puts stringent constraints on the mass
and radius of the star, as shown in
Figure~\ref{fig.sect.obs.sgr.MRsgr1806}. This constraint rules out some stiff
equations of state based on the relativistic mean field theory
proposed by Glendenning~\cite{glendenning-00}. The 28~Hz and 54~Hz
QPOs in the 1998 flare from SGR~1900+14 have been identified with the
$\ell=2$ and $\ell=4$ toroidal modes, respectively. In this case, the
mass and radius are much less constrained, as can be seen in
Figure~\ref{fig.sect.obs.sgr.MRsgr1900}. The identification of 
higher frequency QPOs is more controversial.

\epubtkImage{xg1806.png}{%
\begin{figure}[htbp]
  \centerline{\includegraphics[scale=0.5,angle=0]{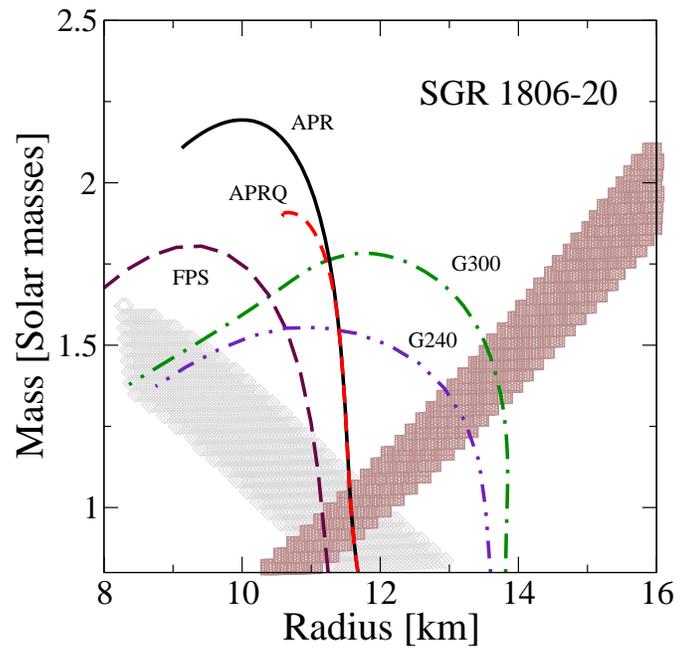}}
  \caption{Constraints on the mass and radius of SGR~1806$-$20
  obtained from the seismic analysis of quasi-periodic
  oscillations in X-ray emission during the December~27, 2004
  giant flare. For comparison, the mass-radius relation for several
  equations of state is shown (see~\cite{samuelsson-07} for further
  details).}
  \label{fig.sect.obs.sgr.MRsgr1806}
\end{figure}}

The effect of rotation on oscillation modes is to split the
frequency of each mode with a given $\ell$ into $2\ell + 1$
frequencies. It has recently been pointed out that some of the
resulting modes might, thus, become secularly unstable, according to the
Chandrasekhar--Friedman--Schutz (CFS)
criterion~\cite{vavoulidis-07}. The study of oscillation modes becomes
even more difficult in the presence of a magnetic field. Roughly
speaking, the effects of the magnetic field increase the mode
frequencies~\cite{duncan-98, messios-01, piro-05b, lee-07, sotani-07}.
Simple Newtonian estimates lead to an increase of the frequencies by a
factor $\sqrt{1+(B/B_\mu)^2}$, where $B_\mu \equiv \sqrt{4\pi\mu}$ is
expressed in terms of the shear modulus
$\mu$~\cite{duncan-98}. Sotani and et al.~\cite{sotani-07} recently
carried out calculations in general relativity with a dipole magnetic
field and found numerically that the frequencies are increased by a
factor $\sqrt{1+\alpha_{n,\ell}(B/B_\mu)^2}$, where $\alpha_{n,\ell}$
is a numerical coefficient. However, it has been
emphasized by Messios et al.~\cite{messios-01} that the
effects of the magnetic field strongly depend on its
configuration. The most important effect is to couple the crust to the core so
that the whole stellar interior vibrates
during a giant flare~\cite{glampedakis-07}. Low frequency QPOs 
could, thus, be associated with
magnetohydrodynamic (MHD) modes in the core~\cite{levin-07}.

\epubtkImage{xg1900.png}{%
  \begin{figure}[htbp]
    \centerline{\includegraphics[scale=0.5,angle=0]{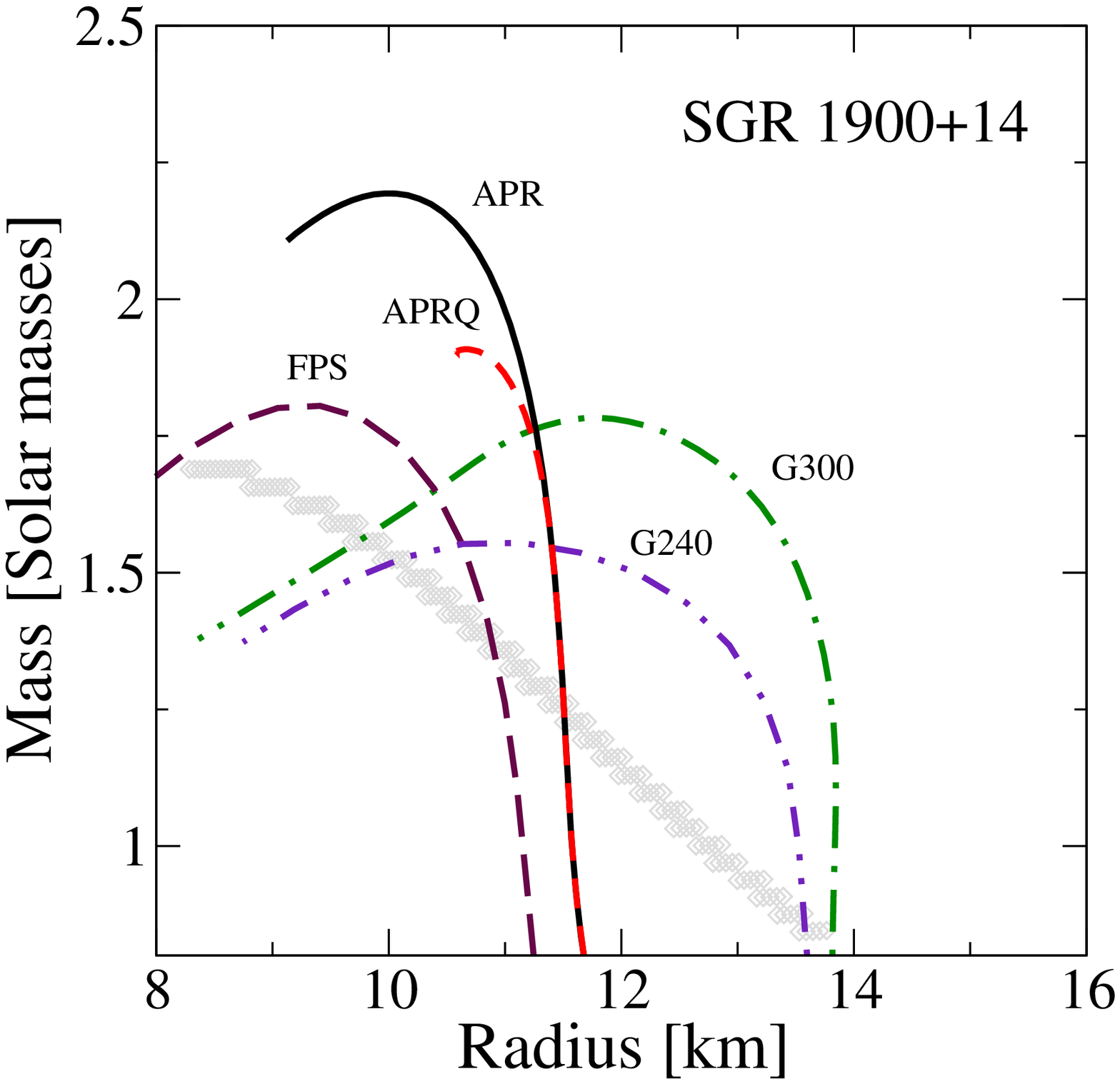}}
    \caption{Constraints on the mass and radius of SGR~1900+14
    obtained from the seismic analysis of the quasi-periodic
    oscillations in the X-ray emission during the August~27,~1998
    giant flare. For comparison, the mass-radius relation for several
    equations of state is shown (see~\cite{samuelsson-07} for further
    details). From~\cite{samuelsson-07}.}
    \label{fig.sect.obs.sgr.MRsgr1900}
\end{figure}}

Another important aspect to be addressed is the
presence of neutron superfluid, which permeates the inner
crust. The formalism for treating a superfluid in a magneto-elastic
medium has been recently developed both in general
relativity~\cite{cartersamuelsson-06} and in the Newtonian
limit~\cite{carter-06, carterchachoua-06}, based on a variational
principle. This formalism has not yet been applied to study
oscillation modes in magnetars. However, we can anticipate 
the effects of the neutron superfluid using the  
two-fluid description of the crust reviewed in Section~\ref{sect.hydro.twofluid}. Following the same arguments as for two-fluid models of neutron star
cores~\cite{andersson-01b}, two classes of oscillations can be 
expected to exist in the inner crust, depending on
whether neutron superfluid is co-moving or
countermoving with the crust. The countermoving modes are predicted to
be very sensitive to entrainment effects, which are very strong in
the crust~\cite{chamel-05, chamel-06}. 

The neutron-star--oscillation problem deserves further theoretical study. The
prospect of probing neutron star crusts by analyzing the X-ray emission 
of giant magnetar flares is very promising.

\subsection{Low mass X-ray binaries}
\label{sect.obs.LMXB}

As reviewed in Section~\ref{sect.accretion}, the accretion of matter
(mainly hydrogen and helium) onto the surface of neutron stars triggers
thermonuclear fusion reactions. Under certain circumstances, these
reactions can become explosive, giving rise to X-ray bursts. The
unstable burning of helium ash produced by the fusion of accreted
hydrogen, is thought to be at the origin of type I X-ray bursts. A new
type of X-ray burst has been recently discovered. These \emph{superbursts} are a thousand times more energetic than normal
bursts and last several hours compared to a few tens of seconds, but occur
much more rarely. These superbursts could be due to the unstable burning
of $^{12}$C accumulated from He burning. The mass of
$^{12}$C fuel has to be as high as $10^{-9}\,M_\odot$ to get
$E_{\mathrm{burst}}\sim 10^{42}$~erg. It can be seen from
Figure~\ref{fig.sect.structure.Mcr-z-GSC-ACC} that $^{12}$C
ignition has to occur at $\rho \sim10^9~\mdens$ at a depth of $\sim
30$~m. At the accretion rates characteristic of typical superbursters
($\dot{M} \sim (1\mbox{\,--\,}3)\times 10^{-9}\,M_\odot/\mathrm{y}$), this
would correspond to recurrence times of a few years. It also seems
that crustal heating might be quite important to getting such a relatively
low ignition density of $^{12}$C~\cite{gupta-07}. The ignition
conditions are very sensitive to the thermal properties of the crust
and core~\cite{cumming-06}. X-ray observations of low-mass X-ray
binaries thus provide another way of probing the interior of neutron
stars, both during thermonuclear bursting episodes and during periods
of quiescence as discussed below.

\subsubsection{Burst oscillations}
\label{sect.obs.LMXB.oscillations}

During the past ten years, millisecond oscillations have been
discovered in X-ray bursts in low mass X-ray
binaries as illustrated in
Figure~\ref{fig.sect.obs.burst.oscillations} (see, for
instance,~\cite{strohmayer-06b} and references therein for a recent
review). Such oscillations have been observed during the rise time of
bursts, as well as at later times in the decay phase. The
observations of burst oscillations in the accreting millisecond
pulsars SAX~J1808.4$-$3658~\cite{zand-01, markwardt-02, chakrabarty-03}
and XTE~J1814$-$338~\cite{strohmayer-03} firmly established that burst
oscillations occur close to the spin frequency of the neutron
star. This conclusion was further supported by the discovery of about
500,000 highly coherent pulsation cycles at 582~Hz during a superburst
from the low mass X-ray binary 4U~1636$-$356~\cite{strohmayer-02}. This
suggests that burst oscillations arise from some nonuniformities on the
neutron star surface. In the burst rise, the oscillations are
likely to be caused by the presence of hot spots induced by the
ignition of nuclear burning. This interpretation naturally explains
why oscillation amplitudes decrease with time as the burning
region spreads over the entire surface~\cite{strohmayer-97}, as shown
in Figure~\ref{fig.sect.obs.burst.oscillations.spreading}. Nevertheless,
this model cannot explain the oscillations detected in the burst tail,
since the duration of the burst (of the order 10\,--\,30~seconds) is much
larger than the spreading time of thermonuclear burning (typically
less than 1~second). Surface inhomogeneities during the cooling phase could be
produced by the dynamic formation of vortices driven by the Coriolis
force~\cite{spitkovsky-02} and by nonradial surface
oscillations~\cite{heyl-04}. The outer envelope of an accreting
neutron star is formed of three distinct regions: a hot bursting
shell, an ocean and the solid outer crust, so that many different
oscillation modes could be excited during a burst. The observed
frequencies and positive frequency drifts are consistent with shallow
surface waves excited in the hot bursting layer changing into crustal
interface waves in the ocean as the surface cools~\cite{piro-05}. This
model can also explain the energy dependence of the burst oscillation
amplitude~\cite{piro-06}. The interface waves resemble shallow surface
waves, but with a large radial displacement at the ocean/crust
boundary due to the elasticity of the crust~\cite{piro-05a}. The
frequency of the interface wave is reduced by a factor
$\sqrt{\mu/P}\sim 0.1$, where $\mu$ is the shear modulus and $P$ the
pressure, as compared to a rigid surface. The frequencies of these
modes depend on the composition of the neutron star surface
layers. This raises the exciting possibility of probing accreting
neutron star crusts with X-ray burst oscillations. Very recently
evidence has come forth for burst oscillations at a frequency of 1122~Hz
in the X-ray transient XTE~J1739$-$285~\cite{kaaret-07}. If 
confirmed, it would imply that this system contains the fastest
spinning neutron star ever discovered. Since the spinning rate
is limited by the mass shedding limit, these  observations would thus
put constraints on the gravitational mass $M$ and circumferential
equatorial radius $R$ of the neutron star in XTE~J1739$-$285~\cite{bejger-07}
\begin{equation}
R< 15.52 \left(\frac{M}{1.4\,M_\odot}\right)^{1/3} \mathrm{\ km} \, .
\end{equation}

\epubtkImage{burst_oscillations.png}{%
\begin{figure}[htbp]
  \centerline{\includegraphics[scale=0.6,angle=0]{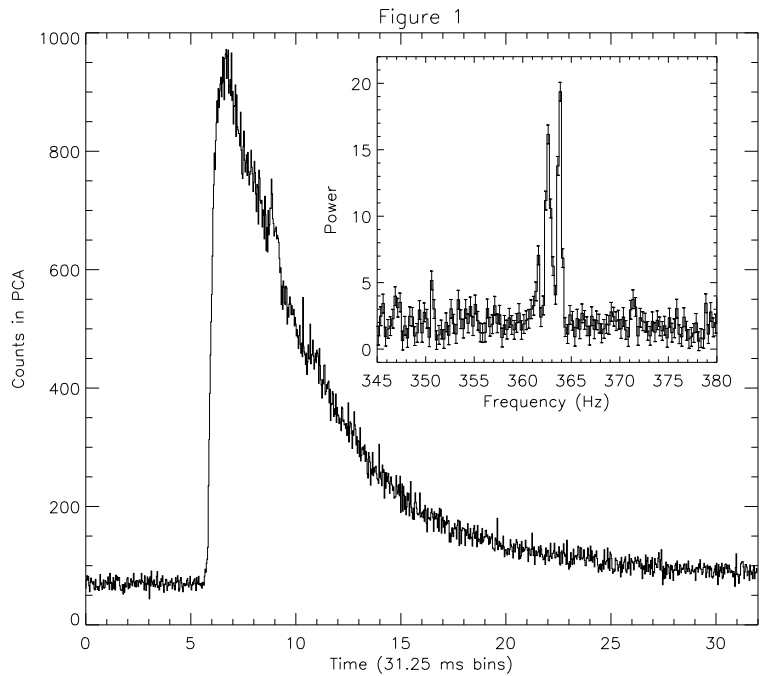}}
  \caption{Oscillations detected in an X-ray burst from 4U~1728$-$34
  at frequency $\sim$~363~Hz~\cite{strohmayer-96}.}
  \label{fig.sect.obs.burst.oscillations}
\end{figure}}

\epubtkImage{burning.png}{%
\begin{figure}[htbp]
  \centerline{\includegraphics[scale=0.5,angle=0]{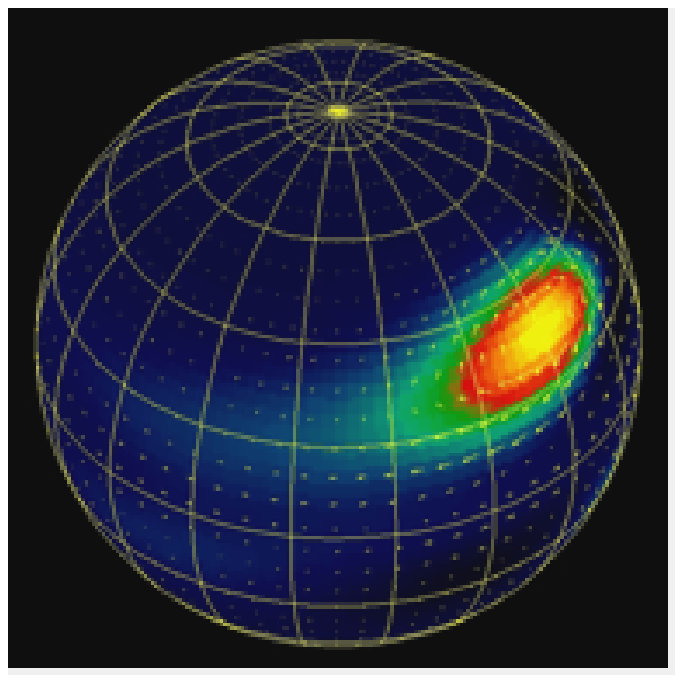}\includegraphics[scale=0.5,angle=0]{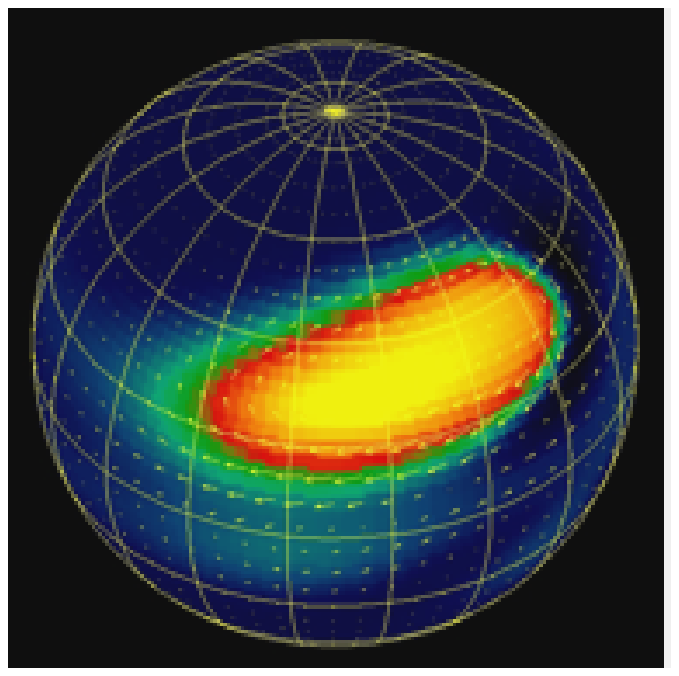}\includegraphics[scale=0.5,angle=0]{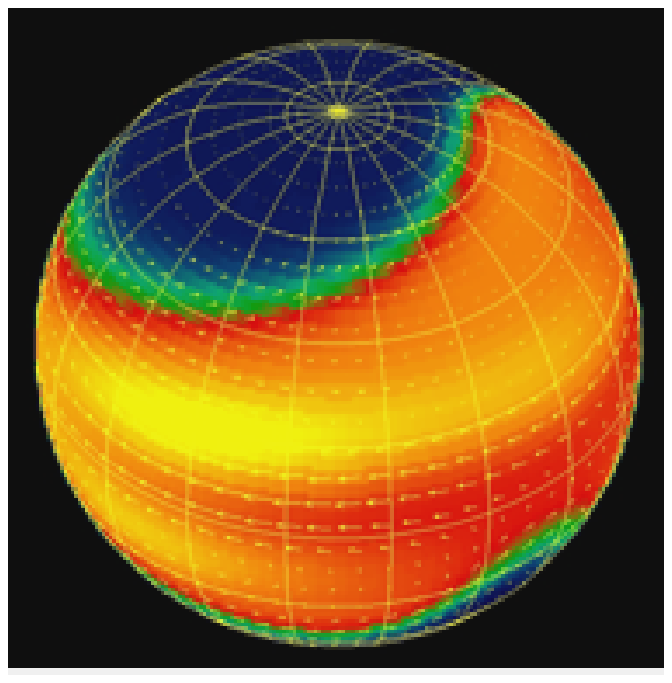}}
  \caption{Spreading of a thermonuclear burning hotspot on the
  surface of a rotating neutron star simulated by
  Spitkovsky~\cite{spitkovsky-02} (from
  \url{http://www.astro.princeton.edu/~anatoly/}).}
  \label{fig.sect.obs.burst.oscillations.spreading}
\end{figure}}

\subsubsection{Soft X-ray transients in quiescence}
\label{sect.obs.LMXB.crust-heat-SXT}

The phenomenon of deep crustal heating appears  to be relevant
for the understanding of the thermal radiation observed in
soft X-ray transients (SXTs) in quiescence, when the accretion
from a disk is switched off or strongly suppressed. Typically,
the quiescent emission is much higher than it would be
in an old cooling neutron star. It has been suggested that
this is because the interiors of neutron stars in
SXTs are heated up during relatively short periods of
accretion and bursting by the nonequilibrium processes
associated with nuclear reactions taking place in the deep
layers of the crust~(\cite{brown-98}, see also
Section~\ref{sect.accretion.crust-heating}). The deep crustal
heating model, combined with appropriate models of the neutron star
atmosphere and interior, is used to explain measured
luminosities of SXTs in quiescence. The luminosity in
quiescence depends on the structure of neutron star cores, and
particularly on the rate of neutrino cooling. This opens up the
new possibility of exploring the internal structure and
equation of state of neutron stars (see~\cite{colpi-01, rutledge-02,
  yakovlev-03, yakovlev-04b} and references therein).

Let us denote the duration of the accretion stage, with
accretion rate $\dot{M}_{\mathrm{a}}$, by $t_{\mathrm{a}}$,
 and the duration  of quiescence
between two active periods by $t_{\mathrm{q}}$, with $t_{\mathrm{q}}\gg
t_{\mathrm{a}}$. After a few thousands of accretion-quiescence
cycles, an SXT reaches a steady thermal state with the well-defined
thermal structure of quiescence. This thermal structure  is fully
determined by the time-averaged accretion rate $\langle \dot{M}\rangle
=t_{\mathrm{a}}\dot{M}_{\mathrm{a}}/(t_{\mathrm{a}}+t_{\mathrm{q}})$.
A steady state in quiescence satisfies the global energy balance ``on
average''. The heat associated with nuclear H and He burning and X-ray
bursting during $t_{\mathrm{a}}$ is nearly completely radiated away,
and therefore does not contribute to the steady-state energy balance.
Therefore, to a good approximation, the sum of the total average
cooling rates (photon surface  and neutrino volume emission) is
balanced by deep crustal heating during an accretion period. Except
for a thin blanketing envelope, the interior of an SXT in
quiescence is isothermal, with temperature $T_{\mathrm{int}}$. A
blanketing envelope separates the isothermal interior from the
surface, where the photons are emitted with a spectrum formed
in a  photosphere of effective temperature $T_{\mathrm{eff}}$. Therefore,
\begin{equation}
  L_\gamma(T_{\mathrm{eff}}) +L_\nu(T_{\mathrm{int}})=L_{\mathrm{dh}}(\langle
  \dot{M}\rangle) \, ,
  \label{eq:accrete-SXT-thermal-balance}
\end{equation}
where the total time-averaged deep-crustal heating rate is
\begin{equation}
  L_{\mathrm{dh}}(\langle \dot{M}\rangle)=Q_{\mathrm{tot}}\; \langle
  \dot{M}\rangle/m_u\approx 6.03\times 10^{33}\; \langle
  \dot{M}_{-10}\rangle {Q_{\mathrm{tot}} \over \mathrm{MeV}} \mathrm{\ erg\ s}^{-1}
  \, ,
  \label{eq:accrete-L_dh}
\end{equation}
and $Q_{\mathrm{tot}}$ is the total heat released per accreted
nucleon.

\epubtkImage{Ksen-1-5.png}{%
  \begin{figure}[htbp]
    \centerline{\includegraphics[scale=0.8]{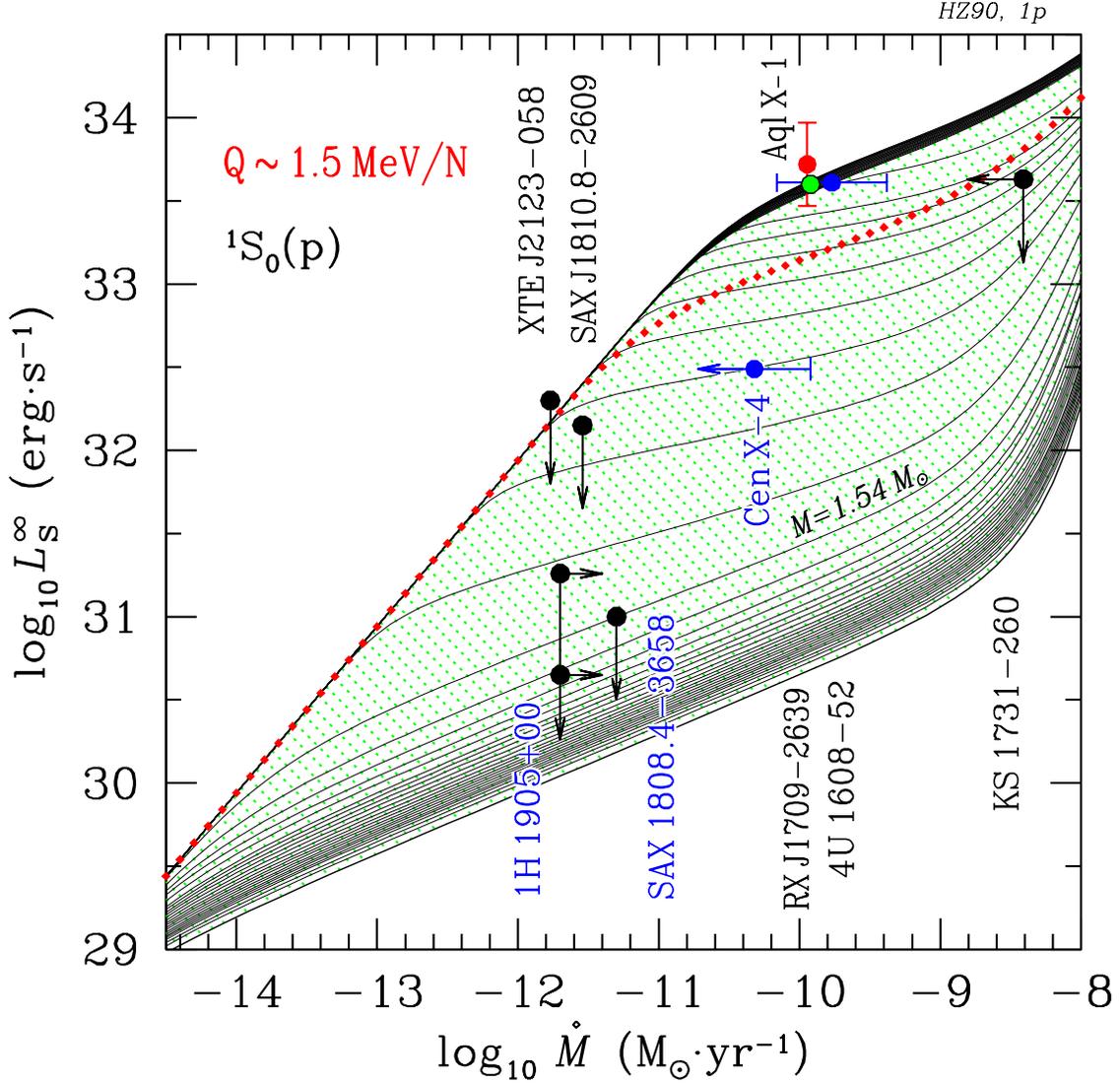}}
    \caption{X-ray luminosities of SXTs in quiescence vs.\ time-averaged accretion rates. The heating curves correspond to
      different neutron star masses, increasing from the top to the
      bottom, with a step of $\Delta M=0.02\,M_\odot$. The EoS of the
      core is moderately stiff, with $M_{\max}=1.977\,M_\odot$~\cite{prakash-88}. The model of a strong
      proton and a weak neutron superfluidity is
      assumed~\cite{levenfish-07}. The highest curve (hottest stars)
      corresponds to $M=M_\odot$, and the lowest one (coldest stars)
      to $M=M_{\max}$. The upper bundle of coalescing curves
      corresponds to masses $M_\odot\le M\le M_{\mathrm{D}}$, where a star of
      mass $M_{\mathrm{D}}$ has a central density equal to the threshold for
      the direct Urca process. The red dotted line represents thermal states
      of a ``basic model'' (nonsuperfluid core, slow cooling via a
      modified Urca process). Total deep heat released per
      accreted nucleon is $Q_{\mathrm{tot}}=1.5$~MeV. Figure made by K.P.\
      Levenfish. For a further description of neutron star models
      and observational data see Levenfish \&
      Haensel~\cite{levenfish-07}.}
    \label{fig.sect.obs.LMXB.Ksen-1-5}
\end{figure}}

\epubtkImage{Ksen-0-15.png}{%
  \begin{figure}[htbp]
    \centerline{\includegraphics[scale=0.4]{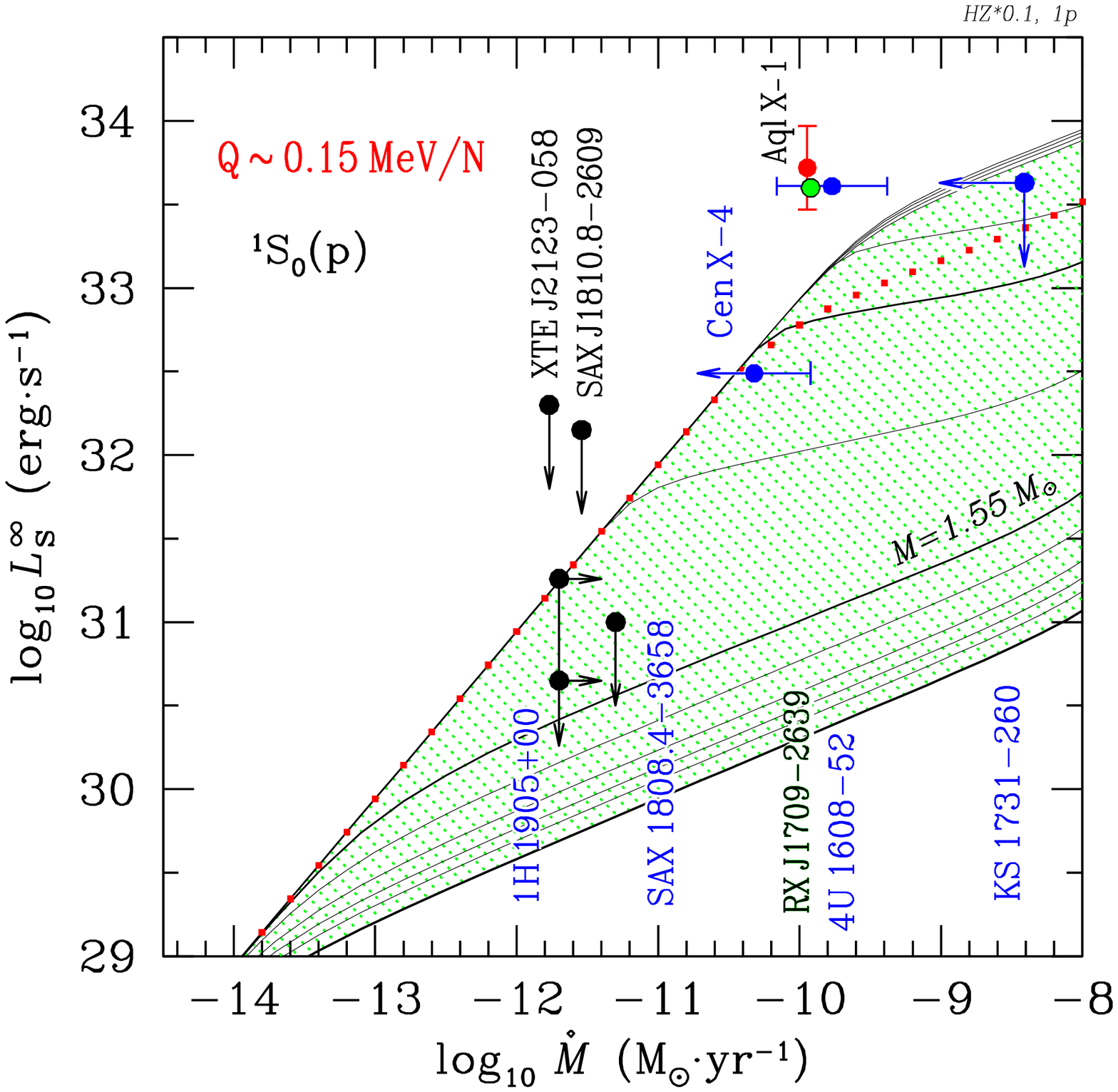}}
    \caption{Same as Figure~\ref{fig.sect.obs.LMXB.Ksen-1-5} but
    assuming ten times smaller $Q_{\mathrm{tot}}=0.15$~MeV. Such a low
    crustal heating is contradicted by the measured luminosities of
    three SXTs: Aql~X$-$1, RXJ1709$-$2639, and 4U~1608$-$52. Figure
    made by K.P.\ Levenfish.}
    \label{fig.sect.obs.LMXB.Ksen-0-15}
\end{figure}}

As can be seen in Figure~\ref{fig.sect.obs.LMXB.Ksen-1-5},
$Q_{\mathrm{tot}}=1.5$~MeV/nucleon is consistent with SXTs
observations. However, different sources require different neutron
star masses $M$. This is because the core neutrino cooling rate
depends on the mass of the inner core, where the direct Urca process
is possible. As shown in Figure~\ref{fig.sect.obs.LMXB.Ksen-0-15},
$Q_{\mathrm{tot}}=0.15$~MeV/nucleon (and {\it a fortiori\/}
$Q_{\mathrm{tot}}=0$) would contradict observations of Aql~X$-$1,
RX~1709$-$2639 and 4U~1608$-$52 in quiescence.

\subsubsection{Initial cooling in quasi-persistent SXTs}
\label{sect.obs.LMXB.initial-cool-SXTs}

Quasi-persistent SXTs, with accretion periods lasting for years --
decades, might be particularly useful for studying the structure of
neutron star crusts. This is because one can observe their 
thermal relaxation between the accreting and quiescent stages. For
standard SXTs, with accretion lasting days -- weeks, such relaxation
cannot be detected, because crustal heating due to accretion is too
small. On the contrary, thermal relaxation toward the quiescent state
for KS~1731$-$260 (after accreting over 12.5~y) and for
MXB~1659$-$29 (after accreting over 2.5~y), called ``initial cooling'',
was observed~\cite{cackett-06}. Let us consider the thermal
relaxation of KS~1731$-$260. After 12.5~y
of accretion  and deep crustal heating, the crust and the surface
became significantly hotter than in the quiescent state.
 The cooling curve depends on crust properties, such as
thermal conductivity (Section~\ref{sect.cond}), thickness
(Section~\ref{sect.structure}), distribution of heat sources
(Section~\ref{sect.accretion}), and neutrino emissivity
(Section~\ref{sect.neutrino}). Some of these properties depend
strongly on the crust structure, as illustrated in
Figure~\ref{fig.sect.obs.LMXB.Shternin-cond}. Modeling of the initial
cooling curve can hopefully constrain the crust
physics~\cite{rutledge-02, cackett-06, shternin-07}. An example of
such modeling is shown in
Figure~\ref{fig.sect.obs.LMXB.KS1731-cool}. The cooling curve is much
more sensitive to the crust physics than to that of the dense
core. For example, an amorphous crust, with its low
thermal conductivity (see Figure~\ref{fig.sect.obs.LMXB.Shternin-cond}),
yields a too slow relaxation  (see
Figure~\ref{fig.sect.obs.LMXB.KS1731-cool} ). Moreover, the star has
to be massive to have a sufficiently thin crust to relax sufficiently
rapidly.

\epubtkImage{Shternin-cond.png}{%
\begin{figure}[htbp]
  \centerline{\includegraphics[scale=0.4]{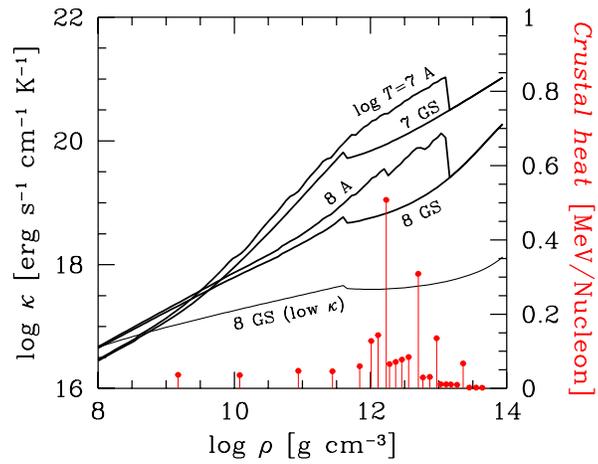}}
  \caption{Left vertical  axis: thermal conductivity of neutron star crust
  vs.\ density, at $\log_{10}(T)[K]=7$  and 8. GS -- pure ground state
  crust. A -- pure accreted crust. GS (low $\kappa$) -- amorphous
  ground state crust. Right vertical axis -- deep crustal heating per
  accreted nucleon in a thin heating shell, vs.\ density, according to
  $A_{\mathrm{i}}=56$ model of Haensel \& Zdunik~\cite{haensel-07}. From
  Shternin et al.~\cite{shternin-07}. Figure made by P.S.\ Shternin.}
\label{fig.sect.obs.LMXB.Shternin-cond}
\end{figure}}

\epubtkImage{KS1731-cool.png}{%
\begin{figure}[htbp]
  \centerline{\includegraphics[scale=1.0]{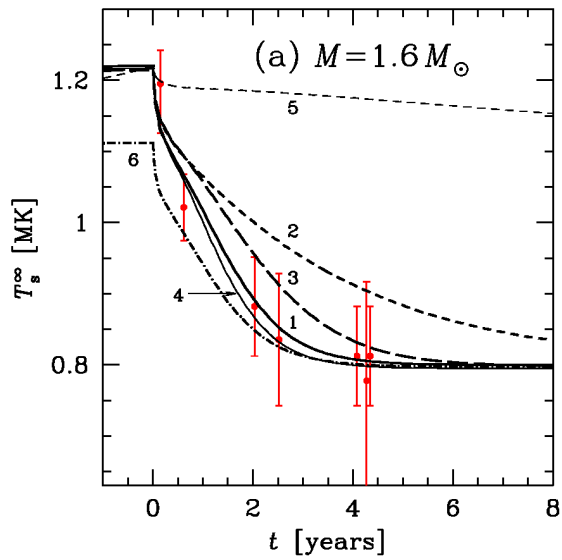}}
  \caption{Theoretical cooling curves for KS~1731$-$260 relaxing
  toward a quiescent state; observations expressed in terms of
  effective surface temperature as measured by a distant observer,
  $T^\infty_{\mathrm{s}}$. Curves 1 and 4 were obtained for pure crystal
  accreted crust with two different models of superfluidity of
  neutrons in the inner crust (strong and weak). Curves 1 and 4 give
  best fit to data points. Line 5 -- amorphous ground state crust; line
  2 -- ground state crust without neutron superfluidity: both are ruled
  out by observations. From Shternin et al.~\cite{shternin-07}. Figure
  made by P.S.\ Shternin.}
\label{fig.sect.obs.LMXB.KS1731-cool}
\end{figure}}

\newpage


\section{Conclusion}
\label{sect.conclusion}

The conditions prevailing inside the crusts of neutron stars are not
so extreme as those encountered in the dense core. Nonetheless, they
are still far beyond those accessible in terrestrial laboratories. The
matter in neutron star crust is subject to very high pressures, as well as
huge magnetic fields, which can attain up to $10^{14}\mbox{\,--\,}10^{15}$~G
in magnetars. For comparison, it is worth reminding ourselves that the strongest
(explosive) magnetic fields ever produced on the Earth reach 
``only'' $3\times10^7$~G~\cite{motokowa-04}. The description of such
environments requires the interplay of many different branches of
physics, from nuclear physics to condensed matter and plasma physics.

Considerable progress in the microscopic modeling of neutron star
crusts has been achieved during the last few years. Yet the structure
and properties of the crust remain difficult to predict,
depending on the formation and subsequent cooling of the star. Even in the ground
state approximation, the structure of the neutron star crust is only
well determined at $\rho\lesssim 10^{11}~\mdens$, for which
experimental data are available. Although all theoretical calculations 
predict the same large-scale picture of the denser layers of the crust, they
do not quantitatively agree, reflecting the uncertainties in the
properties of very exotic nuclei and uniform highly-asymmetric
nuclear matter. The inner crust is expected to be formed of a 
lattice of neutron-rich nuclear clusters coexisting with a degenerate
relativistic electron gas and a neutron liquid. The structure of the
inner crust, its composition and the shape of the clusters are
model dependent, especially in the bottom layers at densities $\sim
10^{14}~\mdens$. However, the structure of the inner crust is crucial
for calculating many properties, such as neutrino emissivities, as
well as transport properties like electric and thermal
conductivities. Superfluidity of unbound neutrons in the crust seems
to be well established, both observationally and
theoretically. However, much remains to be done to understand its
properties in detail and, in particular, the effects of the nuclear
lattice.

The interpretation of many observed neutron star phenomena, like 
 pulsar glitches, X-ray bursts in low-mass X-ray binaries,
initial cooling in soft X-ray transients, or quasi-periodic
oscillations in soft gamma repeaters, can potentially shed light on
the microscopic properties of the crust. However, their description
requires  consistent models of the crust from the
nuclear scale up to the macroscopic scale. In particular,
understanding the evolution of the magnetic field, the thermal
relaxation of the star, the formation of mountains, the occurrence
of starquakes and the propagation of seismic waves, requires the
development of global models combining general relativity, elasticity,
magnetohydrodynamics and superfluid hydrodynamics. Theoretical
modeling of neutron star crusts is very challenging, but not out of
reach. A confrontation of these models with observations
could hopefully help us to unveil the intimate nature of dense matter
at subnuclear densities. The improvement of observational techniques, as
well as the development of gravitational wave astronomy in the
near future, open very exciting perspectives.

We are aware that some aspects of the physics of neutron star crusts
have not been dealt with in this review. Our intention was not to
write an exhaustive monograph, but to give a glimpse of the
great variety of topics that are necessarily addressed by different
communities of scientists. We hope that this review will be useful to
the reader for his/her own research.

\newpage


\section{Acknowledgments}

N.C.\ gratefully acknowledges financial support from a Marie Curie
Intra-European grant (contract number MEIF-CT-2005-024660) and from FNRS (Belgium). 
P.H.\ was partially supported by the Polish MNiSW grant no.\ N20300632/0450.

We express our deep gratitude to D.G.\ Yakovlev for his help in the
preparation of the present review. He read the whole manuscript and,
during many hours of Skype sessions, he went over with one of us (P.H.)
all the sections that required corrections, clarifications
and improvement (they were legion). We thank him for his patience
and expertise shared with us.

We are grateful to A.Y.\ Potekhin, who suggested several improvements 
in the sections referring to the effects of magnetic fields. 
He also suggested appropriate figures and made some new ones, 
which enriched the content of this review. Sometimes a good figure  
is better than hundreds of words.

We also thank Lars Samuelsson for his comments on elasticity and on axial modes
in neutron stars. 

We would like to address our gratitude to our colleagues, who contributed 
to this review through discussions or collaborations. 

\newpage


\begin{appendix}
\section{List of Notations}

\begin{tabular}[t]{lll}

$A$ &--& mass number \\
$\pmb{A}$ &--& electromagnetic potential vector \\
$a_0$ &--& Bohr radius \\
$\pmb{B},B$ &--& magnetic field \\
$c$ &--& speed of light \\
$c_{\mathrm{s}}$ &--& speed of sound \\
$\Delta(\pmb{k})$ &--& wave number dependent pairing gap \\
$\Delta_{\mathrm{F}}$ &--& pairing gap at the Fermi level $\Delta(\pmb{k_{\mathrm{F}}})$ \\
$d_n$ &--& interneutron spacing \\
$d_\upsilon$ &--& intervortex spacing \\
$E\{A,Z\}$ &--& energy of a nucleus of mass number $A$ and charge number $Z$ \\
$e$ &--& proton electric charge \\
$\eta$ &--& shear viscosity \\
$\epsilon$ &--& single particle energy \\
$\epsilon_{\mathrm{F}}$ &--& Fermi energy \\
$\varepsilon$ &--& energy density \\
$\pmb{\cal F}$ &--& force per unit length \\
$\pmb{f}$ &--& force density \\
$f_\mu$ &--& 4-force density \\
$G$ &--& gravitational constant \\
$\gamma$ &--& adiabatic index \\
$\hbar$ &--& Dirac constant \\
$h$ &--& Planck constant \\
$I$ &--& moment of inertia \\
$J$ &--& angular momentum \\
$\pmb{k}$ &--& wave vector \\
$k$ &--& wave number \\
$k_{\mathrm{B}}$ &--& Boltzman constant \\
$k_{\mathrm{F}}$ &--& Fermi wavenumber \\
$\kappa$ &--& thermal conductivity (Section~\ref{sect.cond}) \\
$\pmb{\kappa},\kappa$ &--& circulation of a vortex (Section~\ref{sect.super}) \\
$\Lambda$ &--& Lagrangian density (Section~\ref{sect.hydro})\\
$\mu$ &--& shear modulus \\
$\mu_e$ &--& electron chemical potential \\
$\mu_n$ &--& neutron chemical potential \\
$\mu_p$ &--& proton chemical potential \\
$M$ &--& neutron star gravitational mass \\
$M_\odot$ &--& solar mass \\
$m$ &--& mass \\
$m^\star_{\mathrm{c}}$ &--& dynamic effective mass of confined nucleons in neutron star crust\\
$m_e$ &--& electron mass \\
$m^*_e$ &--& electron (relativistic) effective mass in dense matter \\
$m^\star_{\mathrm{f}}$ &--& dynamic effective mass of the free neutrons in neutron star crust \\
$m_n$ &--& neutron mass \\
$m^*_n$ &--& neutron effective mass in dense matter \\
$m_p$ &--& proton mass \\
$m_{\mathrm{u}}$ &--& atomic mass unit

\end{tabular}

\newpage

\begin{tabular}[t]{lll}

$n$ &--& number density of particles \\
$n_{\mathrm{b}}$ &--& number density of baryons\\
$n_{\mathrm{c}}$ &--& number density of confined nucleons in neutron star crust\\
$n_e$ &--& number density of electrons\\
$n_{\mathrm{f}}$ &--& number density of free neutrons in neutron star crust\\
$n_n$ &--& number density of neutrons\\
$n_{\mathrm{N}}$ &--& number density of nuclei (ions) \\
$n_p$ &--& number density of protons\\
$n_\upsilon$ &--& surface density of vortex lines \\
$\Omega$ &--& angular velocity \\
$\pi_\mu$ &--& momentum 4-covector \\
$P$ &--& pressure \\
$\pmb{p},p$ &--& momentum \\
$p_{\mathrm{F}}$ &--& Fermi momentum \\
$p_{\mathrm{F}e}$ &--& electron Fermi momentum \\
$\varphi$ &--& single particle wave function \\
$\rho$ &--& mass density \\
$\rho_{\mathrm{ND}}$ &--& neutron drip density \\
$r_{\mathrm{g}}$ &--& Schwarzschild radius \\
$r_\upsilon$ &--& radius of a superfluid vortex core \\
$R$ &--& neutron star circumferential radius \\
$R_{\mathrm{cell}}$ &--& radius of a Wigner--Seitz sphere \\
$\sigma$ &--& surface thermodynamic potential~(Section~\ref{sect.groundstate}) \\
$\sigma$ &--& electric conductivity~(Section~\ref{sect.cond}) \\
$\sigma$ &--& Poisson coefficient~(Section~\ref{sect.elast}) \\
$\sigma_{ij}$ &--& elastic stress tensor \\
$\sigma_{\max}$ &--& breaking strain \\
$\sigma_{\mathrm{s}}$ &--& surface tension \\
$T$ &--& temperature \\
$T_{\mathrm{c}}$ &--& critical temperature \\
$T_{\mathrm{F}}$ &--& Fermi temperature \\
$T_{\mathrm{F}e}$ &--& electron Fermi temperature \\
$T_{\mathrm{m}}$ &--& melting temperature \\
$T_{\mathrm{p}e}$ &--& electron plasma temperature \\
$T_{\mathrm{p i}}$ &--& ion plasma temperature \\
$T^\infty_{\mathrm{s}}$ &--& effective surface temperature as seen by a
  distant observer \\
$T^\mu_{\ \nu}$ &--& stress-energy tensor \\
${\cal V}_{\mathrm{cell}}$ &--& volume of the Wigner--Seitz cell \\
${\cal V}_{\mathrm{N}}$ &--& volume of the proton cluster \\
$v_{\mathrm{F}}$ &--& Fermi velocity \\
$v_{\mathrm{F}e}$ &--& electron Fermi velocity \\
$\pmb{v}$ &--& velocity \\
$\pmb{v_\upsilon}$ &--& vortex velocity \\
$\pmb{v_{\mathrm{f}}}$ &--& velocity of neutron superfluid in neutron star crust\\
$\pmb{v_{\mathrm{c}}}$ &--& velocity of confined nucleons in neutron star crust \\
$w$ &--& volume fraction occupied by nuclear clusters \\
$\xi$ &--& coherence length \\
$x_{\mathrm{r}}$ &--& relativity parameter \\
$Z$ &--& charge (proton) number \\
$\zeta$ &--& bulk viscosity \\

\end{tabular}

\newpage

\section{List of Abbreviations}

\begin{tabular}[t]{lll}

bcc &--& body-centered cubic \\
BCS &--& Bardeen--Cooper--Schrieffer \\
BE &--& Boltzman equation \\
BEC &--& Bose--Einstein condensate \\
EoS &--& equation of state \\
LMXB &--& low-mass X-ray binary \\
QED &--& quantum electrodynamics \\
QPO &--& Quasi Periodic Oscillation \\
SGR &--& Soft Gamma Repeater \\
SXT &--& Soft X-ray Transient \\
W-S &--& Wigner--Seitz \\

\end{tabular}

\end{appendix}

\newpage



\bibliography{refs}

\end{document}